\documentclass[11pt,a4paper]{article}
\usepackage[T1]{fontenc}
\usepackage{lmodern}
\usepackage[margin=25mm]{geometry}
\usepackage{amsmath,amssymb,amsthm,mathtools}
\usepackage{mathrsfs}
\usepackage{array,booktabs,tabularx}
\usepackage{microtype}
\usepackage{comment}
\usepackage{xcolor}
\usepackage{cite}
\usepackage[colorlinks=true,allcolors=blue]{hyperref}
\newcommand{\dd}{\mathrm{d}}
\newcommand{\bx}{\mathbf{x}}
\newcommand{\bk}{\mathbf{k}}
\newcommand{\bn}{\mathbf{n}}
\newcommand{\cH}{\mathcal H}
\newcommand{\Eta}{\mathbb H}
\newcommand{\Op}{\mathcal O}
\newcommand{\STF}{\operatorname{STF}}
\newcommand{\eps}{\epsilon}

\numberwithin{equation}{section}
\allowdisplaybreaks
\title{\bfseries Cosmological perturbations from localized sources:\\ a gauge-invariant multipolar theory}
\author{\textbf{Geoffrey Comp\`ere}\textsuperscript{a},
\textbf{Sk Jahanur Hoque}\textsuperscript{b}\\[0.7em]
\small \textsuperscript{a}Universit\'e Libre de Bruxelles, International Solvay Institutes\\
\small and Brussels Laboratory of the Universe (BLU-ULB), CP 231, B-1050 Brussels, Belgium\\[0.4em]
\small \textsuperscript{b}Birla Institute of Technology and Science, Pilani, Hyderabad Campus,\\
\small Jawaharnagar, Hyderabad 500 078, India}
\date{}
\begin{document}
\maketitle
\begin{abstract}
We develop a first-principles description of the linearized gravitational
field of localized sources in spatially flat
Friedmann--Lema\^{\i}tre--Robertson--Walker universes with arbitrary smooth
expansion histories. The construction determines the complete retarded
metric and the induced perturbation of the surrounding cosmic fluid
from prescribed source stresses and a cosmologically stationary past field.
It includes all multipole orders and
accounts for curvature tails, through which the field depends on the
source's past history inside the light cone.
Generalizing the Blanchet--Damour multipolar formalism, we identify
gauge-invariant mass and current moments together with a canonical
cosmic fluid moment that determines an independent scalar response. This response vanishes in
Minkowski and pure de Sitter spacetime, although curvature tails persist
in de Sitter. The gauge-invariant multipoles are obtained directly in position
space. This gives an exterior description without the nonlocal spatial
projection through the source interior required by the usual
scalar--vector--tensor decomposition. We also relate the canonical
multipole metric to the standard variables of cosmological perturbation
theory.
The retarded construction requires neither a slow-motion nor a
geometric-optics approximation.
We define effective multipolar sources by truncating
the moments of slowly varying sub-Hubble extended sources at a prescribed
Cartesian multipole order, recovering the earlier quadrupolar and
octupolar models in de Sitter spacetime.
We recover the known Minkowski and
de Sitter multipolar fields and obtain an analytical representation
for gravitational propagation in the $\Lambda$CDM model including radiation.
\end{abstract}
\tableofcontents

\section{Introduction}
A localized event in an expanding universe disturbs both the geometry
and the surrounding cosmic fluid. Cosmological
curvature generates tails that depend on the source's past history already at linear order,
while coupling to the cosmic fluid induces density and pressure
perturbations. How can
these effects be described together, and which properties of the
resulting exterior field characterize the source independently of the
choice of coordinates?

In this paper, we develop a unified, first-principles multipolar
description of the retarded gravitational field and cosmic fluid
response produced by localized sources in a spatially flat
Friedmann--Lema\^{\i}tre--Robertson--Walker (FLRW) universe with an
arbitrary smooth expansion history. Our aim is to relate the structure
and evolution of a source to the complete disturbance it creates,
including its gravitational radiation and its influence on the
surrounding cosmic fluid. We seek a gauge-invariant description of that
exterior disturbance directly in position space, with its causal
relation to the source kept explicit.

For isolated gravitating systems, the seminal work of Blanchet and
Damour \cite{Blanchet:1985sp,Blanchet:1989ki} laid the foundations of
the multipolar post-Minkowskian/post-Newtonian  approach to gravitational-wave generation. Their
construction organizes the nonlinear exterior gravitational field in
source multipoles and matches it to the post-Newtonian field near the
source, connecting gravitational-wave generation and propagation
(see also \cite{WillWiseman:1996,Thorne:1980ru,Blanchet:2024review}).
At linear order, the vacuum exterior field is characterized by canonical
mass and current moments which are determined by the source stress-energy
tensor \cite{Blanchet:1989ki,Damour:1990gj}.
The Blanchet--Damour construction provides the starting point for our cosmological extension at
linear order. Two physical ingredients must now be incorporated:
propagation through a curved, evolving background and an exterior
which contains the cosmic fluid and its perturbations.

Identifying physical perturbations in such a background is the central
achievement of the seminal papers of Bardeen \cite{Bardeen:1980kt}
and Kodama and Sasaki \cite{Kodama:1984ziu}, which laid the foundations
of modern gauge-invariant cosmological perturbation theory. Bardeen
formulated the scalar, vector and tensor perturbation equations in terms
of gauge-invariant metric and matter variables. Kodama and Sasaki
extended and extensively developed this framework, including its
geometrical interpretation, multicomponent matter systems and
cosmological applications. For reviews of this framework and subsequent
developments, see \cite{Mukhanov:1990me,Malik:2008im,Baumann:AdvCosmo}.

For the localized-source problem, an additional construction is needed
to characterize the exterior field by source multipoles. The usual
scalar--vector--tensor decomposition involves nonlocal spatial
projections defined with a choice of boundary conditions
\cite{Flanagan:2005yc}. The exterior metric alone does not generally
determine its whole-space projection, which also depends on the metric
inside the source \cite{Ashtekar:2017wgq}. Moreover, the separate
gauge-invariant fields need not depend causally on the source: their
scalar, vector and tensor contributions combine to give the causal
local tidal response \cite{Chu:2019ndv}. Our approach constructs
the exterior multipole description directly from the coupled metric
and fluid field equations, and then relates it to the standard
cosmological variables. It brings
together the source-based viewpoint of the multipolar formalism and
the treatment of gauge freedom and matter response central to
cosmological perturbation theory.

Retarded solutions for prescribed sources in cosmological backgrounds
have also been constructed directly in metric components.
Kopeikin, Ram\'irez, Mashhoon and Sazhin \cite{Kopeikin:2001uk}
used a suitable gauge to obtain hyperbolic perturbation equations and
retarded solutions on a spatially flat dust background, see also \cite{Ramirez:2001wp,Kopeikin:2013am} for extensions. Our construction will differ substantially from theirs. % though our approaches share some qualitative features.
Bashinsky and Bertschinger \cite{Bashinsky:2002vx} demonstrated how a
position-space Green-function formulation makes the causal evolution
of coupled scalar metric and matter perturbations explicit, providing
a physical interpretation of acoustic propagation before recombination.
Their treatment is restricted to scalar perturbations of specified
matter species; our construction addresses the complete retarded metric
and its gauge-invariant exterior multipoles for localized sources in
an arbitrary spatially flat FLRW background. For gravitational
radiation in de Sitter spacetime, de Vega, Ram\'irez and S\'anchez
\cite{deVega:1998ia} decoupled the metric wave equations in a suitable
gauge and constructed the retarded field of a generic source,
including its light-cone and tail contributions. Chu
\cite{Chu:2015yua} developed a dimensional-reduction construction
of gravitational-wave Green functions in spatially flat FLRW
backgrounds and obtained explicit power-law tails. These results
provide both methods and checks for our treatment of propagation.
Our multipolar construction relates this propagation to the source
moments and reconstructs the complete metric and cosmic fluid response.

A direct route from source moments to cosmological radiation emerged
in the de Sitter quadrupole analysis of Ashtekar, Bonga and Kesavan
\cite{Ashtekar:2015lxa}. They generalized Einstein's quadrupole
energy-flux formula to a positive cosmological constant, showing why
the changed asymptotic geometry and curvature tail must be treated
explicitly. Our subsequent work \cite{Compere:2023ktn} completed
the quadrupolar metric in generalized harmonic and Bondi gauges,
including both mass and current sectors, and studied displacement
memory. It also corrected the moment truncation used in earlier
calculations: retaining a pressure quadrupole requires stress moments
that had been discarded, as follows from stress-energy conservation.
Our FLRW extension \cite{Compere:2026xoe} then incorporated the
response of the cosmic fluid, obtained the decoupled metric equations
and constructed retarded fields through quadrupolar order for
power-law expansion. Following the de Sitter quadrupolar
construction, Harsh, Shetye and Virmani \cite{Harsh:2026vmh}
extended the calculation to a consistent octupolar truncation and
resolved its electric and magnetic sectors.
The present work pursues
this research line at all multipole orders and for arbitrary smooth
expansion histories, with the gauge-invariant content of the coupled
exterior field as a central objective.
A particular outcome will be a formulation of effective
multipolar sources representing finite multipolar truncations of
sub-Hubble, slowly varying extended sources. {The quadrupolar
construction will account for the cosmic-fluid response while preserving
the earlier de Sitter source class~\cite{Compere:2023ktn,Compere:2026xoe}.
At octupolar order, the formulation will recover the de Sitter model
of \cite{Harsh:2026vmh}.}

The physical interest of keeping track of the complete disturbance is
particularly clear for primordial events, which can produce density
and velocity perturbations as well as gravitational radiation. Three
possible applications are primordial black-hole formation, first-order
phase transitions and unstable cosmic defects.

Primordial black holes can form through the gravitational collapse of
sufficiently large primordial overdensities
\cite{CarrHawking:1974}. Clesse and Garc\'ia-Bellido
\cite{ClesseGarciaBellido:2015} studied a formation scenario in which a
mild waterfall stage of hybrid inflation enhances small-scale curvature
perturbations, producing a broad primordial black-hole mass distribution.
They proposed these black holes as dark-matter candidates and as seeds
of supermassive black holes. Such scenarios motivate studying the
disturbance left in the cosmic fluid by individual collapse events.
The collapse itself is nonlinear; our formalism could describe the
weak-field metric and fluid response outside the collapsing region,
once a collapse model supplies the source history and matching data.

In a first-order phase transition, bubbles nucleate, expand and collide,
transferring released energy to their walls and the surrounding plasma.
Early relativistic analyses of bubble growth established the role of
detonation and deflagration fronts
\cite{Steinhardt:1982bubbles,Gyulassy:1984bubbles}, which could also be supersonic \cite{KurkiSuonioLaine:1995supersonic}.
The gravitational-radiation signal was anticipated by Witten
\cite{Witten:1984phases} and estimated by Hogan
\cite{Hogan:1986phase}, who emphasized acoustic motions in the plasma.
Kamionkowski, Kosowsky and Turner \cite{Kamionkowski:1994phase}
calculated radiation from bubble collisions and estimated the
contribution from hydrodynamic turbulence. The same bubble and fluid
histories could supply localized sources for our construction, allowing
their gravitational radiation and the accompanying metric and cosmic
fluid perturbations to be described together.

Unstable cosmic defects provide another class of localized events.
Zel'dovich, Kobzarev and Okun \cite{Zeldovich:1974domains}
studied the cosmological formation of domain walls and the collapse of
closed walls under their tension. Kibble \cite{Kibble:1976defects}
established the general connection between symmetry breaking, vacuum
topology and cosmic-defect formation. The subsequent dynamics of walls
and strings, including collapsing closed walls and radiating string
loops, are reviewed by Vilenkin \cite{Vilenkin:1985defects}.
These finite objects provide concrete sources whose changing shape
and stresses can be related to an exterior gravitational field and to
the disturbance of the cosmic fluid. A multipolar treatment could
resolve an individual collapse or decay event before averaging over
a population of defects.

The subsequent evolution of these primordial disturbances depends on
the properties of the cosmic medium. For example, Silk
\cite{Silk:1968damping} showed that photon diffusion damps small-scale
density perturbations before recombination. Longer-lived records in
matter clustering, analogous to gravitational-wave fossils
\cite{Masui:2010cz,Schmidt:2013gwa}, provide a further motivation,
although their calculation requires nonlinear couplings beyond the
linear response studied here. Predicting observable signatures
therefore requires the relevant matter dynamics, including diffusion
and thermalization when appropriate. Our purpose is to supply the
common gravitational description on which such calculations can build.

The paper is organized in two stages. Sections~\ref{sec:equations}--\ref{sec:reconstruct}
set out the source assumptions and construct the retarded metric and cosmic 
fluid response. Sections~\ref{sec:cartesian}--\ref{sec:multipoles}
express these fields in terms of  source moments   and
construct effective multipolar sources, and
Section~\ref{sec:octupolar} specializes to de Sitter spacetime and
compares with earlier multipolar results. The second stage identifies
the gauge-invariant content: Section~\ref{sec:direct-gauge-invariant-multipoles}
identifies the invariant multipoles in generalized harmonic gauge
and constructs a simple canonical metric by a temporal gauge transformation,
Section~\ref{sec:minkowski} checks the Minkowski limit against
Blanchet--Damour \cite{Blanchet:1985sp,Blanchet:1989ki} and
Damour--Iyer \cite{Damour:1990gj}, and
Section~\ref{sec:SVT-comparison} relates the construction to the
standard scalar--vector--tensor description. The appendices contain
supporting kernel constructions and Fourier representations.

{\bf Note added:} In the late stages of writing this manuscript, we
received a preprint written by Harsh, Shetye and Virmani
\cite{Harsh:AllMultipolesPreprint}. Their current draft concerns de Sitter spacetime
only and partially overlaps with our de Sitter specialization.

\section{Perturbation equations and admissible sources}\label{sec:equations}
In this section, we will recall the decoupled perturbation equations derived in
Ref.~\cite{Compere:2026xoe} using a generalization of harmonic gauge
and specify the source histories and initial data for the retarded problem.
The residual freedom of the generalized harmonic gauge will also be
identified. We will distinguish which components
of the stress-energy tensor can be prescribed and which must be
reconstructed from conservation. The linearized
McVittie monopole \cite{McVittie:1933zz} will then illustrate the response of the background cosmic fluid
and its description by the shifted stress-energy tensor perturbation.

\subsection{Background, generalized harmonic gauge and linearized perturbations}
We begin by fixing the background conventions.
We allow both metric signatures, with $\mathfrak{s}:=+1$ for
$(-,+,+,+)$ and $\mathfrak{s}:=-1$ for $(+,-,-,-)$. Greek indices run
from $0$ to $3$, Latin indices from $1$ to $3$, and
$x^\mu=(c\eta,x^i)$ are conformal Cartesian coordinates. The Minkowski
metric is $\eta_{\mu\nu}:=\mathfrak{s}\operatorname{diag}(-1,1,1,1)$;
its inverse is $\eta^{\mu\nu}$. Spatial contractions use
$\delta_{ij}$, with $\epsilon_{123}=1$. We use normalized
symmetrization and denote spatial symmetric trace-free (STF) projection
by angle brackets. Our curvature convention is
\begin{equation}\label{eq:curvature-convention}
 R^\rho{}_{\sigma\mu\nu}
 :=\partial_\mu\Gamma^\rho{}_{\nu\sigma}
 -\partial_\nu\Gamma^\rho{}_{\mu\sigma}
 +\Gamma^\rho{}_{\mu\alpha}\Gamma^\alpha{}_{\nu\sigma}
 -\Gamma^\rho{}_{\nu\alpha}\Gamma^\alpha{}_{\mu\sigma},
 \qquad R_{\sigma\nu}:=R^\rho{}_{\sigma\rho\nu}.
\end{equation}
Let $\eta$ and $t$ denote conformal
and cosmological proper time, respectively. We consider a smooth positive scale factor
$a(\eta)$ on the conformal-time interval
$\Eta:=(\eta_-,\eta_+)$, which specifies the background history
covered by the spacetime. The endpoints of the conformal-time interval $\Eta$ may be finite or infinite. In conformal coordinates, the background metric reads
\begin{equation}
 \bar g_{\mu\nu}\dd x^\mu\dd x^\nu\equiv a^2(\eta)\eta_{\mu\nu}\dd x^\mu\dd x^\nu
 =\mathfrak{s}a^2(\eta)(-c^2\dd\eta^2+\delta_{ij}\dd x^i\dd x^j),
 \qquad \dd t=a\,\dd\eta.
\end{equation}
We have $\partial_0=c^{-1}\partial_\eta$. Dots denote conformal-time derivatives, $\dot a:=\partial_\eta a$,
while proper-time derivatives are denoted by $\partial_t$. Spatial
indices are contracted with $\delta_{ij}$. The word ``tensor'' refers
to a spatial tensor and does not imply a transverse-traceless
projection. We define the Hubble rate $H:=a^{-1}\partial_ta$,
the conformal Hubble rate $\cH:=\dot a/a=\partial_\eta\ln a=aH$ and the deceleration parameter
$q:=-\dot{\cH}/\cH^2$, which is finite wherever $\cH\ne0$.
The cosmic fluid supporting the background has the perfect-fluid
stress-energy tensor
$\bar T^\mu{}_\nu=\mathfrak{s}\operatorname{diag}(-\eps,p,p,p)$, where $\eps$ and
$p$ are its energy density and pressure. With Newton's constant $G$
and cosmological constant $\Lambda$ (positive for de Sitter vacuum),
Einstein's equations $G_{\mu\nu}+\mathfrak{s}\Lambda g_{\mu\nu}=8\pi G T_{\mu\nu}/c^4$,
evaluated on the background, determine these fluid variables from the scale factor as
\begin{equation}\label{eq:friedmann}
 \eps=\frac{c^2(3\cH^2-\Lambda c^2a^2)}{8\pi G a^2},
 \qquad p=\frac{c^2(-2\dot{\cH}-\cH^2+\Lambda c^2a^2)}{8\pi G a^2}.
\end{equation}

Let us introduce a perturbative parameter $\lambda$. We expand
$g_{\mu\nu}=\bar g_{\mu\nu}+\lambda h_{\mu\nu}+O(\lambda^2)$.
Indices of $h_{\mu\nu}$ are raised and lowered with $\bar g$, and
its trace is $h:=\bar g^{\mu\nu}h_{\mu\nu}$.
We define the rescaled trace-reversed metric perturbation
$\bar h_{\mu\nu}$ and $\hat h$ by\footnote{For mostly plus signature,
this is the standard trace reversal followed by division by $a$.
For mostly minus signature, we find the additional minus sign convenient:
the components of $\bar h_{\mu\nu}$ and their field equations then
remain unchanged under an overall reversal of the metric signature.}
\begin{equation}\label{eq:field-variables}
 \bar h_{\mu\nu}:=\frac{\mathfrak{s}}a\left(h_{\mu\nu}-\frac12\bar g_{\mu\nu}h\right),
 \qquad \hat h:=\bar h_{00}+\bar h_{kk}.
\end{equation}
We raise and lower indices of the rescaled components
$\bar h_{\mu\nu}$ with $\eta_{\mu\nu}$:
$\bar h^{\mu\nu}:=\eta^{\mu\alpha}\eta^{\nu\beta}\bar h_{\alpha\beta}$.
Thus $\bar h^{0i}=-\bar h_{0i}$ and
$\eta^{\mu\nu}\bar h_{\mu\nu}=\mathfrak{s}(-\bar h_{00}+\bar h_{kk})=-\mathfrak{s}ah$.
The covariant derivative $\bar\nabla$ is associated with $\bar g$,
and its index is raised with the background inverse metric:
$\bar\nabla^\mu:=\bar g^{\mu\alpha}\bar\nabla_\alpha$.
Inverting the trace reversal gives
\begin{equation}\label{eq:metric-inverse-covariant}
 h_{\mu\nu}=\mathfrak{s}a\left(\bar h_{\mu\nu}
 -\frac12\eta_{\mu\nu}\eta^{\alpha\beta}\bar h_{\alpha\beta}\right).
\end{equation}
Equivalently, the components of the original metric perturbation are
\begin{equation}\label{eq:metric-final}
 h_{00}=\frac{\mathfrak{s}a}2\hat h,\qquad h_{0i}=\mathfrak{s}a\bar h_{0i},\qquad
 h_{ij}=\mathfrak{s}a\left[\bar h_{ij}
 +\delta_{ij}\left(\frac{\hat h}2-\bar h_{kk}\right)\right].
\end{equation}
In these variables, the generalized harmonic gauge of
\cite[Eqs.~(2.10), (2.16)]{Compere:2026xoe} takes the weighted divergence form\footnote{Our gauge condition differs from that of Ram\'irez and
Kopeikin \cite[Eqs.~(31)--(32)]{Ramirez:2001wp}. To compare the
conventions, define the ordinary trace-reversed perturbation
$\psi_{\mu\nu}:=h_{\mu\nu}-\tfrac12\bar g_{\mu\nu}h=\mathfrak{s}a\bar h_{\mu\nu}$
and the background fluid four-velocity
$\bar u^\mu:=ca^{-1}\delta^\mu_0$, normalized by $\bar u_\mu\bar u^\mu=-\mathfrak{s}c^2$.
Our condition~\eqref{eq:generalized-harmonic-gauge} is equivalent to
$\bar\nabla^\nu\psi_{\mu\nu}=-(2\mathfrak{s}H/c^2)\bar u^\nu\psi_{\mu\nu}$,
whereas their condition, translated to our signature convention, is
$\bar\nabla^\nu\psi_{\mu\nu}=-(2\mathfrak{s}H/c^2)\bar u^\nu\psi_{\mu\nu}
+(16\pi G/c^4)(\partial_\mu\bar\phi)\delta\phi$.
Here $\bar\phi(\eta)$ is their homogeneous background scalar field and
$\delta\phi$ its perturbation. The additional scalar-field term affects
only the temporal gauge condition and is absent from ours.}:
\begin{equation}\label{eq:generalized-harmonic-gauge}
 \bar\nabla^\mu\left(a^{-1}\bar h_{\mu\nu}\right)=0.
\end{equation}
We define the four gauge residuals $C_\mu := -\mathfrak{s}a^3\bar\nabla^\nu(a^{-1}\bar h_{\mu\nu})$ and impose their
vanishing as the gauge condition:
\begin{subequations}\label{eq:gauge-components}
\begin{align}
 C_0&:=\frac1c(\partial_\eta\bar h_{00}+\cH\bar h_{kk})-\partial_j\bar h_{0j}=0,\label{eq:gauge0}\\
 C_i&:=\frac1c(\partial_\eta+\cH)\bar h_{0i}
       -\partial_j\bar h_{ij}=0.\label{eq:gaugei}
\end{align}
\end{subequations}

\subsection{Shifted stress-energy tensor perturbation and the two potentials}
The linearized Einstein equations couple the metric perturbation to
perturbations of the cosmic fluid. To make the propagation of the metric
explicit, we will separate a local metric-dependent contribution from the
stress-energy tensor perturbation. In the generalized harmonic gauge,
this redefinition will yield wave equations with two time-dependent
potentials. We will then express stress-energy conservation in terms of the
shifted stress-energy tensor perturbation.

We start by expanding the stress-energy tensor as
$T_{\mu\nu}=\bar T_{\mu\nu}+\lambda\delta T_{\mu\nu}+O(\lambda^2)$.
We isolate the local metric-dependent contribution associated with
the background cosmic fluid,
\begin{equation}\label{eq:fluid-metric-contribution}
 \delta_h T^{\mathrm{fluid}}_{\mu\nu}
 :=\mathfrak{s}p h_{\mu\nu}+\frac{\eps+p}{2}h_{00}\eta_{\mu\nu},
\end{equation}
and we define the shifted stress-energy tensor
perturbation by subtracting this contribution:
\begin{equation}\label{eq:shift-definition}
 \tau_{\mu\nu}:=\delta T_{\mu\nu}
                   -\delta_h T^{\mathrm{fluid}}_{\mu\nu}.
\end{equation}
This algebraic subtraction leaves the dynamical fluid
perturbations in $\tau_{\mu\nu}$. We also call $\tau_{\mu\nu}$ the
shifted source when referring to its role in the metric field equations. Using the field definitions
\eqref{eq:field-variables}, its components evaluate to
\begin{subequations}\label{eq:shift-components}
\begin{align}
 \tau_{00}&=\delta T_{00}+\frac{a}{4}(\eps-p)\hat h,
 \label{eq:shift00}\\
 \tau_{0i}&=\delta T_{0i}-pa\bar h_{0i},
 \label{eq:shift0i}\\
 \tau_{ij}&=\delta T_{ij}-pa\big(\bar h_{ij}-\delta_{ij}\bar h_{kk}\big)
 -\frac{a}{4}(\eps+3p)\delta_{ij}\hat h.
 \label{eq:shift}
\end{align}
\end{subequations}
The distinction between a localized perturbing source and the induced
response of the background matter was already made explicitly in
\cite{Kopeikin:2001uk}. Equations~\eqref{eq:shift-components} reproduce
\cite[Eq.~(2.19)]{Compere:2026xoe}; the shifted stress-energy tensor perturbation denoted by
$\delta\tilde T_{\mu\nu}$ in that reference is denoted here by
$\tau_{\mu\nu}$.

Conversely, the stress-energy tensor perturbation is recovered by
\begin{subequations}\label{eq:inverse-source-shift}
\begin{align}
 \delta T_{00}&=\tau_{00}-\frac a4(\eps-p)\hat h,
 \label{eq:inverse-source-shift00}\\
 \delta T_{0i}&=\tau_{0i}+pa\bar h_{0i},
 \label{eq:inverse-source-shift0i}\\
 \delta T_{ij}&=\tau_{ij}+pa\bar h_{ij}
       +a\delta_{ij}\left[\frac{\eps+3p}{4}\hat h-p\bar h_{kk}\right].
 \label{eq:inverse-source-shiftij}
\end{align}
\end{subequations}
These are algebraic inverses of Eqs.~\eqref{eq:shift-components}.

The physical matter variables are the energy-density perturbation
$\delta\eps$, the pressure perturbation $\delta p$, the momentum
density $j_i$ and the anisotropic stress $\pi_{ij}$. We measure them
in orthonormal frames carried by Eulerian observers, whose timelike
frame vector is the future-directed unit normal to the constant-$\eta$
hypersurfaces in the perturbed spacetime
\cite[Sec.~3.3.3]{Gourgoulhon:2007ue}.
These operational definitions distinguish the physical density
$\delta\eps$ from the shifted coordinate component $\tau_{00}$.
We retain $\delta\eps,\delta p,j_i,\pi_{ij}$, so that perturbations
and background quantities use the same density and pressure notation.
Their relations to the shifted stress-energy tensor perturbation are
\begin{subequations}\label{eq:physical-matter-variables}
\begin{align}
 \delta\eps&=\frac{\tau_{00}}{a^2}
                 +\frac{\eps+p}{4a}\hat h,
 \label{eq:physical-density}\\
 \delta p&=\frac{\tau_{kk}}{3a^2}
                 +\frac{\eps+p}{4a}\hat h,
 \label{eq:physical-pressure}\\
 j_i&=-\frac{\tau_{0i}}{a^2c},
 \label{eq:physical-momentum}\\
 \pi_{ij}&=\frac1{a^2}
       \left(\tau_{ij}-\frac13\delta_{ij}\tau_{kk}\right).
 \label{eq:physical-anisotropic-stress}
\end{align}
\end{subequations}
At linear order, $\delta\eps$ and $\delta p$ also equal the
rest-frame density and pressure perturbations. For a perfect-fluid
perturbation, $\pi_{ij}=0$ and $j_i$ is the background enthalpy
density $(\eps+p)/c^2$ times the fluid velocity relative to the
Eulerian observers. 
%When $h_{0i}\ne0$, this velocity differs from the coordinate velocity.
%Thus $\tau_{00}$ and $\tau_{kk}/3$ contain the same metric-dependent shift; neither is by itself the physical density or pressure perturbation. 
No equation of state or perfect-fluid
condition on the perturbation has been imposed in
Eqs.~\eqref{eq:physical-matter-variables}.

The linearized Einstein equations in this gauge involve two wave
operators, denoted by $\Op^-$ and $\Op^+$:
\begin{subequations}\label{eq:metric-wave-equations}
\begin{align}
 \Op^-\bar h_{ij}&=-\frac{16\pi G}{c^4}a\tau_{ij},\label{eq:spatial}\\
 \Op^+\bar h_{0i}&=-\frac{16\pi G}{c^4}a\tau_{0i},\label{eq:mixed}\\
 \Op^+\hat h&=-\frac{16\pi G}{c^4}a(\tau_{00}+\tau_{kk}),\label{eq:scalar}
\end{align}
\end{subequations}
where
\begin{subequations}\label{eq:wave-operator-definitions}
\begin{align}
 \Op^\pm&:=\square+V^\pm(\eta),\qquad
 \square:=\mathfrak{s}\eta^{\mu\nu}\partial_\mu\partial_\nu
 =-c^{-2}\partial_\eta^2+\Delta,\label{eq:ops}\\
 V^-&:=\frac{\cH^2+\dot{\cH}}{c^2}=\frac{\ddot a}{c^2a},\qquad
 V^+:=\frac{\cH^2-\dot{\cH}}{c^2}=\frac1{c^2}\left(2\frac{\dot a^2}{a^2}-\frac{\ddot a}{a}\right).
 \label{eq:potentials}
\end{align}
\end{subequations}
Here $\Delta:=\delta^{ij}\partial_i\partial_j$ is the spatial Laplacian.
We call $V^\pm$ the wave potentials.
When $\cH\ne0$, we have $V^\pm=\cH^2(1\pm q)/c^2$\footnote{The same two potentials occur in the spatial and mixed component
operators of Ram\'irez and Kopeikin
\cite[Eqs.~(35)--(36)]{Ramirez:2001wp}: their
$\Omega=\ln a$ gives $\dot\Omega=\cH$ and
$\dot\Omega^2\pm\ddot\Omega=c^2V^{\mp}$. However, their sources comprise a scalar-field
perturbation and a separately background-conserved perturbing
stress-energy tensor.}. The expressions
\eqref{eq:potentials} also apply at zeros of $\cH$.
Equation~\eqref{eq:spatial} contains the spatial trace and trace-free
equations, and therefore comprises six components. Together with
Eqs.~\eqref{eq:mixed} and \eqref{eq:scalar}, it gives all ten field equations.
These are Eqs.~(2.23)--(2.24) of
Ref.~\cite{Compere:2026xoe}, with $G$ and $c$ restored. Substituting
Eqs.~\eqref{eq:shift-components} into the linearized Einstein tensor verifies the
equations for arbitrary $a$, after using Eq.~\eqref{eq:friedmann}. Using the background Friedmann equations~\eqref{eq:friedmann}, we can
equivalently express the potentials in terms of the cosmic fluid
density and pressure:
\begin{subequations}\label{eq:potentials-fluid}
\begin{align}
 V^-&=\frac{4\pi G a^2}{3c^4}(\eps-3p)+\frac{2\Lambda a^2}{3},
 \label{eq:potential-minus-fluid}\\
 V^+&=\frac{4\pi G a^2}{c^4}(\eps+p).
 \label{eq:potential-plus-fluid}
\end{align}
\end{subequations}

Let us now discuss the conservation equation. Varying
$\nabla^\mu T_{\mu\nu}=0$, including the inverse metric and connection,
gives \cite[Eq.~(2.17)]{Compere:2026xoe}
\begin{equation}\label{eq:linear-conservation}
 \bar\nabla^\mu\delta T_{\mu\nu}
 -h^{\mu\rho}\bar\nabla_\rho\bar T_{\mu\nu}
 -\bar g^{\mu\rho}\left(
 \delta\Gamma^\lambda{}_{\rho\mu}\bar T_{\lambda\nu}
 +\delta\Gamma^\lambda{}_{\rho\nu}\bar T_{\mu\lambda}\right)=0.
\end{equation}
Inserting the background perfect fluid, the gauge conditions
\eqref{eq:gauge-components} and the source-shift relations
\eqref{eq:shift-components} into Eq.~\eqref{eq:linear-conservation} gives the evolution
equations for the shifted energy-density perturbation $\tau_{00}$
and the momentum density $\tau_{0i}$
\cite[Eqs.~(2.18), (4.6)]{Compere:2026xoe}:
\begin{subequations}\label{eq:shifted-conservation}
\begin{align}
 \partial_\eta(a\tau_{00})&=ca\partial_i\tau_{0i}-\dot a\tau_{kk}+a^2s\hat h,
 \label{eq:energy}\\
 (\partial_\eta+\cH)(a\tau_{0i})&=ca\partial_j\tau_{ij},
 \label{eq:momentum}
\end{align}
\end{subequations}
where
\begin{equation}\label{eq:s}
 s:=-\frac1{4a^2}\partial_\eta\big[a^2(\eps+p)\big]
   =-\frac{c^4\dot V^+}{16\pi G a^2}.
\end{equation}
The last equality follows by differentiating Eq.~\eqref{eq:potential-plus-fluid}.
Thus the source term $a^2s\hat h$ describes the coupling of the
matter perturbation to the metric. Covariant conservation of the full
stress-energy tensor remains satisfied.
\subsection{Residual gauge freedom}
\label{sec:residual-gauge}
The generalized harmonic gauge leaves a residual coordinate freedom.
We will identify the diffeomorphisms that preserve it and determine
their action on both the metric and the shifted stress-energy tensor
perturbation $\tau_{\mu\nu}$.

We start with a diffeomorphism generated by $\xi^\mu$, which acts as
\begin{equation}\label{eq:linear-gauge-transformation}
 h_{\mu\nu}\longmapsto h_{\mu\nu}
 +\mathcal L_\xi\bar g_{\mu\nu}
 =h_{\mu\nu}+\bar\nabla_\mu\xi_\nu+\bar\nabla_\nu\xi_\mu,
 \qquad \xi_\mu:=\bar g_{\mu\nu}\xi^\nu.
\end{equation}
Denoting the linear gauge variation by $\delta_\xi$, we define the
rescaled components of the gauge vector by
\begin{equation}\label{eq:rescaled-gauge-vector}
 \zeta^\mu:=a\xi^\mu.
\end{equation}
The metric transformation~\eqref{eq:linear-gauge-transformation} and
field definitions~\eqref{eq:field-variables} give
\begin{subequations}\label{eq:rescaled-metric-gauge-action}
\begin{align}
 \delta_\xi\bar h_{00}&=\frac1c(-\dot\zeta^0+3\cH\zeta^0)+\partial_i\zeta^i,
 \label{eq:rescaled-metric-gauge00}\\
 \delta_\xi\bar h_{0i}&=\frac1c(\partial_\eta-\cH)\zeta^i-\partial_i\zeta^0,
 \label{eq:rescaled-metric-gauge0i}\\
 \delta_\xi\bar h_{ij}&=2\partial_{(i}\zeta^{j)}
 -\delta_{ij}\left[\frac1c(\dot\zeta^0+\cH\zeta^0)+\partial_k\zeta^k\right].
 \label{eq:rescaled-metric-gaugeij}
\end{align}
\end{subequations}
In particular, taking the $00$ component plus the spatial trace gives
\begin{equation}\label{eq:scalar-metric-gauge-action}
 \delta_\xi\hat h=-\frac4c\dot\zeta^0.
\end{equation}

For an arbitrary infinitesimal diffeomorphism, substitution into
the gauge residuals~\eqref{eq:gauge-components} gives
\begin{subequations}\label{eq:direct-gauge-residual-action}
\begin{align}
 \delta_\xi C_0&=(\square-3V^+)\zeta^0,\label{eq:direct-gauge-residual-action-1}\\
 \delta_\xi C_i&=-\Op^-\zeta^i.\label{eq:direct-gauge-residual-action-2}
\end{align}
\end{subequations}
A residual transformation preserves $C_\mu=0$, so it obeys
\begin{subequations}\label{eq:residual-gauge-rescaled}
\begin{align}
 (\square-3V^+)\zeta^0&=0,\label{eq:residual-gauge-rescaled-time}\\
 \Op^-\zeta^i&=0.\label{eq:residual-gauge-rescaled-space}
\end{align}
\end{subequations}
Here $\xi^0$ and $\xi^i$ are the conformal-coordinate components
of the gauge vector. The rescaled spatial components obey the same
homogeneous wave equation as $\bar h_{ij}$ in the absence of shifted
stresses. The rescaled temporal component has potential $-3V^+$,
so its operator is $\square-3V^+=\Op^+-4V^+$.
This operator governs the remaining temporal coordinate freedom.
The source metric is propagated by $\Op^\pm$; in
Section~\ref{sec:canonical-gauge} we will remove the explicit temporal
gauge profile from its canonical representative.
This differs from the temporal residual operator $\square+V^+$
in \cite[Eq.~(39)]{Ramirez:2001wp}; their spatial residual
operator in their Eq.~(40) agrees with $\Op^-$ and therefore with ours. The difference in the temporal component follows
from their gauge conditions which differs from ours.

The same diffeomorphism acts on the matter perturbation. Its linear gauge variation is
\begin{equation}\label{eq:matter-gauge-transformation}
 \delta_\xi(\delta T_{\mu\nu})=\mathcal L_\xi\bar T_{\mu\nu}.
\end{equation}
To obtain the transformation of $\tau_{\mu\nu}$, we vary both terms
in its definition~\eqref{eq:shift-definition}, using
Eqs.~\eqref{eq:fluid-metric-contribution},
\eqref{eq:linear-gauge-transformation} and
\eqref{eq:matter-gauge-transformation}, while keeping the background
fixed. The background continuity equation
$\dot\eps=-3\cH(\eps+p)$, which follows from the Friedmann
equations~\eqref{eq:friedmann}, simplifies the $00$ component.
This gives the transformation laws
\cite[Eqs.~(2.20)--(2.22)]{Compere:2026xoe}, in our notation,
\begin{subequations}\label{eq:shifted-source-gauge}
\begin{align}
 \delta_\xi\tau_{00}
 &=\frac ac(\eps+p)(\partial_\eta-3\cH)\zeta^0,
 \label{eq:shifted-source-gauge00}\\
 \delta_\xi\tau_{0i}
 &=a(\eps+p)\partial_i\zeta^0,
 \label{eq:shifted-source-gauge0i}\\
 \delta_\xi\tau_{ij}
 &=\frac ac\delta_{ij}\big[\dot p\,\zeta^0+(\eps+p)\dot\zeta^0\big].
 \label{eq:shifted-source-gaugeij}
\end{align}
\end{subequations}
Here we have used $\xi^0=\zeta^0/a$ and
$\dot\xi^0=(\dot\zeta^0-\cH\zeta^0)/a$.
These laws hold for any infinitesimal $\zeta^\mu$; residual gauge
transformations additionally satisfy
Eqs.~\eqref{eq:residual-gauge-rescaled}. The terms involving $\xi^i$
cancel in the subtraction defining $\tau_{\mu\nu}$. The remaining
variation of $\tau_{ij}$ is proportional to $\delta_{ij}$, so its
spatial trace-free part $\tau_{\langle ij\rangle}:=\tau_{ij}-\delta_{ij}\tau_{kk}/3$ is gauge
invariant. In contrast, $\tau_{00}$, $\tau_{0i}$ and $\tau_{kk}$
generally change under a temporal gauge transformation. Thus a
classification of the metric perturbation modulo residual gauge
transformations must include the cosmic fluid perturbations.
Using the transformation~\eqref{eq:scalar-metric-gauge-action} of $\hat h$
in the physical definitions~\eqref{eq:physical-matter-variables}
gives the simpler matter transformation laws
\begin{subequations}\label{eq:physical-matter-gauge}
\begin{align}
 \delta_\xi(\delta\eps)&=\frac1c\dot\eps\,\xi^0,
 \label{eq:physical-matter-gauge-scalars}\\
 \delta_\xi(\delta p)&=\frac1c\dot p\,\xi^0,
 \label{eq:physical-matter-gauge-pressure}\\
 \delta_\xi j_i&=-\frac{\eps+p}{c}\partial_i\xi^0,
 \label{eq:physical-matter-gauge-stress}\\
 \delta_\xi\pi_{ij}&=0.
 \label{eq:physical-matter-gauge-anisotropy}
\end{align}
\end{subequations}
The density and pressure variations reflect the change of time
slice in a fluid whose background density and pressure evolve.
The gradient in the momentum transformation reflects the change
of the Eulerian observers associated with those slices.
For a problem with prescribed initial and boundary data, admissible
residual transformations must also preserve those data and any
conditions imposed on the source.

\subsection{Nearly spatially compact sources and retarded data}
\label{sec:source-data}
The retarded problem requires a source history and initial data.
We will specify which shifted source components are confined to the
source region and which are determined by conservation, following the
nearly spatially compact source prescription of
\cite[Sec.~2, discussion following Eq.~(2.22)]{Compere:2026xoe}.

We start by fixing a comoving source radius $d>0$ and an initial time
$\eta_i\in\Eta$.
We reserve $\bx'$ for source positions and $\bx$ for observation
positions. We denote derivatives with respect to $x'^i$ by
$\partial'_i:=\partial/\partial x'^i$ and the source-coordinate
Laplacian by $\Delta':=\delta^{ij}\partial'_i\partial'_j$.
This convention is retained in source-to-field formulas; suppressed
arguments are specified locally when they would clutter an expression.
We prescribe smooth shifted stresses $\tau_{ij}=\tau_{ji}$ supported in
$|\bx'|\le d$. Their physical support radius is bounded by $a(\eta)d$.
This fixed comoving ball, followed through time, defines the source
world tube. The condition $\tau_{ij}=0$ outside this world tube need not
be preserved by a residual gauge transformation. By
Eq.~\eqref{eq:shifted-source-gaugeij}, a temporal transformation preserves
this condition only if
$\dot p\,\zeta^0+(\eps+p)\dot\zeta^0=0$ outside the world tube.
Preserving compact momentum support also requires
$(\eps+p)\partial_i\zeta^0=0$ there. These restrictions do not
completely fix the gauge: the admissible residual transformations
must preserve the source and initial conditions as well as generalized
harmonic gauge. 
%A transformation that fails these conditions changes the source representative and must transform the surrounding fluid together with the metric, as in Section~\ref{sec:direct-gauge-invariant-multipoles}.

We allow a gravitational field already present before $\eta_i$.
We specify it by past metric data $\bar h_{\mu\nu}^{\mathrm{in}}(\bx')$,
with $\hat h^{\mathrm{in}}:=\bar h_{00}^{\mathrm{in}}+
\bar h_{kk}^{\mathrm{in}}$ and $a_i:=a(\eta_i)$. The corresponding
\emph{past field} is
\begin{subequations}\label{eq:past-reference-metric}
\begin{align}
 \bar h_{ij}^{\mathrm{past}}(\eta,\bx')
 &:=\frac{a(\eta)}{a_i}\bar h_{ij}^{\mathrm{in}}(\bx'),
 \label{eq:past-reference-spatial}\\
 \bar h_{0i}^{\mathrm{past}}(\eta,\bx')
 &:=\frac{a_i}{a(\eta)}\bar h_{0i}^{\mathrm{in}}(\bx'),
 \label{eq:past-reference-mixed}\\
 \hat h^{\mathrm{past}}(\eta,\bx')&:=\hat h^{\mathrm{in}}(\bx'),
 \label{eq:past-reference-hat}\\
 \bar h_{00}^{\mathrm{past}}(\eta,\bx')
 &:=\hat h^{\mathrm{in}}(\bx')
       -\frac{a(\eta)}{a_i}\bar h_{kk}^{\mathrm{in}}(\bx').
 \label{eq:past-reference-00}
\end{align}
\end{subequations}
We call agreement with this field before and in a neighbourhood of
$\eta_i$ \emph{cosmological past stationarity}. It specifies the past
in our generalized harmonic gauge, with the expansion factors displayed
explicitly; for constant $a$, all metric components are stationary.
This extends the past-stationary prescription of
\cite[Eqs.~(1.5)--(1.6) and Sec.~4.1]{Blanchet:1985sp}.

We assume that the past data decay at spatial infinity and that their
Laplacians have support in $|\bx'|\le d$. The gauge conditions require
\begin{equation}\label{eq:past-reference-transverse}
 \partial'_j\bar h_{ij}^{\mathrm{in}}=0,\qquad
 \partial'_i\bar h_{0i}^{\mathrm{in}}=0.
\end{equation}
The metric equations~\eqref{eq:metric-wave-equations} then determine
the past shifted stress-energy tensor perturbation:
\begin{subequations}\label{eq:past-reference-source}
\begin{align}
 \tau_{ij}^{\mathrm{past}}(\eta,\bx')
 &:=-\frac{c^4\Delta'\bar h_{ij}^{\mathrm{in}}(\bx')}{16\pi G a_i},
 \label{eq:past-source-spatial}\\
 \tau_{0i}^{\mathrm{past}}(\eta,\bx')
 &:=-\frac{c^4a_i\Delta'\bar h_{0i}^{\mathrm{in}}(\bx')}{16\pi G a^2(\eta)},
 \label{eq:past-source-mixed}\\
 \tau_{00}^{\mathrm{past}}(\eta,\bx')
 &:=-\frac{c^4[\Delta'+V^+(\eta)]\hat h^{\mathrm{in}}(\bx')}{16\pi G a(\eta)}
       +\frac{c^4\Delta'\bar h_{kk}^{\mathrm{in}}(\bx')}{16\pi G a_i}.
 \label{eq:past-source-density}
\end{align}
\end{subequations}
Indeed, $(c^{-2}\partial_\eta^2-V^-)a=0$ and
$(c^{-2}\partial_\eta^2-V^+)(1/a)=0$ remove the time-dependent terms
in the spatial and mixed equations. The equation for $\hat h$
fixes the last line. Equations~\eqref{eq:past-reference-transverse}
and $16\pi G a^2s=-c^4\dot V^+$ then give both conservation laws.
Thus the initial matter perturbation is compatible with the metric.

We extend the past field over the conformal-time interval $\Eta$
and define the dynamical parts by
\begin{subequations}\label{eq:past-dynamical-split}
\begin{align}
 \bar h_{\mu\nu}&=\bar h_{\mu\nu}^{\mathrm{past}}
                    +\bar h_{\mu\nu}^{\mathrm{dyn}},
 &\hat h&=\hat h^{\mathrm{past}}+\hat h^{\mathrm{dyn}},
 \label{eq:past-dynamical-metric}\\
 \tau_{\mu\nu}&=\tau_{\mu\nu}^{\mathrm{past}}
                    +\tau_{\mu\nu}^{\mathrm{dyn}}.
 \label{eq:past-dynamical-source}
\end{align}
\end{subequations}
The prescribed $\tau_{ij}^{\mathrm{dyn}}$ vanish before and near
$\eta_i$. Linearity gives the same field, gauge and conservation
equations for the dynamical parts, with zero past data. This separates
the pre-existing field from the subsequent disturbance without
introducing an impulsive source at $\eta_i$.

We require the full momentum component $\tau_{0i}$ to obey
Eq.~\eqref{eq:momentum}. At fixed $\bx'$, it has the integrating-factor form
\begin{equation}\label{eq:momentum-integrating-factor}
\partial_\eta(a^2\tau_{0i})=ca^2\partial'_j\tau_{ij}.
\end{equation}
Integrating the momentum
and energy equations \eqref{eq:momentum} and \eqref{eq:energy} gives
\begin{subequations}\label{eq:integrated-source-conservation}
\begin{align}
 \tau_{0i}(\eta,\bx')&=\frac{a_i^2\tau_{0i}^{\mathrm{in}}(\bx')}{a^2(\eta)}
 +\frac c{a^2(\eta)}\int_{\eta_i}^{\eta}\dd\zeta\,
 a^2(\zeta)\partial'_j\tau_{ij}(\zeta,\bx'),
 \label{eq:momentumsolution}\\
 \tau_{00}(\eta,\bx')&=\frac{a_i\tau_{00}^{\mathrm{in}}(\bx')}{a(\eta)}
 +\frac1{a(\eta)}\int_{\eta_i}^{\eta}\dd\zeta\,
  \big[ca\partial'_i\tau_{0i}-\dot a\tau_{kk}
       +a^2s\hat h\big](\zeta,\bx').
 \label{eq:energy-solution}
\end{align}
\end{subequations}
Here $\tau_{\mu\nu}^{\mathrm{in}}(\bx'):=
\tau_{\mu\nu}^{\mathrm{past}}(\eta_i,\bx')$ is fixed by
Eqs.~\eqref{eq:past-reference-source}.
The prescribed stresses and past momentum determine $\tau_{0i}$ through
Eq.~\eqref{eq:momentumsolution}, preserving its confinement to the
source ball.
By contrast, reconstructing $\tau_{00}$
requires $\hat h$ and therefore also the metric equations.

The shifted energy-density perturbation $\tau_{00}$ cannot in general be restricted to the same source
region. Indeed, outside $|\bx'|\le d$, Eq.~\eqref{eq:energy} becomes
$\partial_\eta(a\tau_{00})=a^2s\hat h$. Setting $\tau_{00}=0$ there would
suppress the response of the surrounding cosmic fluid: for $s\ne0$,
the same equation would require $\hat h=0$ there, which is generically
inconsistent with Einstein's equations.
We call $\tau_{00}$ outside the region of prescribed stresses the
induced exterior density. It is determined by
Eq.~\eqref{eq:energy-solution}.
For the past field, the same conservation law gives
$\partial_\eta(a\tau_{00}^{\mathrm{past}})=a^2s\hat h^{\mathrm{in}}$,
with initial density $\tau_{00}^{\mathrm{in}}$ fixed by
Eq.~\eqref{eq:past-source-density}.
The past field can already carry an exterior density contribution
at $\eta_i$, so $\tau_{00}^{\mathrm{past}}$ is generally not spatially
compact.

Our construction determines the gravitational field and stress-energy
perturbation compatible with prescribed shifted stresses and covariant
conservation. We do not impose an equation of state nor develop a
dynamical matter model. Determining which prescribed histories arise
from a fluid, scalar field or compact-object model requires additional
matter equations and lies outside the scope of this paper.
For example, the cosmological Einstein--Boltzmann formulation
of Ma and Bertschinger \cite{Ma:1995ey} evolves specified matter
species. Such matter equations supply the additional input needed
to select a physical source history for our retarded construction.

\subsection{The linearized McVittie monopole}
\label{sec:mcvittie-monopole}
The McVittie solution describes a constant central mass $M$ in a
spatially flat cosmological background \cite{McVittie:1933zz}.
Its linearization will provide a benchmark for the perturbation
equations and illustrate the distinction between $\tau_{00}$ and
the physical fluid density.

We start from the exact metric in conformal time and isotropic
spatial coordinates,
\begin{equation}\label{eq:mcvittie-exact}
 \dd s^2=\mathfrak{s}a^2\left[-\left(\frac{1-GM/(2c^2a\rho)}{1+GM/(2c^2a\rho)}\right)^2
 c^2\dd\eta^2+\left(1+\frac{GM}{2c^2a\rho}\right)^4\delta_{ij}\dd x^i\dd x^j\right],
 \qquad \rho:=|\bx|,
\end{equation}
where $a=a(\eta)$. This is the isotropic-coordinate form of
\cite[Eqs.~(1), (11)]{Nolan:2014maa}.
For constant $H>0$, the background obeys $\eps+p=0$, and the exact
metric~\eqref{eq:mcvittie-exact} is Schwarzschild--de Sitter
\cite[Sec.~I]{Nolan:2014maa}. 
For a general expansion history, the existence of a black-hole
horizon is a property of the full McVittie spacetime
\cite[Sec.~I]{Nolan:2014maa}.
%; the present check concerns only its linearized exterior.

Expanding to first order in $M$, in the region $GM/(c^2a\rho)\ll1$,
gives the metric perturbation
\begin{equation}\label{eq:mcvittie-perturbation}
 h_{00}=\frac{2\mathfrak{s}GMa}{c^2\rho},\qquad h_{0i}=0,\qquad
 h_{ij}=\frac{2\mathfrak{s}GMa}{c^2\rho}\delta_{ij}.
\end{equation}
The field definitions~\eqref{eq:field-variables} therefore give
\begin{equation}\label{eq:mcvittie-rescaled}
 \bar h_{00}=\hat h=\frac{4GM}{c^2\rho},\qquad
 \bar h_{0i}=\bar h_{ij}=0.
\end{equation}
These fields satisfy the generalized harmonic gauge conditions
\eqref{eq:gauge-components}: $\bar h_{00}$ is independent
of conformal time and the other components vanish. The gravitational
perturbation is a pure mass monopole, fixed by the single parameter $M$. 
%For $a=1$ and $\eps=p=\Lambda=0$, Eq.~\eqref{eq:mcvittie-perturbation} reduces to the linearized Schwarzschild metric in harmonic coordinates.

The spatial and mixed field equations
\eqref{eq:spatial}--\eqref{eq:mixed} give $\tau_{ij}=\tau_{0i}=0$.
The wave equation~\eqref{eq:scalar} for $\hat h$ fixes $\tau_{00}$. Using
$\Delta(1/\rho)=-4\pi\delta^{(3)}(\bx)$ and the background
identity~\eqref{eq:potential-plus-fluid}, we obtain the distributional
source of the linearized monopole:
\begin{equation}\label{eq:mcvittie-source}
 \tau_{00}=\frac{Mc^2}{a}\delta^{(3)}(\bx)-\frac{Mc^2V^+}{4\pi a\rho}
 =\frac{Mc^2}{a}\delta^{(3)}(\bx)-\frac{GMa(\eps+p)}{c^2\rho},\qquad
 \tau_{0i}=\tau_{ij}=0.
\end{equation}
The delta function represents the central mass in the linearized
equations. The comparison with the exact McVittie metric applies in
the region $GM/(c^2a\rho)\ll1$, where linearization is valid.
The second term in $\tau_{00}$ 
is the induced exterior density: it extends into the surrounding
fluid and is fixed by $M$ and the background, rather than by an
additional source parameter. Its evolution satisfies
the energy equation~\eqref{eq:energy} through the relation between
$\dot V^+$ and $s$ in Eq.~\eqref{eq:s}.

Substituting the field~\eqref{eq:mcvittie-rescaled} and the
source~\eqref{eq:mcvittie-source} into the physical
relations~\eqref{eq:physical-matter-variables} gives, outside the
central mass,
\begin{equation}\label{eq:mcvittie-fluid}
 \delta\eps=0,\qquad \delta p=\frac{GM}{c^2a\rho}(\eps+p),\qquad \rho>0.
\end{equation}
Thus the nonzero shifted energy-density perturbation $\tau_{00}$
coexists with a homogeneous physical fluid energy density and an
inhomogeneous pressure. Equation~\eqref{eq:mcvittie-fluid} agrees
with the linearization of the McVittie density and pressure in
\cite[Eqs.~(7)--(10)]{Nolan:2014maa}.

The McVittie linear solution is the mass-monopole member of
Eqs.~\eqref{eq:past-reference-metric}, with
$\hat h^{\mathrm{in}}=4GM/(c^2\rho)$ and
$\bar h_{ij}^{\mathrm{in}}=\bar h_{0i}^{\mathrm{in}}=0$.
Its source~\eqref{eq:mcvittie-source} is precisely
Eq.~\eqref{eq:past-source-density}. The mass and its accompanying
fluid perturbation are thus retained as past data in our formalism.
%In either case, the linear solution applies where $GM/(c^2a\rho)\ll1$.

\section{Retarded Green functions for an arbitrary scale factor}
\label{sec:green}
The two wave potentials determine how the metric perturbation
propagates through the cosmological background. In this section, we
will first derive analytical representations of their retarded Green functions
for an arbitrary smooth scale factor using the characteristic and dimension-reduction method
of \cite{Chu:2015yua}. We will then define the metric
propagators and derive the identities needed for gauge
propagation and multipolar reconstruction. The de Sitter and
power-law solutions will provide checks before we apply the
construction to the two Green functions in the $\Lambda$CDM model
with radiation.

\subsection{Hadamard form and characteristic construction}
\label{sec:green-definition}

The retarded Green function separates the direct signal on the light
cone from the curvature tail inside it. We will construct the tail
through an auxiliary Green function in one spatial dimension, whose
characteristic equation can be integrated for any smooth potential.

We start with a source event $(\eta',\bx')$, an observation event
$(\eta,\bx)$, and the conformal-time separation $T:=\eta-\eta'$.
For a smooth potential $V(\eta)$ on $\Eta$, denote the retarded Green
function by $G_V(\eta,\bx;\eta',\bx')$: a semicolon separates the
observation coordinates from the source coordinates. Spatial
homogeneity and isotropy allow the reduced notation
\begin{equation}\label{eq:green-argument-convention}
 G_V(\eta,\bx;\eta',\bx')=G_V(\eta;R;\eta'),
 \qquad R:=|\bx-\bx'|.
\end{equation}
The two semicolons isolate the separation $R$, which depends on both
spatial positions. The Green function vanishes for $\eta<\eta'$ and obeys
\begin{equation}\label{eq:green-def}
 (\square+V(\eta))G_V(\eta;R;\eta')
 =-\frac{4\pi}{c}\delta(\eta-\eta')\delta^{(3)}(\bx-\bx').
\end{equation}
Its existence and uniqueness follow from the theory of normally
hyperbolic operators
\cite[Theorem~3.2.11 and Corollary~3.4.3]{Baer:2008qni}.
With $\delta$ and $\theta$ denoting the Dirac distribution and
Heaviside function, its Hadamard form is
\begin{equation}\label{eq:hadamard}
 G_V(\eta;R;\eta')=\frac{\delta(cT-R)}R
       +\theta(cT-R)\mathcal V_V(\eta;R;\eta').
\end{equation}
We denote the light-cone and curvature-tail terms by $G^{\mathrm{LC}}$
and $G_V^{\mathrm{tail}}$, respectively; $\mathcal V_V$ is the smooth
tail coefficient, before multiplication by the step function.
This is the normalization of the reduced Green function $g$ in
\cite[Eqs.~(4.1)--(4.4)]{Burko:2002ge}, whose cosmological examples
have $V=2\beta/(c^2\eta^2)$.
The retarded support is $cT\ge R$, since the conformal FLRW metric
has the same null cones as the Minkowski metric. Source integrals
use the coordinate measure $c\,\dd\eta'\dd^3x'$, with no additional
factor $a^4(\eta')$.

\paragraph{Reduction to one spatial dimension.}
To determine the tail, we use the transverse integration of
\cite[Eqs.~(12)--(14)]{Chu:2015yua}, with $U_{\rm Chu}=-V$.
We define the auxiliary retarded Green function
\begin{equation}\label{eq:green-transverse-reduction}
 G_V^{(1d)}(\eta,R;\eta'):=
 \frac12\int_{|R|}^{\infty}\dd\widetilde R\,\widetilde R\,
 G_V(\eta;\widetilde R;\eta').
\end{equation}
Here $R$ is the signed separation along a chosen spatial axis.
In the auxiliary problem, $(\eta,R)$ are the evaluation coordinates,
grouped before the semicolon; the source time $\eta'$ follows it.
We use the same ordering for the characteristic function $J_V$ introduced
below. The integral is $G_V/(4\pi)$ integrated over the two transverse
coordinates: their angular integration gives
$2\pi\widetilde R\,\dd\widetilde R$ at fixed $R$.
Integrating Eq.~\eqref{eq:green-def} over this plane eliminates
the transverse Laplacian. Indeed, the divergence theorem turns it
into a boundary integral, and the boundary can be placed outside
the finite support of $G_V$ at fixed $\eta,\eta'$.
The remaining derivatives and the potential commute with the
transverse integral, while the spatial delta function integrates
to $\delta(R)$. Hence
\begin{equation}\label{eq:one-dimensional-green-source}
 \left(-c^{-2}\partial_\eta^2+\partial_R^2+V(\eta)\right)
 G_V^{(1d)}(\eta,R;\eta')=-\delta(c(\eta-\eta'))\delta(R).
\end{equation}
This is the response to a unit point impulse in one spatial
dimension, with support $cT\ge|R|$.

Substituting the Hadamard form~\eqref{eq:hadamard} into the
defining integral~\eqref{eq:green-transverse-reduction} gives
\begin{subequations}\label{eq:auxiliary-function-relations}
\begin{equation}\label{eq:one-dimensional-green}
 G_V^{(1d)}(\eta,R;\eta')=
 \tfrac12\theta(cT-|R|)J_V(\eta,R;\eta'),
\end{equation}
and identifies the characteristic function $J_V$ multiplying the step function:
\begin{equation}\label{eq:J-from-tail-integral}
 J_V(\eta,R;\eta'):=
 1+\int_{|R|}^{cT}\dd\widetilde R\,\widetilde R\,
       \mathcal V_V(\eta;\widetilde R;\eta'),
 \qquad |R|\le cT.
\end{equation}
\end{subequations}
The constant one comes from the light-cone term. The tail integral
vanishes at $|R|=cT$, so $J_V=1$ on both null boundaries; its
dependence on $|R|$ makes it even in $R$. Thus the boundary values
and parity of $J_V$ follow directly from the Green function.
Inside the cone, the step function is constant and
Eq.~\eqref{eq:one-dimensional-green-source} gives
\begin{equation}\label{eq:characteristic-problem}
 (-c^{-2}\partial_\eta^2+\partial_R^2+V(\eta))J_V(\eta,R;\eta')=0.
\end{equation}

\paragraph{Characteristic construction and recovery of the tail.}
We will solve this equation with its unit boundary data. Introduce
$U:=(cT-R)/2$ and $V:=(cT+R)/2$ at fixed $\eta'$; the coordinate $V$
is distinct from the potential $V(\eta)$. We use the argument convention
\begin{equation}\label{eq:J-characteristic-arguments}
 J_V(U,V;\eta'):=J_V(\eta'+(U+V)/c,V-U;\eta').
\end{equation}
The left-hand side uses characteristic coordinates; the right-hand
side uses observation time and signed separation, followed by source time.
Since $\partial_\eta=c(\partial_U+\partial_V)/2$ and
$\partial_R=(\partial_V-\partial_U)/2$, the equation and data become
\begin{subequations}\label{eq:characteristic-system}
\begin{align}
 \partial_U\partial_VJ_V(U,V;\eta')
 &=V(\eta'+(U+V)/c)J_V(U,V;\eta'),
 \label{eq:J-characteristic-equation}\\
 J_V(0,V;\eta')&=J_V(U,0;\eta')=1.
 \label{eq:characteristic-data}
\end{align}
\end{subequations}
We now integrate over $[0,U]\times[0,V]$. The mixed derivative gives
the four corner values, whose combination is $J_V(U,V;\eta')-1$
because the other three corners carry unit data. We obtain the
Volterra integral equation
\begin{equation}\label{eq:volterra}
 J_V(U,V;\eta')=1+
 \int_0^U\dd u\int_0^V\dd v\,
 V(\eta'+(u+v)/c)J_V(u,v;\eta').
\end{equation}
This is the characteristic integral construction of
\cite[Appendix~A, Eqs.~(A33)--(A34), (A44)]{Chu:2015yua};
their characteristic displacements are $2V,2U$.
Starting with $J_0(U,V;\eta'):=1$, it determines the solution through
\begin{subequations}\label{eq:green-volterra-iteration}
\begin{align}
 J_{n+1}(U,V;\eta')&:=
 \int_0^U\dd u\int_0^V\dd v\,
 V(\eta'+(u+v)/c)J_n(u,v;\eta'),\qquad n\ge0,
 \label{eq:green-volterra-recurrence}\\
 J_V(U,V;\eta')&=\sum_{n=0}^{\infty}J_n(U,V;\eta').
 \label{eq:green-volterra-sum}
\end{align}
\end{subequations}
The first correction integrates the potential alone; subsequent
corrections use the preceding term. Each correction vanishes on the
initial characteristics, and summing the recurrence gives
Eq.~\eqref{eq:volterra}.
For smooth $V$, the series defines the unique smooth solution on
$U,V\ge0$ with $\eta'+(U+V)/c\in\Eta$. No small-potential assumption
is needed; the convergence and uniqueness arguments are given in
Appendix~\ref{app:volterra-estimates}.

We now recover the three-dimensional Green function. Differentiating
the defining integral~\eqref{eq:green-transverse-reduction} for
$G_V^{(1d)}$ with respect to its lower limit $R>0$, at fixed
$\eta,\eta'$, gives $\partial_R G_V^{(1d)}=-RG_V/2$.
Solving for $G_V$ yields
\begin{align}
 G_V(\eta;R;\eta')&=
 -\frac2R\partial_R G_V^{(1d)}(\eta,R;\eta'),\qquad R>0.
 \label{eq:dimension-raised-green}
\end{align}
Substituting the step-function form~\eqref{eq:one-dimensional-green}
and using $J_V=1$ on the light cone then identifies the tail as
\begin{align}
 \mathcal V_V(\eta;R;\eta')&=
 -\frac1R\partial_RJ_V(\eta,R;\eta'),\qquad R>0.
 \label{eq:tail-from-J}
\end{align}
The derivative of the step function supplies the light-cone term
in Eq.~\eqref{eq:hadamard}; the derivative of $J_V$ supplies
the tail. All radial derivatives hold $\eta,\eta'$ fixed.
At $R=0$, smoothness and evenness of $J_V$ imply that its first
radial derivative starts at order $R$, with coefficient
$\partial_R^2J_V|_{R=0}$. The factor $R$ therefore cancels, giving
\begin{equation}\label{eq:tail-at-origin}
 \mathcal V_V(\eta;0;\eta')
 =-\left.\partial_R^2J_V(\eta,R;\eta')\right|_{R=0},
 \qquad \eta>\eta'.
\end{equation}

The characteristic equation also fixes the tail on the light cone.
On $U=0$, the unit boundary data give $\partial_VJ_V=0$.
Integrating Eq.~\eqref{eq:J-characteristic-equation} along that
boundary from the vertex to $V=cT$, and then using
$\partial_R=(\partial_V-\partial_U)/2$, yields
\begin{subequations}\label{eq:tail-cone-check}
\begin{align}
 \left.\partial_UJ_V(U,V;\eta')\right|_{U=0,\,V=cT}
 &=\int_0^{cT}\dd v\,V(\eta'+v/c),
 \label{eq:transverse-characteristic-cone}\\
 \mathcal V_V(\eta;cT;\eta')&=
 \frac1{2T}\int_{\eta'}^\eta\dd\eta''\,V(\eta''),
 \qquad \lim_{\eta\downarrow\eta'}\mathcal V_V(\eta;cT;\eta')
 =\frac12V(\eta').
 \label{eq:conevalue}
\end{align}
\end{subequations}
Thus the cone value is one half of the average potential along the
propagation interval, as in \cite[Eq.~(23)]{Chu:2015yua}.
Differentiating $T\mathcal V_V(\eta'+T;cT;\eta')$ gives
$V(\eta'+T)/2$, recovering the transport condition of
\cite[Eqs.~(4.5)--(4.9)]{Burko:2002ge} and its inverse-square
specialization in \cite[Eqs.~(3.12)--(3.16)]{Compere:2026xoe}.
For $V=0$, the iteration gives $J_V=1$ and the tail vanishes.
For a general expansion history, the same construction determines
the full tail, and it can also be evolved on the null grid described
in Appendix~\ref{app:green-numerics}.

\subsection{Metric Green functions and propagation identities}
\label{sec:metric-green-propagation}
The general Green-function construction applies to both wave
potentials. We will first define the corresponding metric propagators,
then use the
relation between the two metric operators to derive propagation
identities. These identities will be used to establish gauge
propagation and reconstruct the multipolar metric.

We start by defining the retarded metric Green functions $G^\pm$
and their characteristic functions $J^\pm$ and tail coefficients
$\mathcal V^\pm$ by specialization to the two wave potentials:
\begin{subequations}\label{eq:metric-kernel-definitions}
\begin{align}
 G^\pm(\eta;R;\eta')&:=G_{V^\pm}(\eta;R;\eta'),
 \label{eq:metric-Gpm-definition}\\
 J^\pm(\eta,R;\eta')&:=J_{V^\pm}(\eta,R;\eta'),
 \label{eq:metric-Jpm-definition}\\
 \mathcal V^\pm(\eta;R;\eta')&:=\mathcal V_{V^\pm}(\eta;R;\eta').
 \label{eq:metric-tailpm-definition}
\end{align}
\end{subequations}
The spatial metric components use $G^-$, while the mixed components
and $\hat h$ use $G^+$ in Eqs.~\eqref{eq:metric-wave-equations}.
When the source is fixed at the spatial origin, $\rho:=|\bx|$ is an
observation coordinate. We therefore write $G_V(\eta,\rho;\eta')$ and
$\mathcal V_V(\eta,\rho;\eta')$, grouping $(\eta,\rho)$ before the
semicolon. For each metric operator $\Op^\pm$, we retain a prescribed past
coefficient by writing $f=f_\pm^{\mathrm{past}}+f^{\mathrm{dyn}}$, with
$f_-^{\mathrm{past}}:=f_{\mathrm{in}}a/a_i$ and
$f_+^{\mathrm{past}}:=f_{\mathrm{in}}a_i/a$. Here $f_{\mathrm{in}}$ is constant and
$f^{\mathrm{dyn}}$ has zero past data. We define the retarded integral operator $\mathcal G_V$ for
zero-past histories and the metric propagators $\mathcal G_\pm$,
which include the prescribed past contribution, by
\begin{subequations}\label{eq:canonical-propagator-definitions}
\begin{align}
 \mathcal G_V[f](\eta,\rho)&:=
 \int_\Eta c\,\dd\eta'\,G_V(\eta,\rho;\eta')f(\eta'),
 \label{eq:canonical-propagator}\\
 \mathcal G_\pm[f](\eta,\rho)&:=\frac{f_\pm^{\mathrm{past}}(\eta)}\rho
             +\mathcal G_{V^\pm}[f^{\mathrm{dyn}}](\eta,\rho).
 \label{eq:metric-propagator-definition}
\end{align}
\end{subequations}

The Green function describes the response to a point impulse, whereas
the metric propagator maps a source history, including its prescribed
past part, to a radial field. The past temporal factors satisfy
$(c^{-2}\partial_\eta^2-V^\pm)f_\pm^{\mathrm{past}}=0$, while
$\Delta(1/\rho)=-4\pi\delta^{(3)}(\bx)$.
Applying $\Op^\pm$ to Eq.~\eqref{eq:metric-propagator-definition},
using these identities for the past term and the Green-function
equation~\eqref{eq:green-def} for the retarded integral, gives
\begin{equation}\label{eq:metric-propagator-source}
 \Op^\pm\mathcal G_\pm[f]=-4\pi f(\eta)\delta^{(3)}(\bx).
\end{equation}
Thus each profile solves a homogeneous equation away from the origin.

We will use the relation between the two wave potentials to
transfer time derivatives through the metric propagators. Define
the first-order operators $D_\pm$ and temporal operators $T_\pm$ by
\begin{subequations}\label{eq:canonical-temporal-operators}
\begin{align}
 D_\pm&:=\frac1c(\partial_\eta\pm\cH),
 \label{eq:canonical-Dpm}\\
 T_\pm&:=c^{-2}\partial_\eta^2-V^\pm=D_\mp D_\pm,
 \label{eq:canonical-Tpm}\\
 \Op^\pm&=\Delta-T_\pm.
 \label{eq:canonical-wave-temporal-relation}
\end{align}
\end{subequations}
The factorization of $T_\pm$ follows from Eq.~\eqref{eq:potentials}.
The wave operators therefore form a Darboux
pair \cite{Darboux}: reversing the order of the first-order factors
exchanges their temporal parts. Commutation of $D_\pm$ with $\Delta$
then gives
\begin{subequations}\label{eq:intertwine}
\begin{align}
 \Op^-D_+&=D_+\Op^+,\label{eq:intertwine-minus}\\
 \Op^+D_-&=D_-\Op^-.\label{eq:intertwine-plus}
\end{align}
\end{subequations}
For the prescribed past coefficients and zero-past dynamical parts,
these identities imply
\begin{subequations}\label{eq:canonical-propagation-identities}
\begin{align}
 D_-\mathcal G_-[f]&=\mathcal G_+[D_-f],
 \label{eq:canonical-propagation-minus}\\
 D_+\mathcal G_+[f]&=\mathcal G_-[D_+f],
 \label{eq:canonical-propagation-plus}\\
 \Delta\mathcal G_\pm[f]&=\mathcal G_\pm[T_\pm f],\qquad \rho>0.
 \label{eq:canonical-propagation-trace}
\end{align}
\end{subequations}
For the first equality, the intertwining relation and
Eq.~\eqref{eq:metric-propagator-source} give
\begin{equation}\label{eq:propagation-minus-source-check}
 \Op^+D_-\mathcal G_-[f]
 =D_-\Op^-\mathcal G_-[f]
 =-4\pi(D_-f)(\eta)\delta^{(3)}(\bx).
\end{equation}
This is the source of $\mathcal G_+[D_-f]$. Since $D_-a=0$,
both sides have zero past data and coincide by uniqueness. The
second equality follows in the same way using $D_+(1/a)=0$.
Thus the identities hold for the past terms as well as the dynamical parts.

For the third equality, apply the first two successively:
\begin{equation}\label{eq:propagation-temporal-composition}
 T_-\mathcal G_-[f]=D_+\mathcal G_+[D_-f]
 =\mathcal G_-[D_+D_-f]=\mathcal G_-[T_-f].
\end{equation}
Reversing the order gives $T_+\mathcal G_+[f]=\mathcal G_+[T_+f]$.
The exterior wave equation identifies $T_\pm\mathcal G_\pm[f]$
with $\Delta\mathcal G_\pm[f]$, proving the last identity.
The prescribed past contributions are already included in
Eq.~\eqref{eq:metric-propagator-definition}.

\subsection{de Sitter, power-law cosmologies and \texorpdfstring{$\Lambda$}{Lambda}CDM with radiation}
\label{sec:explicit-backgrounds}
We will now specialize the construction to cosmological backgrounds
for which the retarded Green functions can be checked explicitly. The de Sitter,
radiation and matter solutions of \cite{Ramirez:2001wp} and the
general power-law tails of \cite{Chu:2015yua,Compere:2026xoe}
provide the benchmarks. We will then construct both retarded metric Green functions
for the $\Lambda$CDM model with radiation using the elliptic background
solution of \cite{Coquereaux:1981ya,Coquereaux:2014sva}.

\paragraph{de Sitter spacetime.}
We first recover the known de Sitter retarded Green functions to
check the characteristic construction. We set $a=-1/(H\eta)$
and $\Lambda=3H^2/c^2$, with constant $H>0$, and take
$\Eta:=(-\infty,0)$ as the conformal time range. Assigning the vacuum
energy to $\Lambda$ gives $\eps=p=s=0$. The potentials
\eqref{eq:potentials} reduce to
$V^+=0$ and $V^-=2/(c^2\eta^2)$, with characteristic functions
\begin{equation}
 J^+=1,\qquad
 J^-=1+\frac{c^2T^2-R^2}{2c^2\eta\eta'}.
\end{equation}
They obey Eq.~\eqref{eq:characteristic-problem} and equal one on
the light cone.
Equation~\eqref{eq:tail-from-J} therefore gives
\begin{equation}\label{eq:dSgreen}
 G^+=\frac{\delta(cT-R)}R,\qquad
 G^-=\frac{\delta(cT-R)}R+\frac{\theta(cT-R)}{c^2\eta\eta'}.
\end{equation}
With $\eps=p=0$, the metric-dependent contribution in Eq.~\eqref{eq:shift-definition} vanishes, so the shifted
stress-energy tensor perturbation $\tau_{\mu\nu}$ coincides with
the stress-energy tensor perturbation $\delta T_{\mu\nu}$.
Equation~\eqref{eq:dSgreen} reproduces
\cite[Eq.~(3.35)]{Compere:2026xoe}, with the inverse-square potential
coefficient of that reference set to zero for $G^+$ and two for $G^-$.
The conversion of its causal distributions is
$\delta_+(\sigma)=\delta(cT-R)/R$ and
$\theta_+(-\sigma)=\theta(cT-R)$, with
$\sigma:=(R^2-c^2T^2)/2$. These are the retarded Green functions underlying the de Sitter
solution of Ref.~\cite{Compere:2023ktn}, in the conventions of
the defining equation \eqref{eq:green-def}. Earlier retarded
metric solutions for generic de Sitter sources were obtained in
\cite{deVega:1998ia}. Our light-cone and $C=2$ Green functions also agree
with \cite[Eqs.~(43)--(44)]{Ramirez:2001wp} after multiplying
their Green functions by $4\pi$ to match our source normalization.
\paragraph{Power-law cosmologies.}
We next recover the retarded Green functions expressed in hypergeometric functions in
\cite[Eqs.~(3.35)--(3.36)]{Compere:2026xoe}, which provide explicit
checks for radiation and matter domination.
We take $a(\eta)\propto|\eta|^\alpha$, with a positive proportionality
constant and real exponent $\alpha$, on either $\Eta:=(0,\infty)$ or
$\Eta:=(-\infty,0)$. The expanding branch has $\alpha>0$ in the first
case and $\alpha<0$ in the second. The two potentials are
\begin{equation}\label{eq:power-potentials}
 V^- =\frac{\alpha(\alpha-1)}{c^2\eta^2},\qquad
 V^+ =\frac{\alpha(\alpha+1)}{c^2\eta^2}.
\end{equation}
The variable $\sigma:=(R^2-c^2T^2)/2$ equals $\mathfrak{s}$ times the
Minkowski world function. For a potential $V(\eta)=C/(c^2\eta^2)$ with constant
$C$, denote the characteristic function and tail coefficient by
$J_C$ and $\mathcal V_C$. The characteristic function $J_C$ is given by the Legendre
solution of \cite[Sec.~III]{Chu:2015yua}, written using
\cite[Eq.~14.3.6 with $\mu=0$]{DLMF}:
\begin{equation}\label{eq:Jpower}
 J_C={}_2F_1\!\left(-\nu,\nu+1;1;\frac{\sigma}{2c^2\eta\eta'}\right)
     =P_\nu\!\left(1-\frac{\sigma}{c^2\eta\eta'}\right),
 \qquad \nu(\nu+1)=C.
\end{equation}
Here ${}_2F_1$ is the Gauss hypergeometric function and $P_\nu$ is
the Legendre function regular at argument one.
For the coefficients in Eq.~\eqref{eq:power-potentials},
$J^-=P_{-\alpha}(1-\sigma/(c^2\eta\eta'))$ and
$J^+=P_{\alpha}(1-\sigma/(c^2\eta\eta'))$. The hypergeometric
derivative identity \cite[Eq.~15.5.1]{DLMF}, applied to
Eq.~\eqref{eq:Jpower} in the tail formula \eqref{eq:tail-from-J}, gives
\begin{equation}\label{eq:tailpower}
 \mathcal V_C=\frac{C}{2c^2\eta\eta'}\,
 {}_2F_1\!\left(\frac{3-\sqrt{1+4C}}2,
             \frac{3+\sqrt{1+4C}}2;2;
             \frac{\sigma}{2c^2\eta\eta'}\right).
\end{equation}
Combining Eq.~\eqref{eq:tailpower} with Eq.~\eqref{eq:hadamard} reproduces
\cite[Eq.~(3.35)]{Compere:2026xoe}; setting $C=\alpha(\alpha\pm1)$
gives Eq.~(3.36) of that reference. The normalization and causal
distributions are the same under the conversion given above. This is
the known power-law result, also derived in \cite[Sec.~III]{Chu:2015yua}.
At the light cone, the tail coefficient
$\mathcal V_C$ has value $C/(2c^2\eta\eta')$,
in agreement with Eq.~\eqref{eq:conevalue}.

For expanding radiation ($\alpha=1$), we take $\Eta:=(0,\infty)$.
Then $V^-=0$ and $V^+=2/(c^2\eta^2)$, hence
$\mathcal V^-=0$ and $\mathcal V^+=1/(c^2\eta\eta')$.
The absence of a spatial-tensor tail therefore does not imply the
absence of a tail for every metric component. For expanding matter
($\alpha=2$), we again take $\Eta:=(0,\infty)$. The two potentials
are $2/(c^2\eta^2)$ and $6/(c^2\eta^2)$,
and the terminating hypergeometric functions give
\begin{equation}\label{eq:matter-tails}
 \mathcal V^-=\frac1{c^2\eta\eta'},\qquad
 \mathcal V^+=\frac3{c^2\eta\eta'}
 \left(1-\frac{\sigma}{c^2\eta\eta'}\right).
\end{equation}
These radiation and matter tail coefficients realize the values $C=0,2,6$.
The $C=2$ and $C=6$ expressions reproduce
\cite[Eqs.~(44)--(45)]{Ramirez:2001wp} with the same factor
$4\pi$. At $C=6$, both expressions give
$3[c^2(\eta^2+\eta'^2)-R^2]/(2c^4\eta^2\eta'^2)$ in our normalization.
Their light-cone values agree with $C/(2c^2\eta\eta')$, providing a
check of the sign and normalization in Eq.~\eqref{eq:conevalue}.
The retarded Green functions in Eq.~\eqref{eq:dSgreen} and the tail
coefficient in Eq.~\eqref{eq:tailpower} also provide benchmarks for the
characteristic-grid algorithm in Appendix~\ref{app:green-numerics}.

\paragraph{$\Lambda$CDM with radiation.}
We now construct both retarded metric Green functions for a mixed
radiation--matter--$\Lambda$ background. Exact homogeneous tensor
modes for matter and $\Lambda$ were obtained by Sasaki and Suzuki
\cite{Sasaki:2018oxw}, and the corresponding position-space tensor
tail was studied by Jokela, Kajantie and Sarkkinen
\cite[Sec.~3]{Jokela:2022rhk}. Here radiation is retained and both
potentials enter the retarded metric and fluid reconstruction.
We begin with the elliptic background solution, which will specify
the wave potentials entering the retarded Green functions. Radiation and pressureless
matter are taken to be separately conserved. We normalize $a=1$ today and denote the present
Hubble rate by $H_0$. Let $\Omega_r$, $\Omega_m$ and $\Omega_\Lambda$
be the present density fractions of radiation, matter and the
cosmological constant, respectively. Spatial flatness gives
$\Omega_r+\Omega_m+\Omega_\Lambda=1$.
We keep $\Lambda=3H_0^2\Omega_\Lambda/c^2$ on the geometric side of
Einstein's equations. The fluid variables and conformal-time Friedmann
equations are then
\begin{subequations}\label{eq:lcdm-background}
\begin{align}
 \eps&=\frac{3c^2H_0^2}{8\pi G}
       \left(\frac{\Omega_m}{a^3}+\frac{\Omega_r}{a^4}\right),
 &p&=\frac{c^2H_0^2\Omega_r}{8\pi G a^4},\label{eq:lcdm-fluid}\\
 \dot a^2&=H_0^2(\Omega_r+\Omega_m a+\Omega_\Lambda a^4),
 &\ddot a&=H_0^2\left(\frac{\Omega_m}2+2\Omega_\Lambda a^3\right).
 \label{eq:lcdm-friedmann}
\end{align}
\end{subequations}
For positive density parameters, the expanding branch with the big
bang at $\eta=0$ is fixed by
\begin{equation}\label{eq:lcdm-time}
 \eta(a)=\frac1{H_0}\int_0^a
 \frac{\dd b}{\sqrt{\Omega_r+\Omega_m b+\Omega_\Lambda b^4}},
 \qquad 0<\eta<\eta_\infty:=\lim_{a\to\infty}\eta(a)<\infty.
\end{equation}
We take $\Eta:=(0,\eta_\infty)$, on which the expanding scale factor
is smooth and positive.
The Friedmann equation with radiation admits an elliptic solution
\cite{Coquereaux:1981ya}; see also
\cite[Secs.~3 and 5]{Coquereaux:2014sva} for the spatially flat case.
We use the quartic inversion formula \cite[Sec.~20.6]{WW} with
$a(0)=0$. Introduce the Weierstrass elliptic function $\wp(\eta)$
with invariants
\begin{equation}\label{eq:lcdm-weierstrass-invariants}
 g_2:=H_0^4\Omega_\Lambda\Omega_r,\qquad
 g_3:=-\frac{H_0^6\Omega_\Lambda\Omega_m^2}{16},
\end{equation}
and let a prime on $\wp$ denote a conformal-time derivative.\footnote{In
Mathematica, $\wp(\eta)$ is given by
$\mathtt{WeierstrassP}[\eta,\{g_2,g_3\}]$, and its derivative by
$\mathtt{WeierstrassPPrime}[\eta,\{g_2,g_3\}]$. The second argument
is the pair of invariants defined by Eq.~\eqref{eq:lcdm-weierstrass-invariants}.}
One convenient expression for the scale factor is
\begin{equation}\label{eq:lcdm-elliptic}
 a(\eta)=
 \frac{-H_0\sqrt{\Omega_r}\,\wp'(\eta)
           +\frac12H_0^2\Omega_m\wp(\eta)}
      {2\wp(\eta)^2-\frac12H_0^4\Omega_\Lambda\Omega_r}.
\end{equation}
Using $\wp'^2=4\wp^3-g_2\wp-g_3$ and
$\wp''=6\wp^2-g_2/2$ \cite[Eqs.~23.3.10, 23.3.12]{DLMF}, we verify
the background equations~\eqref{eq:lcdm-friedmann}. The Laurent expansion
\cite[Eq.~23.9.2]{DLMF} gives the early-time behavior
\begin{equation}
 a(\eta)=H_0\sqrt{\Omega_r}\,\eta
       +\frac14H_0^2\Omega_m\eta^2+O(\eta^5).
\end{equation}

Substituting the background equations \eqref{eq:lcdm-friedmann}
into the definitions \eqref{eq:potentials}, we obtain the two
wave potentials:
\begin{equation}\label{eq:lcdm-potentials}
 V^-=\frac{H_0^2}{c^2}\left(\frac{\Omega_m}{2a}+2\Omega_\Lambda a^2\right),
 \qquad
 V^+=\frac{H_0^2}{c^2}\left(\frac{2\Omega_r}{a^2}
                         +\frac{3\Omega_m}{2a}\right).
\end{equation}
Radiation contributes to $V^-$ through $a(\eta)$, although
$\Omega_r$ does not appear explicitly in its numerator. Similarly,
$\Omega_\Lambda$ enters $V^+$ through the expansion history.

Applying the iteration of Eq.~\eqref{eq:volterra} to these elliptic potentials
gives an analytical representation of both exact position-space
retarded Green functions:
\begin{subequations}\label{eq:lcdm-iteration}
\begin{align}
J_0^\pm(U,V;\eta')&:=1,\label{eq:lcdm-iteration-initial}\\
J_{n+1}^\pm(U,V;\eta')
 &:=\int_0^U\dd u\int_0^V\dd v\,
 V^\pm(\eta'+(u+v)/c)J_n^\pm(u,v;\eta'),\label{eq:lcdm-iteration-step}\\
G^\pm(\eta;R;\eta')
 &=\frac{\delta(c(\eta-\eta')-R)}R
 -\frac{\theta(c(\eta-\eta')-R)}R\partial_R
 \sum_{n=0}^{\infty}J_n^\pm(\eta,R;\eta').\label{eq:lcdm-green}
\end{align}
\end{subequations}
Here the arguments of $J_n^\pm$ follow the convention
\eqref{eq:J-characteristic-arguments}: $J_n^\pm(\eta,R;\eta')$ denotes
the same iterate expressed in $\eta,R,\eta'$, with $U=(c(\eta-\eta')-R)/2$ and
$V=(c(\eta-\eta')+R)/2$. The radial derivative is taken at fixed $\eta,\eta'$.

The potentials are smooth on $\Eta$. For
$0<\eta'<\eta<\eta_\infty$, the Volterra series therefore converges
and can be differentiated term by term with respect to $R$, as
established in Appendix~\ref{app:volterra-estimates}.
The light-cone tail values are given by Eq.~\eqref{eq:conevalue}.
At $\Omega_r=0$, $V^-$ in Eq.~\eqref{eq:lcdm-potentials} agrees
with the tensor potential in \cite[Eq.~(3.29)]{Jokela:2022rhk},
and the cone value agrees with their Eq.~(3.11). These are direct
checks of the wave potential and the light-cone tail value.
A further comparison with the homogeneous tensor modes of
\cite{Sasaki:2018oxw} can be made through the Fourier representation
in Appendix~\ref{sec:green-fourier}.

The pure radiation, matter and de Sitter cases follow by setting two
density parameters to zero in Eqs.~\eqref{eq:lcdm-friedmann} and
\eqref{eq:lcdm-potentials}. For pure $\Lambda$, one must also change the
choice of conformal-time origin: Eq.~\eqref{eq:lcdm-time} presupposes a big bang, whereas
$a=[H_0\sqrt{\Omega_\Lambda}(\eta_\infty-\eta)]^{-1}$ gives the de
Sitter solution on $\Eta:=(-\infty,\eta_\infty)$, with a finite future
endpoint $\eta_\infty$. Our earlier de Sitter convention sets
$\eta_\infty=0$.
The radiation--matter--$\Lambda$ solution approaches the same de Sitter
behavior, $a\sim[H_0\sqrt{\Omega_\Lambda}(\eta_\infty-\eta)]^{-1}$,
as $\eta\to\eta_\infty$.

\section{Reconstruction of the complete retarded metric}
\label{sec:reconstruct}
The prescribed shifted stresses $\tau_{ij}$ determine both the metric
and the induced matter perturbation once the initial data are fixed.
We will first derive the evolution equations for the generalized
harmonic gauge residuals. The retarded metric will then be constructed
from the spatial and mixed wave equations~\eqref{eq:spatial}--\eqref{eq:mixed},
the temporal gauge condition~\eqref{eq:gauge0} and the integrated
energy equation~\eqref{eq:energy-solution}. We will separate its
light-cone and tail contributions and check the remaining wave equation
for $\hat h$.

\subsection{Consistency of the generalized harmonic gauge}
\label{sec:gauge-consistency}
The evolution of the gauge residuals $C_0,C_i$ determines whether
initially imposed gauge conditions remain satisfied. We will derive
that evolution using the factorization in Eq.~\eqref{eq:canonical-Tpm}
and the intertwining identities~\eqref{eq:intertwine}.

We begin by assuming that all the reduced field equations hold.
For a solution of Eqs.~\eqref{eq:metric-wave-equations}, the residuals
defined in Eqs.~\eqref{eq:gauge-components} satisfy
\begin{subequations}\label{eq:gauge-residual-equations}
\begin{align}
 \Op^-C_i&=-\frac{16\pi G}{c^4}\big[c^{-1}(\partial_\eta+\cH)(a\tau_{0i})
                                     -a\partial_j\tau_{ij}\big],
 \label{eq:gauge-residual-spatial}\\
 \Op^+C_0&=-\frac{16\pi G}{c^5}\big[\partial_\eta(a\tau_{00})
                    -ca\partial_i\tau_{0i}+\dot a\tau_{kk}\big]
                    -\frac{\dot V^+}{c}\hat h.
 \label{eq:gauge-residual-temporal}
\end{align}
\end{subequations}
To obtain the first identity, we apply $\Op^-$ to $C_i$, use
Eq.~\eqref{eq:intertwine-minus}, and substitute the spatial and mixed field
equations. For the second, we write
$C_0=c^{-1}\partial_\eta\hat h-D_-\bar h_{kk}
-\partial_i\bar h_{0i}$ and use the intertwining identity~\eqref{eq:intertwine-plus} together
with $\Op^+\partial_\eta=\partial_\eta\Op^+-\dot V^+$.
This second step also uses the scalar wave equation \eqref{eq:scalar}.

The momentum conservation equation~\eqref{eq:momentum} sets the source
of $C_i$ in Eq.~\eqref{eq:gauge-residual-spatial} to zero. The energy
conservation equation~\eqref{eq:energy} reduces the source of $C_0$ in
Eq.~\eqref{eq:gauge-residual-temporal}
to $-c^{-1}(16\pi G a^2s/c^4+\dot V^+)\hat h$, which vanishes by
Eq.~\eqref{eq:s}. Both gauge residuals $C_i,C_0$ therefore obey homogeneous
wave equations. If they and their first time derivatives vanish
initially, uniqueness of the hyperbolic Cauchy problem
\cite[Theorem~3.2.11]{Baer:2008qni} implies $C_i=C_0=0$ throughout the
evolution. The past field satisfies the gauge conditions for
all times, and the dynamical part vanishes near $\eta_i$, supplying these data.
Both the background Friedmann equations and the reduced perturbation
equations enter this argument.
\subsection{Retarded reconstruction and matter response}
\label{sec:retarded-reconstruction}
The spatial and mixed wave equations determine $\bar h_{ij}$ and
$\bar h_{0i}$ directly from the shifted sources. We will use the
temporal gauge condition to obtain $\hat h$, and then conservation
to determine the matter response.

We start with a smooth positive scale factor $a$ on $\Eta$ and the
background density and pressure fixed by Eq.~\eqref{eq:friedmann}.
Choose the past data of Eqs.~\eqref{eq:past-reference-metric}--
\eqref{eq:past-reference-source}, and prescribe smooth symmetric
$\tau_{ij}^{\mathrm{dyn}}$ supported in $|\bx'|\le d$, vanishing
before and near $\eta_i$. Equation~\eqref{eq:momentumsolution}
then fixes the full $\tau_{0i}$, including its past contribution.
We use the retarded Green functions $G^\pm=G_{V^\pm}$ satisfying
Eq.~\eqref{eq:green-def}, with the potentials~\eqref{eq:potentials}.

The sources of the dynamical wave equations, $a\tau_{ij}^{\mathrm{dyn}}$ and
$a\tau_{0i}^{\mathrm{dyn}}$, have
\emph{past-compact} support: its intersection with the causal past
of any event is compact. The retarded Green operators of the normally
hyperbolic operators $\Op^\pm$ therefore give unique smooth fields
\cite[Theorem~3.8 and Corollary~3.9]{Bar:2013bpw}, even when the
sources persist indefinitely into the future. With
$R(\bx,\bx'):=|\bx-\bx'|$,
adding these dynamical parts to the past field gives
\begin{subequations}\label{eq:retarded-metric-reconstruction}
\begin{align}
 \bar h_{ij}(\eta,\bx)&=\bar h_{ij}^{\mathrm{past}}(\eta,\bx)
 +\frac{4G}{c^4}\int_{\eta_i}^{\eta}c\,\dd\eta'\,a(\eta')
 \int\dd^3x'\,G^-(\eta;R;\eta')\tau_{ij}^{\mathrm{dyn}}(\eta',\bx'),
 \label{eq:spatial-solution}\\
 \bar h_{0i}(\eta,\bx)&=\bar h_{0i}^{\mathrm{past}}(\eta,\bx)
 +\frac{4G}{c^4}\int_{\eta_i}^{\eta}c\,\dd\eta'\,a(\eta')
 \int\dd^3x'\,G^+(\eta;R;\eta')\tau_{0i}^{\mathrm{dyn}}(\eta',\bx').
 \label{eq:mixed-solution}
\end{align}
For $\hat h=\bar h_{00}+\bar h_{kk}$, the temporal gauge condition
\eqref{eq:gauge0} takes the evolution form
$\partial_\eta\hat h=(\partial_\eta-\cH)\bar h_{kk}
+c\partial_i\bar h_{0i}$. Integrating at fixed $\bx$ yields
\begin{equation}\label{eq:scalar-solution}
 \hat h(\eta,\bx)=\hat h^{\mathrm{in}}(\bx)
 +\int_{\eta_i}^{\eta}\dd\zeta\,
  \big[(\partial_\zeta-\cH)\bar h_{kk}
       +c\partial_i\bar h_{0i}\big](\zeta,\bx).
\end{equation}
\end{subequations}
Substituting this $\hat h$ into Eq.~\eqref{eq:energy-solution}
determines $\tau_{00}$ with the compatible initial value
$\tau_{00}^{\mathrm{in}}$ fixed by Eq.~\eqref{eq:past-source-density}.
The remaining metric component is
$\bar h_{00}=\hat h-\bar h_{kk}$; Eq.~\eqref{eq:metric-final} then
gives $h_{\mu\nu}$, and Eqs.~\eqref{eq:inverse-source-shift} recover
$\delta T_{\mu\nu}$.

We separate the dynamical part into propagation on and inside
the light cone. The Green-function decomposition~\eqref{eq:hadamard}
gives $\bar h_{ij}=\bar h_{ij}^{\mathrm{past}}
+\bar h_{ij}^{\mathrm{LC}}+\bar h_{ij}^{\mathrm{tail}}$.
With the source retarded time $\eta_{\mathrm{ret}}:=\eta-R(\bx,\bx')/c$,
the light-cone delta function fixes $\eta'=\eta_{\mathrm{ret}}$,
while the tail step function restricts $\eta'\le\eta_{\mathrm{ret}}$.
The two contributions are therefore
\begin{subequations}\label{eq:lightcone-tail-fields}
\begin{align}
 \bar h^{\mathrm{LC}}_{ij}(\eta,\bx)
 &=\frac{4G}{c^4}\int\frac{\dd^3x'}{R(\bx,\bx')}\,
    \left.a(\eta')\tau_{ij}^{\mathrm{dyn}}(\eta',\bx')\right|_{\eta'=\eta_{\mathrm{ret}}},
 \label{eq:LC-field}\\
 \bar h^{\mathrm{tail}}_{ij}(\eta,\bx)
 &=\frac{4G}{c^4}\int\dd^3x'\int_{\eta_i}^{\eta_{\mathrm{ret}}}c\,\dd\eta'\,
   \mathcal V^-(\eta;R(\bx,\bx');\eta')a(\eta')\tau_{ij}^{\mathrm{dyn}}(\eta',\bx').
 \label{eq:tail-field}
\end{align}
\end{subequations}
The source $a\tau_{ij}^{\mathrm{dyn}}$ in the spatial wave equation is zero
at retarded times before $\eta_i$, in accordance with the stated
retarded data, and the tail integral vanishes when $\eta_{\mathrm{ret}}<\eta_i$.
The mixed fields $\bar h^{\mathrm{LC}}_{0i}$ and
$\bar h^{\mathrm{tail}}_{0i}$ follow from
Eq.~\eqref{eq:mixed-solution} by the same formulas with $ij$ replaced
by $0i$ and $\mathcal V^-$ by $\mathcal V^+$.

The temporal gauge condition holds by construction. For the spatial
conditions, the spatial and mixed wave equations together with momentum
conservation give $\Op^-C_i=0$ through
Eq.~\eqref{eq:gauge-residual-spatial}. This identity uses only those
two wave equations. Near $\eta_i$ the fields equal the past field,
which satisfies the gauge conditions, so $C_i$ and its first time
derivative initially vanish.
The unique solution of the homogeneous wave equation with these zero
initial data is $C_i=0$ throughout the evolution.

It remains to check the wave equation~\eqref{eq:scalar} for $\hat h$.
We define its residual by
\begin{equation}\label{eq:scalar-residual-definition}
 \mathcal R:=\Op^+\hat h+\frac{16\pi G}{c^4}a(\tau_{00}+\tau_{kk}).
\end{equation}
The temporal gauge condition \eqref{eq:gauge0}, imposed through
Eq.~\eqref{eq:scalar-solution}, and the intertwining identity
\eqref{eq:intertwine-plus} give
\begin{equation}\label{eq:scalar-residual-wave-step}
\begin{aligned}
 \Op^+\dot{\hat h}
 &=(\partial_\eta-\cH)\Op^-\bar h_{kk}
   +c\partial_i\Op^+\bar h_{0i}\\
 &=-\frac{16\pi G}{c^4}\big[(\partial_\eta-\cH)(a\tau_{kk})
                  +ca\partial_i\tau_{0i}\big].
\end{aligned}
\end{equation}
The second line uses the spatial and mixed wave equations
\eqref{eq:spatial}--\eqref{eq:mixed} and $\partial_i a=0$.
Since $\Op^+=-c^{-2}\partial_\eta^2+\Delta+V^+(\eta)$, differentiating
$\Op^+\hat h$ also produces $\dot V^+\hat h$. At fixed $\bx$,
we therefore obtain
\begin{equation}\label{eq:scalar-residual-check}
\begin{aligned}
 \dot{\mathcal R}
 &=\Op^+\dot{\hat h}+\dot V^+\hat h
   +\frac{16\pi G}{c^4}\partial_\eta\big[a(\tau_{00}+\tau_{kk})\big]\\
 &=\dot V^+\hat h+\frac{16\pi G}{c^4}\big[
     \partial_\eta(a\tau_{00})-ca\partial_i\tau_{0i}
     +\dot a\tau_{kk}\big]\\
 &=(\dot V^++16\pi G a^2s/c^4)\hat h=0.
\end{aligned}
\end{equation}
In the second line, the two terms $\partial_\eta(a\tau_{kk})$
cancel and $a\cH=\dot a$. The energy equation~\eqref{eq:energy}
then replaces the bracket by $a^2s\hat h$, and the background identity
\eqref{eq:s} gives the final equality. Thus $\mathcal R$ is constant in time at each spatial point.
It vanishes on the past field by Eq.~\eqref{eq:past-source-density},
and hence throughout the evolution, proving Eq.~\eqref{eq:scalar}.
The reconstructed fields therefore satisfy all ten reduced field
equations, both conservation equations and all four generalized
harmonic gauge conditions, and hence the original linearized
Einstein equations.

\label{sec:initial-gravitational-fields}
The past field and the dynamical part together give the full
metric and matter perturbation. In particular, the mass field of
Section~\ref{sec:mcvittie-monopole} is retained without a limiting
procedure at the past boundary. The support of the dynamical part is
causally bounded: for $\eta\ge\eta_i$, both the additional metric
field and $\tau_{00}^{\mathrm{dyn}}$ vanish outside
$|\bx|\le d+c(\eta-\eta_i)$. The past field may already extend beyond
this region. Point-particle and localized multipole sources are
included by interpreting the same equations distributionally.
\section{Source moments and their symmetric trace-free part}\label{sec:cartesian}
The retarded solution expresses the metric in terms of the full source
history. To relate this field to the source's spatial structure and
motion, we will organize the source information into multipole moments.
We will begin with instantaneous moments of the shifted density,
momentum and stresses, whose balance laws express the relations imposed
by stress-energy conservation. Their symmetric trace-free decomposition
will separate the angular orders while retaining the radial information
carried by traces.

In Section~\ref{sec:multipoles}, we will then account for the different
emission times across the source. The multipolar decomposition of the
Green function will determine how the source history is weighted to
form retarded source moments, from which we will reconstruct the
exterior field, including its curvature tail. The instantaneous moments
introduced here will also provide the quantities used in finite
source-size approximations and comparisons with known solutions.

We apply the moment construction to the dynamical parts in
Eq.~\eqref{eq:past-dynamical-split}, treating every component and
every multipole rank in the same way. To keep the formulas compact,
we suppress the superscript $\mathrm{dyn}$ on fields, sources and moments
throughout this section and Sections~\ref{sec:multipoles} and
\ref{sec:octupolar}. We retain it only where the past and dynamical
contributions appear together or are explicitly contrasted; past
quantities keep their superscript $\mathrm{past}$.
The past source moments will be introduced in
Eqs.~\eqref{eq:past-retarded-source-moments}, and the two contributions
combined in Eqs.~\eqref{eq:full-multipolar-fields}. The canonical
reduction will then use the complete fields and moments.

We begin with the multi-index conventions $L:=i_1\cdots i_\ell$,
$x_L:=x_{i_1}\cdots x_{i_\ell}$ and
$\partial_L:=\partial_{i_1}\cdots\partial_{i_\ell}$, with $|L|=\ell$.
For any multi-index, its length is the number of component indices.
Repeated multi-indices are fully contracted. For $\ell=0$, the
multi-index $L$ is empty: we set $x_L:=1$ and let $\partial_L$ be
the identity operator, so $\partial_L f=f$. Thus the monopole term
contains neither a factor of the source position nor a spatial derivative.
Parentheses denote normalized symmetrization. Angle brackets denote symmetrization and
removal of all traces on the enclosed indices; $\STF_L$ denotes this
projection on $L$ alone. We use the spatial antisymmetric symbol
$\epsilon_{ijk}$ with $\epsilon_{123}=1$.

We reserve $\bx$ for the observation point and $\bx'$ for the source
integration point. The observation radius is $\rho=|\bx|$;
we denote the source radius by $\rho':=|\bx'|$, also measured
from the source origin. The separation
between these points is $R=|\bx-\bx'|$, as defined in
Section~\ref{sec:green}. The multi-index conventions also apply to
primed coordinates, $x'_L:=x'_{i_1}\cdots x'_{i_\ell}$.
\subsection{Source moments and their balance laws}
Stress-energy conservation gives balance laws for the source moments
at every rank. We will derive them following Ref.~\cite{Compere:2026xoe},
retaining the cosmic-fluid contribution $a^2s\hat h$ from the energy
equation~\eqref{eq:energy}.

We start with the instantaneous source moments: the density moments
$Q^{(\rho)}_L$, momentum moments $P_{i|L}$, stress moments $S_{ij|L}$
and pressure moments $Q^{(p)}_L$, defined as in
\cite[Eqs.~(4.1)--(4.4)]{Compere:2026xoe}:
\begin{subequations}\label{eq:source-moment-definitions}
\begin{align}
 Q^{(\rho)}_L(\eta)&:=\frac1{c^2}\int\dd^3x'\,a(\eta)^{\ell+1}\tau_{00}(\eta,\bx')x'_L,
 \label{eq:density-moments}\\
 P_{i|L}(\eta)&:=\frac1c\int\dd^3x'\,a(\eta)^{\ell+1}\tau_{0i}(\eta,\bx')x'_L,
 \label{eq:momentum-moments}\\
 S_{ij|L}(\eta)&:=\int\dd^3x'\,a(\eta)^{\ell+1}\tau_{ij}(\eta,\bx')x'_L,
 \qquad Q^{(p)}_L(\eta):=\frac1{c^2}S_{kk|L}(\eta).
 \label{eq:CHmoments}
\end{align}
\end{subequations}
The vertical bar separates the source component indices from the
symmetric multipole indices. When $L$ is empty, we omit it and the
vertical bar, writing, for example, $S_{ij}$ for the stress monopole. We retain both the component trace
$S_{kk|L}$ and all traces over the multipole indices. The factor
$a^{\ell+1}$ combines the physical volume $a^3\dd^3x'$, positions
$ax'_i$ and orthonormal components $a^{-2}\tau_{\mu\nu}$ on a
constant-time slice. The shifted stress-energy tensor perturbation
$\tau_{\mu\nu}$ is evaluated in the gauge defined by
Eqs.~\eqref{eq:gauge-components}.

The dynamical stress and momentum components have support in $|\bx'|\le d$.
For the dynamical density component, the moments $Q_L^{(\rho)}$
also include the induced exterior density throughout the causally
bounded region $|\bx'|\le d+c(\eta-\eta_i)$ established in
Section~\ref{sec:retarded-reconstruction}. They integrate
$\tau_{00}^{\mathrm{dyn}}=\tau_{00}-\tau_{00}^{\mathrm{past}}$,
as defined in Eq.~\eqref{eq:past-dynamical-source}.
The resulting moments $Q_L^{(\rho)}$ are finite at each finite time
because the dynamical density has bounded spatial support. This
remains true when the initial shifted density has divergent
whole-space moments: for a McVittie past field, the pre-existing
$1/\rho'$ contribution in Eq.~\eqref{eq:mcvittie-source} is subtracted
before integration. 

We derive the balance laws by multiplying Eqs.~\eqref{eq:energy} and
\eqref{eq:momentum} by $a^\ell x'_L$ and integrating. For $\ell\ge1$,
integration by parts with $\partial'_j:=\partial/\partial x'^j$ gives
\begin{equation}
 \int\dd^3x'\,x'_L\partial'_j\tau_{ij}
 =-\ell\int\dd^3x'\,\tau_{i(i_1}x'_{i_2}\cdots x'_{i_\ell)},
\end{equation}
where the boundary term vanishes because the integration boundary
can be placed outside the region carrying the shifted stresses $\tau_{ij}$. To track the powers
of $a$, write
$\partial_\eta(a^{\ell+1}\tau_{\mu\nu})
=a^\ell\partial_\eta(a\tau_{\mu\nu})
+\ell\cH a^{\ell+1}\tau_{\mu\nu}$.
The momentum equation contributes one further term
$-\cH a^{\ell+1}\tau_{0i}$, which changes $\ell$ to $\ell-1$
in its balance law. Using $\partial_t=a^{-1}\partial_\eta$ then gives
\cite[Eqs.~(4.9)--(4.11)]{Compere:2026xoe}
\begin{subequations}\label{eq:source-moment-balances}
\begin{align}
 \partial_tQ^{(\rho)}_L
 &=H\big(\ell Q^{(\rho)}_L-Q^{(p)}_L\big)
   -\ell P_{(i_1|i_2\cdots i_\ell)}
   +\frac{s}{c^2}\int\dd^3x'\,a^{\ell+1}\hat h x'_L,
 \label{eq:momentenergy}\\
 \partial_tP_{i|L}
 &=(\ell-1)HP_{i|L}-\ell S_{i(i_1|i_2\cdots i_\ell)}.
 \label{eq:momentmomentum}
\end{align}
\end{subequations}
The ranks $\ell=0,1$ contain the moments associated with spacetime
symmetries. At these ranks, Eqs.~\eqref{eq:momentenergy} and
\eqref{eq:momentmomentum} give
\begin{subequations}\label{eq:low-moment-balances}
\begin{align}
\partial_tQ^{(\rho)}
 &=-H Q^{(p)}+\frac{s}{c^2}\int\dd^3x'\,a\hat h,\label{eq:energy-monopole-balance}\\
\partial_tQ^{(\rho)}_i
 &=H\big(Q^{(\rho)}_i-Q^{(p)}_i\big)-P_i
   +\frac{s}{c^2}\int\dd^3x'\,a^2\hat h x'_i,\label{eq:energy-dipole-balance}\\
\partial_tP_i&=-HP_i,\label{eq:momentum-monopole-balance}\\
\partial_tP_{i|j}&=-S_{ij}.\label{eq:momentum-dipole-balance}
\end{align}
\end{subequations}
Spatial translations and rotations are isometries of every spatially
flat FLRW background. For an arbitrary scale factor, the spatial
translation charges $\mathcal P_i$ and angular momenta $J_i$ are conserved:
\begin{equation}\label{eq:spatial-isometry-charges}
 \mathcal P_i:=-aP_i,\qquad J_i:=\epsilon_{ijk}P_{j|k},
 \qquad \partial_t\mathcal P_i=\partial_tJ_i=0.
\end{equation}
Their conservation follows from $\partial_ta=aH$ and the symmetry
$S_{ij}=S_{ji}$ in Eqs.~\eqref{eq:momentum-monopole-balance}--
\eqref{eq:momentum-dipole-balance}.

For Minkowski spacetime, setting $a=1$ and
$\Lambda=\eps=p=0$, so $H=s=0$ and
$\tau_{\mu\nu}=\delta T_{\mu\nu}$, time translations and Lorentz
boosts supply four further conserved charges: the energy $E$ and
center-of-mass charges $\mathcal K_i$,
\begin{equation}\label{eq:poincare-charges}
 E:=c^2Q^{(\rho)},\qquad \mathcal K_i:=Q^{(\rho)}_i-t\mathcal P_i,
 \qquad \partial_tE=\partial_t\mathcal K_i=0.
\end{equation}
Indeed, Eqs.~\eqref{eq:energy-monopole-balance}--
\eqref{eq:energy-dipole-balance} reduce to
$\partial_tQ^{(\rho)}=0$ and
$\partial_tQ^{(\rho)}_i=-P_i=\mathcal P_i$.
Together, $E$, $\mathcal P_i$, $J_i$ and $\mathcal K_i$ are the ten
Poincar\'e charges.\footnote{We use the signs of
\cite[Eqs.~(3.16)--(3.17)]{Compere:2019gft} at $c=1$.
Our $E$ and $J_i$ correspond to its $\mathcal E$ and $\mathcal J_i$.
The sign of $\mathcal K_i$ agrees with Eq.~(3.4) and the discussion
following Eq.~(3.2) of that reference: the density dipole
$Q^{(\rho)}_i$ corresponds to its comoving center-of-mass, while
$\mathcal K_i$ subtracts its uniform drift. Here the charges are
conserved at linear order; the gravitational-radiation fluxes in
that reference are quadratic in the perturbations. The energy, density
dipole and angular momentum agree with
\cite[Eqs.~(5.43b), (5.45b), (5.47c)]{Damour:1990gj}.}
In this Minkowski specialization, substituting the source moments
\eqref{eq:source-moment-definitions} into the definition of $\mathcal K_i$ in
Eq.~\eqref{eq:poincare-charges} gives
\begin{equation}\label{eq:center-of-mass-source-charge}
 \mathcal K_i=\int\dd^3x'\,
 \left[\frac{x'_i}{c^2}\delta T^{00}-\frac tc\delta T^{0i}\right],
\end{equation}
where indices are raised by $\eta^{\mu\nu}$. This source expression
and its interpretation as a conserved center-of-mass charge apply here
only to the Minkowski background.
%The density dipole $Q^{(\rho)}_i$ itself is not conserved:for $E\ne0$, the center of energy $Q^{(\rho)}_i/E$ moves with constant velocity $\mathcal P_i/E$.

For a generic FLRW expansion, time translations and Lorentz boosts
are not isometries. The $H$ and $s$ terms in
Eqs.~\eqref{eq:energy-monopole-balance}--\eqref{eq:energy-dipole-balance} prevent the corresponding
Minkowski expressions from being conserved; the six charges in
Eq.~\eqref{eq:spatial-isometry-charges} remain conserved.
%These statements require no multipolar truncation. 
The conserved charges of the dynamical part vanish because
it has zero past data. The total charges include those of the
prescribed past field, in particular its mass in the flat limit
and its angular momentum. Compactness and
$\partial'_i\tau_{0i}^{\mathrm{past}}=0$ give zero total linear
momentum for this past-stationary class: integrating
$\tau_{0i}^{\mathrm{past}}=\partial'_j(x'_i\tau_{0j}^{\mathrm{past}})$ yields
$\int\dd^3x'\,\tau_{0i}^{\mathrm{past}}=0$.
\subsection{STF projection and the traces of source moments}
\label{sec:source-moment-STF}
A Cartesian source moment generally contains several angular multipoles.
Its symmetric trace-free part in the multipole indices isolates the
highest angular order. The traces contain lower angular orders weighted
by powers of the source radius: each contraction of two source positions
gives $\delta^{ij}x'_ix'_j=(\rho')^2$. We will first define repeated
traces and the STF projector, then express each moment in terms of STF
tensors of different ranks while retaining all these contributions.

We start with a symmetric spatial tensor of rank $n$. Its \emph{$k$-fold trace}
is obtained by contracting $k$ pairs of indices with $\delta^{ij}$,
leaving $n-2k$ free indices. Thus $0\le k\le\lfloor n/2\rfloor$,
where $\lfloor n/2\rfloor$ denotes the greatest integer not exceeding
$n/2$. 
With $|L|=n-2k$, our notation is
\begin{equation}\label{eq:k-fold-trace}
 U_{L j_1j_1\cdots j_kj_k}
 :=\delta^{a_1b_1}\cdots\delta^{a_kb_k}
 U_{L a_1b_1\cdots a_kb_k}.
\end{equation}
Each $j_s$ is an independent dummy index summed over the three
spatial directions. For $k=0$, no indices are contracted and the
tensor is unchanged. Using these traces, the STF projection of a symmetric rank-$\ell$
tensor $U_L$ in three spatial dimensions is given by
\cite[Eq.~(2.2)]{Thorne:1980ru}, \cite[Eq.~(A2)]{Blanchet:1985sp},
\cite[Eq.~(2.2)]{Damour:1990gj}:
\begin{equation}\label{eq:STFprojector}
 U_{\langle i_1\cdots i_\ell\rangle}
 =\sum_{k=0}^{\lfloor\ell/2\rfloor}
 \frac{(-1)^k\ell!}{(\ell-2k)!}c_{\ell-k,k}
 \delta_{(i_1i_2}\cdots\delta_{i_{2k-1}i_{2k}}
 U_{i_{2k+1}\cdots i_\ell)j_1j_1\cdots j_kj_k},
\end{equation}
where we define the coefficients
\begin{equation}\label{eq:tracecoefficients}
 c_{\ell k}:=\frac{(2\ell-1)!!}{(2k)!!(2\ell+2k-1)!!},
 \qquad \ell,k\ge0.
\end{equation}
The indices of $c_{\ell k}$ are integers. With $0!!=(-1)!!=1$,
we have $c_{\ell 0}=1$, and the scalar case $\ell=0$ in
Eq.~\eqref{eq:STFprojector} corresponds to $c_{00}=1$.
Parentheses denote normalized symmetrization over the free indices,
and the repeated $j$ indices are contracted.
In particular,
\begin{equation}
 U_{\langle ij\rangle}=U_{ij}-\frac13\delta_{ij}U_{kk},
 \qquad
 U_{\langle ijk\rangle}=U_{ijk}
 -\frac35\delta_{(ij}U_{k)mm}.
\end{equation}
For source moments such as $S_{ij|L}$, the STF projection acts only
on the position indices $L$, leaving the component indices $i,j$
unchanged.

The projector removes traces from a tensor of fixed rank. To retain
the information carried by those traces, we also need the inverse
decomposition into STF tensors of lower ranks. The monomial identity
of \cite[Eq.~(A21a)]{Blanchet:1985sp}, expressed in
source positions and normalized symmetrization, gives
\begin{align}\label{eq:STFdecomp}
 x'_{i_1}\cdots x'_{i_n}
 &=\sum_{k=0}^{\lfloor n/2\rfloor}
 \frac{n!}{(n-2k)!}c_{n-2k+1,k}(\rho')^{2k}
 \delta_{(i_1i_2}\cdots\delta_{i_{2k-1}i_{2k}}
 x'_{\langle i_{2k+1}\cdots i_n\rangle)}.
\end{align}
The coefficient in Eq.~\eqref{eq:STFdecomp} follows from the
conversion of their unnormalized braces in
\cite[Eq.~(A19)]{Blanchet:1985sp}; multiplying their unit-vector
identity by $(\rho')^n$ supplies the radial factors.
The inner projection acts on the remaining $n-2k$ indices. At fixed
Cartesian degree $n$, this identity separates the irreducible angular
orders $n,n-2,\ldots$ and retains every trace. Integration against
$a^{n+1}\tau_{ij}$ gives the corresponding decomposition of
$S_{ij|N}$, where $N:=i_1\cdots i_n$. Replacing $\tau_{ij}$ by $\tau_{0i}/c$ gives that of
$P_{i|N}$.

Thus a fixed angular order $\ell$ receives radial information from
Cartesian moments of degrees $\ell,\ell+2,\ldots$; their traces
supply this information without introducing new source quantities.
For example,
$S_{ij|mn}=S_{ij|\langle mn\rangle}+\delta_{mn}S_{ij|kk}/3$:
the STF part has angular order two, whereas the trace has angular
order zero.

The retarded integrals~\eqref{eq:spatial-solution} and
\eqref{eq:mixed-solution} contain the sources of the wave equations,
$a\tau_{ij}$ and $a\tau_{0i}$. To match this normalization, we keep a single
factor $a$ (and $1/c$ for momentum) in their radial integrals and express them in terms of
the source moments:
\begin{subequations}\label{eq:radialmoments-traces}
\begin{align}
a\int\dd^3x'\,(\rho')^{2k}x'_{\langle L\rangle}\tau_{ij}
 &=a^{-\ell-2k}\STF_L S_{ij|L j_1j_1\cdots j_kj_k},\label{eq:spatial-radialmoments-traces}\\
\frac ac\int\dd^3x'\,(\rho')^{2k}x'_{\langle L\rangle}\tau_{0i}
 &=a^{-\ell-2k}\STF_L P_{i|L j_1j_1\cdots j_kj_k}.\label{eq:mixed-radialmoments-traces}
\end{align}
\end{subequations}
The time argument $\eta$ is suppressed throughout these identities;
the spatial integrals are taken at fixed conformal time.
To see why they hold, start from a Cartesian moment of rank
$\ell+2k$. Each contraction of two multipole indices replaces a
pair of source positions by
$\delta^{ab}x'_a x'_b=(\rho')^2$. The $k$ traces therefore supply
$(\rho')^{2k}$, and the subsequent projection $\STF_L$ replaces
the remaining product $x'_L$ by $x'_{\langle L\rangle}$.
The moment definition contains $a^{\ell+2k+1}$; multiplying by
$a^{-\ell-2k}$ leaves precisely the factor $a$ in the
source integral on the left, with the momentum normalization $1/c$ retained. The component indices $ij$ or $0i$ are
unaffected by either operation.

\section{Multipolar expansion of the retarded fields}\label{sec:multipoles}
The retarded integrals of Section~\ref{sec:reconstruct} sum the
contributions from every point of the source. Our aim is to express
the exterior field in terms of source moments that depend only on
conformal time, with propagation to the observer described by the Green
functions of Section~\ref{sec:green}. These moments must retain the
effects of the source's finite extent, including the different emission
times across it.

We will first derive the Green-function multipolar expansion by
constructing the source-time kernels. Integrating this expansion
against the dynamical shifted stresses and momentum density will
then identify the retarded source moments. We will separate the
light-cone and curvature-tail fields and add the prescribed past
field to obtain the complete exterior metric. Finally, we will
derive source-size approximations for sub-Hubble sources that evolve
slowly during their light-crossing time, while retaining exact
propagation to the observer. We will then construct a consistent
quadrupolar truncation on a general FLRW background{
with vanishing Cartesian density octupole, and compare its light-cone
and curvature-tail fields with the known quadrupolar expressions}.

As in Section~\ref{sec:cartesian}, unmarked fields, sources and moments
denote their dynamical parts. We retain $\mathrm{dyn}$ only where past
and dynamical contributions appear together or are explicitly
contrasted, and keep $\mathrm{past}$ on past quantities.

\subsection{Green-function multipolar expansion}
\label{sec:source-time-kernels}
We will first decompose the response to a single source event. The
construction will use a general smooth potential $V$, so that the same
result applies to the spatial and mixed metric fields upon choosing
$V^-$ and $V^+$, respectively.

We start with $G_V(\eta,\rho;\eta')$, the retarded response to a point
impulse at the spatial origin. Because $V$ depends only on time,
spatial derivatives commute with $\square+V$. Derivatives of this Green
function therefore generate multipolar fields centred at the origin.
For a source event $(\eta',\bx')$ with $\rho':=|\bx'|<\rho$, we seek
the Green-function multipolar expansion in STF form, with source-time kernels
$K^V_\ell$ to be determined:
\begin{equation}\label{eq:curved-green-STF-translation}
 G_V(\eta;|\bx-\bx'|;\eta')
 \overset{\rho>\rho'}{=}\sum_{\ell\ge0}\frac{(-1)^\ell}{\ell!}
 x'_{\langle L\rangle}\partial_L
 \int_{\Eta}c\,\dd\eta''\,G_V(\eta,\rho;\eta'')
 \frac{K^V_\ell(\rho',\eta';\eta'')}{\rho'^{\ell+1}}.
\end{equation}
Here $\partial_L$ acts on the observation position at fixed $\eta$,
and $x'_{\langle L\rangle}$ displays the angular dependence on the source
position. 
The source-time kernel $K^V_\ell(\rho',\eta';\eta'')$, divided by $\rho'^{\ell+1}$,
weights the response of the Green function $G_V(\eta,\rho;\eta'')$  over the auxiliary source time $\eta''$.
It accounts for the source radius $\rho'$ and emission time $\eta'$,
independently of the observation position $\bx$ and time $\eta$.
In $K^V_\ell(\rho',\eta';\eta'')$, the semicolon separates the auxiliary
evaluation variables $(\rho',\eta')$ from the time $\eta''$ labelling
data at zero radius. We will determine this kernel from the source
wave equation and the response to a source at the origin.

\paragraph{Radial equation and data at the origin.}
The Green function defined by Eq.~\eqref{eq:green-def} also satisfies
the source-variable equation, since its wave operator is formally
self-adjoint \cite[Eq.~(10)]{Chu:2015yua}:
\begin{equation}\label{eq:source-green-definition}
 \bigl[-c^{-2}\partial_{\eta'}^2+\Delta'+V(\eta')\bigr]
 G_V(\eta;|\bx-\bx'|;\eta')
 =-\frac{4\pi}{c}\delta(\eta-\eta')\delta^3(\bx-\bx').
\end{equation}
Its right-hand side vanishes for $\rho'<\rho$. The polynomial
$x'_{\langle L\rangle}$ is harmonic and homogeneous of degree $\ell$.
The product rule therefore gives, for any radial factor $F$,
\begin{equation}\label{eq:source-STF-radial-operator}
 \Delta'\bigl[x'_{\langle L\rangle}F(\rho')\bigr]
 =x'_{\langle L\rangle}
 \left(\partial_{\rho'}^2+\frac{2\ell+2}{\rho'}\partial_{\rho'}\right)F(\rho').
\end{equation}
The $2\ell$ term uses $x'^i\partial'_i x'_{\langle L\rangle}
=\ell x'_{\langle L\rangle}$, and the remaining $2$ comes from the
three-dimensional radial Laplacian. Substituting
$F=K^V_\ell/\rho'^{\ell+1}$, as in
Eq.~\eqref{eq:curved-green-STF-translation}, cancels the terms
proportional to $\partial_{\rho'}K^V_\ell$ and gives
\begin{equation}\label{eq:source-reduced-radial-operator}
 \left(\partial_{\rho'}^2+\frac{2\ell+2}{\rho'}\partial_{\rho'}\right)
 \frac{K^V_\ell}{\rho'^{\ell+1}}
 =\frac1{\rho'^{\ell+1}}
 \left(\partial_{\rho'}^2-\frac{\ell(\ell+1)}{\rho'^2}\right)K^V_\ell.
\end{equation}

Using Taylor expansion around $\bx'=0$ and the STF decomposition
in Eq.~\eqref{eq:STFdecomp}, we have
\begin{equation}\label{eq:source-STF-central-translation}
 \left.G_V(\eta;|\bx-\bx'|;\eta')\right|_{\text{angular rank }\ell}
 =\frac{(-1)^\ell}{\ell!}x'_{\langle L\rangle}
   \partial_LG_V(\eta,\rho;\eta')+O(\rho'^{\ell+2}).
\end{equation}
The factor $(-1)^\ell/\ell!$ follows from the translation
$\bx-\bx'$: each source coordinate contributes $-\partial_i$, while
$1/\ell!$ is the Taylor coefficient.
The STF decomposition of a Cartesian term of degree $n$ contains only
angular ranks $n,n-2,\ldots$, since each trace removes two indices.
Thus rank $\ell$ receives contributions from degrees $\ell,\ell+2,\ldots$.
The absence of degree $\ell+1$ can also be seen from its angular
projection onto rank $\ell$: multiply that Taylor term by
$x'_{\langle L\rangle}$ and integrate over a sphere at fixed $\rho'$.
Under $\bx'\mapsto-\bx'$, the two factors acquire signs
$(-1)^{\ell+1}$ and $(-1)^\ell$. Their product changes sign, so
opposite source directions cancel in the angular integral.

Factoring out the degree-$\ell$ polynomial $x'_{\langle L\rangle}$
leaves a radial coefficient with allowed powers
$1,\rho'^2,\rho'^4,\ldots$. Its first possible correction to the
value at $\rho'=0$ is therefore of order $\rho'^2$; there is no term
linear in $\rho'$. To reproduce
Eq.~\eqref{eq:source-STF-central-translation}, we will require the
integral in Eq.~\eqref{eq:curved-green-STF-translation} to have value
$G_V(\eta,\rho;\eta')$ and vanishing first radial derivative at
$\rho'=0$, with $\eta,\eta'$ and the observation point fixed.
Together with the source wave equation and the radial identity
\eqref{eq:source-reduced-radial-operator}, these requirements lead us to
construct $K^V_\ell$ as the regular solution of
\begin{subequations}\label{eq:curved-average-radial-problem}
\begin{align}
 \left[-\partial_{\rho'}^2+c^{-2}\partial_{\eta'}^2-V(\eta')
       +\frac{\ell(\ell+1)}{\rho'^2}\right]
 K^V_\ell(\rho',\eta';\eta'')&=0,\qquad\rho'>0,
 \label{eq:curved-average-radial-equation}\\
 \lim_{\rho'\to0}\frac{K^V_\ell(\rho',\eta';\eta'')}{\rho'^{\ell+1}}
 &=\delta(c(\eta'-\eta'')),
 \label{eq:curved-average-radial-data}\\
 \lim_{\rho'\to0}\partial_{\rho'}
 \frac{K^V_\ell(\rho',\eta';\eta'')}{\rho'^{\ell+1}}&=0.
 \label{eq:curved-average-radial-derivative}
\end{align}
\end{subequations}
Equation~\eqref{eq:curved-average-radial-equation}
has the form of an auxiliary wave equation with $\rho'/c$ taking the
role of time and $c\eta'$ the role of a spatial coordinate;
$c\eta''$ labels the initial impulse location.
The limits hold after integration over $\eta''$ against a smooth
temporal profile $f(\eta'')$: the integral of
$K^V_\ell/\rho'^{\ell+1}$ tends to $f(\eta')$, and its first radial
derivative tends to zero. These data and the radial equation uniquely
determine the regular kernel; the resulting integral is smooth and
even in $\rho'$.

\paragraph{Reciprocity.}

The kernels obey reciprocity,
\begin{equation}\label{eq:curved-average-reciprocity}
 K^V_\ell(\rho',\eta';\eta'')=K^V_\ell(\rho',\eta'';\eta').
\end{equation}
Reciprocity means that exchanging the impulse location $\eta''$
and the response point $\eta'$ gives the same response at fixed
$\rho'$.
To establish this property, continue the real potential smoothly as
a bounded function outside the interval sampled by a compact
characteristic diamond. The self-adjoint spatial operator then has
a complete set of real modes $\psi_n$, with eigenvalues $\lambda_n$,
obeying
\begin{subequations}\label{eq:source-kernel-spatial-modes}
\begin{align}
 [c^{-2}\partial_{\eta'}^2-V(\eta')]\psi_n(\eta')
 &=\lambda_n\psi_n(\eta'),
 \label{eq:source-kernel-mode-eigenvalue}\\
 \sum_n\psi_n(\eta')\psi_n(\eta'')
 &=\delta(c(\eta'-\eta'')).
 \label{eq:source-kernel-mode-completeness}
\end{align}
\end{subequations}
The mode sum includes integration over any continuous spectrum.
For each mode, define the regular radial amplitude $k_{\ell n}$ by
\begin{subequations}\label{eq:source-kernel-radial-modes}
\begin{align}
 \left[-\partial_{\rho'}^2+\lambda_n
       +\frac{\ell(\ell+1)}{\rho'^2}\right]k_{\ell n}(\rho')&=0,
 \label{eq:source-kernel-mode-radial-equation}\\
 \lim_{\rho'\to0}\frac{k_{\ell n}(\rho')}{\rho'^{\ell+1}}&=1,
 \qquad
 \lim_{\rho'\to0}\partial_{\rho'}
 \frac{k_{\ell n}(\rho')}{\rho'^{\ell+1}}=0.
 \label{eq:source-kernel-mode-radial-data}
\end{align}
\end{subequations}
The centrifugal term enters this radial equation but is independent
of $\eta'$, so it does not couple different spatial modes.
Combining the spatial modes with these radial amplitudes gives
\begin{equation}\label{eq:source-kernel-mode-expansion}
 K^V_\ell(\rho',\eta';\eta'')
 =\sum_n k_{\ell n}(\rho')\psi_n(\eta')\psi_n(\eta'').
\end{equation}
The eigenvalue equation ensures that each term satisfies
Eq.~\eqref{eq:curved-average-radial-equation}, while completeness
and the radial initial data reproduce the kernel's data at $\rho'=0$.
Uniqueness therefore identifies this expansion with the kernel.
The factor $\psi_n(\eta'')$ specifies how strongly the impulse excites
the mode, $k_{\ell n}(\rho')$ evolves its amplitude, and
$\psi_n(\eta')$ gives its value at the response point.
Exchanging $\eta'$ and $\eta''$ leaves every term unchanged,
proving Eq.~\eqref{eq:curved-average-reciprocity}.
Finite propagation makes the result independent of the continuation
of $V$ whenever the characteristic diamond lies within $\Eta$.
Reciprocity also implies that the kernel satisfies
Eq.~\eqref{eq:curved-average-radial-equation} in $\eta''$.
\par

\paragraph{Characteristic construction of the monopole kernel.}
At $\ell=0$, the centrifugal term in
Eq.~\eqref{eq:curved-average-radial-equation} vanishes.
The data at zero source radius
follow from Eq.~\eqref{eq:curved-average-radial-data} alone:
\begin{equation}\label{eq:source-kernel-monopole-data}
 K^V_0(0,\eta';\eta'')=0,\qquad
 \partial_{\rho'}K^V_0(0,\eta';\eta'')=\delta(c(\eta'-\eta'')).
\end{equation}
These are the initial data for Eq.~\eqref{eq:curved-average-radial-equation}
at $\ell=0$, namely
\begin{equation}\label{eq:source-kernel-monopole-equation}
 \left[-\partial_{\rho'}^2+c^{-2}\partial_{\eta'}^2-V(\eta')\right]
 K^V_0(\rho',\eta';\eta'')=0,\qquad\rho'>0.
\end{equation}
We evolve this equation in the auxiliary time $\rho'/c$ from $\rho'=0$.
Compared with the auxiliary Green-function equation
\eqref{eq:one-dimensional-green-source}, $\rho'/c$ and $c\eta'$ take
the roles played there by $\eta$ and $R$, respectively.
Here the characteristics $c(\eta'-\eta'')=\pm\rho'$ bound the support
$c|\eta'-\eta''|\le\rho'$. We will apply the same characteristic
construction, with $V(\eta')$ now depending on the auxiliary spatial
coordinate.

We introduce $u_\pm:=(\rho'\pm c(\eta'-\eta''))/2$ and define the smooth
characteristic function $\mathscr J_V$ multiplying the step function,
analogously to $J_V$ in Eq.~\eqref{eq:one-dimensional-green}:
\begin{equation}\label{eq:curved-average-monopole}
 K^V_0(\rho',\eta';\eta'')
 =\frac12\theta(\rho'-c|\eta'-\eta''|)
 \mathscr J_V(u_+,u_-;\eta'').
\end{equation}
Substituting this form into the homogeneous wave equation
\eqref{eq:source-kernel-monopole-equation}, derivatives of the step
function produce delta distributions on the two characteristics.
For $\rho'>0$, their coefficients are proportional to
$\partial_{u_-}\mathscr J_V(0,u_-;\eta'')$ on $u_+=0$ and
$\partial_{u_+}\mathscr J_V(u_+,0;\eta'')$ on $u_-=0$.
The wave equation has zero right-hand side there, so both tangential
derivatives vanish: $\mathscr J_V$ is constant along each
characteristic. Smoothness makes these constants equal at the vertex.

To fix their common value, integrate $K^V_0/\rho'$ over $\eta''$
against a smooth profile $f(\eta'')$. The step function restricts the
integral to $[\eta'-\rho'/c,\eta'+\rho'/c]$, and $\mathscr J_V$ tends
to its vertex value as this interval shrinks. The limiting result is
$\mathscr J_V(0,0;\eta')f(\eta')$. The initial derivative in
Eq.~\eqref{eq:source-kernel-monopole-data} requires this to equal
$f(\eta')$, fixing the vertex value to one. With this vertex value,
the shrinking-interval average gives
\begin{equation}\label{eq:source-kernel-monopole-delta-limit}
 \lim_{\rho'\to0}\int_\Eta c\,\dd\eta''\,
 \frac{K^V_0(\rho',\eta';\eta'')}{\rho'}f(\eta'')
 =\lim_{\rho'\to0}\frac1{2\rho'}\int_{\eta'-\rho'/c}^{\eta'+\rho'/c}c\,\dd\eta''\,f(\eta'')
 =f(\eta').
\end{equation}
This is the distributional limit
$K^V_0/\rho'\to\delta(c(\eta'-\eta''))$ required by
Eq.~\eqref{eq:curved-average-radial-data} at $\ell=0$.

Inside the support, the step function is constant. Substituting
Eq.~\eqref{eq:curved-average-monopole} into the radial equation and
using $\partial_{\rho'}^2-c^{-2}\partial_{\eta'}^2
=\partial_{u_+}\partial_{u_-}$ gives the characteristic problem
\begin{subequations}\label{eq:source-kernel-characteristic-problem}
\begin{align}
 \partial_{u_+}\partial_{u_-}\mathscr J_V(u_+,u_-;\eta'')
 &=-V(\eta''+(u_+-u_-)/c)\mathscr J_V(u_+,u_-;\eta''),
 \label{eq:source-kernel-characteristic-equation}\\
 \mathscr J_V(u_+,0;\eta'')&=\mathscr J_V(0,u_-;\eta'')=1.
 \label{eq:source-kernel-characteristic-data}
\end{align}
\end{subequations}
Integrating over $[0,u_+]\times[0,u_-]$ combines the four corner
values into $\mathscr J_V-1$, exactly as in the derivation of
Eq.~\eqref{eq:volterra}. We therefore obtain the Volterra integral equation
\begin{equation}\label{eq:curved-average-volterra}
 \mathscr J_V(u_+,u_-;\eta'')
 =1-\int_0^{u_+}\dd\xi\int_0^{u_-}\dd\zeta\,
 V(\eta''+(\xi-\zeta)/c)\mathscr J_V(\xi,\zeta;\eta'').
\end{equation}
The same successive substitution, starting from $\mathscr J_V=1$,
constructs the solution. Radial evolution gives the minus sign.
The integral samples $V(\eta''+(\xi-\zeta)/c)$ throughout the
characteristic rectangle, over conformal times in
$[\eta''-u_-/c,\eta''+u_+/c]$. The time
$\eta'=\eta''+(u_+-u_-)/c$ corresponds  to its corner
$(\xi,\zeta)=(u_+,u_-)$; determining $\mathscr J_V$ requires the
potential over the sampled interval.
The minus sign leaves the absolute-value estimates of
Appendix~\ref{app:volterra-estimates} unchanged. With the potential
bounded over this sampled interval, the same argument establishes
convergence and uniqueness on each compact characteristic rectangle.

\paragraph{Higher multipoles.}
The same characteristic function determines higher multipolar orders. Starting from
Eq.~\eqref{eq:curved-average-monopole}, define successively
\begin{equation}\label{eq:curved-average-recursion}
 K^V_{\ell+1}(\rho',\eta';\eta'')
 =\frac{2\ell+3}{\rho'^{\ell+1}}
   \int_0^{\rho'}\dd r\,r^{\ell+1}K^V_\ell(r,\eta';\eta'').
\end{equation}
Differentiation gives
\begin{equation}\label{eq:source-kernel-rank-first-order}
 \left(\partial_{\rho'}+\frac{\ell+1}{\rho'}\right)K^V_{\ell+1}
 =(2\ell+3)K^V_\ell.
\end{equation}

Applying $c^{-2}\partial_{\eta'}^2-V(\eta')$ to the recurrence
\eqref{eq:curved-average-recursion}, we use the rank-$\ell$ radial
equation and integrate by parts. The regular lower boundary term
vanishes, leaving
\begin{equation}\label{eq:source-kernel-rank-time-identity}
 [c^{-2}\partial_{\eta'}^2-V(\eta')]K^V_{\ell+1}
 =(2\ell+3)\left(\partial_{\rho'}-\frac{\ell+1}{\rho'}\right)K^V_\ell.
\end{equation}
Applying $\partial_{\rho'}-(\ell+1)/\rho'$ to
Eq.~\eqref{eq:source-kernel-rank-first-order} gives the same
right-hand side; the product of radial operators on the left is
$\partial_{\rho'}^2-(\ell+1)(\ell+2)/\rho'^2$.
This proves the radial equation at rank $\ell+1$.
\par

The monopole initial condition~\eqref{eq:curved-average-radial-data} follows from
Eq.~\eqref{eq:source-kernel-monopole-delta-limit}. We now show that
the recurrence preserves it at higher ranks.
Assuming this initial condition
at rank $\ell$, the integral of $K^V_\ell$ over $\eta''$ against
a smooth profile $f(\eta'')$ has leading term $r^{\ell+1}f(\eta')$. The radial integral in
Eq.~\eqref{eq:curved-average-recursion} then has leading term
$\rho'^{2\ell+3}f(\eta')/(2\ell+3)$. Its prefactor gives
$\rho'^{\ell+2}f(\eta')$, establishing the initial condition at rank
$\ell+1$.

To obtain an explicit formula, insert the monopole solution in the
recurrence and interchange the finite radial integrals. At each
subsequent step the remaining elementary integral is
\begin{equation}\label{eq:curved-average-rank-integral}
 \int_s^{\rho'}\dd r\,r(r^2-s^2)^{\ell-1}
 =\frac{(\rho'^2-s^2)^\ell}{2\ell},\qquad\ell\ge1.
\end{equation}
The recurrence factors multiply to $(2\ell+1)!!$, while these
integrations supply $2^{\ell-1}(\ell-1)!$. Including the monopole
factor $1/2$ and setting $z:=r/\rho'$ gives, for $\ell\ge1$ and
$\rho'>0$,

\begin{equation}\label{eq:curved-average-explicit}
\begin{aligned}
 K^V_\ell(\rho',\eta';\eta'')
 &=(2\ell+1)\rho'^\ell
   \theta\!\left(\rho'-c|\eta'-\eta''|\right)\\
 &\quad\times\int_{c|\eta'-\eta''|/\rho'}^1\dd z\,z\delta_{\ell-1}(z)
   \mathscr J_V\!\left(\frac{\rho'z+c(\eta'-\eta'')}{2},
                      \frac{\rho'z-c(\eta'-\eta'')}{2};\eta''\right),
\end{aligned}
\end{equation}

where we define the normalized polynomial weight
\cite[Appendix~B]{Blanchet:1989ki},
\cite[Eqs.~(3.4)--(3.6)]{Damour:1990gj}:
\begin{subequations}\label{eq:time-weight}
\begin{align}
 \delta_\ell(z)&:=\frac{(2\ell+1)!!}{2^{\ell+1}\ell!}(1-z^2)^\ell,
 \quad -1\le z\le1,\quad\ell\ge0,\label{eq:time-weight-definition}\\
 \int_{-1}^1\dd z\,\delta_\ell(z)&=1.\label{eq:time-weight-normalization}
\end{align}
\end{subequations}
The step function is inherited from the monopole
kernel~\eqref{eq:curved-average-monopole}: the radial recurrence
\eqref{eq:curved-average-recursion} integrates only over $r\le\rho'$
and preserves this support at every rank. The integral in
Eq.~\eqref{eq:curved-average-explicit} is evaluated only for
$c|\eta'-\eta''|\le\rho'$.
Integrating
$K^V_\ell/\rho'^{\ell+1}$ against a temporal profile $f(\eta'')$
then uses its values throughout $[\eta'-\rho'/c,\eta'+\rho'/c]$.

\paragraph{Proof of the multipolar expansion.}

To establish Eq.~\eqref{eq:curved-green-STF-translation}, we first
compare its two sides at a fixed angular rank. At an arbitrary
observation position $\bx$, integrate over the observation time $\eta$
against an arbitrary smooth weight $w(\eta)$ of compact support in
$\Eta$. For $\rho'<\rho$, this integration smooths the light-cone
distributions. It preserves the source wave equation because the
derivatives in that equation act on $\eta'$ and $\bx'$.
Define the exact radial coefficient $F_L$ by
\begin{equation}\label{eq:translation-exact-tested-coefficient}
 \left.\int_\Eta c\,\dd\eta\,w(\eta)
 G_V(\eta;|\bx-\bx'|;\eta')\right|_{\text{angular rank }\ell}
 =\frac{(-1)^\ell}{\ell!}x'_{\langle L\rangle}F_L(\rho',\eta').
\end{equation}
The central profile $f_L$ and the radial
coefficient $\widetilde F_L$ constructed from the kernel are
\begin{subequations}\label{eq:translation-constructed-coefficient}
\begin{align}
 f_L(\eta')&:=\int_\Eta c\,\dd\eta\,w(\eta)
             \partial_{\langle L\rangle}G_V(\eta,\rho;\eta'),
 \label{eq:translation-tested-central-profile}\\
 \widetilde F_L(\rho',\eta')&:=\int_\Eta c\,\dd\eta''\,
 \frac{K^V_\ell(\rho',\eta';\eta'')}{\rho'^{\ell+1}}f_L(\eta'').
 \label{eq:translation-tested-kernel-coefficient}
\end{align}
\end{subequations}
The STF projection of $\partial_L$ leaves its contraction with
$x'_{\langle L\rangle}$ unchanged. Thus $\widetilde F_L$ is precisely
the coefficient obtained from the right-hand side of
Eq.~\eqref{eq:curved-green-STF-translation} after the same integration.

To compare the two coefficients, define the difference field
\begin{equation}\label{eq:translation-reduced-difference}
 Z_{\ell L}(\rho',\eta'):=\rho'^{\ell+1}
 \bigl[F_L(\rho',\eta')-\widetilde F_L(\rho',\eta')\bigr].
\end{equation}
The source equation~\eqref{eq:source-green-definition} and the radial
identity~\eqref{eq:source-reduced-radial-operator} show that
$\rho'^{\ell+1}F_L$ obeys the same equation as $K^V_\ell$.
The term $\rho'^{\ell+1}\widetilde F_L$ is an integral of that kernel
against $f_L$, so it obeys the same equation directly.
At the origin, Eq.~\eqref{eq:source-STF-central-translation} gives
$F_L(0,\eta')=f_L(\eta')$ and $\partial_{\rho'}F_L(0,\eta')=0$;
the kernel data~\eqref{eq:curved-average-radial-data} and
\eqref{eq:curved-average-radial-derivative} give these same values
for $\widetilde F_L$. Thus the difference satisfies
\begin{subequations}\label{eq:translation-zero-data-problem}
\begin{align}
 \left[-\partial_{\rho'}^2+c^{-2}\partial_{\eta'}^2-V(\eta')
       +\frac{\ell(\ell+1)}{\rho'^2}\right]
 Z_{\ell L}(\rho',\eta')&=0,
 \label{eq:translation-difference-equation}\\
 \lim_{\rho'\to0}\frac{Z_{\ell L}(\rho',\eta')}{\rho'^{\ell+1}}&=0,
 \label{eq:translation-difference-value}\\
 \lim_{\rho'\to0}\partial_{\rho'}
 \frac{Z_{\ell L}(\rho',\eta')}{\rho'^{\ell+1}}&=0.
 \label{eq:translation-difference-derivative}
\end{align}
\end{subequations}
Both coefficients are smooth and even in $\rho'$, so the difference
belongs to the regular class of
Eqs.~\eqref{eq:curved-average-radial-problem}.
The regular mode construction~\eqref{eq:source-kernel-radial-modes}
already determines the radial evolution from these data: each
spatial mode is proportional to $k_{\ell n}$, with its amplitude
fixed by the leading coefficient at $\rho'=0$.
The zero data in Eqs.~\eqref{eq:translation-zero-data-problem}
therefore set every mode amplitude to zero, giving $Z_{\ell L}=0$
and hence $F_L=\widetilde F_L$.
This establishes equality at every angular rank. Completeness of
the STF expansion and the arbitrariness of $w$ then recover
Eq.~\eqref{eq:curved-green-STF-translation} as an equality of
distributions for $\rho'<\rho$.
\par

\paragraph{Limiting cases.}
For $V=0$, Eq.~\eqref{eq:curved-average-volterra} gives $\mathscr J_0=1$.
For $\ell\ge1$, we evaluate the integral in
Eq.~\eqref{eq:curved-average-explicit} using
$\partial_z\delta_\ell(z)=-(2\ell+1)z\delta_{\ell-1}(z)$ and
$\delta_\ell(1)=0$. Using the evenness of
$\delta_\ell$ and including the monopole
formula~\eqref{eq:curved-average-monopole} gives, at every rank,
\begin{equation}\label{eq:flat-average-kernel}
 K^0_\ell(\rho',\eta';\eta'')
 =\rho'^\ell\theta(\rho'-c|\eta'-\eta''|)
       \delta_\ell\!\left(\frac{c(\eta'-\eta'')}{\rho'}\right).
\end{equation}

For the inverse-square potential $V(\eta')=C/(c^2\eta'^2)$ with
$C=\nu(\nu+1)$, the regular kernel is
\begin{equation}\label{eq:power-retarded-kernel}
 K^V_\ell(\rho',\eta';\eta'')
 =K^0_\ell(\rho',\eta';\eta'')\,
 {}_2F_1\!\left(-\nu,\nu+1;\ell+1;
       \frac{\rho'^2-c^2(\eta''-\eta')^2}{4c^2\eta''\eta'}\right).
\end{equation}
The hypergeometric equation verifies the radial equation. The factor
multiplying $K^0_\ell$ tends to one as $\rho'\to0$ on the kernel
support, preserving the initial condition~\eqref{eq:curved-average-radial-data}.
The sampled diamond must remain on one
side of $\eta=0$. Setting $\nu=1$ gives the de Sitter kernel
\eqref{eq:dS-source-time-kernel}.
\par

\subsection{Retarded source moments and the exterior metric}
\label{sec:lightconemoments}
\label{sec:time-averaged-moments}
The Green-function multipolar expansion~\eqref{eq:curved-green-STF-translation}
expresses the response to a source at $\bx'$ through spatial derivatives
of $G_V(\eta,\rho;\eta'')$, whose source is at the origin. The source
position enters through $x'_{\langle L\rangle}$ and the source-time
kernel $K^V_\ell(\rho',\eta';\eta'')$. We will integrate these factors
against the source to define the retarded source moments, then use
the central Green function to reconstruct the dynamical exterior
metric. We will identify its light-cone and curvature-tail parts
and restore the prescribed past field.

We start from the spatial source integral~\eqref{eq:spatial-solution}.
The dynamical source vanishes before $\eta_i$, and the Green function
vanishes for $\eta'>\eta$, so the time integral can be taken over
$\Eta$. For $\rho:=|\bx|>d$, insert the Green-function multipolar expansion
\eqref{eq:curved-green-STF-translation} with $V=V^-$. Taking the
observation derivatives $\partial_L$ outside the source integrals
and exchanging the integrations over $\eta'$ and $\eta''$ gives
\begin{equation}\label{eq:retarded-field-moment-identification}
\begin{aligned}
 \bar h_{ij}(\eta,\bx)
 &\overset{\rho>d}{=}\frac{4G}{c^4}\sum_{\ell\ge0}\frac{(-1)^\ell}{\ell!}\partial_L
   \int_{\Eta}c\,\dd\eta''\,G^-(\eta,\rho;\eta'')\\
 &\qquad\times\left[
   \int\dd^3x'\,x'_{\langle L\rangle}\int_{\Eta}c\,\dd\eta'\,
   \frac{K^{V^-}_\ell(\rho',\eta';\eta'')}{\rho'^{\ell+1}}
   a(\eta')\tau_{ij}(\eta',\bx')\right].
\end{aligned}
\end{equation}
The bracket depends only on the central conformal time $\eta''$.
Repeating the construction for the mixed field with $V^+$ identifies
the second family of retarded source moments. The spatial and mixed
moments use $V^-$ and $V^+$, respectively. For the prescribed scale
factor $a$, we define the retarded source moments
\begin{subequations}\label{eq:curved-retarded-moments}
\begin{align}
 \mathcal F_{ij|L}(\eta'';a)
 &:=\int\dd^3x'\,x'_{\langle L\rangle}\int_{\Eta}c\,\dd\eta'\,
 \frac{K^{V^-}_\ell(\rho',\eta';\eta'')}{\rho'^{\ell+1}}a(\eta')\tau_{ij}(\eta',\bx'),
 \label{eq:curved-retarded-spatial-moment}\\
 \mathcal F_{0i|L}(\eta'';a)
 &:=\frac1c\int\dd^3x'\,x'_{\langle L\rangle}\int_{\Eta}c\,\dd\eta'\,
 \frac{K^{V^+}_\ell(\rho',\eta';\eta'')}{\rho'^{\ell+1}}a(\eta')\tau_{0i}(\eta',\bx').
 \label{eq:curved-retarded-mixed-moment}
\end{align}
\end{subequations}
The semicolon separates the central conformal time from the prescribed
scale-factor history. The dependence on $a$ enters through the
potentials in the kernels and the sources of the wave equations,
$a\tau_{ij}$ and $a\tau_{0i}$; the source histories are implicit in the notation.
The moments are STF in $L$, while the source component indices
remain unprojected. The weights $K^{V^\pm}_\ell/\rho'^{\ell+1}$ have the
regular distributional limit~\eqref{eq:curved-average-radial-data}
at the source origin. Replacing the bracket in
Eq.~\eqref{eq:retarded-field-moment-identification} by the spatial
moment, and using the mixed moment in the same way, gives
\begin{subequations}\label{eq:curved-retarded-fields}
\begin{align}
 \bar h_{ij}(\eta,\bx)
 &\overset{\rho>d}{=}\frac{4G}{c^4}\sum_{\ell\ge0}\frac{(-1)^\ell}{\ell!}\partial_L
   \int_{\Eta}c\,\dd\eta''\,G^-(\eta,\rho;\eta'')
       \mathcal F_{ij|L}(\eta'';a),
 \label{eq:curved-retarded-spatial-field}\\
 \bar h_{0i}(\eta,\bx)
 &\overset{\rho>d}{=}\frac{4G}{c^3}\sum_{\ell\ge0}\frac{(-1)^\ell}{\ell!}\partial_L
   \int_{\Eta}c\,\dd\eta''\,G^+(\eta,\rho;\eta'')
       \mathcal F_{0i|L}(\eta'';a).
 \label{eq:curved-retarded-mixed-field}
\end{align}
\end{subequations}
The time integrations are rearranged at each angular rank, with
the kernels acting on smooth, spatially compact source histories;
their support bounds the contributing times. For these sources,
the reconstructed field is smooth. Its spherical-harmonic
coefficients decay faster than any inverse power of the angular
rank, including after any fixed number of time and radial
derivatives, uniformly on compact exterior regions $\rho>d$
\cite[Chs.~2 and 4]{AtkinsonHan},
\cite[Appendix~B]{Blanchet:1985sp}. At each rank,
Eq.~\eqref{eq:curved-green-STF-translation} gives the angular
projection of this field. This decay therefore ensures convergence
of Eqs.~\eqref{eq:curved-retarded-fields}, also after any fixed
number of spacetime derivatives.
The same potential enters each moment and its Green function:
$V^-$ for the spatial field and $V^+$ for the mixed field. The
remaining dynamical metric components and exterior density follow
from Section~\ref{sec:retarded-reconstruction}. These formulas are
exact for $\rho>d$, without a slow-motion approximation.

\paragraph{Source history.}
At fixed $\eta''$ and source radius $\rho'$, the kernel has support
$c|\eta'-\eta''|\le\rho'$, and the dynamical source vanishes for
$\eta'<\eta_i$. The contributing source times therefore satisfy
$\max(\eta_i,\eta''-\rho'/c)\le\eta'\le\eta''+\rho'/c$.
Across the source $\rho'\le d$, these times lie within
$[\eta''-d/c,\eta''+d/c]$. The Green functions in
Eqs.~\eqref{eq:curved-retarded-fields} restrict $\eta''\le\eta-\rho/c$,
so every sampled emission time satisfies
$\eta'\le\eta''+\rho'/c\le\eta-\rho/c+d/c<\eta$. Thus the reconstructed
field is retarded even though a moment samples times on both sides
of its central time. Every nonempty source-time interval lies in
$\Eta$, since its lower endpoint is at least $\eta_i$ and its
upper endpoint is below $\eta$.

The dynamical moments can begin at $\eta_i-d/c$. Their representation
therefore requires $\eta_i-d/c\in\Eta$: the background must be defined
for a conformal-time interval $d/c$ before the prescribed source
history begins. The propagation integrals in
Eqs.~\eqref{eq:curved-retarded-fields} include this earlier interval.
The original source integrals~\eqref{eq:spatial-solution}--\eqref{eq:mixed-solution}
require no such temporal margin.

For the central times contributing to
Eqs.~\eqref{eq:curved-retarded-fields}, the spatial and mixed source
moments~\eqref{eq:curved-retarded-moments} therefore take the
finite-interval forms
\begin{subequations}\label{eq:retarded-moment-finite-time-interval}
\begin{align}
 \mathcal F_{ij|L}(\eta'';a)
 &=\int\dd^3x'\,\frac{x'_{\langle L\rangle}}{\rho'^{\ell+1}}
   \int_{\max(\eta_i,\eta''-\rho'/c)}^{\eta''+\rho'/c}c\,\dd\eta'\,
   K^{V^-}_\ell(\rho',\eta';\eta'')a(\eta')\tau_{ij}(\eta',\bx'),
 \label{eq:retarded-spatial-moment-finite-time-interval}\\
 \mathcal F_{0i|L}(\eta'';a)
 &=\frac1c\int\dd^3x'\,\frac{x'_{\langle L\rangle}}{\rho'^{\ell+1}}
   \int_{\max(\eta_i,\eta''-\rho'/c)}^{\eta''+\rho'/c}c\,\dd\eta'\,
   K^{V^+}_\ell(\rho',\eta';\eta'')a(\eta')\tau_{0i}(\eta',\bx').
 \label{eq:retarded-mixed-moment-finite-time-interval}
\end{align}
\end{subequations}
Each time integral is zero when its upper limit lies below its lower
limit.

\paragraph{Light-cone source moments and Minkowski limit.}
The Hadamard form~\eqref{eq:hadamard} fixes the light-cone term
$\delta(c(\eta-\eta')-R)/R$ independently of the potential and hence
of the scale factor. For $V=0$, the Volterra equation~\eqref{eq:volterra}
gives $J_V=1$, so the tail vanishes by Eq.~\eqref{eq:tail-from-J}.
Thus, on every background,
\begin{equation}\label{eq:light-cone-zero-potential-kernel}
 G^{\mathrm{LC}}(\eta;R;\eta')
 =G_0(\eta;R;\eta')=\frac{\delta(c(\eta-\eta')-R)}R.
\end{equation}
To obtain the multipolar form of the light-cone field~\eqref{eq:LC-field},
we now expand $G_0$ using Eq.~\eqref{eq:curved-green-STF-translation}
with $K^0_\ell$ from Eq.~\eqref{eq:flat-average-kernel}.
Integrating against $a\tau_{ij}$ and setting
$\eta':=\eta''+\rho'z/c$ identifies the light-cone source moments
\begin{equation}\label{eq:timemoment}
 \mathcal F^{\mathrm{LC}}_{ij|L}(\eta'';a)
 :=\int\dd^3x'\,x'_{\langle L\rangle}\int_{-1}^1\dd z\,
 \delta_\ell(z)a(\eta''+\rho'z/c)\tau_{ij}(\eta''+\rho'z/c,\bx').
\end{equation}
The normalized weight $\delta_\ell$ is defined in
Eq.~\eqref{eq:time-weight-definition}; it also appears in the
flat-space multipole formulas of \cite[Eq.~(B.3)]{Blanchet:1989ki} and
\cite[Eqs.~(3.4)--(3.6)]{Damour:1990gj}.
The argument $a$ in $\mathcal F^{\mathrm{LC}}_{ij|L}(\eta'';a)$
refers to the prescribed cosmological background; its dependence
enters through $a\tau_{ij}$.
The mixed light-cone moments $\mathcal F^{\mathrm{LC}}_{0i|L}(\eta'';a)$
follow by replacing $\tau_{ij}$ with $\tau_{0i}/c$.

For propagation from the origin, the delta function in
Eq.~\eqref{eq:light-cone-zero-potential-kernel} evaluates each moment
at $\eta''=\eta-\rho/c$, giving
\begin{subequations}\label{eq:LCSTF}
\begin{align}
 \bar h^{\mathrm{LC}}_{ij}(\eta,\bx)
 &\overset{\rho>d}{=}\frac{4G}{c^4}\sum_{\ell\ge0}\frac{(-1)^\ell}{\ell!}\partial_L
       \left[\frac{\mathcal F^{\mathrm{LC}}_{ij|L}(\eta-\rho/c;a)}\rho\right],
 \label{eq:LCSTF-spatial}\\
 \bar h^{\mathrm{LC}}_{0i}(\eta,\bx)
 &\overset{\rho>d}{=}\frac{4G}{c^3}\sum_{\ell\ge0}\frac{(-1)^\ell}{\ell!}\partial_L
       \left[\frac{\mathcal F^{\mathrm{LC}}_{0i|L}(\eta-\rho/c;a)}\rho\right].
 \label{eq:LCSTF-mixed}
\end{align}
\end{subequations}
The spatial derivatives also act on the central retarded time
$\eta-\rho/c$.

In Minkowski spacetime, $a=1$ and $V^\pm=0$, so both metric
source-time kernels reduce to $K^0_\ell$. Consequently,
$\mathcal F_{ij|L}(\eta'';1)=\mathcal F^{\mathrm{LC}}_{ij|L}(\eta'';1)$,
and likewise for the mixed moments. The curvature tail vanishes,
so Eqs.~\eqref{eq:LCSTF} give the complete dynamical spatial and
mixed fields and recover
the classical exterior multipole theorem of
\cite[Eqs.~(B.2)--(B.3)]{Blanchet:1989ki} and
\cite[Eqs.~(3.3)--(3.6)]{Damour:1990gj}. The theorem was first
derived in spherical harmonics by Campbell, Macek and Morgan
\cite{Campbell:1977jf} and in STF form by Blanchet and Damour
\cite[Appendix~B]{Blanchet:1989ki}. The outgoing-frequency form of Eq.~\eqref{eq:LCSTF-spatial} is given in
Appendix~\ref{app:retarded-STF}.

\paragraph{Curvature-tail contribution.}
We insert the Hadamard decomposition~\eqref{eq:hadamard} of $G^-$
and $G^+$ into the spatial and mixed fields~\eqref{eq:curved-retarded-fields}.
The light-cone term of each central Green function evaluates its
metric source moment at $\eta''=\eta-\rho/c$, while the tail term
integrates its earlier values. Subtracting the extended source's
light-cone fields~\eqref{eq:LCSTF} then gives
\begin{subequations}\label{eq:curved-retarded-tail-field}
\begin{equation}\label{eq:curved-retarded-spatial-tail-field}
\begin{aligned}
 \bar h^{\mathrm{tail}}_{ij}(\eta,\bx)
 &\overset{\rho>d}{=}\frac{4G}{c^4}\sum_{\ell\ge0}\frac{(-1)^\ell}{\ell!}\partial_L
 \biggl[\frac{\mathcal F_{ij|L}(\eta-\rho/c;a)
                 -\mathcal F^{\mathrm{LC}}_{ij|L}(\eta-\rho/c;a)}{\rho}\\
 &\hspace{4em}+\int_{\Eta\cap(-\infty,\eta-\rho/c]}c\,\dd\eta''\,
   \mathcal V^-(\eta,\rho;\eta'')
   \mathcal F_{ij|L}(\eta'';a)\biggr],
\end{aligned}
\end{equation}
\begin{equation}\label{eq:curved-retarded-mixed-tail-field}
\begin{aligned}
 \bar h^{\mathrm{tail}}_{0i}(\eta,\bx)
 &\overset{\rho>d}{=}\frac{4G}{c^3}\sum_{\ell\ge0}\frac{(-1)^\ell}{\ell!}\partial_L
 \biggl[\frac{\mathcal F_{0i|L}(\eta-\rho/c;a)
                 -\mathcal F^{\mathrm{LC}}_{0i|L}(\eta-\rho/c;a)}{\rho}\\
 &\hspace{4em}+\int_{\Eta\cap(-\infty,\eta-\rho/c]}c\,\dd\eta''\,
   \mathcal V^+(\eta,\rho;\eta'')
   \mathcal F_{0i|L}(\eta'';a)\biggr].
\end{aligned}
\end{equation}
\end{subequations}
The integration domain combines the background interval $\Eta$ with
the retarded support $\eta''\le\eta-\rho/c$ of the central Green
function. In each sector, the difference between the metric and
light-cone source moments accounts for the curvature dependence of
the source-time kernel, $K_\ell^{V^-}$ or $K_\ell^{V^+}$, relative
to $K_\ell^0$. The integral accounts for the tail in propagation
from the origin to the observer. Their sum gives the Hadamard tail
of the extended source.

\paragraph{Restoring the past field.}
The prescribed past sources $a\tau_{ij}^{\mathrm{past}}$ and
$a\tau_{0i}^{\mathrm{past}}$ in the wave equations are proportional
to $a$ and $1/a$, respectively. Define their retarded source moments by
\begin{subequations}\label{eq:past-retarded-source-moments}
\begin{align}
 \mathcal F^{\mathrm{past}}_{ij|L}(\eta;a)
 &:=\frac{a(\eta)}{a_i}\int\dd^3x'\,x'_{\langle L\rangle}
                             a_i\tau_{ij}^{\mathrm{in}}(\bx'),
 \label{eq:past-retarded-spatial-moment}\\
 \mathcal F^{\mathrm{past}}_{0i|L}(\eta;a)
 &:=\frac{a_i}{ca(\eta)}\int\dd^3x'\,x'_{\langle L\rangle}
                             a_i\tau_{0i}^{\mathrm{in}}(\bx').
 \label{eq:past-retarded-mixed-moment}
\end{align}
\end{subequations}
Here the initial shifted source data
$\tau_{\mu\nu}^{\mathrm{in}}(\bx')=\tau_{\mu\nu}^{\mathrm{past}}(\eta_i,\bx')$
are defined after Eqs.~\eqref{eq:integrated-source-conservation}.
These expressions agree with the retarded source-moment definitions
\eqref{eq:curved-retarded-moments} evaluated on the past sources.
Their time integrals simplify because the corresponding kernels
preserve the two prescribed temporal factors: for
$[\eta-\rho'/c,\eta+\rho'/c]\subset\Eta$,
\begin{subequations}\label{eq:past-source-kernel-averages}
\begin{align}
 \int_\Eta c\,\dd\eta'\,
 \frac{K^{V^-}_\ell(\rho',\eta';\eta)}{\rho'^{\ell+1}}
 \frac{a(\eta')}{a_i}&=\frac{a(\eta)}{a_i},
 \label{eq:past-spatial-kernel-average}\\
 \int_\Eta c\,\dd\eta'\,
 \frac{K^{V^+}_\ell(\rho',\eta';\eta)}{\rho'^{\ell+1}}
 \frac{a_i}{a(\eta')}&=\frac{a_i}{a(\eta)}.
 \label{eq:past-mixed-kernel-average}
\end{align}
\end{subequations}
Indeed, $f=a/a_i$ obeys $c^{-2}f''-V^-f=0$, and $f=a_i/a$ obeys
$c^{-2}f''-V^+f=0$. For either temporal factor, $\rho'^{\ell+1}f(\eta)$
therefore solves the radial equation~\eqref{eq:curved-average-radial-equation}
and, after division by $\rho'^{\ell+1}$, has value $f(\eta)$ and
zero radial derivative at $\rho'=0$. These are precisely the data
\eqref{eq:curved-average-radial-data}--\eqref{eq:curved-average-radial-derivative}
integrated against $f$.
Uniqueness and reciprocity~\eqref{eq:curved-average-reciprocity}
give Eqs.~\eqref{eq:past-source-kernel-averages}.
Thus the past moments require only the spatial integrals of the
initial sources in Eqs.~\eqref{eq:past-retarded-source-moments}.

The past metric perturbations $\bar h_{ij}^{\mathrm{past}}(\eta,\bx)$
and $\bar h_{0i}^{\mathrm{past}}(\eta,\bx)$ are fixed by the Poisson equations
\eqref{eq:past-source-spatial} and \eqref{eq:past-source-mixed}, with
the assumed boundary condition that they vanish at spatial infinity,
$\rho\to\infty$ at fixed $\eta$.
Their harmonic multipole expansions are
\begin{subequations}\label{eq:past-multipolar-fields}
\begin{align}
 \bar h_{ij}^{\mathrm{past}}(\eta,\bx)
 &\overset{\rho>d}{=}\frac{4G}{c^4}\sum_{\ell\ge0}\frac{(-1)^\ell}{\ell!}\partial_L
 \left[\frac{\mathcal F^{\mathrm{past}}_{ij|L}(\eta;a)}{\rho}\right],
 \label{eq:past-multipolar-spatial}\\
 \bar h_{0i}^{\mathrm{past}}(\eta,\bx)
 &\overset{\rho>d}{=}\frac{4G}{c^3}\sum_{\ell\ge0}\frac{(-1)^\ell}{\ell!}\partial_L
 \left[\frac{\mathcal F^{\mathrm{past}}_{0i|L}(\eta;a)}{\rho}\right].
 \label{eq:past-multipolar-mixed}
\end{align}
\end{subequations}

\paragraph{Complete exterior metric.}
Adding the past fields~\eqref{eq:past-multipolar-fields} to the
dynamical fields~\eqref{eq:curved-retarded-fields} gives
\begin{subequations}\label{eq:full-multipolar-fields}
\begin{align}
 \bar h_{ij}^{\mathrm{past}}(\eta,\bx)+\bar h_{ij}^{\mathrm{dyn}}(\eta,\bx)
 &\overset{\rho>d}{=}\frac{4G}{c^4}\sum_{\ell\ge0}\frac{(-1)^\ell}{\ell!}\partial_L
 \biggl[\frac{\mathcal F^{\mathrm{past}}_{ij|L}(\eta;a)}{\rho}
 +\int_\Eta c\,\dd\eta''\,G^-(\eta,\rho;\eta'')
 \mathcal F^{\mathrm{dyn}}_{ij|L}(\eta'';a)\biggr]
 \nonumber\\
 &=\frac{4G}{c^4}\sum_{\ell\ge0}\frac{(-1)^\ell}{\ell!}\partial_L
 \mathcal G_-\!\left[
 \mathcal F^{\mathrm{past}}_{ij|L}(\cdot;a)
 +\mathcal F^{\mathrm{dyn}}_{ij|L}(\cdot;a)\right](\eta,\rho),
 \label{eq:full-multipolar-spatial}\\
 \bar h_{0i}^{\mathrm{past}}(\eta,\bx)+\bar h_{0i}^{\mathrm{dyn}}(\eta,\bx)
 &\overset{\rho>d}{=}\frac{4G}{c^3}\sum_{\ell\ge0}\frac{(-1)^\ell}{\ell!}\partial_L
 \biggl[\frac{\mathcal F^{\mathrm{past}}_{0i|L}(\eta;a)}{\rho}
 +\int_\Eta c\,\dd\eta''\,G^+(\eta,\rho;\eta'')
 \mathcal F^{\mathrm{dyn}}_{0i|L}(\eta'';a)\biggr]
 \nonumber\\
 &=\frac{4G}{c^3}\sum_{\ell\ge0}\frac{(-1)^\ell}{\ell!}\partial_L
 \mathcal G_+\!\left[
 \mathcal F^{\mathrm{past}}_{0i|L}(\cdot;a)
 +\mathcal F^{\mathrm{dyn}}_{0i|L}(\cdot;a)\right](\eta,\rho).
 \label{eq:full-multipolar-mixed}
\end{align}
\end{subequations}
The second equalities use the retarded integral
\eqref{eq:canonical-propagator} and the metric propagator definition
\eqref{eq:metric-propagator-definition} to combine each past
Poisson term and dynamical integral into $\mathcal G_\pm$.
This compact form will be used in
Section~\ref{sec:direct-gauge-invariant-multipoles} to apply the
propagation identities~\eqref{eq:canonical-propagation-identities}
when decomposing the metric into irreducible STF coefficients.

\subsection{Sub-Hubble, slowly varying sources}
\label{sec:finite-source-size}
The exact retarded source moments defined in
Eqs.~\eqref{eq:curved-retarded-moments} retain the source history over
its light-crossing time. We will now expand the complete spatial and
mixed sources, including their prescribed past parts, for sub-Hubble,
slowly varying sources. After specifying the source scales,
we will expand the source-time average about zero radius and
use its coefficients to construct finite source-size approximations
to the complete exterior metric.

\paragraph{Physical assumptions and the scales being expanded.}
Let $d$ bound the comoving source radius, so $D_{\rm src}(\eta'):=a(\eta')d$
bounds its physical size at conformal time $\eta'$. Let
$T_{\rm src}(\eta')$ denote a proper-time
scale controlling the variation of the prescribed dynamical stresses and their
time derivatives. The conditions
\begin{equation}\label{eq:small-source-physical-scales}
 D_{\rm src}|H|/c\ll1,\qquad \frac{D_{\rm src}}{cT_{\rm src}}\ll1
\end{equation}
mean that cosmic expansion and source evolution are both slow over
the light-crossing time. For a source whose changing multipoles are
governed by internal motion, $T_{\rm src}\sim D_{\rm src}/v$ reduces the
second condition to a nonrelativistic characteristic speed $v\ll c$.
Slow motion at the initial time alone does not establish this
condition at later times: it is an assumption on the source history
that contributes to the observation. We do not supply a matter
equation of state or dynamics that would enforce it.

In the Volterra equation~\eqref{eq:curved-average-volterra}, the two
characteristic integrations each extend over a length bounded by $d$.
The first curvature correction is therefore of order
$d^2|V^\pm|$; the radial recurrence~\eqref{eq:curved-average-recursion}
preserves this scaling at higher multipolar ranks.
Using Eq.~\eqref{eq:potentials}, we estimate its size,
up to numerical factors of order one, as
\begin{equation}\label{eq:small-source-curvature-scale}
 d^2|V^\pm|\lesssim \frac{D_{\rm src}^2}{c^2}\bigl(H^2+|\partial_tH|\bigr)\ll1.
\end{equation}
When $|\partial_tH|$ is of order $H^2$ or smaller, these corrections
are at most of order $(D_{\rm src}H/c)^2$, and the sub-Hubble condition suffices.
During a rapid change of the background, we also require
$D_{\rm src}^2|\partial_tH|/c^2\ll1$.
For approximations at arbitrary finite order, we assume that the
dynamical stresses and $H$ are smooth, with their $n$th proper-time
derivatives bounded by their characteristic amplitudes divided by the
$n$th power of $T_{\rm src}$ or the background variation timescale, respectively.
These timescales are assumed long compared with $D_{\rm src}/c$.
At each order, the bounds are uniform over the full sampled history
and independent of source size; convergence of the infinite series
is not assumed.

The approximation concerns the size and evolution of the emitter.
It does not require the propagation
time to be small compared with the curvature scale:
a sub-Hubble source can produce a substantial accumulated tail.

\paragraph{Expanding the moments while retaining exact propagation.}
We start with the source-time average of a smooth profile $f$.
The data in Eq.~\eqref{eq:curved-average-radial-data} fix its value
at zero radius to $f(\eta')$. Smoothness and evenness in $\rho'$
then give
\begin{equation}\label{eq:source-time-even-expansion}
 \int_\Eta c\,\dd\eta''\,
 \frac{K^V_\ell(\rho',\eta';\eta'')}{\rho'^{\ell+1}}f(\eta'')
 =f(\eta')+\rho'^2C_\ell[f](\eta')+O(\rho'^4).
\end{equation}
Multiplying by $\rho'^{\ell+1}$ gives the integral of $K^V_\ell$
against $f$, which obeys the radial
equation~\eqref{eq:curved-average-radial-equation}.
Substituting the expansion, the terms of order $\rho'^{\ell-1}$
cancel between the second radial derivative and the centrifugal
term. The terms of order $\rho'^{\ell+1}$ give
\begin{equation}\label{eq:source-time-first-correction}
 2(2\ell+3)C_\ell[f](\eta')
 =c^{-2}f''(\eta')-V(\eta')f(\eta').
\end{equation}
Indeed, the radial derivative and centrifugal terms acting on
$\rho'^{\ell+3}$ give the coefficient
$\ell(\ell+1)-(\ell+3)(\ell+2)=-2(2\ell+3)$.
We therefore obtain
\begin{equation}\label{eq:source-time-central-check}
 \int_{\Eta}c\,\dd\eta''\,
 \frac{K^V_\ell(\rho',\eta';\eta'')}{\rho'^{\ell+1}}f(\eta'')
 =f(\eta')+\frac{\rho'^2}{2(2\ell+3)}
\bigl[c^{-2}f''(\eta')-V(\eta')f(\eta')\bigr]+O(\rho'^4).
\end{equation}
For $f=1$ on the sampled interval, the first correction is
$-\rho'^2V(\eta')/[2(2\ell+3)]$: the integral of
$K^V_\ell/\rho'^{\ell+1}$ therefore generally differs from one
at finite radius.

To apply this expansion to the complete sources, we first evaluate
their prescribed past parts. Their time dependence in
Eqs.~\eqref{eq:past-source-spatial}--\eqref{eq:past-source-mixed}
is fixed by $a\tau_{ij}^{\mathrm{past}}=a\tau_{ij}^{\mathrm{in}}$ and
$a\tau_{0i}^{\mathrm{past}}=a_i^2\tau_{0i}^{\mathrm{in}}/a$.
The potentials~\eqref{eq:potentials} give $T_-a=0$ and
$T_+(1/a)=0$. Thus, with $T_\pm$ acting on $\eta'$ at fixed $\bx'$,
\begin{subequations}\label{eq:past-source-time-annihilation}
\begin{align}
 T_-\!\left[a(\eta')\tau_{ij}^{\mathrm{past}}(\eta',\bx')\right]
 &=\tau_{ij}^{\mathrm{in}}(\bx')T_-a(\eta')=0,
 \label{eq:past-spatial-time-annihilation}\\
 T_+\!\left[a(\eta')\tau_{0i}^{\mathrm{past}}(\eta',\bx')\right]
 &=a_i^2\tau_{0i}^{\mathrm{in}}(\bx')T_+\!\left[\frac1{a(\eta')}\right]=0.
 \label{eq:past-mixed-time-annihilation}
\end{align}
\end{subequations}
For $V=V^\pm$, the quadratic correction in
Eq.~\eqref{eq:source-time-central-check} is
$\rho'^2T_\pm f/[2(2\ell+3)]$, by the definition~\eqref{eq:canonical-Tpm}.
At fixed $\bx'$, substitute $f:=a\tau_{ij}^{\mathrm{past}}$ in the
minus sector and $f:=a\tau_{0i}^{\mathrm{past}}/c$ in the plus sector.
Equations~\eqref{eq:past-source-time-annihilation} then give
$T_\pm f=0$, or $c^{-2}f''=V^\pm f$: the time-derivative and
potential terms cancel. This agrees with the exact kernel
averages~\eqref{eq:past-source-kernel-averages}.

For the dynamical spatial source moment~\eqref{eq:curved-retarded-spatial-moment},
set $V=V^-$ and use reciprocity~\eqref{eq:curved-average-reciprocity}
to exchange $\eta'$ and $\eta''$ in
Eq.~\eqref{eq:source-time-central-check}. Apply the resulting expansion
to $f(\eta'):=a(\eta')\tau_{ij}^{\mathrm{dyn}}(\eta',\bx')$ at fixed source position.
Multiplication by $x'_{\langle L\rangle}$ and spatial integration then
turn powers of $\rho'^2$ into Cartesian traces by
Eq.~\eqref{eq:spatial-radialmoments-traces}. The leading term and its
first correction are
\begin{equation}\label{eq:small-source-leading-moment}
\begin{aligned}
 \mathcal F^{\mathrm{dyn}}_{ij|L}(\eta'';a)\simeq{}
 &a(\eta'')^{-\ell}S^{\mathrm{dyn}}_{ij|\langle L\rangle}(\eta'')\\
 &+\frac{[c^{-2}\partial_{\eta''}^2-V^-(\eta'')]
       [a(\eta'')^{-\ell-2}\STF_L S^{\mathrm{dyn}}_{ij|Lkk}(\eta'')]}{2(2\ell+3)}.
\end{aligned}
\end{equation}
The two corrections have different origins. The time derivative
accounts for evolution during the source light-crossing time;
the potential term accounts for curvature across the source.
The derivative acts also on the scale factor. Since
$\partial_{\eta'}=a\partial_t$, the expansion is controlled by the
source and background scales stated above.

For a truncation at Cartesian degree $m$, the radial equation generates
successive even powers of the source radius. At fixed rank $\ell$,
each trace increases the Cartesian degree by two, so only
$p\le\lfloor(m-\ell)/2\rfloor$ is retained. Matching powers in the
radial equation gives the coefficient at degree $2p+2$ by applying
$T_\pm$ to the coefficient at degree $2p$ and dividing by
$2(p+1)(2p+2\ell+3)$. Iteration generates the coefficients
$c_{\ell+1,p}$ of Eq.~\eqref{eq:tracecoefficients}. Hence
Eqs.~\eqref{eq:past-source-time-annihilation} make every past-source
correction with $p\ge1$ vanish, including after spatial integration.
The term $p=0$ gives exactly the past moments
\eqref{eq:past-retarded-source-moments}. We use the superscript $[m]$
for quantities obtained by retaining stress and momentum moments through
Cartesian degree $m$, including fields reconstructed from those moments.
For $0\le\ell\le m$, the complete spatial moment at this finite order is
\begin{equation}\label{eq:curved-retarded-moment-finite}
\begin{aligned}
 \mathcal F^{[m]}_{ij|L}(\eta'';a)
 :={}&\mathcal F^{\mathrm{past}}_{ij|L}(\eta'';a)\\
 &+\sum_{p=0}^{\lfloor(m-\ell)/2\rfloor}c_{\ell+1,p}
 [c^{-2}\partial_{\eta''}^2-V^-(\eta'')]^p
 \left[a(\eta'')^{-\ell-2p}\STF_L
 S^{\mathrm{dyn}}_{ij|L j_1j_1\cdots j_pj_p}(\eta'')\right].
\end{aligned}
\end{equation}
The coefficients $c_{\ell+1,p}$ are defined in
Eq.~\eqref{eq:tracecoefficients}. The power denotes operator composition:
at higher orders derivatives act on $V^-$ as well as on the moment.
The complete mixed moment $\mathcal F^{[m]}_{0i|L}(\eta'';a)$ follows
by replacing $\mathcal F^{\mathrm{past}}_{ij|L}$, $V^-$ and
$S^{\mathrm{dyn}}_{ij|L}$ with $\mathcal F^{\mathrm{past}}_{0i|L}$,
$V^+$ and $P^{\mathrm{dyn}}_{i|L}$, respectively.
For $a=1$, this reduces to the Cartesian trace expansion of
Blanchet--Damour \cite[Eqs.~(B.14a)--(B.14b)]{Blanchet:1989ki}
and Damour--Iyer \cite[Eq.~(3.10)]{Damour:1990gj}.

Using the metric propagator~\eqref{eq:metric-propagator-definition}
to include the prescribed past data, we obtain the complete field
approximation for $\rho>d$:
\begin{equation}\label{eq:curved-retarded-field-finite}
 \bar h^{[m]}_{ij}(\eta,\bx)
 :=\frac{4G}{c^4}\sum_{\ell=0}^m\frac{(-1)^\ell}{\ell!}\partial_L
 \mathcal G_-\!\left[\mathcal F^{[m]}_{ij|L}(\cdot;a)\right](\eta,\rho).
\end{equation}
The mixed field $\bar h^{[m]}_{0i}$ follows with $\mathcal G_+$,
$\mathcal F^{[m]}_{0i|L}$ and prefactor $4G/c^3$. The past contribution at each retained
STF rank is exact; its truncation error comes only from the omitted
angular orders $\ell>m$. The Green functions within $\mathcal G_\pm$
remain unexpanded, so propagation to the observer, including
curvature tails, is exact for the retained source moments.
The dynamical parts of the finite moments depend only on
instantaneous source traces and derivatives and vanish before
$\eta_i$; the exact dynamical retarded source moments can instead
start at $\eta_i-d/c$.

Equivalently, a finite spatial Taylor expansion of the dynamical
source integral~\eqref{eq:spatial-solution}, after time integration,
gives the unprojected Cartesian form, with $|N|=n$,
\begin{equation}\label{eq:cartesian-general}
 \bar h_{ij}(\eta,\bx)\overset{\rho>d}{=}\frac{4G}{c^4}\sum_{n=0}^{m}\frac{(-1)^n}{n!}\partial_N
 \int_{\eta_i}^{\eta}c\,\dd\eta'\,
 G^-(\eta,\rho;\eta')a(\eta')^{-n}S_{ij|N}(\eta')
 +\mathcal R_m(\eta,\bx).
\end{equation}
Here $\mathcal R_m$ is the finite Taylor remainder, and the unmarked
fields and moments denote their dynamical parts. The finite sum is the
dynamical part of $\bar h^{[m]}_{ij}$. The mixed expression uses $P_{i|N}$, $G^+$ and
prefactor $4G/c^3$. The fields $\hat h^{[m]}$ and $\tau_{00}^{[m]}$
are reconstructed from the retained metric through
Eqs.~\eqref{eq:scalar-solution} and \eqref{eq:energy-solution}.

Equation~\eqref{eq:cartesian-general}, including its remainder,
holds throughout the exterior, $\rho>d$. The accuracy of a low
multipolar truncation also depends on the omitted angular orders.
Spatial derivatives of the light-cone field act both on the inverse
distance and on the retarded emission time. Derivatives of the inverse
distance produce powers of $d/\rho$, so their contribution to the
truncation error is more strongly suppressed for distant observers,
$\rho\gg d$. This improves the geometric accuracy of the truncation;
errors associated with time variation and curvature tails still depend
on the contributing source history.

In de Sitter spacetime, the source-size and slow-motion assumptions
are those used by Ashtekar, Bonga and Kesavan
\cite[Sec.~III~A, Eqs.~(3.1)--(3.9)]{Ashtekar:2015lxa}.
Their leading approximation at future conformal infinity has
corrections of order $HD_0/c$ and $v/c$, including the replacement of a
source-dependent tail endpoint by the central retarded time.
Those specific estimates are not assumed for an arbitrary FLRW
background. Here the exact retarded Green function retains the history dependence;
the error in approximating its input moments must be propagated over
the actual history. A long accumulated tail can
preclude a uniform relative-error estimate based on local scale
ratios alone.

\subsection{Quadrupolar truncation}
\label{sec:general-quadrupolar-check}
We will construct the complete metric and induced matter response
of effective quadrupolar sources for an arbitrary smooth scale factor.
The second Cartesian order of the source-size expansion in
Section~\ref{sec:finite-source-size} provides a finite source
representation. {We will impose momentum conservation on
its retained coefficients and reconstruct its density by energy
conservation. We will define the quadrupolar truncation by requiring
the full Cartesian density octupole to vanish and derive the
corresponding condition on the retained momentum moments.}
The superscript $[2]$ denotes this model; as in Section~\ref{sec:cartesian},
$\mathrm{dyn}$ is suppressed when only dynamical quantities appear.

\paragraph{Effective quadrupolar source.}
An effective multipolar source of order $m$ is represented
by delta-function derivatives through Cartesian rank $m$ in its stress
and momentum components. Momentum conservation restricts their
coefficients, and energy conservation reconstructs its density{,
including the cosmic-fluid response}.
We start with the finite Cartesian expansion~\eqref{eq:cartesian-general}
at $m=2$. We define the stress and momentum components of
the effective quadrupolar source by
\begin{subequations}\label{eq:quadrupolar-effective-sources}
\begin{align}
 a(\eta)\tau^{[2]}_{ij}(\eta,\bx')
 &:=\sum_{n=0}^{2}\frac{(-1)^n}{n!}a(\eta)^{-n}
 S_{ij|N}(\eta)\partial'_N\delta^{(3)}(\bx'),
 \label{eq:quadrupolar-effective-stress}\\
 \frac{a(\eta)}c\tau^{[2]}_{0i}(\eta,\bx')
 &:=\sum_{n=0}^{2}\frac{(-1)^n}{n!}a(\eta)^{-n}
 P_{i|N}(\eta)\partial'_N\delta^{(3)}(\bx').
 \label{eq:quadrupolar-effective-momentum}
\end{align}
\end{subequations}
Here $|N|=n$ and $\partial'_N$ acts on the delta function.
These two components are supported at the origin.
{For an extended source under the assumptions of
Section~\ref{sec:finite-source-size}, these two finite sums give the
second-order stress and momentum expansion, with the metric
remainder $\mathcal R_2$ of Eq.~\eqref{eq:cartesian-general} and its
mixed counterpart. The quadrupolar model additionally requires
the retained moments to obey the conservation and octupole
cancellation conditions derived below.}

Their moments follow from the identity
\begin{equation}\label{eq:quadrupolar-delta-moments}
 \int\dd^3x'\,x'_L\partial'_N\delta^{(3)}(\bx')
 =(-1)^n\left.\partial'_Nx'_L\right|_{\bx'=0}.
\end{equation}
The right-hand side vanishes unless $|L|=n$. At equal rank,
contraction with the symmetric coefficient cancels the factorial
in Eqs.~\eqref{eq:quadrupolar-effective-sources}. Evaluating the
source moments~\eqref{eq:source-moment-definitions} therefore gives
$S_{ij|L}$ and $P_{i|L}$ for $|L|\le2$, while all higher moments
of these two components vanish. With $S_{ij|L}$ and $P_{i|L}$ denoting
these retained moments, we obtain the {second-order Cartesian cutoff}
\begin{equation}\label{eq:quadrupolar-compact-cutoff}
 S_{ij|L}=P_{i|L}=0,\qquad |L|>2.
\end{equation}
{The rank-two Cartesian traces are retained.}

\paragraph{Conservation of the effective quadrupolar source.}
We impose both conservation laws~\eqref{eq:shifted-conservation}
on the effective quadrupolar source. At $(\eta,\bx')$, they read
\begin{subequations}\label{eq:quadrupolar-source-conservation}
\begin{align}
 (\partial_\eta+\cH)(a\tau^{[2]}_{0i}/c)
 &=\partial'_j(a\tau^{[2]}_{ij}),
 \label{eq:quadrupolar-source-momentum-conservation}\\
 \partial_\eta(a\tau^{[2]}_{00})
 &=ca\partial'_i\tau^{[2]}_{0i}-\dot a\tau^{[2]}_{kk}
 +a^2s\hat h^{[2]}.
 \label{eq:quadrupolar-source-energy-conservation}
\end{align}
\end{subequations}
The left-hand side of Eq.~\eqref{eq:quadrupolar-source-momentum-conservation}
contains delta-function derivatives through
order two, whereas the stress quadrupole on the right produces
third-order derivatives proportional to $S_{i(j|kl)}$.
Matching all derivative orders gives the momentum balance laws
and the closure condition
\begin{subequations}\label{eq:quadrupolar-momentum-closure}
\begin{align}
 (\partial_t+H)P_i&=0,
 \label{eq:quadrupolar-momentum-monopole}\\
 \partial_tP_{i|j}&=-S_{ij},
 \label{eq:quadrupolar-momentum-dipole}\\
 (\partial_t-H)P_{i|jk}&=-2S_{i(j|k)},
 \label{eq:quadrupolar-momentum-evolution}\\
 S_{i(j|kl)}&=0.
 \label{eq:quadrupolar-stress-closure}
\end{align}
\end{subequations}
The first three relations evolve the retained momentum moments;
the closure condition~\eqref{eq:quadrupolar-stress-closure}
prevents the stress quadrupole from generating a momentum octupole.
Contracting Eq.~\eqref{eq:quadrupolar-stress-closure} first on two
symmetrized indices and then on the free index and a symmetrized
index gives
\begin{equation}\label{eq:quadrupolar-stress-traces}
 S_{ij|kk}=S_{kk|ij}=c^2Q^{(p)}_{ij},\qquad
 S_{ik|kj}=-\frac{c^2}{2}Q^{(p)}_{ij}.
\end{equation}
Thus a pressure quadrupole requires the corresponding rank-two
stress moments; their traces must be retained.

Energy conservation determines the density moments. The divergence
of the retained momentum quadrupole produces third-order derivatives
of the delta function and can therefore generate a density octupole.
Since $Q^{(p)}_{ijk}=0$ by Eq.~\eqref{eq:quadrupolar-compact-cutoff},
the rank-three energy balance~\eqref{eq:momentenergy} becomes
\begin{equation}\label{eq:quadrupolar-density-octupole-balance}
 (\partial_t-3H)Q^{(\rho)}_{ijk}
 =-3P_{(i|jk)}
 +\frac{s}{c^2}\int\dd^3x'\,a^4\hat h^{[2]}x'_ix'_jx'_k.
\end{equation}
{The temporal gauge condition~\eqref{eq:scalar-solution}
determines $\hat h^{[2]}$ from the spatial and mixed metric fields,
with zero past data:
\begin{equation}\label{eq:quadrupolar-hat-reconstruction}
 \hat h^{[2]}(\eta,\bx)
 \overset{\rho>d}{=}\int_{\eta_i}^{\eta}\dd\zeta\,
 \big[(\partial_\zeta-\cH)\bar h^{[2]}_{kk}
       +c\partial_i\bar h^{[2]}_{0i}\big](\zeta,\bx).
\end{equation}
This identity also holds distributionally at the source.}
{To solve the density balance entirely in terms of the
retained momentum moments,} define the Cartesian moments of $\hat h^{[2]}$ for
$\ell=1,3$ and the momentum trace combination
\begin{subequations}\label{eq:quadrupolar-octupole-auxiliaries}
\begin{align}
 \mathcal U_L(\eta)&:=\int\dd^3x'\,
       \hat h^{[2]}(\eta,\bx')x'_L,
 \label{eq:quadrupolar-hat-moments}\\
 \Pi_i(\eta)&:=P_{i|ll}(\eta)-2P_{l|il}(\eta).
 \label{eq:quadrupolar-momentum-trace}
\end{align}
\end{subequations}
The dynamical fields have bounded spatial support at each finite
time, so spatial integration by parts produces no boundary terms.
Equation~\eqref{eq:quadrupolar-momentum-monopole} with zero past
data gives $P_i=0$; the mixed wave equation~\eqref{eq:mixed} then
also gives $\int\dd^3x'\,\bar h^{[2]}_{0i}=0$.
Integrate the spatial gauge condition $C_l=0$ in Eq.~\eqref{eq:gaugei} against
$x'^2\delta_{il}-2x'_ix'_l$ and the temporal condition~\eqref{eq:gauge0}
against $x'_i$. Using $T_+=D_-D_+$ from
Eq.~\eqref{eq:canonical-Tpm}, followed by the mixed wave equation,
gives
\begin{equation}\label{eq:quadrupolar-hat-dipole}
\begin{aligned}
 \partial_\eta\mathcal U_i(\eta)
 &=\frac c2T_+\int\dd^3x'\,
 \big[\bar h^{[2]}_{0i}x'^2-2\bar h^{[2]}_{0l}x'_ix'_l\big](\eta,\bx')\\
 &=\frac{8\pi G}{c^2a(\eta)^2}\Pi_i(\eta).
\end{aligned}
\end{equation}
Next, integrate the scalar wave equation~\eqref{eq:scalar} against
$x'_ix'_jx'_k$. Since
$\Delta'(x'_ix'_jx'_k)=6\delta_{(ij}x'_{k)}$, this yields
\begin{equation}\label{eq:quadrupolar-hat-octupole-wave}
 [\partial_\eta^2-c^2V^+(\eta)]\mathcal U_{ijk}(\eta)
 -6c^2\delta_{(ij}\mathcal U_{k)}(\eta)
 =\frac{16\pi G}{a(\eta)^3}Q^{(\rho)}_{ijk}(\eta).
\end{equation}
Differentiate this equation and substitute the energy
balance~\eqref{eq:quadrupolar-density-octupole-balance}, with
$\partial_t=a^{-1}\partial_\eta$. The terms proportional to
$\dot V^+\mathcal U_{ijk}$ cancel by Eq.~\eqref{eq:s}; using
Eq.~\eqref{eq:quadrupolar-hat-dipole} leaves
\begin{equation}\label{eq:quadrupolar-hat-octupole-evolution}
 [\partial_\eta^2-c^2V^+(\eta)]\partial_\eta\mathcal U_{ijk}(\eta)
 =-\frac{48\pi G}{a(\eta)^2}
 \big[P_{(i|jk)}(\eta)-\delta_{(ij}\Pi_{k)}(\eta)\big].
\end{equation}
For $\eta\ge\zeta$, introduce the temporal Green kernel $g_+$
and the associated background weight $W_+$:
\begin{subequations}\label{eq:quadrupolar-octupole-kernels}
\begin{align}
 g_+(\eta,\zeta)&:=\frac1{a(\eta)a(\zeta)}
                  \int_\zeta^\eta\dd\lambda\,a(\lambda)^2,
 \label{eq:quadrupolar-temporal-kernel}\\
 W_+(\eta,\zeta)&:=\partial_\eta g_+(\eta,\zeta)
 -c^2V^+(\eta)\int_\zeta^\eta\dd\sigma\,g_+(\sigma,\zeta).
 \label{eq:quadrupolar-octupole-weight}
\end{align}
\end{subequations}
The factorization~\eqref{eq:canonical-Tpm} gives
$(\partial_\eta^2-c^2V^+)g_+=0$ for $\eta>\zeta$, with
$g_+(\zeta,\zeta)=0$ and
$\left.\partial_\eta g_+(\eta,\zeta)\right|_{\eta=\zeta}=1$.
Solving Eq.~\eqref{eq:quadrupolar-hat-octupole-evolution}
with this kernel and zero past data determines
$\partial_\eta\mathcal U_{ijk}$. Integrating once in time and
substituting into Eq.~\eqref{eq:quadrupolar-hat-octupole-wave},
with $\mathcal U_i$ obtained from
Eq.~\eqref{eq:quadrupolar-hat-dipole}, gives the full Cartesian
density octupole:
\begin{equation}\label{eq:quadrupolar-density-octupole-solution}
\begin{aligned}
 Q^{(\rho)}_{ijk}(\eta)
 &=-3a(\eta)^3\int_{\eta_i}^{\eta}\frac{\dd\zeta}{a(\zeta)^2}
 \Big[W_+(\eta,\zeta)P_{(i|jk)}(\zeta)\\
 &\hspace{43mm}
 +\big(1-W_+(\eta,\zeta)\big)\delta_{(ij}\Pi_{k)}(\zeta)\Big].
\end{aligned}
\end{equation}
The Kronecker-delta term retains the Cartesian traces generated by
the coupling to $\mathcal U_i$. Energy conservation therefore fixes
the density octupole without an additional constraint on the retained
stress and momentum moments.

If the density octupole is required to vanish throughout the dynamical
history, the STF projection of
Eq.~\eqref{eq:quadrupolar-hat-octupole-wave} is homogeneous with zero
past data. Its solution vanishes, and the energy
balance~\eqref{eq:quadrupolar-density-octupole-balance} then requires
the corresponding STF momentum moment to vanish. Together with the
trace of Eq.~\eqref{eq:quadrupolar-density-octupole-solution}, this
gives the necessary and sufficient conditions
{
\begin{subequations}\label{eq:quadrupolar-octupole-cancellation}
\begin{equation}\label{eq:quadrupolar-octupole-cancellation-STF}
 P_{\langle i|jk\rangle}(\eta)=0,
\end{equation}
\begin{equation}\label{eq:quadrupolar-octupole-cancellation-trace}
 \int_{\eta_i}^{\eta}\frac{\dd\zeta}{a(\zeta)^2}
 \Big\{2W_+(\eta,\zeta)P_{i|ll}(\zeta)
 +[5-6W_+(\eta,\zeta)]\Pi_i(\zeta)\Big\}=0.
\end{equation}
\end{subequations}
The angle brackets project on the STF part over $i,j,k$; the second
equation uses the definition~\eqref{eq:quadrupolar-momentum-trace}
of $\Pi_i$.

We define the quadrupolar truncation by the effective sources
\eqref{eq:quadrupolar-effective-sources}, their conservation laws,
and a vanishing full Cartesian density octupole throughout the
dynamical history:
\begin{equation}\label{eq:quadrupolar-defining-conditions}
 Q^{(\rho)}_{ijk}(\eta)=0.
\end{equation}
For zero past data, this is equivalent to the cancellation conditions
\eqref{eq:quadrupolar-octupole-cancellation}, including the trace condition.

These conditions admit a constructive implementation with a
prescribed $\Pi_i$. First integrate
Eq.~\eqref{eq:quadrupolar-hat-dipole} with $\mathcal U_i(\eta_i)=0$.
The scalar-wave moment equation~\eqref{eq:quadrupolar-hat-octupole-wave}
then determines $\mathcal U_{ijk}$ from
\begin{equation}\label{eq:quadrupolar-zero-density-hat-moment}
 [\partial_\eta^2-c^2V^+(\eta)]\mathcal U_{ijk}(\eta)
 =6c^2\delta_{(ij}\mathcal U_{k)}(\eta),
\end{equation}
with zero initial value and first derivative. Energy conservation
\eqref{eq:quadrupolar-density-octupole-balance} fixes the symmetric
momentum moment:
\begin{equation}\label{eq:quadrupolar-compensating-momentum}
 P_{(i|jk)}(\eta)
 =\frac{a(\eta)^4s(\eta)}{3c^2}\mathcal U_{ijk}(\eta).
\end{equation}
The forcing in Eq.~\eqref{eq:quadrupolar-zero-density-hat-moment}
is pure trace, so $\mathcal U_{\langle ijk\rangle}=0$ and
$P_{\langle i|jk\rangle}=0$. The trace of
Eq.~\eqref{eq:quadrupolar-compensating-momentum}, together with
the definition~\eqref{eq:quadrupolar-momentum-trace} of $\Pi_i$,
determines both momentum traces without setting $\Pi_i$ to zero.
Differentiating Eq.~\eqref{eq:quadrupolar-zero-density-hat-moment}
and substituting Eqs.~\eqref{eq:quadrupolar-hat-dipole} and
\eqref{eq:quadrupolar-compensating-momentum}, with Eq.~\eqref{eq:s},
recovers Eq.~\eqref{eq:quadrupolar-hat-octupole-evolution}, verifying
the compatibility of this construction.

Contracting the momentum balance~\eqref{eq:quadrupolar-momentum-evolution}
gives
\begin{equation}\label{eq:quadrupolar-momentum-trace-balance}
 (\partial_t-H)\Pi_i=2S_{ll|i}=2c^2Q^{(p)}_i.
\end{equation}
Thus the momentum traces and pressure dipole remain allowed.
On a general FLRW background, the symmetric momentum moment in
Eq.~\eqref{eq:quadrupolar-compensating-momentum} compensates the
cosmic-fluid contribution to the density octupole.}

\paragraph{Metric and matter reconstruction.}
Inserting Eqs.~\eqref{eq:quadrupolar-effective-sources} into the
source integrals~\eqref{eq:spatial-solution}--\eqref{eq:mixed-solution}
and integrating by parts transfers the derivatives to $G^\pm$.
Spatial homogeneity gives $\partial'_NG^\pm=(-1)^n\partial_NG^\pm$,
which yields the dynamical spatial and mixed fields
\begin{subequations}\label{eq:quadrupolar-metric-integrals}
\begin{align}
 \bar h^{[2]}_{ij}(\eta,\bx)
 &\overset{\rho>d}{=}\frac{4G}{c^4}\sum_{n=0}^{2}\frac{(-1)^n}{n!}\partial_N
 \int_{\eta_i}^{\eta}c\,\dd\eta'\,
 G^-(\eta,\rho;\eta')a(\eta')^{-n}S_{ij|N}(\eta'),
 \label{eq:quadrupolar-spatial-integral}\\
 \bar h^{[2]}_{0i}(\eta,\bx)
 &\overset{\rho>d}{=}\frac{4G}{c^3}\sum_{n=0}^{2}\frac{(-1)^n}{n!}\partial_N
 \int_{\eta_i}^{\eta}c\,\dd\eta'\,
 G^+(\eta,\rho;\eta')a(\eta')^{-n}P_{i|N}(\eta').
 \label{eq:quadrupolar-mixed-integral}
\end{align}
\end{subequations}
These are the $m=2$ finite sum in Eq.~\eqref{eq:cartesian-general}
and its mixed counterpart, exact for the retained model.
{Their moments obey the momentum balance and closure
conditions~\eqref{eq:quadrupolar-momentum-closure} and the octupole
cancellation conditions~\eqref{eq:quadrupolar-octupole-cancellation}.}
The spatial gauge conditions follow from momentum conservation,
by the argument of Section~\ref{sec:retarded-reconstruction}.
{Substitution into
Eq.~\eqref{eq:quadrupolar-hat-reconstruction} determines $\hat h^{[2]}$
and hence $\bar h^{[2]}_{00}=\hat h^{[2]}-\bar h^{[2]}_{kk}$.
The momentum quadrupole contributes to the divergence
$\partial_i\bar h^{[2]}_{0i}$ through three spatial derivatives
contracted with $P_{(i|jk)}$, as fixed by
Eq.~\eqref{eq:quadrupolar-compensating-momentum}. Its STF part
vanishes, but its traces and the pressure dipole in
$\bar h^{[2]}_{kk}$ can contribute to $\hat h^{[2]}$.

Energy conservation~\eqref{eq:energy-solution} reconstructs the
shifted density. Since the effective stress and momentum are
supported at the origin, the exterior response reduces to
\begin{equation}\label{eq:quadrupolar-density-reconstruction}
 \tau_{00}^{[2]}(\eta,\bx)
 \overset{\rho>d}{=}\frac1{a(\eta)}\int_{\eta_i}^{\eta}\dd\zeta\,
 a(\zeta)^2s(\zeta)\hat h^{[2]}(\zeta,\bx).
\end{equation}
The full density octupole vanishes by the cancellation conditions
\eqref{eq:quadrupolar-octupole-cancellation}, although the cosmic fluid
response can still carry higher density moments.}
The scalar-wave consistency argument of
Section~\ref{sec:retarded-reconstruction} applies to these conserved
sources and establishes the remaining field equation. The retained
moments therefore determine the full dynamical metric and its
induced matter response.

The prescribed past field is added through
Eqs.~\eqref{eq:past-multipolar-fields}, with its stress and momentum moments
retained through the same order and obeying the closure conditions.
The compatible initial $\hat h^{\mathrm{in}}$ and
$\tau_{00}^{\mathrm{in}}$ enter
Eqs.~\eqref{eq:scalar-solution} and \eqref{eq:energy-solution}, respectively.

{In de Sitter, $s=0$ reduces
Eq.~\eqref{eq:quadrupolar-compensating-momentum} to $P_{(i|jk)}=0$,
while leaving $\Pi_i$ and the associated pressure dipole available.
The definition therefore recovers the Comp\`ere--Hoque quadrupolar
truncation \cite{Compere:2023ktn,Compere:2026xoe}, as detailed in
Section~\ref{sec:quadrupolar-map}. On a general FLRW background,
Eq.~\eqref{eq:quadrupolar-compensating-momentum} includes the
cosmic-fluid response required for a vanishing density octupole.}

\paragraph{Light-cone and curvature-tail fields.}
Applying the Hadamard decomposition~\eqref{eq:hadamard} to
Eqs.~\eqref{eq:quadrupolar-metric-integrals} separates each retained
dynamical field into its light-cone and curvature-tail parts.
The spatial and mixed expressions have the same form: replacing
$S_{ij|L}$ by $P_{i|L}$, $\mathcal V^-$ by $\mathcal V^+$ and the
prefactor $4G/c^4$ by $4G/c^3$ converts the spatial field into the
mixed field. It therefore suffices to evaluate the spatial field
explicitly.

The light-cone part of Eq.~\eqref{eq:quadrupolar-spatial-integral}
is obtained by differentiating the central field $f(\eta-\rho/c)/\rho$
through order two. Evaluating the source moments and scale factors
at $\eta'=\eta-\rho/c$, after taking their $\eta'$ derivatives, gives
\begin{equation}\label{eq:quadrupolar-LC-map}
\begin{aligned}
 \frac{c^4\bar h^{[2],\mathrm{LC}}_{ij}}{4G}
 &\overset{\rho>d}{=}\frac1\rho\left[S_{ij}
       +\frac{n_k}{c}\partial_{\eta'}(a^{-1}S_{ij|k})
       +\frac1{2c^2}n_kn_l\partial_{\eta'}^2(a^{-2}S_{ij|kl})\right]
 \\
 &\quad+\frac1{\rho^2}\left[a^{-1}n_kS_{ij|k}
       +\frac1{2c}(3n_kn_l-\delta_{kl})
                          \partial_{\eta'}(a^{-2}S_{ij|kl})\right]
 +\frac{3n_kn_l-\delta_{kl}}{2\rho^3a^2}S_{ij|kl}.
\end{aligned}
\end{equation}

For the tail, derivatives act on both its coefficient
$\mathcal V^-(\eta,\rho;\eta')$ and its moving upper limit.
For a smooth source-time profile $f(\eta')$, the Leibniz rule at
fixed observation time gives
\begin{equation}\label{eq:tail-upper-limit-derivative}
\begin{aligned}
 \partial_\rho^2\int_{\eta_i}^{\eta-\rho/c}c\,\dd\eta'\,\mathcal V^- f
 &=\int_{\eta_i}^{\eta-\rho/c}c\,\dd\eta'\,(\partial_\rho^2\mathcal V^-)f
 -2(\partial_\rho\mathcal V^-)f\big|_{\eta'=\eta-\rho/c}
 +\frac1c\partial_{\eta'}(\mathcal V^- f)\big|_{\eta'=\eta-\rho/c}.
\end{aligned}
\end{equation}
Together with
$\partial_{kl}=n_kn_l\partial_\rho^2+
(\delta_{kl}-n_kn_l)\partial_\rho/\rho$ on radial functions,
this separates the retained tail into a history contribution
\begin{equation}\label{eq:quadrupolar-tail-history-map}
\begin{aligned}
 \frac{c^4\bar h^{[2],\mathrm{tail,history}}_{ij}}{4G}
 &\overset{\rho>d}{=}\int_{\eta_i}^{\eta-\rho/c}c\,\dd\eta'\,
 \bigg[\mathcal V^- S_{ij}
 -a^{-1}n_k\partial_\rho\mathcal V^-\,S_{ij|k}\\
 &\quad+\frac{a^{-2}}2\left(
 n_kn_l\partial_\rho^2\mathcal V^-
 +\frac{\delta_{kl}-n_kn_l}{\rho}\partial_\rho\mathcal V^-
 \right)S_{ij|kl}\bigg]
\end{aligned}
\end{equation}
and an upper-endpoint contribution
\begin{equation}\label{eq:quadrupolar-tail-endpoint-map}
\begin{aligned}
 \frac{c^4\bar h^{[2],\mathrm{tail,endpoint}}_{ij}}{4G}
 &\overset{\rho>d}{=}\bigg[\mathcal V^- a^{-1}n_kS_{ij|k}
 -\frac{\mathcal V^-}{2\rho}a^{-2}(\delta_{kl}-n_kn_l)S_{ij|kl}
 \\
 &\quad+\frac12 n_kn_l\left\{
 \frac1c\partial_{\eta'}(\mathcal V^- a^{-2}S_{ij|kl})
 -2(\partial_\rho\mathcal V^-)a^{-2}S_{ij|kl}\right\}\bigg]_{\eta'=\eta-\rho/c}.
\end{aligned}
\end{equation}
The moments and scale factors in the history integral are evaluated
at $\eta'$; endpoint values are at $\eta'=\eta-\rho/c$. Derivatives of
$\mathcal V^-$ hold its other arguments fixed.

To compare these fields with the quadrupolar solution of
\cite{Compere:2026xoe}, use $\partial_{\eta'}=a(\eta')\partial_t$
and $\partial_ta=Ha$ in the light-cone field~\eqref{eq:quadrupolar-LC-map}.
The derivatives in its $1/\rho$ term become
\begin{subequations}\label{eq:quadrupolar-time-map}
\begin{align}
\partial_{\eta'}(a^{-1}S_{ij|k})
   &=(\partial_t-H)S_{ij|k},\label{eq:dipole-time-map}\\
\partial_{\eta'}^2(a^{-2}S_{ij|kl})
   &=\left[\partial_t^2-3H\partial_t+2H^2-2\partial_tH\right]
        S_{ij|kl}.\label{eq:quadrupole-time-map}
\end{align}
\end{subequations}
Using $\partial_tH=-(1+q)H^2$ where $\cH\ne0$ then recovers
\cite[Eqs.~(5.30), (5.33)]{Compere:2026xoe}, with the STF projection
on the component indices $i,j$ for the latter equation.
For a power-law background, the tail coefficient $\mathcal V_C$
of Eq.~\eqref{eq:tailpower} and its first derivatives on the cone are
\begin{equation}\label{eq:powerlaw-cone-derivatives-map}
\begin{aligned}
 \mathcal V_C&=\frac{C}{2c^2\eta(\eta-\rho/c)},\qquad
 \partial_\rho\mathcal V_C=-\frac{C(C-2)\rho}{8c^4(\eta(\eta-\rho/c))^2},
 \\
 \partial_{\eta'}\mathcal V_C
 &=-\frac{C}{2c^2\eta(\eta-\rho/c)^2}-\frac{C(C-2)\rho}{8c^3(\eta(\eta-\rho/c))^2}.
\end{aligned}
\end{equation}
With $C=\alpha(\alpha-1)$, substituting the tail coefficient and these
cone values into the STF parts on
$i,j$ of Eqs.~\eqref{eq:quadrupolar-tail-history-map} and
\eqref{eq:quadrupolar-tail-endpoint-map} recovers
\cite[Eqs.~(5.34)--(5.35)]{Compere:2026xoe}, including the endpoint
terms. The mixed sector uses $C=\alpha(\alpha+1)$.

\section{Multipolar fields in de Sitter spacetime}
\label{sec:octupolar}
The de Sitter background gives an explicit realization of the
construction in Section~\ref{sec:multipoles}. We will first evaluate
the exact retarded source moments and metric, retaining the source's
finite light-crossing time. We will then specialize the conserved
effective multipolar sources to de Sitter, reduce their curvature tails to
endpoint terms, and compare their quadrupolar and octupolar fields
with earlier solutions. The spatial tail is present already in the
generic-source solution of \cite{deVega:1998ia} and the quadrupole
analysis of \cite{Ashtekar:2015lxa}. {We retain the
finite-distance fields and first consider the general finite-order
source construction, which allows the additional density moments
generated by conservation. We will then impose the quadrupolar
conditions of Section~\ref{sec:general-quadrupolar-check} and recover
the model of \cite{Compere:2023ktn,Compere:2026xoe}.
At octupolar order, we will require the rank-four density moment
to vanish and recover the model of \cite{Harsh:2026vmh}.}

As in Sections~\ref{sec:cartesian} and \ref{sec:multipoles}, unmarked
fields, sources and moments denote their dynamical parts, with zero
past data. We retain $\mathrm{dyn}$ only where past and dynamical
contributions appear together or are explicitly contrasted, and keep
$\mathrm{past}$ on past quantities. The superscript $[m]$ identifies
fields generated by an effective multipolar source of order $m$. The full solution includes the
prescribed past field of Eqs.~\eqref{eq:past-reference-metric}.

\subsection{Exact retarded source moments and metric}
\label{sec:dS-retarded-moments}
We begin in the expanding de Sitter patch, setting $a(\eta)=-1/(H\eta)$
with constant $H>0$ and $\eta<0$, and assign the vacuum energy to
$\Lambda$. Then $\eps=p=s=0$, $\tau_{\mu\nu}=\delta T_{\mu\nu}$,
$V^-=2/(c^2\eta^2)$ and $V^+=0$. The spatial tail coefficient is
$1/(c^2\eta\eta')$, while the mixed
field has no tail.

To evaluate the de Sitter source-time weight, we solve the radial
equation~\eqref{eq:curved-average-radial-equation} with
$V^-(\eta')=2/(c^2\eta'^2)$ and the regular central data in
Eqs.~\eqref{eq:curved-average-radial-data} and
\eqref{eq:curved-average-radial-derivative}, obtaining
\begin{equation}\label{eq:dS-source-time-kernel}
 K^{V^-}_\ell(\rho',\eta';\eta'')
 =K^0_\ell(\rho',\eta';\eta'')\left[
 1-\frac{\rho'^2-c^2(\eta''-\eta')^2}{2c^2(\ell+1)\eta''\eta'}\right].
\end{equation}
Here $K^0_\ell$ is the flat-potential kernel in
Eq.~\eqref{eq:flat-average-kernel}. The sampled characteristic diamond
lies entirely at negative conformal times.

We obtain the de Sitter retarded source moments by inserting the
kernel~\eqref{eq:dS-source-time-kernel}, with $K^0_\ell$ given by
Eq.~\eqref{eq:flat-average-kernel}, into the general
definition~\eqref{eq:curved-retarded-spatial-moment}.
The substitution $\eta'=\eta''+\rho'z/c$ gives
$c\,\dd\eta'=\rho'\dd z$ and maps the kernel support to $-1\le z\le1$.
As functions of the central conformal time $\eta$, the source-time
weight and de Sitter retarded moment are
\begin{subequations}\label{eq:dS-retarded-moments}
\begin{align}
 \delta^{\mathrm{dS}}_\ell(\eta,\rho';z)
 &:=\delta_\ell(z)\left[1-
 \frac{\rho'^2(1-z^2)}{2c^2(\ell+1)\eta(\eta+\rho'z/c)}\right],
 \label{eq:dS-retarded-weight}\\
 \mathcal F_{ij|L}(\eta;a)
 &=\int\dd^3x'\,x'_{\langle L\rangle}\int_{-1}^1\dd z\,
 \delta^{\mathrm{dS}}_\ell(\eta,\rho';z)
 a(\eta+\rho'z/c)\tau_{ij}(\eta+\rho'z/c,\bx').
 \label{eq:dS-retarded-moment}
\end{align}
\end{subequations}
In the exterior field, these moments are evaluated at central times
$\eta''\le\eta-\rho/c$. Since $\rho>d$ and $\eta<0$, every sampled
source time $\eta'=\eta''+\rho'z/c$ is negative.
The correction to the flat weight
accounts for curvature during the source's light-crossing time.
The mixed potential is $V^+=0$, so
$\mathcal F_{0i|L}(\eta;a)=\mathcal F^{\mathrm{LC}}_{0i|L}(\eta;a)$
for this de Sitter scale factor. This light-cone source moment still
averages the source over its finite light-crossing time, as in
Eq.~\eqref{eq:timemoment}.

Substituting the full de Sitter Green function~\eqref{eq:dSgreen}
into Eq.~\eqref{eq:curved-retarded-spatial-field} gives, with
$u:=\eta-\rho/c$,
\begin{equation}\label{eq:dS-exact-retarded-STF}
 \bar h_{ij}(\eta,\bx)
 \overset{\rho>d}{=}\frac{4G}{c^4}\sum_{\ell\ge0}\frac{(-1)^\ell}{\ell!}\partial_L
 \left[\frac{\mathcal F_{ij|L}(u;a)}{\rho}
 +\frac1{c\eta}\int_{-\infty}^{u}\frac{\dd\eta''}{\eta''}
       \mathcal F_{ij|L}(\eta'';a)\right].
\end{equation}
The lower limit uses the retarded source history:
$\mathcal F_{ij|L}(\eta'';a)=0$ for $\eta''<\eta_i-d/c$.
This expression retains every angular order and the complete source-time
weight; it involves no finite source-size or slow-motion approximation.
As in Eq.~\eqref{eq:curved-retarded-spatial-tail-field}, its original Hadamard
tail consists of the history integral together with the
correction
$[\mathcal F_{ij|L}(u;a)-\mathcal F^{\mathrm{LC}}_{ij|L}(u;a)]/\rho$
inside the STF derivatives.

For the stated zero-past data, the energy equation~\eqref{eq:energy}
with $s=0$ leaves $\tau_{00}$ supported in the same source region as
$\tau_{0i}$ and $\tau_{ij}$. The wave equation for $\hat h$ therefore also has a compact source.
Define $\mathcal F^{\mathrm{LC}}_{00|L}(\eta;a)$ by the time average
\eqref{eq:timemoment} with $\tau_{ij}$ replaced by $\tau_{00}/c^2$;
$\mathcal F^{\mathrm{LC}}_{kk|L}(\eta;a):=
\delta^{ij}\mathcal F^{\mathrm{LC}}_{ij|L}(\eta;a)$ is the spatial
trace of the light-cone stress moment. Since $G^+=G^{\mathrm{LC}}$,
the remaining fields are
\begin{subequations}\label{eq:dS-exact-mixed-scalar}
\begin{align}
 \bar h_{0i}(\eta,\bx)
 &\overset{\rho>d}{=}\frac{4G}{c^3}\sum_{\ell\ge0}\frac{(-1)^\ell}{\ell!}\partial_L
       \left[\frac{\mathcal F_{0i|L}(u;a)}{\rho}\right],
 \label{eq:dS-exact-mixed}\\
 \hat h(\eta,\bx)
 &\overset{\rho>d}{=}\frac{4G}{c^2}\sum_{\ell\ge0}\frac{(-1)^\ell}{\ell!}\partial_L
       \left[\frac{\mathcal F^{\mathrm{LC}}_{00|L}(u;a)+\mathcal F^{\mathrm{LC}}_{kk|L}(u;a)/c^2}{\rho}\right],
 \label{eq:dS-exact-scalar}
\end{align}
\end{subequations}
where $u=\eta-\rho/c$. The moments of $\tau_{00}$ are
computed from its conserved history; they are not additional prescribed
data. Equations~\eqref{eq:dS-exact-retarded-STF} and
\eqref{eq:dS-exact-mixed-scalar}, with
$\bar h_{00}=\hat h-\bar h_{kk}$ and Eq.~\eqref{eq:metric-final},
give all components of the dynamical exterior metric. Adding the past
field~\eqref{eq:past-reference-metric} gives the full solution.
Although $\bar h_{0i}$ and $\hat h$
have no propagation tail, $\bar h_{00}$ and the physical spatial
metric also involve $\bar h_{kk}$ and can inherit its tail.

\subsection{Finite multipolar expansions and tail reduction}
\label{sec:dS-tail-endpoints}
{We will evaluate the metric of an effective multipolar
source of arbitrary Cartesian order $m$ using the de Sitter Green
functions.} Momentum conservation will restrict the retained stresses,
while energy conservation will determine the density moments.

\paragraph{Effective multipolar sources and conservation.}
We obtain the effective multipolar source of order $m$
by replacing the upper limit $2$ by $m$ in the stress and momentum
components~\eqref{eq:quadrupolar-effective-sources}. Its source moments obey
\begin{equation}\label{eq:dS-compact-source-cutoff}
 S_{ij|L}=P_{i|L}=0,\qquad |L|>m.
\end{equation}
The momentum balance law~\eqref{eq:momentmomentum} evolves the
retained $P_{i|L}$. At rank $m+1$, it gives the closure condition
\begin{equation}\label{eq:dS-compact-source-closure}
 S_{i(j_1|j_2\cdots j_{m+1})}=0.
\end{equation}
It prevents the retained stress moments from generating a momentum
moment beyond the cutoff, as in
Eq.~\eqref{eq:quadrupolar-stress-closure}.

The density moments are determined by the energy balance
law~\eqref{eq:momentenergy} with $s=0$ and zero past data.
Since $Q^{(p)}_L=S_{kk|L}/c^2$ vanishes for $|L|>m$, its rank-$m+1$
equation is driven only by the symmetric part of the retained
momentum moment. At ranks $\ell>m+1$, both driving terms
vanish, leaving $(\partial_t-\ell H)Q^{(\rho)}_L=0$.
The zero past prescription sets $Q^{(\rho)}_L(t(\eta_i))=0$
for these dynamical density moments, so they remain zero.
The density balance therefore gives
\begin{subequations}\label{eq:dS-generated-density-moments}
\begin{align}
 [\partial_t-(m+1)H]Q^{(\rho)}_{i_1\cdots i_{m+1}}
 &=-(m+1)P_{(i_1|i_2\cdots i_{m+1})},
 \label{eq:dS-generated-density-balance}\\
 Q^{(\rho)}_L&=0,\qquad |L|>m+1.
 \label{eq:dS-generated-density-cutoff}
\end{align}
\end{subequations}
Thus the reconstructed density can contain moments of Cartesian
rank $m+1$, fixed by the retained stress and momentum moments. Imposing a density cutoff at
rank $m$ would instead restrict their symmetric momentum moment.
{The quadrupolar and octupolar conditions
\eqref{eq:quadrupolar-defining-conditions} and
\eqref{eq:octupolar-defining-conditions} remove this generated
density moment at $m=2$ and $m=3$, respectively.}

\paragraph{Field and time conventions.}
\label{sec:octupolar-dS-map}
To express the fields in the variables used in the de Sitter
comparisons, we apply the same rescaling to all dynamical components:
\begin{subequations}\label{eq:HSV-dictionary}
\begin{align}
 \chi_{\mu\nu}&:=a^{-1}\bar h_{\mu\nu},\label{eq:dS-chi}\\
 \hat\chi&:=\chi_{00}+\chi_{kk}=a^{-1}\hat h.\label{eq:dS-hatchi}
\end{align}
\end{subequations}
In \cite[Eq.~(5.40)]{Compere:2026xoe}, $\tilde\chi_{ij}$ corresponds
to our $\bar h_{ij}$. The dynamical matter perturbation obeys
$\delta T_{\mu\nu}=\tau_{\mu\nu}$, with the source moments defined
in Eqs.~\eqref{eq:source-moment-definitions}.
The physical observation radius satisfies
\begin{subequations}\label{eq:HSV-radius}
\begin{align}
 r&:=a(\eta)\rho,\label{eq:dS-radius-definition}\\
 \frac1{a(u)\rho}&=-\frac{H(\eta-\rho/c)}{\rho}
 =\frac Hc+\frac1r.\label{eq:dS-radius-identity}
\end{align}
\end{subequations}
At the retarded conformal time $u=\eta-\rho/c$, the moments are
functions of the source proper time $t(u)$. Repeated use of
$\partial_u=a(u)\partial_t$ and $\partial_ta=Ha$ gives
\begin{equation}\label{eq:dS-ordered-time-identity}
 \partial_u^p[a(u)^{-\ell}S_{ij|L}(u)]
 =a(u)^{p-\ell}\prod_{j=0}^{p-1}
       [\partial_t-(\ell-j)H]S_{ij|L}(t(u)).
\end{equation}
The constant-$H$ factors commute.

\paragraph{Conservation and tail endpoints.}
The de Sitter tail coefficient $\mathcal V^-=1/(c^2\eta\eta')$
is independent of the spatial separation. Substituting it into
the finite Cartesian expansion~\eqref{eq:cartesian-general} and
using $a(\eta)=-1/(H\eta)$ gives the retained spatial tail
\begin{equation}\label{eq:dS-finite-tail-integral}
 \frac{c^4\chi_{ij}^{[m],\mathrm{tail}}(\eta,\bx)}{4G}
 \overset{\rho>d}{=}\frac{H^2}{c}
 \sum_{\ell=0}^{m}\frac{(-1)^\ell}{\ell!}\partial_L
 \int_{\eta_i}^{u}\dd\eta'\,a(\eta')^{1-\ell}S_{ij|L}(\eta').
\end{equation}
For $\ell=0$, the rank-two energy and momentum balance laws give
\begin{equation}\label{eq:dS-stress-map}
 S_{ij}=\frac12\partial_t\left[
       (\partial_t-2H)Q^{(\rho)}_{ij}+HQ^{(p)}_{ij}\right].
\end{equation}
This identity holds without a multipolar cutoff. Using
$\dd t=a\dd\eta$ in the rank-zero tail therefore gives
\begin{equation}\label{eq:dS-tail-stress-endpoints}
\begin{aligned}
 \frac{4GH^2}{c^5}\int_{\eta_i}^{u}\dd\eta'\,a(\eta')S_{ij}(\eta')
 &=\frac{4GH^2}{c^5}\int_{t(\eta_i)}^{t(u)}\dd t'\,S_{ij}(t')\\
 &=\frac{2GH^2}{c^5}\Big[(\partial_t-2H)Q^{(\rho)}_{ij}
                       +HQ^{(p)}_{ij}\Big]_{t(\eta_i)}^{t(u)}.
\end{aligned}
\end{equation}
The lower endpoint vanishes for the dynamical part. If the integral
is written for the full source, that endpoint is retained together
with the prescribed past field.

For $\ell\ge1$, one spatial derivative of the
history integral in Eq.~\eqref{eq:dS-finite-tail-integral} gives
\begin{equation}\label{eq:dS-tail-multipole-endpoints}
 \partial_L\int_{\eta_i}^{u}\dd\eta'\,a(\eta')^{1-\ell}S_{ij|L}(\eta')
 =-\frac1c\partial_{L-1}\!\left[n_{i_\ell}a(u)^{1-\ell}S_{ij|L}(u)\right].
\end{equation}
Here $L-1=i_1\cdots i_{\ell-1}$, and the remaining derivatives act at
fixed $\eta$ on both $n_{i_\ell}$ and $u=\eta-\rho/c$. The identity uses
$\partial_i u=-n_i/c$ and the spatial independence of the lower limit.
Thus every nonzero-rank tail is an upper-endpoint contribution,
while Eq.~\eqref{eq:dS-tail-stress-endpoints} reduces the stress-monopole
tail by conservation. The entire retained tail consists of endpoint
terms, which can remain nonzero. The exact retarded source moments
in Eq.~\eqref{eq:dS-exact-retarded-STF} retain the finite source-time
spread; their monopole cannot simply be replaced by $S_{ij}$.

\paragraph{The retained de Sitter field through order $m$.}
Combining the light-cone part of the finite Cartesian
expansion~\eqref{eq:cartesian-general} with the endpoint identities
\eqref{eq:dS-tail-stress-endpoints} and
\eqref{eq:dS-tail-multipole-endpoints} gives
\begin{equation}\label{eq:dS-finite-order-field}
\begin{aligned}
 \frac{c^4\chi^{[m]}_{ij}(\eta,\bx)}{4G}
 &\overset{\rho>d}{=}\frac1{a(\eta)}\sum_{\ell=0}^{m}\frac{(-1)^\ell}{\ell!}\partial_L
       \left[\frac{a(u)^{-\ell}S_{ij|L}(u)}{\rho}\right]\\
 &\quad+\frac{H^2}{2c}\Big[(\partial_t-2H)Q^{(\rho)}_{ij}
                          +HQ^{(p)}_{ij}\Big]_{t(\eta_i)}^{t(u)}\\
 &\quad+\frac{H^2}{c^2}\sum_{\ell=1}^{m}\frac{(-1)^{\ell+1}}{\ell!}\partial_{L-1}
       \left[n_{i_\ell}a(u)^{1-\ell}S_{ij|L}(u)\right].
\end{aligned}
\end{equation}
The second and third lines give the tail contribution. An empty sum
vanishes, and spatial derivatives act at fixed observation time,
including on $u$ and $n_i$.
Since $G^+=G^{\mathrm{LC}}$, the mixed field and the solution of the
$\hat h$ wave equation with the reconstructed density give
\begin{subequations}\label{eq:dS-finite-mixed-scalar}
\begin{align}
 \chi^{[m]}_{0i}(\eta,\bx)
 &\overset{\rho>d}{=}\frac{4G}{c^3a(\eta)}
 \sum_{\ell=0}^{m}\frac{(-1)^\ell}{\ell!}\partial_L
 \left[\frac{a(u)^{-\ell}P_{i|L}(u)}\rho\right],
 \label{eq:dS-finite-mixed}\\
 \hat\chi^{[m]}(\eta,\bx)
 &\overset{\rho>d}{=}\frac{4G}{c^2a(\eta)}
 \sum_{\ell=0}^{m+1}\frac{(-1)^\ell}{\ell!}\partial_L
 \left[\frac{a(u)^{-\ell}[Q^{(\rho)}_L(u)+Q^{(p)}_L(u)]}\rho\right].
 \label{eq:dS-finite-scalar}
\end{align}
\end{subequations}
The last term in $\hat\chi^{[m]}$ has Cartesian
rank $m+1$. At that rank, $Q^{(p)}_L=S_{kk|L}/c^2=0$ by the
cutoff~\eqref{eq:dS-compact-source-cutoff}; only the density moment
$Q^{(\rho)}_L$ contributes, as determined by
Eq.~\eqref{eq:dS-generated-density-balance}.
This solution of the $\hat h$ wave equation is the temporal-gauge
reconstruction of Section~\ref{sec:retarded-reconstruction}, since
both obey the same field equation and zero past data. Together
with $\chi^{[m]}_{00}=\hat\chi^{[m]}-\chi^{[m]}_{kk}$, these formulas
give the full dynamical metric generated by the effective multipolar source.

The condition $\ell+2p\le m$ in
Eq.~\eqref{eq:curved-retarded-moment-finite} retains Cartesian traces
together with their STF parts. When the finite Cartesian sum is
used as a source-size approximation to an extended source under the
assumptions of Section~\ref{sec:finite-source-size}, the spatial
field has the remainder $\mathcal R_m/a(\eta)$ of
Eq.~\eqref{eq:cartesian-general}. The prescribed past field is added
as in Section~\ref{sec:general-quadrupolar-check}; the density-rank
bound~\eqref{eq:dS-generated-density-cutoff} concerns the dynamical
part with zero past data.

\subsection{Quadrupolar truncation}
\label{sec:quadrupolar-map}
{We will specialize the quadrupolar truncation of
Section~\ref{sec:general-quadrupolar-check} to de Sitter and recover
the model previously studied by the authors in
\cite{Compere:2023ktn,Compere:2026xoe}.

We start with the effective sources
\eqref{eq:quadrupolar-effective-sources}, the conservation conditions
\eqref{eq:quadrupolar-momentum-closure}, and the vanishing density octupole
\eqref{eq:quadrupolar-defining-conditions}.
Since $s=0$, Eq.~\eqref{eq:quadrupolar-compensating-momentum} gives
$P_{(i|jk)}=0$. Symmetrizing
Eq.~\eqref{eq:quadrupolar-momentum-evolution} then gives
$S_{(ij|k)}=0$. Together with $S_{i(j|kl)}=0$, these are the
conservation constraints of
\cite[Eqs.~(4.21)--(4.22)]{Compere:2026xoe}.
Equation~\eqref{eq:dS-generated-density-cutoff} also eliminates
all higher density moments for zero past data, so the density obeys
\begin{equation}\label{eq:dS-quadrupolar-generated-density}
 Q^{(\rho)}_L=0,\qquad \ell>2.
\end{equation}
Tracing $P_{(i|jk)}=0$ relates the momentum traces without
eliminating them:
\begin{equation}\label{eq:dS-quadrupolar-momentum-traces}
 P_{l|il}=-\frac12P_{i|ll},\qquad \Pi_i=2P_{i|ll}.
\end{equation}
The momentum trace $P_{i|ll}$ and the pressure dipole related to it
by Eq.~\eqref{eq:quadrupolar-momentum-trace-balance} remain allowed,
as in \cite{Compere:2023ktn,Compere:2026xoe}.}
The stress monopole remains related to the density and pressure
quadrupoles by Eq.~\eqref{eq:dS-stress-map}.

Setting $\partial_tH=0$ in the light-cone field and time conversions
\eqref{eq:quadrupolar-LC-map} and \eqref{eq:quadrupolar-time-map},
and applying the rescaling~\eqref{eq:dS-chi}, gives
\begin{equation}\label{eq:dS-quadrupolar-LC}
\begin{aligned}
 \frac{c^4\chi^{[2],\mathrm{LC}}_{ij}}{4G}
 &\overset{\rho>d}{=}\frac1{a(\eta)\rho}\left[S_{ij}
 +\frac{n_k}{c}(\partial_t-H)S_{ij|k}
 +\frac1{2c^2}n_kn_l(\partial_t-H)(\partial_t-2H)S_{ij|kl}\right]\\
 &\quad+\frac1{a(\eta)a(u)\rho^2}\left[n_kS_{ij|k}
 +\frac1{2c}(3n_kn_l-\delta_{kl})(\partial_t-2H)S_{ij|kl}\right]\\
 &\quad+\frac{3n_kn_l-\delta_{kl}}
                  {2a(\eta)a(u)^2\rho^3}S_{ij|kl}.
\end{aligned}
\end{equation}
Here the moments are evaluated at $u=\eta-\rho/c$, and $\partial_t$
acts at the corresponding source proper time $t(u)$.
For $\mathcal V^-=1/(c^2\eta\eta')$, the tail formulas
\eqref{eq:quadrupolar-tail-history-map} and
\eqref{eq:quadrupolar-tail-endpoint-map} reduce to the endpoint terms
in Eq.~\eqref{eq:dS-finite-order-field}. At $m=2$, they give
\begin{equation}\label{eq:dS-quadrupolar-tail}
\begin{aligned}
 \frac{c^4\chi^{[2],\mathrm{tail}}_{ij}}{4G}
 \overset{\rho>d}{=}\frac{H^2}{c^2}\Bigg\{&\frac c2\Big[(\partial_t-2H)Q^{(\rho)}_{ij}
                         +HQ^{(p)}_{ij}\Big]_{t(\eta_i)}^{t(u)}
 +n_kS_{ij|k}(u)\\
 &+\frac1{2c}n_kn_l(\partial_t-H)S_{ij|kl}(u)
 -\frac{\delta_{kl}-n_kn_l}{2a(u)\rho}S_{ij|kl}(u)\Bigg\}.
\end{aligned}
\end{equation}
The first term uses Eq.~\eqref{eq:dS-tail-stress-endpoints}; the
remaining terms follow from Eq.~\eqref{eq:dS-tail-multipole-endpoints}
at ranks one and two. The angular part of the second spatial
derivative supplies the last term. No history integral remains,
and the prescribed past field retains the pre-existing source contribution.

Replacing the stress moments in Eq.~\eqref{eq:dS-quadrupolar-LC}
by $P_{i|L}$ and its prefactor $4G/c^4$ by $4G/c^3$ gives
$\chi^{[2]}_{0i}$. The temporal field is given by
{Eq.~\eqref{eq:dS-finite-scalar} with $m=2$ and the
density cutoff~\eqref{eq:dS-quadrupolar-generated-density}:
\begin{equation}\label{eq:dS-quadrupolar-scalar}
 \hat\chi^{[2]}(\eta,\bx)
 \overset{\rho>d}{=}\frac{4G}{c^2a(\eta)}
 \sum_{\ell=0}^{2}\frac{(-1)^\ell}{\ell!}\partial_L
 \left[\frac{a(u)^{-\ell}[Q^{(\rho)}_L(u)+Q^{(p)}_L(u)]}\rho\right].
\end{equation}
Both the mixed field and $\hat\chi^{[2]}$ use the light-cone Green
function $G^+$. Their source moments obey the conservation laws
above, and $\chi^{[2]}_{00}=\hat\chi^{[2]}-\chi^{[2]}_{kk}$
completes the metric. The exterior shifted density vanishes by
Eq.~\eqref{eq:quadrupolar-density-reconstruction}, since $s=0$.

The spatial light-cone and tail fields recover the de Sitter specialization of
\cite[Eqs.~(5.30), (5.33)--(5.35)]{Compere:2026xoe}:}
$q=-1$, the inverse-square spatial potential coefficient is $C=2$,
and the tail coefficient is $1/(c^2\eta\eta')$. Their sum gives
\cite[Eq.~(5.40)]{Compere:2026xoe} after using
$S_{ij}=-\partial_tP_{(i|j)}$ and Eq.~\eqref{eq:dS-radius-identity},
{recovering also the construction of \cite{Compere:2023ktn}.}

{Equations~\eqref{eq:dS-quadrupolar-LC}--\eqref{eq:dS-quadrupolar-tail},
together with the corresponding $\chi^{[2]}_{0i}$ and
$\chi^{[2]}_{00}$, give the complete metric for the quadrupolar
truncation defined by Eq.~\eqref{eq:quadrupolar-defining-conditions}.}

\subsection{Octupolar truncation}
We will define the octupolar truncation by truncating the stress,
momentum and density moments at Cartesian octupolar order ($\ell=3$).
We will derive the resulting conservation
constraints, evaluate the cubic field contributions and recover the
model of Harsh, Shetye and Virmani~\cite{Harsh:2026vmh}.

\paragraph{Conservation constraints and octupolar source content.}
\label{sec:octupolar-constraints}
We start from an effective multipolar source of Cartesian order
$m=3$, whose stress and momentum moments obey
\begin{equation}\label{eq:octupolar-cutoff}
 S_{ij|L}=P_{i|L}=0,\qquad |L|>3.
\end{equation}
Its momentum-conservation closure condition is
\begin{equation}\label{eq:octupolar-closure}
 S_{i(j|klm)}=0.
\end{equation}
The energy balance law~\eqref{eq:momentenergy} at rank four gives
\begin{equation}\label{eq:dS-octupolar-generated-density}
 (\partial_t-4H)Q^{(\rho)}_{ijkl}=-4P_{(i|jkl)}.
\end{equation}
We define the octupolar truncation by the cutoff
\eqref{eq:octupolar-cutoff}, the conservation laws and a vanishing
full Cartesian rank-four density moment throughout the dynamical history:
\begin{equation}\label{eq:octupolar-defining-conditions}
 Q^{(\rho)}_{ijkl}=0.
\end{equation}
With zero past data, Eq.~\eqref{eq:dS-octupolar-generated-density}
makes this condition equivalent to the vanishing of the fully
symmetric momentum moment. The fully symmetric part of the momentum
balance law~\eqref{eq:momentmomentum} at rank three then sets the
corresponding stress moment to zero:
\begin{equation}\label{eq:octupolar-comparison-constraints}
 P_{(i|jkl)}=0,\qquad S_{(ij|kl)}=0.
\end{equation}
Higher density moments already vanish by
Eq.~\eqref{eq:dS-generated-density-cutoff}. Thus all Cartesian
source moments above rank three vanish, including their traces.

The closure condition~\eqref{eq:octupolar-closure} fixes contractions
of the rank-three stress moments:
\begin{equation}\label{eq:octupolar-stress-traces}
 n_i n_k n_l n_m S_{ij|klm}=0,\qquad
 S_{(ij|k)ll}=\frac{c^2}{3}Q^{(p)}_{ijk},
\end{equation}
as in \cite[Eqs.~(5.14)--(5.15)]{Harsh:2026vmh}. The first identity is a
contraction of $S_{j(i|klm)}=0$. To derive the second, trace
$S_{i(j|klm)}=0$ on two symmetrized indices and then symmetrize the
remaining $i,j,k$; this gives
$S_{l(i|jk)l}=-S_{(ij|k)ll}$.
Tracing instead the free index against one symmetrized index gives
$3S_{l(i|jk)l}+c^2Q^{(p)}_{ijk}=0$.
These two equations prove the stated trace relation. Thus the trace
of a rank-three stress moment is in general nonzero.
We use \emph{electric} for mass-type irreducible multipoles, built
from density and pressure moments, and \emph{magnetic} for current-type
multipoles, obtained by contracting momentum moments with
$\epsilon_{ijk}$ and taking their STF part. At lower ranks, the energy
and momentum balance laws give
\begin{subequations}\label{eq:octupolar-conservation-relations}
\begin{align}
 P_{(i|jk)}
 &=-\frac13\left[(\partial_t-3H)Q^{(\rho)}_{ijk}
                         +H Q^{(p)}_{ijk}\right],
 \label{eq:octupolar-electric-P}\\
 S_{(ij|k)}
 &=\frac16(\partial_t-H)
       \left[(\partial_t-3H)Q^{(\rho)}_{ijk}
                         +H Q^{(p)}_{ijk}\right],
 \label{eq:octupolar-electric-S}\\
 S_{i(j|kl)}&=-\frac13(\partial_t-2H)P_{i|jkl}.
 \label{eq:octupolar-magnetic-S}
\end{align}
\end{subequations}
All three right-hand sides can be nonzero. We recover the relations
of \cite[Eqs.~(5.6), (5.10)--(5.13)]{Harsh:2026vmh} directly from
the balance laws. In particular, $S_{(ij|k)}$ contains electric
octupolar information despite having only one source-position index.

The rank-three electric tensors are the STF parts of
$Q^{(\rho)}_{ijk}$ and $Q^{(p)}_{ijk}$. We define the magnetic octupole
and obtain its balance law as
\begin{subequations}\label{eq:current-octupole-relations}
\begin{align}
 J_{ijk}&:=\frac32\STF_{ijk}\!\left[\epsilon_{iab}P_{a|bjk}\right],
 \label{eq:current-octupole}\\
 (\partial_t-2H)J_{ijk}
 &=-3\STF_{ijk}\!\left[\epsilon_{iab}S_{aj|bk}\right].
 \label{eq:current-octupole-balance}
\end{align}
\end{subequations}
Equation~\eqref{eq:current-octupole-balance} follows by differentiating
the definition~\eqref{eq:current-octupole} and substituting the momentum
balance law~\eqref{eq:momentmomentum} at $|L|=3$.
Symmetry of $S_{ab|jk}$ removes its
contraction with $\epsilon_{iab}$, and the two other terms coincide
after STF projection. Our magnetic octupole is $3/2$ times that of
\cite[Eq.~(5.3)]{Harsh:2026vmh}; with this conversion, its balance law
reproduces their Eq.~(5.8). The magnetic octupole $J_{ijk}$ is thus present
in both $P_{i|jkl}$ and $S_{ij|kl}$. Its definition in
Eq.~\eqref{eq:current-octupole} contracts a component index with a
multipole index before taking the STF projection. This illustrates
the decomposition of component and multipole indices together into
irreducible representations of total angular momentum, used to define
mass and current multipoles in \cite[Secs.~II and V]{Damour:1990gj}.
It differs from the projection on $L$ alone in
Eqs.~\eqref{eq:radialmoments-traces}. The exterior component-field
expansions in Eqs.~\eqref{eq:curved-retarded-fields} already include
both parity sectors without requiring this further decomposition.

\paragraph{Explicit cubic terms.}
We now set $m=3$ in Eq.~\eqref{eq:dS-finite-order-field} and display
the terms with three source-position indices. These must be combined
with the lower-rank contributions, subject to the momentum closure
condition~\eqref{eq:octupolar-closure}, the constraints
\eqref{eq:octupolar-comparison-constraints} and the balance laws
\eqref{eq:octupolar-conservation-relations}. We denote this selection
by $[\cdot]_{|L|=3}$; it is distinct from the complete angular
$\ell=3$ sector.

The cubic light-cone contribution follows by differentiating the
$n=3$ term in the first line of Eq.~\eqref{eq:dS-finite-order-field}.
After the rescaling~\eqref{eq:dS-chi}, we use the derivative
identity~\eqref{eq:dS-ordered-time-identity} and the radius
relations~\eqref{eq:HSV-radius} to obtain
\begin{equation}\label{eq:octupolar-LC-block}
\begin{aligned}
 [\chi^{[3],\mathrm{LC}}_{ij}]_{|L|=3}
 &\overset{\rho>d}{=}\frac{2G}{c^7r}\left[
 \frac13n_kn_ln_m(\partial_t^3-4H^2\partial_t)S_{ij|klm}
 -H n_k(\partial_t^2-2H\partial_t)S_{ij|kll}\right]\\
 &\quad+\frac{2G}{c^6r^2}\left[
 n_kn_ln_m(2\partial_t^2-3H^2)S_{ij|klm}
 -n_k(\partial_t^2+H\partial_t-3H^2)S_{ij|kll}\right]\\
 &\quad+\frac{2G}{c^5r^3}\left[
 5n_kn_ln_m\partial_t S_{ij|klm}-3n_k\partial_t S_{ij|kll}\right]\\
 &\quad+\frac{2G}{c^4r^4}\left[5n_kn_ln_mS_{ij|klm}-3n_kS_{ij|kll}\right].
\end{aligned}
\end{equation}
The moments in Eq.~\eqref{eq:octupolar-LC-block} are evaluated at
source proper time $t(u)$, with $u=\eta-\rho/c$.
The derivatives $\partial_t$ act only on the moments, holding the
physical observation radius $r=a(\eta)\rho$ of
Eq.~\eqref{eq:dS-radius-definition} and the direction $n_i$ fixed.

The mixed Green function $G^+$ in Eq.~\eqref{eq:dSgreen} has no tail.
Replacing $S_{ij|L}$ by $P_{i|L}$ in
Eq.~\eqref{eq:octupolar-LC-block} therefore gives the complete cubic
mixed field, with moments evaluated at the same source time $t(u)$:
\begin{equation}\label{eq:HSV-vector-octupole-map}
\begin{aligned}
 [\chi^{[3]}_{0i}]_{|L|=3}
 &\overset{\rho>d}{=}\frac{2G}{c^6r}\left[
 \frac13(\partial_t^3-4H^2\partial_t)n_kn_ln_m P_{i|klm}
 -H(\partial_t^2-2H\partial_t)n_kP_{i|kll}\right]\\
 &\quad+\frac{2G}{c^5r^2}\left[
 (2\partial_t^2-3H^2)n_kn_ln_mP_{i|klm}
 -(\partial_t^2+H\partial_t-3H^2)n_kP_{i|kll}\right]\\
 &\quad+\frac{2G}{c^4r^3}\partial_t
          \left[5n_kn_ln_mP_{i|klm}-3n_kP_{i|kll}\right]
 +\frac{2G}{c^3r^4}\left[5n_kn_ln_mP_{i|klm}-3n_kP_{i|kll}\right].
\end{aligned}
\end{equation}
The replacement $P_{i|L}\mapsto Q^{(\rho)}_L+Q^{(p)}_L$ gives
$[\hat\chi^{[3]}]_{|L|=3}$ after multiplying the prefactors by $c$.
The complete $\hat\chi^{[3]}$ follows from
Eq.~\eqref{eq:dS-finite-scalar} with the sum ending at $\ell=3$,
since the rank-four density moment vanishes by
Eq.~\eqref{eq:octupolar-defining-conditions}.

For the spatial sector, the cubic tail is already reduced to an
endpoint by Eq.~\eqref{eq:dS-finite-order-field}:
\begin{equation}\label{eq:dS-cubic-tail-endpoint}
 [\chi^{[3],\mathrm{tail}}_{ij}]_{|L|=3}
 \overset{\rho>d}{=}\frac{2GH^2}{3c^6}\partial_{kl}
       \left[n_m a(u)^{-2}S_{ij|klm}(u)\right].
\end{equation}
Evaluating these two derivatives and adding the light-cone
contribution~\eqref{eq:octupolar-LC-block} gives, with moments at $t(u)$,
\begin{equation}\label{eq:HSV-tensor-octupole-map}
\begin{aligned}
 [\chi^{[3]}_{ij}]_{|L|=3}
 &\overset{\rho>d}{=}\frac{2G}{3c^8}\left(H^2+\frac{c\partial_t}{r}\right)
 (\partial_t-H)
 \left[(\partial_t+H)n_kn_ln_mS_{ij|klm}
                         -3H n_kS_{ij|kll}\right]\\
 &\quad+\frac{2G}{c^6r^2}(\partial_t-H)
 \left[2(\partial_t+H)n_kn_ln_mS_{ij|klm}
                  -(\partial_t+2H)n_kS_{ij|kll}\right]\\
 &\quad+\frac{2G}{c^5r^3}\partial_t
       \left[5n_kn_ln_mS_{ij|klm}-3n_kS_{ij|kll}\right]
 +\frac{2G}{c^4r^4}\left[5n_kn_ln_mS_{ij|klm}-3n_kS_{ij|kll}\right].
\end{aligned}
\end{equation}
\paragraph{Complete octupolar metric.}
Adding the cubic terms in Eqs.~\eqref{eq:HSV-tensor-octupole-map}
and \eqref{eq:HSV-vector-octupole-map} to the lower-rank terms in
Eqs.~\eqref{eq:dS-finite-order-field} and \eqref{eq:dS-finite-mixed}
gives $\chi^{[3]}_{ij}$ and $\chi^{[3]}_{0i}$.
The temporal component $\chi^{[3]}_{00}=\hat\chi^{[3]}-\chi^{[3]}_{kk}$
follows from Eq.~\eqref{eq:dS-finite-scalar} with the sum ending at
$\ell=3$. These fields reproduce the octupolar solution of Harsh,
Shetye and Virmani~\cite{Harsh:2026vmh}, whose source truncation in
their Eq.~(4.1) coincides with the definition adopted here.

Equation~\eqref{eq:HSV-vector-octupole-map} and the
corresponding cubic contribution to $\hat\chi^{[3]}$ recover
\cite[Eqs.~(4.21)--(4.22)]{Harsh:2026vmh}, with $G$ and $c$ restored.
The first line of Eq.~\eqref{eq:HSV-tensor-octupole-map} combines the
$r^0$ and $r^{-1}$ contributions of
\cite[Eqs.~(4.36)--(4.37)]{Harsh:2026vmh}; the remaining lines give
their $r^{-2}$, $r^{-3}$ and $r^{-4}$ terms.
The electric and magnetic octupoles in
Eqs.~\eqref{eq:octupolar-electric-S} and
\eqref{eq:current-octupole-balance} also enter the lower-rank terms
of Eq.~\eqref{eq:dS-finite-order-field}. Projecting the complete
retained field gives the spherical-harmonic organization of
\cite[Sec.~5 and Appendices~B--D]{Harsh:2026vmh}.

Setting the Cartesian rank-three stress, momentum and density
moments to zero reduces this model to the quadrupolar truncation of
Section~\ref{sec:quadrupolar-map}. The momentum and energy balance
laws then give $S_{i(j|kl)}=0$ and $P_{(i|jk)}=0$, respectively,
while allowing the momentum traces in
Eq.~\eqref{eq:dS-quadrupolar-momentum-traces}.

\paragraph{Global tail and comparison conventions.}
The light-cone and tail split used in \cite{Harsh:2026vmh}
redistributes an endpoint term relative to our Hadamard decomposition.
With $R:=|\bx-\bx'|$ and $u=\eta-\rho/c$, the two terms of the
Green function~\eqref{eq:dSgreen} give the light-cone and tail
contributions to the rescaled spatial field:
\begin{align}\label{eq:dS-pre-IBP}
 \chi_{ij}\overset{\rho>d}{=}\frac{4G}{c^4}\int\dd^3x'\bigg[
 \frac{\eta}{R(\eta-R/c)}\delta T_{ij}(\eta-R/c,\bx')
 +\frac1c\int_{\eta_i}^{\eta-R/c}\frac{\dd\eta'}{\eta'^2}
                        \delta T_{ij}(\eta',\bx')\bigg].
\end{align}
For the zero-past source used here, integration by parts yields
\begin{align}\label{eq:dS-post-IBP}
 \chi_{ij}\overset{\rho>d}{=}\frac{4G}{c^4}\int\dd^3x'\bigg[
 \frac{\delta T_{ij}(\eta-R/c,\bx')}{R}
 +\frac1c\int_{\eta_i}^{\eta-R/c}\frac{\dd\eta'}{\eta'}
                     \partial_{\eta'}\delta T_{ij}(\eta',\bx')
 \bigg].
\end{align}
Integration by parts in Eq.~\eqref{eq:dS-pre-IBP} gives the boundary term
$-\delta T_{ij}(\eta-R/c,\bx')/[c(\eta-R/c)]$ at the upper limit
$\eta'=\eta-R/c$. Adding it to the light-cone integrand replaces
the factor $\eta/[R(\eta-R/c)]$ by $1/R$. The lower endpoint vanishes
for $\delta T^{\mathrm{dyn}}_{ij}$. Applying the same integration by parts
to the full source $\delta T^{\mathrm{past}}_{ij}+\delta T^{\mathrm{dyn}}_{ij}$
retains the initial contribution
$(4G/c^5)\int\dd^3x'\,\delta T^{\mathrm{past}}_{ij}(\eta_i,\bx')/\eta_i$.
Integration by parts thus redistributes an endpoint term between
the two contributions while leaving their sum unchanged.

Keeping the cubic terms in Eq.~\eqref{eq:dS-post-IBP} reproduces their separate
Eqs.~(4.27) and (4.34), whose sum is
Eq.~\eqref{eq:HSV-tensor-octupole-map}. Splitting its last integral
at $u$ isolates the contribution with common upper limit:

\begin{align}\label{eq:dS-global-tail-map}
 \frac{4G}{c^5}\int\dd^3x'\int_{\eta_i}^{u}\frac{\dd\eta'}{\eta'}
          \partial_{\eta'}\delta T_{ij}(\eta',\bx')
 =-\frac{2GH}{c^5}\bigg[(\partial_t-H)
 \left((\partial_t-2H)Q^{(\rho)}_{ij}+H Q^{(p)}_{ij}\right)
          \bigg]_{t(\eta_i)}^{t(u)}.
\end{align}
Here we used $\int\dd^3x'\partial_{\eta'}\delta T_{ij}
=(\partial_t-H)S_{ij}$, $\dd\eta'/\eta'=-H\dd t$ and
Eq.~\eqref{eq:dS-stress-map}.
No octupolar or quadrupolar closure conditions were required for this identity.
Taking $\eta_i\to-\infty$ and identifying
$u=\eta_{\rm ret}$, Eq.~\eqref{eq:dS-global-tail-map} agrees with
their global-tail expression \cite[Eqs.~(2.79)--(2.80)]{Harsh:2026vmh},
with our source moments and the factors of $G$ and $c$ restored.
The lower-endpoint contribution is their $\chi_{ij}(-\infty)$,
which they explicitly retain; it vanishes for the zero-past dynamical source.
This moment boundary term is distinct from the term proportional to
$T_{ij}/\eta'$ discarded in the earlier integration by parts,
under the assumption that $T_{ij}$ remains finite as $\eta'\to-\infty$
\cite[Sec.~2.4]{Harsh:2026vmh}.
Splitting Eq.~\eqref{eq:dS-post-IBP} at $u$ then recovers their three
terms $\chi^{(I)}_{ij}$, $\chi^{(II)}_{ij}$ and $\chi^{(III)}_{ij}$
\cite[Eqs.~(2.62)--(2.65)]{Harsh:2026vmh}.\footnote{This
splitting was already used for the quadrupolar truncation in
\cite[Sec.~2.5, Eqs.~(2.81)--(2.82)]{Compere:2023ktn}.}
Their $\chi^{(I)}_{ij}$ includes the upper boundary term of the tail
integral and therefore differs from our rescaled light-cone field
alone.

Their fields $\chi_{\mu\nu},\hat\chi$
\cite[Eqs.~(2.2)--(2.8)]{Harsh:2026vmh} use the rescaling in
Eqs.~\eqref{eq:HSV-dictionary}. Applied to our dynamical source,
their moments $Q^{(\rho)}_L,P_{i|L},S_{ij|L}$ agree with ours at $c=1$, and their
$Q^{(\rho+p)}_L$ is $Q^{(\rho)}_L+Q^{(p)}_L$.
We retain Newton's constant, which they set to one from Section~2.3
onwards. Their $\eta_{\rm ret}$ is our $u$, and their physical
radius is $r$ in Eqs.~\eqref{eq:HSV-radius}. In their Bondi-coordinate
expressions, differentiation with respect to Bondi retarded time
agrees with $\partial_t$ on the source moments; that coordinate
is distinct from our conformal retarded time $u$.

Pure de Sitter spacetime separates two effects present in a general
FLRW background: curvature produces a tail even though the background
enthalpy density vanishes. With the vacuum energy assigned to
$\Lambda$, the reconstructed density remains confined to the compact
source and no exterior cosmic fluid perturbation is induced. The
prescribed past field retains pre-existing mass and current multipoles.
The canonical reduction in the next section will identify the
invariant mass, current and cosmic fluid content of the general FLRW
solution and its de Sitter specialization.

\section{Canonical metric multipoles and cosmic fluid response}
\label{sec:direct-gauge-invariant-multipoles}
We start from the complete retarded metric and cosmic fluid
perturbations constructed from the source history and past data of
Section~\ref{sec:source-data}. We call this solution the source
representative; it satisfies generalized harmonic gauge and has
$\tau_{ij}=\tau_{0i}=0$ outside the source world tube. We will first
express its metric coefficients in the Cartesian-derivative STF basis
of Blanchet--Damour \cite[Eqs.~(2.25)--(2.33)]{Blanchet:1985sp}
in terms of the retarded source moments. We will then form six
gauge-invariant combinations and derive their relations from the
harmonic conditions and exterior equations. This determines the
profiles entering the canonical metric without a spatial
scalar--vector--tensor decomposition.

The temporal residual freedom involves the operator
$\square-3V^+$, whereas the sourced metric is reconstructed with
$\Op^\pm$. Our aim is to express the canonical fields through their
invariant profiles and the two metric propagators, without solving
a separate propagation problem for the temporal gauge freedom.
We will obtain the canonical representative by an explicit gauge
transformation of the retarded solution, acting on both the metric and
cosmic fluid perturbations according to
Eqs.~\eqref{eq:rescaled-metric-gauge-action} and
\eqref{eq:shifted-source-gauge}. Its metric retains the three spatial
harmonic conditions and has $\bar h_{kk}=0$. We call its gauge the
canonical harmonic gauge.

The invariant data will be expressed through the canonical mass
moments $M_L$, current moments $J_L$ and canonical cosmic fluid
moment $N_L$. The latter generates the fluid response $R_L$. This response is
controlled by the background enthalpy density and vanishes in
Minkowski and pure de Sitter spacetime. We will derive the source formulas and past
contributions before constructing the canonical metric in
Section~\ref{sec:canonical-gauge} and deriving its cosmic fluid
perturbations in Section~\ref{sec:canonical-fluid}. Finally,
the lowest multipoles in Section~\ref{sec:canonical-fluid-and-lowest-multipoles} will determine
the conserved linear and angular momentum charges, as well as the mass
and center of mass, which generally evolve on an FLRW background.

From this section onward, fields and profiles include both parts of
Eq.~\eqref{eq:past-dynamical-split}. The spatial and mixed source moments also
denote the sums of their past and dynamical contributions in
Eqs.~\eqref{eq:full-multipolar-fields}. All coefficient families
inherit the same split, and $\mathcal G_\pm$ include their past
terms by Eq.~\eqref{eq:metric-propagator-definition}.
The reduction below is established at every STF rank and for finite
multipolar sums; convergence of the infinite multipolar sums will not be addressed. 

\subsection{Irreducible coefficients of the existing metric expansion}
{
We start from the STF decomposition of the exterior metric, which
defines the metric coefficient profiles $A_L,\ldots,I_L,K_L$:
\begin{subequations}\label{eq:BD-FLRW-metric}
\begin{align}
 \frac{c^2\bar h_{00}}{4G}&=\sum_{\ell\ge0}\partial_L A_L,
 \label{eq:BD-FLRW-00}\\
 \frac{c^3\bar h_{0i}}{4G}&=\sum_{\ell\ge0}\partial_{iL}B_L
 +\sum_{\ell\ge1}\big(\partial_{L-1}C_{iL-1}
                  +\epsilon_{iab}\partial_{aL-1}D_{bL-1}\big),
 \label{eq:BD-FLRW-0i}\\
 \frac{c^4\bar h_{ij}}{4G}&=\begin{aligned}[t]
 &\sum_{\ell\ge0}\big(\partial_{ijL}E_L+\delta_{ij}\partial_L F_L\big)\\
 &+\sum_{\ell\ge1}\big(\partial_{(i|L-1}G_{|j)L-1}
       +\epsilon_{ab(i}\partial_{j)aL-1}H_{bL-1}\big)\\
 &+\sum_{\ell\ge2}\big(\partial_{L-2}I_{ijL-2}
       +\partial_{aL-2}\big(\epsilon_{ab(i}K_{j)bL-2}\big)\big),
 \end{aligned}\label{eq:BD-FLRW-ij}
\end{align}
\end{subequations}
where all coefficient profiles depend on $(\eta,\rho)$ and their
free multipole indices are STF. We use the Blanchet--Damour ordering
$A_L,\ldots,I_L,K_L$, replacing their last letter by $K_L$ to reserve
$J_L$ for the canonical current moments.
We will now relate these metric coefficients to the source moments
defined in Eqs.~\eqref{eq:curved-retarded-moments}, using the complete
retarded fields in Eqs.~\eqref{eq:full-multipolar-fields}.

The spatial field in Eq.~\eqref{eq:full-multipolar-spatial} uses
$\mathcal G_-$ and the Green function $G^-$, whereas the mixed field
in Eq.~\eqref{eq:full-multipolar-mixed} uses $\mathcal G_+$ and $G^+$.
Their coefficients therefore form two families, which we collect as
$\boldsymbol P^-_L(\eta,\rho):=(E_L,F_L,G_L,H_L,I_L,K_L)$ evaluated on $(\eta,\rho)$ and
$\boldsymbol P^+_L(\eta,\rho):=(B_L,C_L,D_L)$ evaluated on $(\eta,\rho)$.
At the lowest ranks, we omit $C_L,D_L,G_L,H_L$ for $\ell=0$ and
$I_L,K_L$ for $\ell=0,1$.

To identify these coefficients, we decompose the component and
multipole indices of the source moments together. The algebraic
projections use symmetrization, the STF projector~\eqref{eq:STFprojector},
and contractions with $\delta_{ij}$ and $\epsilon_{ijk}$ to select
the derivative structures in Eqs.~\eqref{eq:BD-FLRW-metric}.
Because $G^\pm$ carry no tensor indices, these operations commute
with the source-time integrals. Uncontracted observation derivatives
remain outside the integrals. Each contracted pair gives a Laplacian;
repeated application of Eq.~\eqref{eq:canonical-propagation-trace} gives
\begin{equation}\label{eq:BD-repeated-trace-propagation}
 \Delta^k\mathcal G_\pm[f](\eta,\rho)
 =\mathcal G_\pm[T_\pm^k f](\eta,\rho),
 \qquad k=0,1,\ldots,\quad\rho>0.
\end{equation}
Thus a derivative trace changes the source history through $T_\pm$
while retaining the same metric propagator.

For each derivative structure, we collect its projected source moments
with the normalization of Eqs.~\eqref{eq:full-multipolar-fields} and
the operators $T_\pm$ generated by traces.
Linearity combines them into a single history propagated by
$\mathcal G_\pm$. The point-source equation~\eqref{eq:metric-propagator-source}
therefore gives the exterior wave equations for both families in
the source representative:
\begin{equation}\label{eq:BD-profile-wave-equations}
 \Op^\pm\boldsymbol P^\pm_L(\eta,\rho)
 \overset{\rho>d}{=}0.
\end{equation}
They hold componentwise, including the prescribed past and dynamical
contributions, without evaluating the source coefficients.

The propagated history is precisely the coefficient of $1/\rho$
in the corresponding radial profile. To see this, the Hadamard
form~\eqref{eq:hadamard} gives, for a smooth retarded history $f$
and a time $\eta_0$ before its support,
\begin{equation}\label{eq:radial-propagator-small-radius}
 \mathcal G_V[f](\eta,\rho)
 =\frac{f(\eta-\rho/c)}{\rho}
 +\int_{\eta_0}^{\eta-\rho/c}c\,\dd\eta'\,
       \mathcal V_V(\eta,\rho;\eta')f(\eta').
\end{equation}
The tail and its fixed-order time derivatives remain bounded as
$\rho\to0$ on compact time intervals within $\Eta$. Hence
\begin{equation}\label{eq:radial-propagator-residue}
 \lim_{\rho\to0}\rho\mathcal G_V[f](\eta,\rho)=f(\eta).
\end{equation}
Adding the prescribed past term $f_\pm^{\mathrm{past}}(\eta)/\rho$
gives the same identity for $\mathcal G_\pm$ with the total history
$f=f_\pm^{\mathrm{past}}+f^{\mathrm{dyn}}$.
We therefore define the single-time source coefficients by
\begin{subequations}\label{eq:BD-source-profile-definitions}
\begin{align}
 \boldsymbol P_L^{\pm,\mathrm{src}}(\eta)
 &:=\lim_{\rho\to0}\rho\boldsymbol P_L^\pm(\eta,\rho),
 \label{eq:BD-source-profile-residue}\\
 \boldsymbol P_L^{-,\mathrm{src}}(\eta)
 &=\big(E_L^{\mathrm{src}}(\eta),F_L^{\mathrm{src}}(\eta),G_L^{\mathrm{src}}(\eta),
         H_L^{\mathrm{src}}(\eta),I_L^{\mathrm{src}}(\eta),K_L^{\mathrm{src}}(\eta)\big),
 \label{eq:BD-spatial-source-profile-definition}\\
 \boldsymbol P_L^{+,\mathrm{src}}(\eta)
 &=\big(B_L^{\mathrm{src}}(\eta),C_L^{\mathrm{src}}(\eta),D_L^{\mathrm{src}}(\eta)\big).
 \label{eq:BD-mixed-source-profile-definition}
\end{align}
\end{subequations}
The limit is taken on each radial profile before applying the
observation derivatives. It uses the continuation supplied by the
metric propagator, rather than the metric inside the physical source.
We will now compute these coefficients from the retarded source
moments. We use the normalization factors
\begin{subequations}\label{eq:direct-BD-source-factors}
\begin{align}
 \kappa_\ell&:=\frac{(-1)^\ell}{\ell!},\label{eq:canonical-kappa}\\
 \alpha_\ell&:=\frac{2\ell+1}{(\ell+1)(2\ell+3)},\label{eq:direct-BD-source-factors-1}\\
 \beta_\ell&:=\frac{2\ell+1}{2(\ell+1)(\ell+2)(2\ell+5)}.
 \label{eq:direct-BD-source-factors-2}
\end{align}
\end{subequations}
For the mixed field, contracting the component index with a moment
index selects $B_L^{\mathrm{src}}$, while an $\epsilon_{ijk}$ contraction
selects $D_L^{\mathrm{src}}$. The fully symmetric STF projection
selects $C_L^{\mathrm{src}}$, with a correction from the derivative
trace of the $B_L$ term. For any STF profile $U_L$, the STF
projection of the observation derivatives gives
\begin{equation}\label{eq:BD-vector-harmonic-projection}
 \partial_{\langle iL\rangle}U_L
 =\partial_{iL}U_L-\frac{\ell}{2\ell+1}
       \Delta\partial_{L-1}U_{iL-1},\qquad\ell\ge1.
\end{equation}
With $U_L=B_L$, the Laplacian is propagated from
$T_+B_L^{\mathrm{src}}$ by Eq.~\eqref{eq:canonical-propagation-trace}.
Applying the STF projector~\eqref{eq:STFprojector} to the three
component projections therefore gives
\begin{subequations}\label{eq:BD-mixed-source-coefficients}
\begin{align}
 B_L^{\mathrm{src}}(\eta)&=-\kappa_\ell\alpha_\ell
                      \mathcal F_{0a|aL}(\eta;a),
 \label{eq:direct-BD-source-coefficients-1}\\
 C_L^{\mathrm{src}}(\eta)&=-\ell\kappa_\ell
      \STF_L\mathcal F_{0i_\ell|L-1}(\eta;a)
                  -\frac\ell{2\ell+1}\big(T_+B_L^{\mathrm{src}}\big)(\eta),
 \label{eq:direct-BD-electric-source-1}\\
 D_L^{\mathrm{src}}(\eta)&=-\kappa_\ell\frac\ell{\ell+1}
       \STF_L\big(\epsilon_{ab i_\ell}\mathcal F_{0b|aL-1}(\eta;a)\big).
 \label{eq:direct-BD-source-coefficients-3}
\end{align}
\end{subequations}
The gradient coefficient exists for $\ell\ge0$; the other two
start at $\ell=1$.

For the spatial field, contracting both component indices with
moment indices selects the double-gradient coefficient
$E_L^{\mathrm{src}}$. One such contraction followed by an
$\epsilon_{ijk}$ contraction selects $H_L^{\mathrm{src}}$. The same
STF projector gives
\begin{subequations}\label{eq:direct-BD-source-coefficients}
\begin{align}
 E_L^{\mathrm{src}}(\eta)&=2\kappa_\ell\beta_\ell
                      \mathcal F_{ab|abL}(\eta;a),
 \label{eq:direct-BD-source-coefficients-2}\\
 H_L^{\mathrm{src}}(\eta)&=2\kappa_\ell\frac\ell{\ell+2}\alpha_\ell
       \STF_L\big(\epsilon_{ab i_\ell}\mathcal F_{bc|acL-1}(\eta;a)\big).
 \label{eq:direct-BD-source-coefficients-4}
\end{align}
The electric coefficient $G_L^{\mathrm{src}}$ follows by contracting
one component index with one moment index and symmetrizing the
remaining free indices. Subtracting one third of the component trace
separates it from $F_L^{\mathrm{src}}$; the derivative trace of the
$E_L$ term contributes the additional $T_-E_L^{\mathrm{src}}$ term.
The component trace of the full spatial decomposition then fixes
$F_L^{\mathrm{src}}$. These operations give
\begin{align}
 G_L^{\mathrm{src}}(\eta)&=
 \frac{6\ell(2\ell-1)}{2\ell+1}\kappa_\ell\alpha_\ell
 \left(\STF_L\mathcal F_{a i_\ell|aL-1}(\eta;a)
                     -\frac13\mathcal F_{aa|L}(\eta;a)\right)
 -\frac{2\ell}{2\ell+3}\big(T_-E_L^{\mathrm{src}}\big)(\eta),
 \label{eq:direct-BD-electric-source-2}\\
 F_L^{\mathrm{src}}(\eta)&=\frac13\left(
 \kappa_\ell\mathcal F_{aa|L}(\eta;a)
                  -\big(T_-E_L^{\mathrm{src}}\big)(\eta)-G_L^{\mathrm{src}}(\eta)\right).
 \label{eq:BD-source-F}
\end{align}
Finally, the fully symmetric STF projection of the two component
indices with the moment indices selects $I_L^{\mathrm{src}}$.
The corresponding projection with one $\epsilon_{ijk}$ contraction
selects $K_L^{\mathrm{src}}$. Removing the derivative traces already
carried by $E_L,G_L,H_L$ yields
\begin{align}
 I_L^{\mathrm{src}}(\eta)&=\ell(\ell-1)\kappa_\ell
       \STF_L\mathcal F_{i_{\ell-1}i_\ell|L-2}(\eta;a)
       -\frac{\ell-1}{2\ell-1}\big(T_-G_L^{\mathrm{src}}\big)(\eta)
       -\frac{\ell(\ell-1)}{(2\ell-1)(2\ell+1)}\big(T_-^2E_L^{\mathrm{src}}\big)(\eta),
 \label{eq:BD-source-I}\\
 K_L^{\mathrm{src}}(\eta)&=\frac{2\ell(\ell-1)}{\ell+1}\kappa_\ell
       \STF_L\big(\epsilon_{ab i_\ell}
                   \mathcal F_{i_{\ell-1}b|aL-2}(\eta;a)\big)
       -\frac{\ell-1}{2\ell+1}\big(T_-H_L^{\mathrm{src}}\big)(\eta).
 \label{eq:BD-source-K}
\end{align}
\end{subequations}
Here $E_L^{\mathrm{src}},F_L^{\mathrm{src}}$ start at $\ell=0$,
$G_L^{\mathrm{src}},H_L^{\mathrm{src}}$ at $\ell=1$, and
$I_L^{\mathrm{src}},K_L^{\mathrm{src}}$ at $\ell=2$; the absent
$G_L^{\mathrm{src}}$ term is omitted from the monopole formula for
$F_L^{\mathrm{src}}$.
These are algebraic STF projections of the existing source moments,
with each derivative trace replaced by $T_\pm$. The numerical factors
in $B_L^{\mathrm{src}},E_L^{\mathrm{src}},D_L^{\mathrm{src}},H_L^{\mathrm{src}}$
are those of the flat recoupling in
\cite[Eqs.~(5.10)--(5.11)]{Damour:1990gj}; the distinct moments and
operators retain the two cosmological wave potentials.

The coefficients inherit the past and dynamical parts of the source
moments. Applying the metric propagator to either family gives the
single reconstruction formula
\begin{equation}\label{eq:BD-profile-propagation}
 \boldsymbol P^\pm_L(\eta,\rho)
 =\frac{\boldsymbol P_L^{\pm,\mathrm{src},\mathrm{past}}(\eta)}{\rho}
   +\int_\Eta c\,\dd\eta'\,G^\pm(\eta,\rho;\eta')
       \boldsymbol P_L^{\pm,\mathrm{src},\mathrm{dyn}}(\eta').
\end{equation}
This fixes all spatial and mixed radial profiles in the source
representative. For the remaining coefficient $A_L$, substitute
Eqs.~\eqref{eq:BD-FLRW-metric} into the temporal
reconstruction~\eqref{eq:scalar-solution} and use
$\bar h_{00}=\hat h-\bar h_{kk}$. The divergence of the mixed field
has coefficient $\Delta B_L+C_L$, while the spatial trace has
coefficient $\Delta E_L+3F_L+G_L$. Hence
\begin{equation}\label{eq:BD-temporal-reconstruction}
 A_L(\eta,\rho)=A_L(\eta_i,\rho)
 +\int_{\eta_i}^{\eta}\dd \eta'\,
 \big[\Delta B_L+C_L-\frac{\cH}{c^2}(\Delta E_L+3F_L+G_L)\big](\eta',\rho).
\end{equation}
The initial coefficient is fixed by the prescribed past metric, and
$C_L,G_L$ are absent for $\ell=0$. No density moment is required.

In the flat limit, both metric propagators give
$f(\eta-\rho/c)/\rho$, and the trace identity reduces to
$\Delta[f(\eta-\rho/c)/\rho]=f''(\eta-\rho/c)/(c^2\rho)$.
The single-time source coefficients then appear inside explicit
$1/\rho$ factors, as in \cite[Eq.~(2.25)]{Blanchet:1985sp}.
}

\subsection{Invariant combinations and harmonic constraints}
\label{sec:invariant-harmonic-reduction}
We will first separate gauge-invariant combinations of the ten
coefficient families, then derive the relations they obey for the
exterior solution. We start by expanding the rescaled gauge vector:
\begin{subequations}\label{eq:BD-FLRW-gauge-vector}
\begin{align}
 \frac{c^3\zeta^0}{4G}&=\sum_{\ell\ge0}\partial_L W_L,
 \label{eq:BD-FLRW-gauge-time}\\
 \frac{c^4\zeta^i}{4G}&=\sum_{\ell\ge0}\partial_{iL}X_L
 +\sum_{\ell\ge1}\big(\partial_{L-1}Y_{iL-1}
                 +\epsilon_{iab}\partial_{aL-1}Z_{bL-1}\big).
 \label{eq:BD-FLRW-gauge-space}
\end{align}
\end{subequations}
All profiles depend on $(\eta,\rho)$. Substitution of Eqs.~\eqref{eq:BD-FLRW-gauge-vector} and
\eqref{eq:BD-FLRW-metric} into Eqs.~\eqref{eq:rescaled-metric-gauge-action} gives
\begin{subequations}\label{eq:BD-FLRW-gauge-action}
\begin{align}
 \delta_\xi A_L&=\frac1{c^2}\big[-(\partial_\eta-3\cH)W_L+\Delta X_L+Y_L\big],\label{eq:BD-FLRW-gauge-action-1}\\
 \delta_\xi B_L&=c^{-1}D_-X_L-W_L,\label{eq:BD-FLRW-gauge-action-2}\\
 \delta_\xi C_L&=c^{-1}D_-Y_L,\label{eq:BD-FLRW-gauge-action-3}\\
 \delta_\xi D_L&=c^{-1}D_-Z_L,\label{eq:BD-FLRW-gauge-action-4}\\
 \delta_\xi E_L&=2X_L,\label{eq:BD-FLRW-gauge-action-5}\\
 \delta_\xi F_L&=-cD_+W_L-\Delta X_L-Y_L,\label{eq:BD-FLRW-gauge-action-6}\\
 \delta_\xi G_L&=2Y_L,\label{eq:BD-FLRW-gauge-action-7}\\
 \delta_\xi H_L&=2Z_L,\label{eq:BD-FLRW-gauge-action-8}\\
 \delta_\xi I_L&=0,\label{eq:BD-FLRW-gauge-action-9}\\
 \delta_\xi K_L&=0.
 \label{eq:BD-FLRW-gauge-action-10}
\end{align}
\end{subequations}
These transformation laws hold for unrestricted gauge profiles. Families absent at the lowest ranks are omitted:
$C_L,D_L,G_L,H_L,Y_L,Z_L$ start at $\ell=1$, and $I_L,K_L$ at
$\ell=2$.

Using Eqs.~\eqref{eq:BD-FLRW-gauge-action}, we obtain the
gauge-invariant combinations
\begin{subequations}\label{eq:direct-BD-invariants}
\begin{align}
 \mathfrak M_L&:=A_L-\frac{\Delta E_L+G_L}{2c^2}
 +\frac1{c^2}(\partial_\eta-3\cH)\big(\tfrac1{2c}D_-E_L-B_L\big),
 \label{eq:direct-BD-M}\\
 \mathfrak C_L&:=C_L-\tfrac1{2c}D_-G_L,
 \label{eq:direct-BD-C}\\
 \mathfrak J_L&:=D_L-\tfrac1{2c}D_-H_L,
 \label{eq:direct-BD-J}\\
 \mathfrak F_L&:=F_L+\tfrac12\Delta E_L+\tfrac12G_L
 +cD_+\big(\tfrac1{2c}D_-E_L-B_L\big).
 \label{eq:direct-BD-F}
\end{align}
\end{subequations}
Together with $I_L,K_L$, which are already invariant in
Eqs.~\eqref{eq:BD-FLRW-gauge-action}, these give six invariant
combinations at $\ell\ge2$. Their invariance follows algebraically,
without fixing a gauge or imposing residual wave equations on the
gauge profiles.
The combinations $\mathfrak M_L,\mathfrak F_L$ also have monopoles,
while $\mathfrak C_L,\mathfrak J_L$ start at $\ell=1$.

We now impose the generalized harmonic gauge conditions on the
linearized perturbations generated by the prescribed source.
Substituting Eqs.~\eqref{eq:BD-FLRW-metric} into
Eqs.~\eqref{eq:gauge-components} gives relations among the metric coefficients:
\begin{subequations}\label{eq:BD-FLRW-constraints}
\begin{align}
 cD_+B_L&=\Delta E_L+F_L+\tfrac12G_L,\qquad\ell\ge0,\label{eq:BD-FLRW-constraints-1}\\
 cD_+C_L&=\tfrac12\Delta G_L+I_L,\qquad\ell\ge1,\label{eq:BD-FLRW-constraints-2}\\
 cD_+D_L&=\tfrac12\Delta H_L+\tfrac12K_L,\qquad\ell\ge1,\label{eq:BD-FLRW-constraints-3}\\
 \partial_\eta A_L&=\Delta B_L+C_L
                   -\frac{\cH}{c^2}(\Delta E_L+3F_L+G_L),\qquad\ell\ge0.
 \label{eq:BD-FLRW-temporal-constraint}
\end{align}
\end{subequations}
The first three equalities come from $C_i=0$, and the last from
$C_0=0$.

The gauge transformation in Eqs.~\eqref{eq:BD-FLRW-gauge-vector}
preserves generalized harmonic gauge when
$\Op^-X_L=\Op^-Y_L=\Op^-Z_L=0$ and $(\square-3V^+)W_L=0$,
with compatible past data.
Using $D_+D_-=T_-$ and $\Op^-=\Delta-T_-$ from
Eqs.~\eqref{eq:canonical-Tpm} and~\eqref{eq:canonical-wave-temporal-relation},
the family wave equation~\eqref{eq:BD-profile-wave-equations}
allows $D_+D_-$ to be
replaced by $\Delta$ when acting on $E_L,G_L,H_L$; the three spatial
constraints therefore give
\begin{subequations}\label{eq:direct-BD-reduced-equations}
\begin{align}
 \mathfrak F_L&=-\tfrac12\Op^-E_L=0,
 \label{eq:direct-BD-reduced-equations-1}\\
 I_L&=cD_+\mathfrak C_L,
 \label{eq:direct-BD-reduced-equations-2}\\
 K_L&=2cD_+\mathfrak J_L.
 \label{eq:direct-BD-reduced-equations-3}
\end{align}
\end{subequations}
The first identity follows from Eqs.~\eqref{eq:direct-BD-F}
and~\eqref{eq:BD-FLRW-constraints-1}.
For the second and third equalities, the absent $I_L,K_L$ at
$\ell=1$ give $D_+\mathfrak C_i=D_+\mathfrak J_i=0$.
These are the constraints between the gauge-invariant quantities
imposed by the spatial generalized harmonic gauge and spatially compact source profile.

Applying $\Op^+$ to the definitions of $\mathfrak C_L$ and
$\mathfrak J_L$ in Eqs.~\eqref{eq:direct-BD-C} and~\eqref{eq:direct-BD-J},
we use the intertwining identity~\eqref{eq:intertwine-plus} and
the family wave equation~\eqref{eq:BD-profile-wave-equations} to obtain,
for $\rho>d$ and $\ell\ge1$,
\begin{subequations}\label{eq:invariant-mixed-wave-equations}
\begin{align}
 \Op^+\mathfrak C_L&=\Op^+C_L-\tfrac1{2c}D_-\Op^-G_L=0,
 \label{eq:direct-BD-reduced-equations-5}\\
 \Op^+\mathfrak J_L&=\Op^+D_L-\tfrac1{2c}D_-\Op^-H_L=0.
 \label{eq:direct-BD-reduced-equations-6}
\end{align}
\end{subequations}

To express the temporal constraint, define the auxiliary profile
\begin{equation}\label{eq:direct-BD-Omega}
 \Omega_L:=\tfrac1{2c}D_-E_L-B_L,\qquad \delta_\xi\Omega_L=W_L.
\end{equation}
Its transformation law follows from
Eqs.~\eqref{eq:BD-FLRW-gauge-action-2} and
\eqref{eq:BD-FLRW-gauge-action-5}. Differentiating
$\mathfrak M_L$ and using $\mathfrak F_L=0$ yields
\begin{equation}\label{eq:temporal-subtraction-obstruction}
 \partial_\eta\mathfrak M_L-\mathfrak C_L
 =-(\square-3V^+)\Omega_L.
\end{equation}
This expresses the difference of two invariant profiles in terms of the
metric coefficients of the harmonic solution.
For the prescribed-source solution, we define the radial response
profile by the invariant difference
\begin{equation}\label{eq:direct-BD-mass-balance}
 \mathfrak N_L:=\partial_\eta\mathfrak M_L-\mathfrak C_L,
 \qquad \mathfrak C=0\quad(\ell=0).
\end{equation}
The six combinations have thus reduced to three profile families,
$\mathfrak M_L,\mathfrak C_L,\mathfrak J_L$, obeying the displayed
relations. Equivalently, $\mathfrak M_L,\mathfrak J_L,\mathfrak N_L$
characterize the solution. The temporal harmonic condition does not
require the additional response $\mathfrak N_L$ to vanish.

Applied to Eq.~\eqref{eq:direct-BD-Omega}, the same family wave
equation~\eqref{eq:BD-profile-wave-equations} and intertwining
identity~\eqref{eq:intertwine-plus} give $\Op^+\Omega_L=0$
in the exterior of the source representative.
Substituting $\square-3V^+=\Op^+-4V^+$ into
Eq.~\eqref{eq:temporal-subtraction-obstruction} therefore yields
\begin{equation}\label{eq:direct-BD-response-compact}
 \mathfrak N_L=4V^+\Omega_L.
\end{equation}
This relation identifies the coupling to the background enthalpy
density, $V^+=4\pi Ga^2(\eps+p)/c^4$. It holds in the source
representative and after spatial residual transformations, which
leave $\Omega_L$ unchanged. The definition
\eqref{eq:direct-BD-mass-balance} applies also after a temporal change of gauge.

\begingroup
\paragraph{Counting the independent moment families.}
The full metric--matter system retains four gauge functions, but
not all four preserve the compact-source representative. In particular,
Eq.~\eqref{eq:shifted-source-gauge0i} gives
$\delta_\xi\tau_{0i}=a(\eps+p)\partial_i\zeta^0$.
When $\eps+p\ne0$, preserving $\tau_{0i}=0$ in the exterior requires
$\partial_i\zeta^0=0$. Spatial decay of the multipolar gauge vector
then gives $\zeta^0=0$ in the exterior, which also preserves
$\tau_{ij}=0$ by Eq.~\eqref{eq:shifted-source-gaugeij}.
Thus no temporal residual multipolar profile remains.
Equivalently, a temporal residual profile $W_L$ must preserve both
the source-representative wave equations~\eqref{eq:BD-profile-wave-equations}
and harmonic gauge~\eqref{eq:residual-gauge-rescaled-time}:
\begin{subequations}\label{eq:temporal-source-preserving-gauge}
\begin{align}
 \Op^+W_L&=0,\label{eq:temporal-source-preserving-wave}\\
 (\Op^+-4V^+)W_L&=0.
 \label{eq:temporal-source-preserving-harmonic}
\end{align}
\end{subequations}
Subtracting the two equations gives $V^+W_L=0$.
Thus, where $V^+\ne0$, only the three spatial residual gauge
families are consistent with $W_L=0$. At $\ell\ge2$, the ten
coefficient families subject to the four harmonic relations in
Eqs.~\eqref{eq:BD-FLRW-constraints} therefore leave three invariant
families after removing these three residual families: $10-4-3=3$.
The temporal relation has not been removed: by
Eqs.~\eqref{eq:direct-BD-mass-balance} and~\eqref{eq:direct-BD-response-compact},
it fixes $4V^+\Omega_L=\partial_\eta\mathfrak M_L-\mathfrak C_L$
in terms of the invariant profiles.

When $V^+\equiv0$, as in Minkowski and pure de Sitter spacetime,
the two temporal wave equations coincide and the fourth residual
gauge family also preserves the source representative. The temporal
relation then becomes $\mathfrak C_L=\partial_\eta\mathfrak M_L$,
leaving the two invariant mass and current families: $10-4-4=2$. On a general FLRW background the additional invariant
family describes the cosmic fluid response.
\endgroup

\begingroup
\paragraph{Dependence on the spatial frame.}
The gauge-invariant multipole families are defined relative to a
chosen spatial origin and Cartesian axes. Proper rotations transform
their free indices as STF tensors, preserving each family and its
multipole rank, whereas translations change the expansion origin
and mix multipole orders. A background isometry $\phi$, whose
parameters are independent of the perturbative parameter $\lambda$,
preserves $\bar g_{\mu\nu}$ but acts on the perturbation as
$h_{\mu\nu}\mapsto(\phi^*h)_{\mu\nu}$.
This action is distinct from the perturbative gauge variation
$\delta_\xi h_{\mu\nu}=\mathcal L_\xi\bar g_{\mu\nu}$ in
Eq.~\eqref{eq:linear-gauge-transformation}: for a background Killing
vector $K^\mu$, $\mathcal L_K\bar g_{\mu\nu}=0$, while
$\mathcal L_Kh_{\mu\nu}$ need not vanish.
The moments are therefore gauge invariant and transform under
changes of the spatial frame. The six rigid translation and rotation
parameters specify that frame.
\endgroup

\subsection{The source moments and the canonical cosmic fluid moment}
We will now use the source coefficients computed in
Eqs.~\eqref{eq:BD-mixed-source-coefficients} and
\eqref{eq:direct-BD-source-coefficients} to determine the invariant
profiles and their canonical moment histories. Combining
$B_L^{\mathrm{src}}$ and $E_L^{\mathrm{src}}$ in
Eq.~\eqref{eq:direct-BD-Omega}, we define the canonical cosmic fluid
moment $N_L$ and obtain
\begin{subequations}\label{eq:direct-BD-U-source}
\begin{align}
 N_L(\eta)&:=\alpha_\ell\mathcal F_{0a|aL}(\eta;a)
              +\frac{\beta_\ell}{c}(D_-\mathcal F_{ab|abL})(\eta;a),\label{eq:direct-BD-U-source-1}\\
 \Omega_L(\eta,\rho)&=\kappa_\ell\mathcal G_+[N_L](\eta,\rho).
 \label{eq:direct-BD-U-source-2}
\end{align}
\end{subequations}
The second identity follows from
Eqs.~\eqref{eq:direct-BD-Omega}, \eqref{eq:BD-profile-propagation}
and~\eqref{eq:canonical-propagation-minus}, using the definition
\eqref{eq:direct-BD-U-source-1}. The canonical cosmic fluid
moment $N_L(\eta)$ is a source history depending only on time, whereas
$\mathfrak N_L(\eta,\rho)$ is its invariant radial response profile.
The metric propagator $\mathcal G_+$ determines this profile and
$\Omega_L$ in the source representative.

We apply the propagation identity~\eqref{eq:canonical-propagation-minus}
to the definitions~\eqref{eq:direct-BD-C}--\eqref{eq:direct-BD-J}
of $\mathfrak C_L$ and $\mathfrak J_L$. For $\mathfrak N_L$, we
substitute Eq.~\eqref{eq:direct-BD-U-source-2} into
Eq.~\eqref{eq:direct-BD-response-compact} and use
$\kappa_\ell N_L=\tfrac1{2c}D_-E_L^{\mathrm{src}}-B_L^{\mathrm{src}}$,
which follows from Eqs.~\eqref{eq:direct-BD-U-source-1},
\eqref{eq:direct-BD-source-coefficients-1}
and~\eqref{eq:direct-BD-source-coefficients-2}. These substitutions give
\begin{subequations}\label{eq:direct-BD-electric-source}
\begin{align}
 \mathfrak C_L&=\mathcal G_+
               [C_L^{\mathrm{src}}-\tfrac1{2c}D_-G_L^{\mathrm{src}}],
 \label{eq:direct-BD-electric-source-3}\\
 \mathfrak J_L&=\mathcal G_+
               [D_L^{\mathrm{src}}-\tfrac1{2c}D_-H_L^{\mathrm{src}}],
 \label{eq:BD-invariant-current-source-profile}\\
 \mathfrak N_L&=4V^+\mathcal G_+
               [\tfrac1{2c}D_-E_L^{\mathrm{src}}-B_L^{\mathrm{src}}].
 \label{eq:BD-invariant-response-source-profile}
\end{align}
\end{subequations}
The first two equalities hold for $\ell\ge1$, and the third for
$\ell\ge0$. In the third equality, $V^+(\eta)$ multiplies the
propagated profile at the observation time.

The small-radius result~\eqref{eq:radial-propagator-residue}
applies to the propagated invariant profiles $\mathfrak J_L,\mathfrak C_L$
and, through Eqs.~\eqref{eq:direct-BD-response-compact}
and~\eqref{eq:direct-BD-U-source-2}, to $\mathfrak N_L$.
For $\mathfrak M_L$, integrate
$\partial_\eta\mathfrak M_L=\mathfrak C_L+\mathfrak N_L$ from
Eq.~\eqref{eq:direct-BD-mass-balance} with the prescribed initial
data. The same uniform bounds allow the limit to pass through this
finite time integral. Hence its leading coefficient exists as well.

We define the canonical mass and current moments $M_L(\eta),J_L(\eta)$ and
the fluid response $R_L(\eta)$ by the leading $1/\rho$
coefficients of their gauge-invariant radial profiles:
\begin{subequations}\label{eq:direct-BD-canonical-moments}
\begin{align}
 M_L(\eta)&:=\kappa_\ell^{-1}
           \lim_{\rho\to0}\rho\mathfrak M_L(\eta,\rho),
           \qquad\ell\ge0,\label{eq:direct-BD-canonical-moments-1}\\
 J_L(\eta)&:=\frac{2a(\eta)}{\kappa_\ell}
           \lim_{\rho\to0}\rho\mathfrak J_L(\eta,\rho),
           \qquad\ell\ge1,\label{eq:direct-BD-canonical-moments-2}\\
 R_L(\eta)&:=\kappa_\ell^{-1}
           \lim_{\rho\to0}\rho\mathfrak N_L(\eta,\rho),
           \qquad\ell\ge0.
 \label{eq:direct-BD-canonical-moments-3}
\end{align}
\end{subequations}
These limits refer to the radial profiles before
the observation derivatives are applied. They use their mathematical
continuation to $\rho>0$ established above. They do not continue the
physical exterior metric through the source.
The factor $a(\eta)$ in $J_L$ makes its dipole equal to the conserved
angular momentum, as verified below. In the flat limit, $M_L$ has the
Blanchet--Damour normalization, while $J_L$ is $2\ell/(\ell+1)$ times
their current moment $S_L$. This choice simplifies the current contribution
to the canonical metric and leaves the angular-momentum dipole unchanged.
The fluid response $R_L$ vanishes in that limit. Its gauge invariance follows
directly from that of its defining radial response profile $\mathfrak N_L$.
Writing $\mathsf M$, $\mathsf L$ and $\mathsf T$ for mass, length and
time dimensions, respectively, the moments and response have dimensions
\begin{equation}\label{eq:canonical-moment-dimensions}
 [M_L]=\mathsf M\mathsf L^\ell,\qquad
 [J_L]=\mathsf M\mathsf L^{\ell+1}\mathsf T^{-1},\qquad
 [N_L]=\mathsf M\mathsf L^{\ell+2}\mathsf T^{-1},\qquad
 [R_L]=\mathsf M\mathsf L^\ell\mathsf T^{-1}.
\end{equation}

For the prescribed-source solution, the metric propagator $\mathcal G_+$
therefore suffices to reconstruct all invariant radial profiles.
Equations~\eqref{eq:direct-BD-electric-source} give
$\mathfrak C_L,\mathfrak J_L,\mathfrak N_L$, with the local factor
$4V^+(\eta)$ multiplying the propagated profile in the last case.
Integrating the mass balance~\eqref{eq:direct-BD-mass-balance} at fixed
$\rho$ then determines $\mathfrak M_L$ from its prescribed initial
profile. The remaining invariant coefficients $I_L$ and $K_L$ follow
by applying $cD_+$ and $2cD_+$ to $\mathfrak C_L$ and $\mathfrak J_L$,
respectively, by Eqs.~\eqref{eq:direct-BD-reduced-equations-2}
and~\eqref{eq:direct-BD-reduced-equations-3}, while
$\mathfrak F_L=0$ by Eq.~\eqref{eq:direct-BD-reduced-equations-1}.
Thus the construction uses a common metric propagator, time derivatives,
one time integral and the prescribed initial data.

The corresponding $\mathcal G_-$ representations follow from
Eq.~\eqref{eq:canonical-propagation-plus},
$D_+\mathcal G_+[f]=\mathcal G_-[D_+f]$, including the prescribed
past contributions. In particular, $I_L$ and $K_L$ can equivalently
be obtained by propagating the differentiated source histories with
$\mathcal G_-$. These coefficients are gauge invariant by
Eqs.~\eqref{eq:BD-FLRW-gauge-action-9}
and~\eqref{eq:BD-FLRW-gauge-action-10} and obey the $\mathcal O^-$
wave equation~\eqref{eq:BD-profile-wave-equations}. The two metric
propagators therefore provide related representations of physical fields.

The mass balance~\eqref{eq:direct-BD-mass-balance} and the source-coefficient
formulas in Eqs.~\eqref{eq:BD-mixed-source-coefficients} and
\eqref{eq:direct-BD-source-coefficients} now give the source map:
\begin{subequations}\label{eq:direct-BD-final-source-map}
\begin{align}
 \partial_\eta M_L(\eta)&=\kappa_\ell^{-1}
       [C_L^{\mathrm{src}}(\eta)-\tfrac1{2c}(D_-G_L^{\mathrm{src}})(\eta)]+R_L(\eta),
 \label{eq:direct-BD-mass-source}\\
 J_L(\eta)&=-\frac{2\ell a(\eta)}{\ell+1}\STF_L\!\left[\epsilon_{ab i_\ell}
       \left(\mathcal F_{0b|aL-1}(\eta;a)
              +\frac{\ell+1}{\ell+2}\frac{\alpha_\ell}{c}(D_-\mathcal F_{bc|acL-1})(\eta;a)\right)\right],
 \label{eq:direct-BD-current-source}\\
 R_L(\eta)&=4V^+(\eta)\left[\alpha_\ell\mathcal F_{0a|aL}(\eta;a)
                    +\frac{\beta_\ell}{c}(D_-\mathcal F_{ab|abL})(\eta;a)\right]
             =4V^+(\eta)N_L(\eta).
 \label{eq:direct-BD-fluid-moment-source}
\end{align}
\end{subequations}
In the mass equation the
$C_L^{\mathrm{src}},G_L^{\mathrm{src}}$ terms are absent for $\ell=0$.
The mass integration data are fixed by the prescribed past metric,
as made explicit below.

The canonical cosmic fluid moment $N_L(\eta)$ determines the fluid
response $R_L(\eta)$ and the radial response profile
$\mathfrak N_L(\eta,\rho)$, which describe the additional scalar
contribution beyond the mass and current moments. They do not
represent every induced fluid perturbation: the mass and current metric
can itself induce a density response even when $\mathfrak N_L=0$.
For perfect-fluid backgrounds with constant equation of state, Chu
\cite{Chu:2016ngc} studied the scalar retarded response.

When $V^+\ne0$ throughout the relevant source-history interval,
Eq.~\eqref{eq:direct-BD-fluid-moment-source} fixes
$N_L=R_L/(4V^+)$, so $N_L$ inherits the gauge invariance of $R_L$.
There are then three independent gauge-invariant histories at
$\ell\ge2$: $M_L,J_L,N_L$.
Using $N_L$ in Eqs.~\eqref{eq:direct-BD-response-compact}
and~\eqref{eq:direct-BD-U-source-2}
also determines the radial response when $V^+$ has zeros, without
dividing by $V^+$. If $V^+$ vanishes on a source-time interval,
$R_L$ alone need not determine $N_L$ there. Histories $N_L$
producing the same radial response profile carry the same invariant
scalar information. When $V^+$ vanishes identically, both $R_L$ and
$\mathfrak N_L$ vanish and this additional family drops out.
This includes Minkowski and pure de Sitter spacetime.

The past moments~\eqref{eq:past-retarded-source-moments} and the
reconstruction~\eqref{eq:BD-profile-propagation} give spatial
profiles proportional to $a/\rho$ and mixed profiles proportional
to $1/(a\rho)$. Their radial Laplacians vanish for $\rho>0$.
Substituting these time dependences into the definitions
\eqref{eq:direct-BD-invariants} and~\eqref{eq:direct-BD-Omega},
and using the mixed-field transversality condition in
Eq.~\eqref{eq:past-reference-transverse}, gives
$\mathfrak C_L^{\mathrm{past}}=0$,
$D_+\mathfrak J_L^{\mathrm{past}}=D_+\Omega_L^{\mathrm{past}}=0$.
The response relation~\eqref{eq:direct-BD-response-compact} and
mass balance~\eqref{eq:direct-BD-mass-balance} then yield
\begin{subequations}\label{eq:past-invariant-profile-relations}
\begin{align}
 \mathfrak N_L^{\mathrm{past}}&=4V^+\Omega_L^{\mathrm{past}},
 \label{eq:past-invariant-response}\\
 \partial_\eta\big(\mathfrak M_L^{\mathrm{past}}
                         +\frac{4\cH}{c^2}\Omega_L^{\mathrm{past}}\big)&=0.
 \label{eq:past-invariant-mass}
\end{align}
\end{subequations}
For the second equality, the potential definition~\eqref{eq:potentials}
gives $\partial_\eta(\cH/a)=-c^2V^+/a$, while
$D_+\Omega_L^{\mathrm{past}}=0$ fixes $\Omega_L^{\mathrm{past}}\propto1/a$.
These relations fix the past part of the canonical reconstruction.

The source formulas~\eqref{eq:direct-BD-U-source-1}
and~\eqref{eq:direct-BD-current-source} therefore give
$N_L^{\mathrm{past}}(\eta)=a_iN_L^{\mathrm{in}}/a(\eta)$ and
$J_L^{\mathrm{past}}(\eta)=J_L^{\mathrm{in}}$.
Here, and for $M_L^{\mathrm{in}}$ below, the superscript
$\mathrm{in}$ denotes the value of the corresponding past part
at $\eta_i$. The source-time averages of the dynamical part can begin
earlier, at $\eta_i-d/c$. With $\cH_i:=\cH(\eta_i)$,
integrating Eq.~\eqref{eq:past-invariant-mass} and using the moment
definitions~\eqref{eq:direct-BD-canonical-moments} gives
\begin{subequations}\label{eq:past-canonical-moments}
\begin{align}
 M_L^{\mathrm{past}}(\eta)
 &=M_L^{\mathrm{in}}
       +\frac4{c^2}\left[\cH_i-\frac{a_i\cH(\eta)}{a(\eta)}\right]N_L^{\mathrm{in}},
 \label{eq:past-canonical-mass}\\
 J_L^{\mathrm{past}}(\eta)&=J_L^{\mathrm{in}},
 \label{eq:past-canonical-current}\\
 R_L^{\mathrm{past}}(\eta)&=4V^+(\eta)\frac{a_i}{a(\eta)}N_L^{\mathrm{in}}.
 \label{eq:past-canonical-response}
\end{align}
\end{subequations}
Their radial profiles are
$\mathfrak M_L^{\mathrm{past}}=\kappa_\ell M_L^{\mathrm{past}}/\rho$,
$\mathfrak J_L^{\mathrm{past}}=
\kappa_\ell J_L^{\mathrm{in}}/(2a\rho)$ and
$\mathfrak N_L^{\mathrm{past}}=\kappa_\ell R_L^{\mathrm{past}}/\rho$,
with $\mathfrak C_L^{\mathrm{past}}=0$.
The constants come from the given past field and its compact sources.
To determine the mass data, use $\mathfrak F_L=0$ from
Eq.~\eqref{eq:direct-BD-reduced-equations-1} in
Eqs.~\eqref{eq:direct-BD-invariants} and take the spatial trace of
Eqs.~\eqref{eq:BD-FLRW-metric}. With
$\hat h=\bar h_{00}+\bar h_{kk}$, this gives
\begin{equation}\label{eq:direct-BD-scalar-mass}
 \frac{c^2\hat h}{4G}=\sum_{\ell\ge0}\partial_L
            \big(\mathfrak M_L-\frac4{c^2}\partial_\eta\Omega_L\big).
\end{equation}
For the past field, $D_+\Omega_L^{\mathrm{past}}=0$ converts the
coefficient to $\mathfrak M_L^{\mathrm{past}}
+\frac{4\cH}{c^2}\Omega_L^{\mathrm{past}}$.
Using the Green function of the spatial Laplacian and the imposed
decay at spatial infinity, we expand $\hat h^{\mathrm{in}}$ in
exterior multipoles. Identifying their coefficients through
Eqs.~\eqref{eq:direct-BD-canonical-moments-1}
and~\eqref{eq:direct-BD-U-source-2} then gives
\begin{equation}\label{eq:past-mass-source-data}
 M_L^{\mathrm{in}}+\frac{4\cH_i}{c^2}N_L^{\mathrm{in}}
 =-\frac{c^2}{16\pi G}\int\dd^3x'\,x'_{\langle L\rangle}
                                      \Delta'\hat h^{\mathrm{in}}(\bx').
\end{equation}
The current and $N_L^{\mathrm{in}}$ follow from
Eqs.~\eqref{eq:direct-BD-current-source},
\eqref{eq:direct-BD-U-source-1} and
\eqref{eq:past-retarded-source-moments}.
At $\ell=0$, Eqs.~\eqref{eq:past-reference-transverse}
and~\eqref{eq:past-source-mixed} imply
$\partial'_i\tau_{0i}^{\mathrm{in}}=0$.
Integration by parts, using compactness, gives
$\int\dd^3x'\,x'_a\tau_{0a}^{\mathrm{in}}=0$, hence
$N^{\mathrm{in}}=R^{\mathrm{past}}=0$ and the past mass monopole
is constant. Every canonical family is the sum of its past part
and its zero-past dynamical part.

Combining the propagated profiles~\eqref{eq:direct-BD-electric-source}
with the source map~\eqref{eq:direct-BD-final-source-map} reconstructs
$\mathfrak C_L$ and $\mathfrak J_L$. For the mass profile, we substitute
$\mathfrak N_L=4\kappa_\ell V^+\mathcal G_+[N_L]$ from
Eqs.~\eqref{eq:direct-BD-response-compact} and~\eqref{eq:direct-BD-U-source-2}
into the mass balance~\eqref{eq:direct-BD-mass-balance} and integrate
with the prescribed past data. This gives the reconstruction:
\begin{subequations}\label{eq:direct-BD-profiles-from-moments}
\begin{align}
 \mathfrak C_L&=\kappa_\ell\mathcal G_+
                   [\partial_\eta M_L-R_L],\label{eq:direct-BD-profiles-from-moments-1}\\
 \mathfrak J_L&=\frac{\kappa_\ell}{2}
                   \mathcal G_+[J_L/a],\label{eq:direct-BD-profiles-from-moments-2}\\
 \mathfrak M_L(\eta,\rho)&=\frac{\kappa_\ell M_L^{\mathrm{past}}(\eta_0)}\rho
 +\int_{\eta_0}^{\eta}\dd\eta'\,
 \big(\mathfrak C_L+4\kappa_\ell V^+\mathcal G_+[N_L]\big)(\eta',\rho).
 \label{eq:direct-BD-profiles-from-moments-3}
\end{align}
\end{subequations}
The first two equalities apply for $\ell\ge1$; $\mathfrak C=0$ in
the monopole mass equation. We take $\eta_0:=\eta_i-d/c\in\Eta$,
before the support of the dynamical retarded source moments.
The metric propagator $\mathcal G_+$ includes the prescribed past terms
in all three equalities. In the last integrand, $V^+$ and the propagated
profiles are evaluated at $\eta'$, with observation radius $\rho$.
The initial mass profile is fixed by Eq.~\eqref{eq:past-canonical-mass},
and the time integral implements its subsequent mass balance.
The mass equation~\eqref{eq:direct-BD-mass-source}
ensures its leading coefficient is $\kappa_\ell M_L$.
The radial response profile $\mathfrak N_L$ follows directly from
$N_L$ through Eqs.~\eqref{eq:direct-BD-response-compact}
and~\eqref{eq:direct-BD-U-source-2}, including at
zeros of $V^+$.

\subsection{Direct construction of the canonical metric}
\label{sec:canonical-gauge}
We will now obtain the canonical metric from the source representative
by a single gauge transformation. Its profiles will be fixed by the
metric coefficients already computed, so the construction requires
no separate solution of the temporal residual equation
$(\square-3V^+)W_L=0$.

We start by subtracting the transformation of
Eqs.~\eqref{eq:BD-FLRW-gauge-vector} with profiles
\begin{equation}\label{eq:canonical-gauge-subtraction}
 X_L:=\tfrac12E_L,\qquad Y_L:=\tfrac12G_L,\qquad
 Z_L:=\tfrac12H_L,\qquad W_L:=\Omega_L.
\end{equation}
Here $E_L,G_L,H_L,\Omega_L$ are evaluated in the source
representative. Equations~\eqref{eq:BD-FLRW-gauge-action} set
$E_L^{\mathrm{can}}=G_L^{\mathrm{can}}=H_L^{\mathrm{can}}=0$ and
$B_L^{\mathrm{can}}=B_L-\tfrac1{2c}D_-E_L+\Omega_L=0$.
The invariant definitions then give
$A_L^{\mathrm{can}}=\mathfrak M_L$,
$C_L^{\mathrm{can}}=\mathfrak C_L$,
$D_L^{\mathrm{can}}=\mathfrak J_L$ and
$F_L^{\mathrm{can}}=\mathfrak F_L=0$.
The coefficients $I_L,K_L$ are unchanged. Substituting their
expressions from Eqs.~\eqref{eq:direct-BD-reduced-equations} into
the metric decomposition~\eqref{eq:BD-FLRW-metric} gives the
canonical representative:
\begin{subequations}\label{eq:canonical-FLRW-metric}
\begin{align}
 \frac{c^2\bar h^{\rm can}_{00}}{4G}
 &=\sum_{\ell\ge0}\partial_L\mathfrak M_L,
 \label{eq:canonical-FLRW-metric-1}\\*
 \frac{c^3\bar h^{\rm can}_{0i}}{4G}
 &=\sum_{\ell\ge1}\big(\partial_{L-1}\mathfrak C_{iL-1}
              +\epsilon_{iab}\partial_{aL-1}\mathfrak J_{bL-1}\big),
 \label{eq:canonical-FLRW-metric-2}\\*
 \frac{c^4\bar h^{\rm can}_{ij}}{4G}
 &=\sum_{\ell\ge2}\Big[c\partial_{L-2}D_+\mathfrak C_{ijL-2}
           +2c\partial_{aL-2}D_+\big(\epsilon_{ab(i}\mathfrak J_{j)bL-2}\big)\Big].
 \label{eq:canonical-FLRW-metric-3}
\end{align}
\end{subequations}
All profiles are evaluated at $(\eta,\rho)$, and observation
derivatives act on that dependence. The spatial part of the
transformation is residual harmonic by
Eq.~\eqref{eq:BD-profile-wave-equations}; its temporal part leaves $C_i$
unchanged by Eq.~\eqref{eq:direct-gauge-residual-action-2}.
Thus the spatial harmonic conditions are preserved, while STF
symmetry makes the spatial trace vanish:
\begin{subequations}\label{eq:canonical-gauge-conditions}
\begin{align}
 C_i&=D_+\bar h_{0i}-\partial_j\bar h_{ij}=0,
 \label{eq:canonical-gauge-spatial}\\
 \bar h_{kk}&=0.
 \label{eq:canonical-gauge-trace}
\end{align}
\end{subequations}
To evaluate the temporal gauge residual, we substitute
Eqs.~\eqref{eq:canonical-FLRW-metric-1} and~\eqref{eq:canonical-FLRW-metric-2}
into the definition of $C_0$ in Eq.~\eqref{eq:gauge0}, with
$\bar h_{kk}^{\mathrm{can}}=0$. The current term has zero divergence
by antisymmetry of $\epsilon_{iab}$; the mass-balance relation
\eqref{eq:direct-BD-mass-balance} then gives
\begin{equation}\label{eq:canonical-response-momentum}
 C_0^{\mathrm{can}}(\eta,\bx)=\frac{4G}{c^3}
 \sum_{\ell\ge0}\partial_L\mathfrak N_L(\eta,\rho)
 \eqqcolon\frac{4G}{c^3}\mathscr N(\eta,\bx).
\end{equation}
Equivalently, the temporal part of the subtraction changes $C_0$
by $-(4G/c^3)\sum_\ell\partial_L(\square-3V^+)\Omega_L$.
Equations~\eqref{eq:temporal-subtraction-obstruction}
and~\eqref{eq:direct-BD-mass-balance} give
$-(\square-3V^+)\Omega_L=\mathfrak N_L$.
Since $C_0=0$ in the source representative, this change gives
$C_0^{\mathrm{can}}$ in Eq.~\eqref{eq:canonical-response-momentum}.
For nonzero $\mathfrak N_L$ this is a change of temporal gauge;
the invariant profiles themselves are unchanged. In particular,
$\mathfrak N_L$ is still defined by
Eq.~\eqref{eq:direct-BD-mass-balance}, even though the auxiliary $\Omega_L$
computed from the canonical coefficients vanishes.

When $\mathfrak N_L=0$, the canonical metric also satisfies the
temporal harmonic condition $C_0^{\mathrm{can}}=0$, and hence all four
harmonic conditions. This includes the cases $V^+\equiv0$, such as
Minkowski and pure de Sitter spacetime. In general, the canonical
metric is reconstructed from $M_L,J_L,N_L$ using
$\mathcal G_\pm$ and the mass integral in
Eqs.~\eqref{eq:direct-BD-profiles-from-moments}.

\subsection{Cosmic fluid perturbations in the canonical representative}
\label{sec:canonical-fluid}
We will derive the cosmic fluid perturbations accompanying the
canonical metric, verify their conservation laws and exhibit the
independent contribution of the canonical cosmic fluid moment $N_L$.

We start with the linearized Einstein equations for $C_i=0$ and
arbitrary $C_0$:
\begin{subequations}\label{eq:canonical-gauge-field-equations}
\begin{align}
 \Op^-\bar h_{ij}&=-\frac{16\pi G}{c^4}a\tau_{ij}-\delta_{ij}D_+C_0,
 \label{eq:canonical-gauge-spatial-field}\\
 \Op^+\bar h_{0i}&=-\frac{16\pi G}{c^4}a\tau_{0i}-\partial_iC_0,
 \label{eq:canonical-gauge-mixed-field}\\
 \Op^+\hat h&=-\frac{16\pi G}{c^4}a(\tau_{00}+\tau_{kk})-\frac4c\partial_\eta C_0.
 \label{eq:canonical-gauge-scalar-field}
\end{align}
\end{subequations}
We now apply these equations to the canonical metric
$\bar h_{\mu\nu}^{\mathrm{can}}$ in Eqs.~\eqref{eq:canonical-FLRW-metric},
with $C_0^{\mathrm{can}}$ given by Eq.~\eqref{eq:canonical-response-momentum}.
In the exterior, its mixed components obey
$\Op^+\bar h_{0i}^{\mathrm{can}}=0$, and the intertwining identities give
$\Op^-\bar h_{ij}^{\mathrm{can}}=0$ for the spatial components.
Substituting these expressions into
Eqs.~\eqref{eq:canonical-gauge-field-equations} therefore gives
\begin{subequations}\label{eq:canonical-FLRW-fluid}
\begin{align}
 a\tau^{\rm can}_{0i}&=-\frac c{4\pi}\sum_{\ell\ge0}
                           \partial_{iL}\mathfrak N_L,
 \label{eq:canonical-FLRW-fluid-1}\\
 a\tau^{\rm can}_{ij}&=-\frac{c\delta_{ij}}{4\pi}
                    \sum_{\ell\ge0}\partial_LD_+\mathfrak N_L,
 \label{eq:canonical-FLRW-fluid-2}\\
 a\tau^{\rm can}_{00}&=-\frac1{4\pi}\sum_{\ell\ge0}\partial_L
       \big[c^2\Op^+\mathfrak M_L+(\partial_\eta-3\cH)\mathfrak N_L\big].
 \label{eq:canonical-FLRW-fluid-3}
\end{align}
\end{subequations}
Only the temporal part of the gauge transformation acts on the shifted
stress-energy tensor perturbation, by
Eqs.~\eqref{eq:shifted-source-gauge}. The first equality also follows directly from
Eq.~\eqref{eq:shifted-source-gauge0i}: the source representative
has $\tau_{0i}=0$ in the exterior, and subtracting the temporal
component $\zeta^0=(4G/c^3)\sum_\ell\partial_L\Omega_L$ gives
$a\tau_{0i}^{\mathrm{can}}=-(4G/c^3)a^2(\eps+p)\sum_\ell
\partial_{iL}\Omega_L=-c(4\pi)^{-1}\partial_i\mathscr N$.
Thus the gradient form of the exterior momentum follows from the
source prescription and the explicit transformation, including the
past field.
For the $00$ component, the scalar equation first fixes
$a(\tau^{\rm can}_{00}+\tau^{\rm can}_{kk})$
as $-(4\pi)^{-1}\sum_\ell\partial_L
(c^2\Op^+\mathfrak M_L+4\partial_\eta\mathfrak N_L)$;
subtracting the spatial trace gives the displayed result.
These noncompact exterior terms describe the cosmic fluid
perturbations in the canonical representative. The prescribed compact
source remains specified in its source representative.

Using Eq.~\eqref{eq:potential-plus-fluid} in
Eqs.~\eqref{eq:physical-matter-variables}, and then the mass balance
in Eq.~\eqref{eq:direct-BD-mass-balance},
gives the density, pressure and momentum measured by the
Eulerian observers:
\begin{subequations}\label{eq:canonical-physical-fluid}
\begin{align}
 \delta\eps^{\rm can}&=\frac1{4\pi a^3}\sum_{\ell\ge0}\partial_L
       \big(\partial_\eta\mathfrak C_L-c^2\Delta\mathfrak M_L
                                      +3\cH\mathfrak N_L\big),
 \label{eq:canonical-physical-density}\\
 \delta p^{\rm can}&=\frac1{4\pi a^3}\sum_{\ell\ge0}\partial_L
                       \big(c^2V^+\mathfrak M_L-cD_+\mathfrak N_L\big),
 \label{eq:canonical-physical-pressure}\\
 j_i^{\rm can}&=\frac1{4\pi a^3}\partial_i\mathscr N,
 \label{eq:canonical-physical-momentum}\\
 \pi_{ij}^{\rm can}&=0.
 \label{eq:canonical-physical-anisotropic}
\end{align}
\end{subequations}
We define the momentum potential $\delta q^{\mathrm{can}}$ by
$j_i^{\mathrm{can}}=\partial_i\delta q^{\mathrm{can}}$ and fix its
additive function of time by requiring
$\delta q^{\mathrm{can}}(\eta,\bx)\to0$ as $\rho\to\infty$ at fixed $\eta$.
Since $\mathscr N$ also vanishes in this limit,
Eq.~\eqref{eq:canonical-physical-momentum} gives
\begin{subequations}\label{eq:canonical-gauge-fluid-interpretation}
\begin{align}
 \partial_iC_0^{\mathrm{can}}
 &=-\frac{16\pi G}{c^4}a\tau_{0i}^{\mathrm{can}}
   =\frac{16\pi G}{c^3}a^3j_i^{\mathrm{can}},
 \label{eq:canonical-gauge-momentum}\\
 \delta q^{\mathrm{can}}
 &=\frac{\mathscr N}{4\pi a^3}
   =\frac{c^3C_0^{\mathrm{can}}}{16\pi Ga^3}.
 \label{eq:canonical-gauge-momentum-potential}
\end{align}
\end{subequations}
The temporal harmonic residual $C_0^{\mathrm{can}}$ is therefore
proportional to the cosmic fluid momentum potential in the canonical
representative.

We check that the canonical shifted stress-energy tensor in
Eqs.~\eqref{eq:canonical-FLRW-fluid} satisfies the conservation laws
in the exterior. For $C_i=0$, Eq.~\eqref{eq:linear-conservation} becomes
\begin{subequations}\label{eq:canonical-gauge-conservation}
\begin{align}
 D_+(a\tau_{0i})&=a\partial_j\tau_{ij},
 \label{eq:canonical-gauge-momentum-conservation}\\
 \partial_\eta(a\tau_{00})&=ca\partial_i\tau_{0i}-\dot a\tau_{kk}
       -\frac{c^4\dot V^+}{16\pi G}\hat h-\frac{c^5V^+}{4\pi G}C_0.
 \label{eq:canonical-gauge-energy-conservation}
\end{align}
\end{subequations}
For momentum conservation, substituting
Eqs.~\eqref{eq:canonical-FLRW-fluid-1} and~\eqref{eq:canonical-FLRW-fluid-2}
into the two sides of Eq.~\eqref{eq:canonical-gauge-momentum-conservation}
gives the same expression:
\begin{equation}\label{eq:canonical-momentum-conservation-check}
 D_+(a\tau_{0i}^{\mathrm{can}})
 =a\partial_j\tau_{ij}^{\mathrm{can}}
 =-\frac c{4\pi}\sum_{\ell\ge0}\partial_{iL}D_+\mathfrak N_L.
\end{equation}

For energy conservation, we substitute all three components in
Eqs.~\eqref{eq:canonical-FLRW-fluid}, together with
$\hat h^{\mathrm{can}}=\bar h_{00}^{\mathrm{can}}$ from
Eqs.~\eqref{eq:canonical-FLRW-metric} and $C_0^{\mathrm{can}}$ from
Eq.~\eqref{eq:canonical-response-momentum}. Using
$c^2V^+=\cH^2-\dot\cH$ from Eq.~\eqref{eq:potentials}, the left-hand
side minus the right-hand side of
Eq.~\eqref{eq:canonical-gauge-energy-conservation} reduces to
\begin{equation}\label{eq:canonical-energy-conservation-check}
 \frac{c^2}{4\pi}\sum_{\ell\ge0}\partial_L
 \big[\Op^+\mathfrak N_L+\dot V^+\mathfrak M_L
       -\partial_\eta(\Op^+\mathfrak M_L)\big].
\end{equation}
The mass balance~\eqref{eq:direct-BD-mass-balance} and the wave
equation $\Op^+\mathfrak C_L=0$ in
Eq.~\eqref{eq:direct-BD-reduced-equations-5} imply
\begin{equation}\label{eq:canonical-mass-operator-identity}
 \begin{aligned}
 \partial_\eta(\Op^+\mathfrak M_L)
 &=\Op^+(\mathfrak C_L+\mathfrak N_L)+\dot V^+\mathfrak M_L\\
 &=\Op^+\mathfrak N_L+\dot V^+\mathfrak M_L.
 \end{aligned}
\end{equation}
Here $\dot V^+\mathfrak M_L$ comes from differentiating the potential
in $\Op^+$. The expression in
Eq.~\eqref{eq:canonical-energy-conservation-check} therefore vanishes,
which verifies energy conservation, including the term proportional
to $C_0^{\mathrm{can}}$.

The canonical cosmic fluid moment $N_L$ supplies data independent
of the mass and current moments. To exhibit this independence in
a general fluid background, we set
$M_L=J_L=0$ at $\ell\ge2$ and choose a nonzero $N_L$ of finite
past support where $V^+\ne0$. Then $R_L=4V^+N_L\ne0$.
Equations~\eqref{eq:direct-BD-profiles-from-moments}
give $\mathfrak C_L=-4\kappa_\ell\mathcal G_+[V^+N_L]$ and a
nonzero radial response profile $\mathfrak N_L$. The corresponding
source representative has $B_L=-\kappa_\ell\mathcal G_+[N_L]$,
$F_L=cD_+B_L$, $I_L=cD_+\mathfrak C_L$, and all other spatial
coefficients zero. These fields are generated by distributions at
the source origin and satisfy the exterior equations and constraints.
Thus the additional invariant is present already in the allowed
multipole-source class.

\subsection{Lowest multipoles and their balance laws}
\label{sec:canonical-fluid-and-lowest-multipoles}
We will relate the lowest multipoles to the mass, momentum,
center of mass and angular momentum through their balance laws
and prescribed past data.

At the lowest ranks, absent tensor families give the balance laws
\begin{subequations}\label{eq:canonical-FLRW-low-ranks}
\begin{align}
 \partial_\eta M&=R,\label{eq:canonical-FLRW-low-ranks-1}\\
 D_+(\partial_\eta M_i-R_i)&=0,\label{eq:canonical-FLRW-low-ranks-2}\\
 \partial_\eta J_i&=0.
 \label{eq:canonical-FLRW-low-ranks-3}
\end{align}
\end{subequations}
Using $D_+=c^{-1}(\partial_\eta+\cH)$, the mass-dipole balance
\eqref{eq:canonical-FLRW-low-ranks-2} becomes
\begin{equation}\label{eq:canonical-mass-dipole-balance}
 \partial_\eta\!\left[a(\eta)
       \big(\partial_\eta M_i(\eta)-R_i(\eta)\big)\right]=0.
\end{equation}
Integrating Eqs.~\eqref{eq:canonical-FLRW-low-ranks-1},
\eqref{eq:canonical-FLRW-low-ranks-3} and~\eqref{eq:canonical-mass-dipole-balance}
from $\eta_i$ gives
\begin{subequations}\label{eq:canonical-low-rank-solutions}
\begin{align}
 M(\eta)&=M(\eta_i)+\int_{\eta_i}^{\eta}\dd\eta'\,R(\eta'),
 \label{eq:canonical-mass-monopole-solution}\\
 M_i(\eta)&=K_i+\mathcal P_i\int_{\eta_i}^{\eta}\frac{\dd\eta'}{a(\eta')}
                   +\int_{\eta_i}^{\eta}\dd\eta'\,R_i(\eta'),
 \label{eq:canonical-mass-dipole-solution}\\
 J_i(\eta)&=J_i^{\mathrm{in}}.
 \label{eq:canonical-current-dipole-solution}
\end{align}
\end{subequations}
Here $K_i:=M_i(\eta_i)$ and $J_i^{\mathrm{in}}=J_i(\eta_i)$.
The ten integration constants are thus
$M(\eta_i),K_i,\mathcal P_i,J_i^{\mathrm{in}}$, evaluated for the
full canonical moments reconstructed in
Eqs.~\eqref{eq:direct-BD-profiles-from-moments}.
The constants $\mathcal P_i$ and $J_i^{\mathrm{in}}$ are the conserved
linear and angular momenta associated with spatial translations and
rotations, with the conventions of Eq.~\eqref{eq:spatial-isometry-charges}.

For a stationary mass $M$ at comoving position $X_i$, $K_i=MX_i$,
so $K_i/M$ is its center-of-mass position. On a general FLRW
background, $R$ and $R_i$ contribute to the evolution of the mass
and mass dipole in Eqs.~\eqref{eq:canonical-low-rank-solutions}.
The prescribed past data in Eqs.~\eqref{eq:past-canonical-mass}
and~\eqref{eq:past-canonical-response} give $\mathcal P_i=0$;
when $R_i=0$, the dipole remains $M_i=K_i$.

The angular-momentum normalization can be checked directly with
the point-spin source
\begin{subequations}\label{eq:canonical-FLRW-spin-source}
\begin{align}
 \tau_{0i}^{\rm spin}(\eta,\bx)
 &=\frac{c\epsilon_{iab}J_a}{2a^2(\eta)}\partial_b\delta^3(\bx),\label{eq:canonical-FLRW-spin-source-1}\\
 \tau_{ij}^{\rm spin}&=0.
 \label{eq:canonical-FLRW-spin-source-2}
\end{align}
\end{subequations}
Insertion into Eq.~\eqref{eq:direct-BD-current-source} gives $J_i$
itself, with the charge sign of Eq.~\eqref{eq:spatial-isometry-charges}.
This conserved spin is carried by the past field. Similarly,
the McVittie field has $\mathfrak M=M/\rho$ and
$R=\mathfrak N=0$, recovering its mass parameter.
Equations~\eqref{eq:canonical-physical-fluid} then give
$\delta\eps^{\rm can}=0$ and
$\delta p^{\rm can}=GM(\eps+p)/(c^2a\rho)$ in the exterior,
in agreement with Section~\ref{sec:mcvittie-monopole}.

\section{Minkowski limit and canonical flat-space multipoles}
\label{sec:minkowski}
The flat limit tests the canonical FLRW construction against the
established multipolar theory. We will first recover the
Blanchet--Damour metric, then identify the component moments,
residual gauge families and canonical source formulas in the
conventions of Damour--Iyer \cite{Blanchet:1985sp,Damour:1990gj}.
The lowest moments will provide a further check through their relation
to the mass, momentum, angular momentum and center-of-mass charges.

\subsection{Recovery of the canonical metric}
We begin by setting $a=1$ and choosing the vacuum background
$\Lambda=\eps=p=0$. Then $t=\eta$, and the shifted stress-energy
tensor perturbation defined in Eq.~\eqref{eq:shift-definition}
reduces to $\tau_{\mu\nu}=\delta T_{\mu\nu}$.
Equations~\eqref{eq:s} and~\eqref{eq:potentials} give $s=0$ and
$V^\pm=0$. Both metric Green functions $G^\pm$, defined in
Eq.~\eqref{eq:metric-Gpm-definition}, reduce to the flat Green function,
\begin{equation}\label{eq:flat-kernel}
 G_0(t;R;t')=\frac{\delta(c(t-t')-R)}R.
\end{equation}
For a history $f=f^{\mathrm{in}}+f^{\mathrm{dyn}}$ with constant past
part $f^{\mathrm{in}}$, the metric propagators defined in
Eq.~\eqref{eq:metric-propagator-definition} give
\begin{equation}\label{eq:flat-completed-propagator}
 \mathcal G_\pm[f](t,\rho)
 =\frac{f^{\mathrm{in}}}{\rho}
  +\frac{f^{\mathrm{dyn}}(t-\rho/c)}{\rho}
 =\frac{f(t-\rho/c)}{\rho}.
\end{equation}
The temporal residual gauge profiles have the same form after adding
their stationary harmonic contribution. The operators $D_\pm$ and
$T_\pm$ in Eqs.~\eqref{eq:canonical-Dpm} and~\eqref{eq:canonical-Tpm}
reduce to $D_\pm=c^{-1}\partial_t$ and $T_\pm=c^{-2}\partial_t^2$.
The field definitions~\eqref{eq:field-variables} reduce to
\begin{equation}\label{eq:flat-field-dictionary}
 \bar h_{\mu\nu}
 =\mathfrak{s}\left(h_{\mu\nu}-\frac12\eta_{\mu\nu}\eta^{\alpha\beta}h_{\alpha\beta}\right),
 \qquad \hat h=\bar h_{00}+\bar h_{kk}.
\end{equation}

The fluid response $R_L$ and its radial response profile
$\mathfrak N_L$ vanish by
Eqs.~\eqref{eq:direct-BD-fluid-moment-source} and~\eqref{eq:direct-BD-response-compact}.
The transformation of
Eq.~\eqref{eq:canonical-gauge-subtraction} is therefore residual
in the flat limit, and the canonical metric satisfies all four
harmonic conditions. The canonical cosmic fluid moments $N_L$
defined in Eq.~\eqref{eq:direct-BD-U-source-1} now parametrize only
this temporal residual freedom.

The invariant profiles of Eqs.~\eqref{eq:direct-BD-invariants} are
expressed in terms of the canonical mass and current moments
$M_L,J_L$ defined in Eqs.~\eqref{eq:direct-BD-canonical-moments}.
With $u:=t-\rho/c$ and dots on retarded-time
moments denoting $\partial_u$,
Eqs.~\eqref{eq:direct-BD-profiles-from-moments} give
\begin{subequations}\label{eq:BD-flat-profile-limit}
\begin{align}
 \mathfrak M_L(t,\rho)&=\kappa_\ell\frac{M_L(t-\rho/c)}\rho,\\
 \mathfrak C_L(t,\rho)&=\kappa_\ell\frac{\dot M_L(t-\rho/c)}\rho,\\
 \mathfrak J_L(t,\rho)&=\frac{\kappa_\ell}{2}
                                   \frac{J_L(t-\rho/c)}\rho.
\end{align}
\end{subequations}
Here $\kappa_\ell=(-1)^\ell/\ell!$ is defined in
Eq.~\eqref{eq:canonical-kappa}. In particular, the mass integral
in Eq.~\eqref{eq:direct-BD-profiles-from-moments-3} gives
\begin{equation}\label{eq:flat-mass-past-restoration}
 \mathfrak M_L(t,\rho)
 =\frac{\kappa_\ell}{\rho}
 \left[M_L^{\mathrm{in}}+
       \int_{t_0}^{t}\dd t'\,\dot M_L(t'-\rho/c)\right]
 =\kappa_\ell\frac{M_L(t-\rho/c)}{\rho},
 \qquad t_0:=\eta_i-d/c.
\end{equation}
The past mass moment $M_L^{\mathrm{in}}$ is specified in
Eq.~\eqref{eq:past-canonical-mass}, and the lower endpoint samples
the stationary past. Thus the mass and
current profiles include arbitrary compatible past constants, as in
\cite[Theorem~2.1, Eqs.~(2.31)--(2.33)]{Blanchet:1985sp}.
Substitution into
Eqs.~\eqref{eq:canonical-FLRW-metric} yields
\begin{subequations}\label{eq:canonical-flat-metric}
\begin{align}
 \bar h^{\rm can}_{00}
 &\overset{\rho>d}{=}\frac{4G}{c^2}\sum_{\ell\ge0}\frac{(-1)^\ell}{\ell!}
       \partial_L\left[\frac{M_L(u)}{\rho}\right],
 \label{eq:canonical-flat00}\\
 \bar h^{\rm can}_{0i}
 &\overset{\rho>d}{=}\frac{4G}{c^3}\sum_{\ell\ge1}\frac{(-1)^\ell}{\ell!}
   \left\{\partial_{L-1}\left[\frac{\dot M_{iL-1}(u)}{\rho}\right]
   +\frac12\epsilon_{iab}\partial_{aL-1}
       \left[\frac{J_{bL-1}(u)}{\rho}\right]\right\},
 \label{eq:canonical-flat0i}\\
 \bar h^{\rm can}_{ij}
 &\overset{\rho>d}{=}\frac{4G}{c^4}\sum_{\ell\ge2}\frac{(-1)^\ell}{\ell!}
   \left\{\partial_{L-2}\left[\frac{\ddot M_{ijL-2}(u)}{\rho}\right]
   +\partial_{aL-2}
       \left[\frac{\epsilon_{ab(i}\dot J_{j)bL-2}(u)}{\rho}\right]\right\}.
 \label{eq:canonical-flatij}
\end{align}
\end{subequations}
The factors $c^{-2}$, $c^{-3}$ and $c^{-4}$ agree with those in
Blanchet's review~\cite[Eqs.~(65a)--(65c)]{Blanchet:2024review}, with the
current-moment conversion given below.
These are the canonical mass and current fields of
\cite[Eqs.~(2.32a)--(2.32c)]{Blanchet:1985sp}, equivalently
\cite[Eq.~(5.32)]{Damour:1990gj}, in our lower-index
trace-reversed variables.\footnote{At $c=1$, the gothic metric
coefficient of Blanchet--Damour obeys
$G h^{\mu\nu}_{1,\mathrm{BD}}=-\bar h^{\mu\nu}$, and its current
moment obeys $S_L^{\rm BD}=(\ell+1)J_L/(2\ell)$.
Damour--Iyer instead use the raised
trace-reversed field and absorb Newton's constant into their moments:
$M_L^{\rm DI}=GM_L$ and $S_L^{\rm DI}=G(\ell+1)J_L/(2\ell)$.
The reference fields retain their mostly plus convention; these
dictionaries hold for either value of $\mathfrak{s}$ in our rescaled fields.
Raising both indices gives $\bar h^{0i}=-\bar h_{0i}$ in either signature.}

The lowest-rank balance laws~\eqref{eq:canonical-FLRW-low-ranks}
reduce to
\begin{equation}\label{eq:canonical-low-conservation}
 \dot M=0,\qquad \ddot M_i=0,\qquad \dot J_i=0.
\end{equation}
These relations also follow from the harmonic conditions on
Eqs.~\eqref{eq:canonical-flat-metric}: the divergences cancel between
successive components except for the lowest ranks, where the
corresponding higher-component terms are absent. We thus recover
both the canonical metric and the constraints on its lowest moments
directly from the coupled FLRW construction.

\subsection{Component source moments and the Damour--Iyer dictionary}
The source formulas require a comparison before the residual gauge
terms are removed. We start with the retarded solution: subtracting
the spatial trace equation from the wave equation for $\hat h$ gives
$\square\bar h_{00}=-16\pi G\delta T_{00}/c^4$; all ten components
therefore propagate with $G_0$. With $V^+=s=0$, the past
$\tau_{00}$ in Eq.~\eqref{eq:past-source-density} is compactly
supported, and Eq.~\eqref{eq:energy-solution} keeps the dynamical part
inside the source region as well. In this flat-space section all
component source moments, including density moments, refer to the
full source. Its stationary Poisson field plus its dynamical part
combine into
\begin{equation}\label{eq:flat-retarded-field}
 \bar h_{\mu\nu}(t,\bx)
 =\frac{4G}{c^4}\int\dd^3x'\,\frac{\delta T_{\mu\nu}(t-R/c,\bx')}{R},
 \qquad R=|\bx-\bx'|.
\end{equation}
This recovers \cite[Eqs.~(5.1)--(5.2)]{Damour:1990gj}.

The flat limit of the source-time weight gives the component moments
of \cite[Eqs.~(5.3)--(5.4)]{Damour:1990gj}. At $a=1$, the metric
and light-cone moments coincide. Applying
Eq.~\eqref{eq:timemoment} to every component of the full source gives
\begin{equation}\label{eq:flat-source-STF}
 \begin{pmatrix}c^2\mathcal F_{00|L}\\c\mathcal F_{0i|L}\\\mathcal F_{ij|L}\end{pmatrix}(u;1)
 :=\int\dd^3x'\,x'_{\langle L\rangle}
   \int_{-1}^{1}\dd z\,\delta_\ell(z)
   \begin{pmatrix}\delta T_{00}\\\delta T_{0i}\\\delta T_{ij}\end{pmatrix}
       (u+\rho'z/c,\bx').
\end{equation}
Here $\rho'=|\bx'|$ and the normalized weight $\delta_\ell$ is
defined in Eq.~\eqref{eq:time-weight-definition}. The STF projection acts on $L$ only.
At each finite order in source size, their expansion in Cartesian
moment traces follows from Eq.~\eqref{eq:curved-retarded-moment-finite}
at $a=1$, applied to each component. The Fourier representation of
the exact metric is given in Appendix~\ref{app:Cartesian-convergence}.

Specializing the complete exterior fields~\eqref{eq:full-multipolar-fields}
and the reconstruction of $\hat h$ to this common metric propagator gives
\begin{subequations}\label{eq:flat-STF-reconstruction}
\begin{align}
 \begin{pmatrix}\bar h_{00}\\\bar h_{0i}\\\bar h_{ij}\end{pmatrix}
 &\overset{\rho>d}{=}\frac{4G}{c^4}\sum_{\ell\ge0}\frac{(-1)^\ell}{\ell!}\partial_L
 \left[\frac1\rho\begin{pmatrix}c^2\mathcal F_{00|L}\\c\mathcal F_{0i|L}\\\mathcal F_{ij|L}\end{pmatrix}(u;1)\right],
 \label{eq:flat-STF-fields}\\
 \hat h
 &\overset{\rho>d}{=}\frac{4G}{c^2}\sum_{\ell\ge0}\frac{(-1)^\ell}{\ell!}\partial_L
          \left[\frac{\mathcal F_{00|L}(u;1)+\mathcal F_{kk|L}(u;1)/c^2}\rho\right].
 \label{eq:flat-STF-scalar}
\end{align}
\end{subequations}
Undoing trace reversal with Eq.~\eqref{eq:flat-field-dictionary} gives
the explicit metric
$g_{\mu\nu}=\eta_{\mu\nu}+h_{\mu\nu}$ at linear order:
\begin{subequations}\label{eq:flat-component-metric}
\begin{align}
 h_{00}
 &\overset{\rho>d}{=}\frac{2\mathfrak{s}G}{c^2}\sum_{\ell\ge0}\frac{(-1)^\ell}{\ell!}\partial_L
       \left[\frac{\mathcal F_{00|L}+\mathcal F_{kk|L}/c^2}\rho\right],
 \label{eq:flat-metric00}\\
 h_{0i}
 &\overset{\rho>d}{=}\frac{4\mathfrak{s}G}{c^3}\sum_{\ell\ge0}\frac{(-1)^\ell}{\ell!}\partial_L
       \left[\frac{\mathcal F_{0i|L}}\rho\right],
 \label{eq:flat-metric0i}\\
 h_{ij}
 &\overset{\rho>d}{=}\frac{4\mathfrak{s}G}{c^4}\sum_{\ell\ge0}\frac{(-1)^\ell}{\ell!}\partial_L
       \left[\frac{\mathcal F_{ij|L}
          +\tfrac12\delta_{ij}(c^2\mathcal F_{00|L}-\mathcal F_{kk|L})}
                    {\rho}\right].
 \label{eq:flat-metricij}
\end{align}
\end{subequations}
All moments are evaluated at $(u;1)$. These expressions retain the full
time averages and every multipolar order, in both parity sectors.
They give the metric in harmonic gauge without a radiation-zone or
slow-motion approximation.
The stationary mass monopole reproduces the Schwarzschild limit of
Section~\ref{sec:mcvittie-monopole}.

We can now compare directly with Damour--Iyer
\cite[Eqs.~(5.1), (5.3)--(5.4)]{Damour:1990gj}. Their trace-reversed
field carries raised indices and we denote it by $\bar h^{\mu\nu}_{\rm DI}$;
its source coefficients are $F_L,G_{iL},H_{ijL}$. With $c$ explicit,
the dictionary is
\begin{subequations}\label{eq:DI-convention-dictionary}
\begin{align}
 \bar h^{00}_{\rm DI}&=\bar h_{00}=\hat h-\bar h_{kk},&
 \bar h^{0i}_{\rm DI}&=-\bar h_{0i},&
 \bar h^{ij}_{\rm DI}&=\bar h_{ij},\label{eq:DI-field-map}\\
 F_L&=c^2\mathcal F_{00|L},&
 G_{iL}&=-c\mathcal F_{0i|L},&
 H_{ijL}&=\mathcal F_{ij|L}.
 \label{eq:DI-source-map}
\end{align}
\end{subequations}
The relative minus sign in the mixed component is fixed by
$T^{0i}=-T_{0i}$ in either signature. This identifies the
component source moments before any residual gauge transformation.

\subsection{Residual gauge families and canonical source moments}
\label{sec:canonical-flat-moments}
The flat specialization of the gauge reduction in
Section~\ref{sec:direct-gauge-invariant-multipoles} separates the
component field into its canonical part and four residual harmonic
families. We now express this separation in the retarded-time
normalization used in the flat literature:

\begin{equation}\label{eq:flat-canonical-gauge-split}
 \bar h_{\mu\nu}
 =\bar h^{\rm can}_{\mu\nu}
 +\mathfrak{s}\big(2\partial_{(\mu}w_{\nu)}-\eta_{\mu\nu}\partial_\alpha w^\alpha\big),
 \qquad \square w^\mu=0\quad(\rho>d).
\end{equation}
Here $w_\mu:=\eta_{\mu\nu}w^\nu$. This uses the gauge transformation in
Eqs.~\eqref{eq:rescaled-metric-gauge-action}
at $a=1$; the transformation generated by $-w^\mu$ removes the last
two terms. We use the gauge-vector decomposition
\eqref{eq:BD-FLRW-gauge-vector} at $a=1$, with $\zeta^\mu=w^\mu$,
and introduce the four STF retarded-time gauge moments
$W_L^0$, $W_L^\nabla$, $W_L^{\mathrm E}$ and $W_L^{\mathrm B}$
through the profile identifications
\begin{subequations}\label{eq:BD-flat-gauge-profile-map}
\begin{align}
 W_L&=-\kappa_\ell W_L^0(u)/\rho,\\
 X_L&=\kappa_\ell W_L^\nabla(u)/\rho,\\
 Y_L&=\kappa_\ell W_L^{\mathrm E}(u)/\rho,\\
 Z_L&=\kappa_\ell\frac\ell{\ell+1}W_L^{\mathrm B}(u)/\rho.
\end{align}
\end{subequations}
These moments parametrize the temporal, spatial-gradient, electric and
magnetic terms of the four outgoing residual gauge families in
\cite[Eq.~(2.33)]{Blanchet:1985sp}.\footnote{The moments
$(W_L^0,W_L^\nabla,W_L^{\mathrm E},W_L^{\mathrm B})$ have the
normalizations of $(W_L,X_L,Y_L,Z_L)$ in
\cite[Eq.~(4.13)]{Blanchet:1998in}. That reference uses the
contravariant gothic perturbation, whose linear term is
$-\bar h^{\mu\nu}$; hence our source-minus-canonical gauge vector
is minus its gauge vector, with $G$ restored.}

The irreducible decomposition of the component moments in
\cite[Eqs.~(5.10)--(5.12)]{Damour:1990gj}, together with its
conservation relations~(5.27), fixes the gauge moments. Substituting
the component dictionary~\eqref{eq:DI-source-map} gives
\begin{subequations}\label{eq:flat-gauge-source-moments}
\begin{align}
 W_L^\nabla
 &=\frac{2\ell+1}{2(\ell+1)(\ell+2)(2\ell+5)}
       \mathcal F_{ab|abL},\label{eq:flat-gauge-gradient-source}\\
 W_L^0
 &=-\frac{2\ell+1}{(\ell+1)(2\ell+3)}
       \mathcal F_{0a|aL}-\frac1{c^2}\dot W_L^\nabla,
       \label{eq:flat-gauge-time-source}\\
 W_L^{\mathrm E}
 &=-\mathcal F_{aa|L}+3\dot W_L^0-\frac1{c^2}\ddot W_L^\nabla,
       \label{eq:flat-gauge-electric-source}\\
 W_L^{\mathrm B}
 &=\frac{2\ell+1}{(\ell+2)(2\ell+3)}
       \STF_L\!\left(\epsilon_{ab i_\ell}
               \mathcal F_{bc|acL-1}\right).
       \label{eq:flat-gauge-magnetic-source}
\end{align}
\end{subequations}
The first three equations hold for $\ell\ge0$ and the last for
$\ell\ge1$. All moments are evaluated at $u$, with the scale-factor
argument of $\mathcal F$ fixed to $1$. Conservation gives
$W^{\mathrm E}=0$ at $\ell=0$, consistent with the $Y_L$ term
starting at $\ell=1$ in Eq.~\eqref{eq:BD-FLRW-gauge-space}. The multipole
indices after the vertical bar are STF as in
Eq.~\eqref{eq:flat-source-STF}; the contractions with the component
indices before the bar need not vanish.

The temporal source relation also gives $N_L=-W_L^0$ by the flat
limit of Eq.~\eqref{eq:direct-BD-U-source-1}. Since $V^+=0$, this
history supplies only a residual gauge transformation.

We now extract the canonical moments. Adding the $00$ component
and the spatial trace of Eq.~\eqref{eq:flat-canonical-gauge-split}
isolates the mass moments: the canonical spatial trace vanishes,
and the gauge contribution to this sum is $-4\partial_t w^0/c$.
The magnetic projection of the mixed component similarly isolates
the current moments and the time derivative of $W_L^{\mathrm B}$.
Using Eqs.~\eqref{eq:BD-FLRW-gauge-vector} with the profile
identifications~\eqref{eq:BD-flat-gauge-profile-map} gives
\begin{subequations}\label{eq:canonical-moments-with-gauge}
\begin{align}
 M_L&=\mathcal F_{00|L}+\frac1{c^2}\mathcal F_{aa|L}-\frac4{c^2}\dot W_L^0,
       \qquad \ell\ge0,\\
 J_L&=-\frac{2\ell}{\ell+1}\left[\STF_L\!\left(\epsilon_{ab i_\ell}
                 \mathcal F_{0b|aL-1}\right)+\frac1{c^2}\dot W_L^{\mathrm B}\right],
       \qquad \ell\ge1.
\end{align}
\end{subequations}
Eliminating the gauge moments with
Eqs.~\eqref{eq:flat-gauge-source-moments} then gives the explicit
dictionary
\begin{subequations}\label{eq:canonical-source-dictionary}
\begin{align}
 M_L
 &=\mathcal F_{00|L}+\frac1{c^2}\mathcal F_{aa|L}
 +\frac{4(2\ell+1)}{c^2(\ell+1)(2\ell+3)}\dot{\mathcal F}_{0a|aL}
 +\frac{2(2\ell+1)}{c^4(\ell+1)(\ell+2)(2\ell+5)}
       \ddot{\mathcal F}_{ab|abL},
 \label{eq:canonical-mass-source}\\
 J_L
 &=-\frac{2\ell}{\ell+1}\STF_L\!\left[\epsilon_{ab i_\ell}
   \left(\mathcal F_{0b|aL-1}
   +\frac{2\ell+1}{c^2(\ell+2)(2\ell+3)}
       \dot{\mathcal F}_{bc|acL-1}\right)\right].
 \label{eq:canonical-current-source}
\end{align}
\end{subequations}
The ranges and time arguments are those of
Eqs.~\eqref{eq:canonical-moments-with-gauge}. Thus the canonical
moments combine component source moments of several ranks and their
time derivatives. In particular, the pressure term
$\mathcal F_{aa|L}/c^2$ and the momentum and stress corrections are fixed
by the harmonic reduction.

These source formulas also check the general FLRW reduction:
setting $a=1$ in Eqs.~\eqref{eq:direct-BD-final-source-map} gives
Eq.~\eqref{eq:canonical-current-source} directly and the time
derivative of Eq.~\eqref{eq:canonical-mass-source}. The same
initial mass data fix the integration constant.

Inserting the weighted definition~\eqref{eq:flat-source-STF} into
Eq.~\eqref{eq:canonical-mass-source} reproduces
\cite[Eqs.~(5.33)--(5.34)]{Damour:1990gj}: the density-plus-pressure, momentum and stress terms
carry the weights $\delta_\ell$, $\delta_{\ell+1}$ and
$\delta_{\ell+2}$, respectively. Similarly,
Eq.~\eqref{eq:canonical-current-source} reproduces its Eq.~(5.35)
after the conversion $S_L^{\rm DI}=G(\ell+1)J_L/(2\ell)$.
The sign of the momentum contribution in each equation follows
from $T^{0i}=-T_{0i}$. These are exact linearized expressions with
the full time averages; no slow-motion expansion has been used.

At ranks zero and one, source conservation reduces the weighted
integrals in Eqs.~\eqref{eq:canonical-source-dictionary} to the
instantaneous charge integrals, as in
\cite[Eqs.~(5.43b), (5.45b), (5.47c)]{Damour:1990gj}. With the
charge definitions~\eqref{eq:spatial-isometry-charges}
and~\eqref{eq:poincare-charges}, this gives
\begin{equation}\label{eq:canonical-charge-dictionary}
 M=E/c^2,\qquad M_i(u)=\mathcal K_i+u\mathcal P_i,\qquad
 J_i=\epsilon_{ijk}P_{j|k}.
\end{equation}

\section{Comparison with scalar--vector--tensor perturbation theory}
\label{sec:SVT-comparison}
The canonical moments characterize an exterior metric, whereas the
usual scalar--vector--tensor (SVT) projections use a metric on an entire
spatial slice. In this section, we will specify a continuation of the
canonical field and derive its SVT variables from the mass and current
moments $M_L,J_L$ and the canonical cosmic fluid moment $N_L$.
The corresponding cosmic fluid variables and coupled Einstein equations will
then allow a comparison with the standard cosmological formulation
\cite{Bardeen:1980kt,Kodama:1984ziu,Mukhanov:1990me,Malik:2008im}.
This map requires spatial inverse Laplacians but no further evolution
equations. Finally, we will explain which interior information is
needed to obtain the SVT fields of an extended source.

\subsection{From canonical moments to SVT metric fields}
We start from the exterior canonical metric derived in
Eqs.~\eqref{eq:canonical-FLRW-metric}. To define its spatial
projections without specifying the physical metric inside the source,
we continue its radial profiles to the spatial origin
and take the observation derivatives as distributional derivatives.
This defines a canonical multipole field on the whole spatial slice.

We will express this field in Poisson gauge, where the scalar, vector
and tensor metric fields enter the line element as
\cite{Bardeen:1980kt,Kodama:1984ziu}:
\begin{equation}\label{eq:invariant-line-element}
 \dd s^2=\mathfrak{s}a^2\big[-c^2(1+2\lambda\Psi/c^2)\dd\eta^2
       +2\lambda\Sigma_i\dd\eta\dd x^i
       +\big((1-2\lambda\Phi/c^2)\delta_{ij}+\lambda\gamma_{ij}\big)
                   \dd x^i\dd x^j\big]+O(\lambda^2).
\end{equation}
Here $\Psi$ and $\Phi$ are the Bardeen potentials describing the lapse
and spatial curvature, respectively.\footnote{The scalar conventions
agree with \cite[Eq.~(2.3)]{Baumann:AdvCosmo}, which uses
$\mathfrak{s}=-1$ and $c=1$; our dimensional potentials correspond to
$c^2$ times those in that reference. We denote its tensor perturbation by
$\gamma_{ij}$ to distinguish it from our metric perturbation $h_{\mu\nu}$.
The normalization of the vector perturbation $\Sigma_i$ is fixed by
Eq.~\eqref{eq:invariant-line-element}.}
The vector perturbation $\Sigma_i$ and tensor perturbation $\gamma_{ij}$
obey
\begin{subequations}\label{eq:invariant-transversality}
\begin{align}
 \partial_i\Sigma_i&=0,\label{eq:SVT-vector-transverse}\\
 \partial_j\gamma_{ij}&=0,\label{eq:SVT-tensor-transverse}\\
 \gamma_{ii}&=0.\label{eq:SVT-tensor-tracefree}
\end{align}
\end{subequations}

We fix the inverse Laplacian by the decaying Newton kernel,
\begin{equation}\label{eq:canonical-SVT-inverse-Laplacian}
 (\Delta^{-1}f)(\eta,\bx):=-\frac1{4\pi}
       \int\frac{\dd^3y}{|\bx-\mathbf y|}\,f(\eta,\mathbf y),
\end{equation}
with its distributional interpretation for multipole sources.\footnote{We
use finite multipolar sums, or infinite sums for which the spatial
projections converge.}

Let us collect the three sums of invariant radial profiles that enter
the canonical metric:
\begin{subequations}\label{eq:canonical-SVT-sums}
\begin{align}
 \mathscr M(\eta,\bx)&:=\sum_{\ell\ge0}\partial_L
                    \mathfrak M_L(\eta,\rho),
 \label{eq:canonical-SVT-mass-sum}\\
 \mathscr V_i(\eta,\bx)&:=\sum_{\ell\ge1}
       \big[\partial_{L-1}\mathfrak C_{iL-1}(\eta,\rho)
          +\epsilon_{iab}\partial_{aL-1}\mathfrak J_{bL-1}(\eta,\rho)\big],
 \label{eq:canonical-SVT-vector-sum}\\
 \mathscr H_{ij}(\eta,\bx)&:=\sum_{\ell\ge2}
       \big[c\partial_{L-2}D_+\mathfrak C_{ijL-2}(\eta,\rho)
          +2c\partial_{aL-2}D_+\big(\epsilon_{ab(i}\mathfrak J_{j)bL-2}(\eta,\rho)\big)\big].
 \label{eq:canonical-SVT-tensor-sum}
\end{align}
\end{subequations}
Here the radial profiles $\mathfrak M_L(\eta,\rho)$,
$\mathfrak C_L(\eta,\rho)$ and $\mathfrak J_L(\eta,\rho)$ depend
on position only through $\rho=|\bx|$; the spatial derivatives in
these sums generate the angular dependence. Their STF symmetry gives
$\mathscr H_{ii}=0$.
Equations~\eqref{eq:direct-BD-profiles-from-moments} reconstruct these
profiles from $M_L,J_L$ and the radial response profile $\mathfrak N_L$.
The latter is computed from $N_L$ through
Eqs.~\eqref{eq:direct-BD-response-compact} and~\eqref{eq:direct-BD-U-source-2}.

Let $\mathcal X^\mu$ denote the rescaled gauge vector from the
canonical gauge of Section~\ref{sec:canonical-gauge} to Poisson gauge,
with spatial components $\mathcal X_i:=\delta_{ij}\mathcal X^j$.
Starting from $\bar h^{\rm can}_{\mu\nu}$, we subtract the metric
variation in Eqs.~\eqref{eq:rescaled-metric-gauge-action} with
$\zeta^\mu=\mathcal X^\mu$. Requiring the transformed spatial trace-free
metric and mixed component to be transverse gives
\begin{subequations}\label{eq:canonical-SVT-gauge-equations}
\begin{align}
 \Delta\mathcal X_i+\tfrac13\partial_i\partial_j\mathcal X_j
      &=\frac{4G}{c^4}\partial^j\mathscr H_{ij},
 \label{eq:canonical-SVT-spatial-gauge-equation}\\
 \Delta\mathcal X^0&=D_-\partial_i\mathcal X_i
                             -\partial_i\bar h^{\rm can}_{0i}.
 \label{eq:canonical-SVT-temporal-gauge-equation}
\end{align}
\end{subequations}
Taking a divergence of the first equation fixes
$\partial_i\mathcal X_i$; inserting it back fixes $\mathcal X_i$.
The second equation then determines $\mathcal X^0$. With the same
inverse Laplacian throughout, this gives
\begin{subequations}\label{eq:canonical-SVT-compensators}
\begin{align}
 \mathcal X_i&:=\frac{4G}{c^4}\Delta^{-1}\!\left[
       \partial^j\mathscr H_{ij}
       -\tfrac14\partial_i\Delta^{-1}\partial^j\partial^k\mathscr H_{jk}\right],
 \label{eq:canonical-SVT-spatial-compensator}\\*
 {\mathcal X^0}&:=\Delta^{-1}\partial^i
                     \big(D_-\mathcal X_i-\frac{4G}{c^3}\mathscr V_i\big).
 \label{eq:canonical-SVT-temporal-compensator}
\end{align}
\end{subequations}
Here  $\mathcal X^\mu$, $\mathscr V_i$ and $\mathscr H_{ij}$ depend
on $(\eta,\bx)$.

Substituting Eqs.~\eqref{eq:canonical-SVT-compensators} into these
gauge-transformation formulas, and reading off the temporal, mixed
and spatial components, now yields the explicit map
\begin{subequations}\label{eq:canonical-SVT-map}
\begin{align}
 \Psi&=\frac{-G\mathscr M-c\partial_\eta{\mathcal X^0}}{a},
 \label{eq:canonical-SVT-Psi}\\
 \Phi&=\frac{-G\mathscr M+c\cH{\mathcal X^0}
                              +\tfrac{c^2}3\partial_i\mathcal X_i}{a},
 \label{eq:canonical-SVT-Phi}\\
 \Sigma_i&=\frac{4G\mathscr V_i/c^2-cD_-\mathcal X_i+c\partial_i{\mathcal X^0}}{a},
 \label{eq:canonical-SVT-Sigma}\\
 \gamma_{ij}&=\frac{4G\mathscr H_{ij}/c^4-2\partial_{(i}\mathcal X_{j)}
                       +\tfrac23\delta_{ij}\partial_k\mathcal X_k}{a}.
 \label{eq:canonical-SVT-gamma}
\end{align}
\end{subequations}
Their divergences and trace give
Eqs.~\eqref{eq:invariant-transversality} directly. To see gauge
invariance explicitly, let $\mathcal D_{\mu\nu}[\zeta^\sigma]$ denote the metric
variation in Eqs.~\eqref{eq:rescaled-metric-gauge-action}. For a general
input metric $\bar h_{\alpha\beta}$, denote by
$\mathcal X^\mu[\bar h_{\alpha\beta}]$ the corresponding rescaled vector
whose metric variation is subtracted to reach Poisson gauge.
For gauge vectors obeying the same spatial boundary prescription, uniqueness
of the rescaled gauge vector to Poisson gauge gives
\begin{subequations}\label{eq:SVT-dictionary-gauge-invariance}
\begin{align}
 \mathcal X^\mu[\bar h_{\alpha\beta}+\mathcal D_{\alpha\beta}[\zeta^\sigma]]
 &=\mathcal X^\mu[\bar h_{\alpha\beta}]+\zeta^\mu,
 \label{eq:SVT-compensator-covariance}\\
 \bar h^{\rm Pois}_{\mu\nu}[\bar h_{\alpha\beta}+\mathcal D_{\alpha\beta}[\zeta^\sigma]]
 &=\bar h_{\mu\nu}+\mathcal D_{\mu\nu}[\zeta^\sigma]
   -\mathcal D_{\mu\nu}[\mathcal X^\sigma[\bar h_{\alpha\beta}]+\zeta^\sigma]\notag\\
 &=\bar h^{\rm Pois}_{\mu\nu}[\bar h_{\alpha\beta}],
 \label{eq:SVT-representative-invariance}
\end{align}
\end{subequations}
where $\bar h^{\rm Pois}_{\mu\nu}[\bar h_{\alpha\beta}]
:=\bar h_{\mu\nu}-\mathcal D_{\mu\nu}[\mathcal X^\sigma[\bar h_{\alpha\beta}]]$.
Thus Eqs.~\eqref{eq:canonical-SVT-map} give gauge-invariant quantities
from the canonical metric. These quantities are unchanged
under gauge transformations obeying the stated boundary prescription,
including transformations that do not preserve generalized harmonic gauge. The prescribed whole-space
continuation and inverse Laplacian remain part of the comparison;
gauge invariance does not supply missing interior information.
The fluid response remains in these invariants:
$R_L$ enters $\mathfrak C_L$ through $\partial_\eta M_L-R_L$, and
its radial response profile $\mathfrak N_L$ also enters the mass balance
determining $\mathfrak M_L$. The FLRW profiles carry the cosmological tails into this map.

The lowest moments illustrate the physical content. For the McVittie
monopole, $\mathscr M=M/\rho$ and
$\mathscr V_i=\mathscr H_{ij}=0$, so
\begin{equation}\label{eq:canonical-SVT-monopole}
 \Psi=\Phi=-\frac{GM}{a\rho},\qquad \Sigma_i=\gamma_{ij}=0.
\end{equation}
For the conserved spin $J_i$ of
Eqs.~\eqref{eq:canonical-FLRW-spin-source},
$\mathfrak J_i=-J_i/(2a\rho)$ and
$\mathscr V_i=-(\mathbf J\times\bx)_i/(2a\rho^3)$.
Its divergence vanishes and $D_+\mathfrak J_i=0$, giving
\begin{equation}\label{eq:canonical-SVT-spin}
 \Sigma_i=-\frac{2G(\mathbf J\times\bx)_i}{c^2a^2\rho^3},
 \qquad \Psi=\Phi=0,\qquad \gamma_{ij}=0.
\end{equation}
Thus canonical mass and current moments include the scalar Coulomb
field and vector frame dragging. Their higher multipoles also
contribute to the tensor field; the moment families do not correspond
one for one to the scalar, vector and tensor sectors.

\subsection{Cosmic fluid perturbations and coupled equations in Poisson gauge}
The comparison must also transform the cosmic fluid perturbations, since
Poisson gauge changes the time slicing and its associated Eulerian observers.
We will use the same temporal component $\mathcal X^0$ to obtain the fluid variables
and state their coupled Einstein equations.

We begin by evaluating the physical density, pressure, momentum and
anisotropic stress of Eqs.~\eqref{eq:physical-matter-variables} in the
canonical representative, denoting them by a superscript
$\mathrm{can}$. Subtracting the gauge transformation with
$\xi^\mu=\mathcal X^\mu/a$ in
Eqs.~\eqref{eq:physical-matter-gauge} gives
\begin{subequations}\label{eq:canonical-SVT-matter}
\begin{align}
 \delta\eps^{\rm Pois}&=\delta\eps^{\rm can}
                                  -\frac{\dot\eps}{ac}\mathcal X^0,
 \label{eq:canonical-SVT-density}\\
 \delta p^{\rm Pois}&=\delta p^{\rm can}
                                  -\frac{\dot p}{ac}\mathcal X^0,
 \label{eq:canonical-SVT-pressure}\\
 j_i^{\rm Pois}&=j_i^{\rm can}
                          +\frac{\eps+p}{ac}\partial_i\mathcal X^0,
 \label{eq:canonical-SVT-momentum}\\
 \pi_{ij}^{\rm Pois}&=\pi_{ij}^{\rm can}=\pi_{ij}.
 \label{eq:canonical-SVT-anisotropic-stress}
\end{align}
\end{subequations}
The density and pressure differences are the changes in the background
fluid between the time slices selected by the two representatives.
They do not add independent matter data. In the exterior,
Eqs.~\eqref{eq:canonical-physical-fluid} give these perturbations in the
canonical representative explicitly;
for their distributional continuation, the Einstein equations also
include the multipole sources at the origin.

In Poisson gauge the coupled equations take the standard form
\cite{Kodama:1984ziu,Malik:2008im}, with physical momentum
$j_i^{\rm Pois}$ and anisotropic stress $\pi_{ij}$ defined above:
\begin{subequations}\label{eq:invariant-Einstein}
\begin{align}
 \Delta\Phi-\frac{3\cH}{c^2}(\dot\Phi+\cH\Psi)
 &=\frac{4\pi Ga^2}{c^2}\delta\eps^{\rm Pois},
 \label{eq:invariant-density-constraint}\\
 \partial_i(\dot\Phi+\cH\Psi)-\frac{c^2}4\Delta\Sigma_i
 &=-4\pi Ga^2 j_i^{\rm Pois},
 \label{eq:invariant-momentum-constraint}\\
 \ddot\Phi+\cH(\dot\Psi+2\dot\Phi)
 +(2\dot\cH+\cH^2)\Psi-\frac{c^2}3\Delta(\Phi-\Psi)
 &=4\pi Ga^2\delta p^{\rm Pois},
 \label{eq:invariant-pressure-equation}\\
 \partial_{\langle i}\partial_{j\rangle}(\Phi-\Psi)
 -\partial_{(i}(\partial_\eta+2\cH)\Sigma_{j)}
 -\frac{c^2}{2a}\Op^-(a\gamma_{ij})
 &=\frac{8\pi Ga^2}{c^2}\pi_{ij}.
 \label{eq:invariant-anisotropic-equation}
\end{align}
\end{subequations}
The scalar density equation and, when $\Phi=\Psi$, the pressure
equation reproduce \cite[Eqs.~(2.45), (2.50)]{Baumann:AdvCosmo}.
The density and momentum equations are constraints. The
transverse-traceless projection of the last equation supplies the
two tensor wave equations, while a matter law closes the scalar
fluid evolution. Thus the six invariant metric functions consist
of two scalar potentials, a transverse vector and a
transverse-traceless tensor.

\subsection{Exterior data and spatial nonlocality}
The inverse Laplacian in
Eq.~\eqref{eq:canonical-SVT-inverse-Laplacian} explains why this map
must specify a whole-space field. Knowledge of the metric only in
an exterior region does not determine its usual SVT projections
there. This limitation, and the distinction between a global TT
projection and a local transverse projection, are discussed in
\cite[Secs.~2.2--2.3]{Flanagan:2005yc} and
\cite[Sec.~I]{Ashtekar:2017wgq}.
Chu and Liu \cite{Chu:2019ndv} analyze the nonlocal support of
the TT field and its combination with the other sectors in local
curvature observables, in Minkowski and cosmological backgrounds.
The example below isolates the consequence for the present source
dictionary: identical exterior canonical moments need not determine
identical whole-space SVT projections.

A compact perturbation illustrates the missing information already
in Minkowski spacetime. Let $e_{ij}$ be a constant STF tensor and
$f(\eta,\bx)$ a smooth function of compact spacetime support. Set
\begin{subequations}\label{eq:SVT-compact-example}
\begin{align}
 \bar h^{\rm c}_{00}&:=0,\label{eq:SVT-compact-example-temporal}\\
 \bar h^{\rm c}_{0i}&:=0,\label{eq:SVT-compact-example-mixed}\\
 \bar h^{\rm c}_{ij}&:=\epsilon_{iab}\epsilon_{jcd}e_{bd}
                                      \partial_a\partial_c f.
 \label{eq:SVT-compact-example-spatial}
\end{align}
\end{subequations}
The double curl is symmetric and divergence free, so
$\bar h^{\rm c}_{\mu\nu}$ obeys harmonic gauge. It is the retarded
metric of the compact conserved source
$\tau^{\rm c}_{\mu\nu}:=-c^4\square\bar h^{\rm c}_{\mu\nu}/(16\pi G)$.
Its exterior metric and all its exterior canonical moments vanish.
Nevertheless, define
\begin{equation}\label{eq:SVT-compact-trace-potential}
 \Upsilon:=\Delta^{-1}\bar h^{\rm c}_{kk}
              =-\Delta^{-1}e_{ab}\partial_a\partial_b f.
\end{equation}
The whole-space SVT projection gives, outside the support of $f$,
\begin{subequations}\label{eq:SVT-compact-exterior}
\begin{align}
 \Psi&=\tfrac14\partial_\eta^2\Upsilon,
 \label{eq:SVT-compact-exterior-Psi}\\
 \Phi&=\Sigma_i=0,
 \label{eq:SVT-compact-exterior-scalar-vector}\\
 \gamma_{ij}&=\tfrac12\partial_i\partial_j\Upsilon.
 \label{eq:SVT-compact-exterior-tensor}
\end{align}
\end{subequations}
The harmonicity statement applies only in this exterior region.
Globally the definition gives
\begin{equation}\label{eq:SVT-trace-potential-source}
 \Delta\Upsilon=\bar h^{\rm c}_{kk}
              =-e_{ab}\partial_a\partial_b f,
\end{equation}
which generally does not vanish inside the support of $f$.
For illustration, choose $f$ to be spatially spherically symmetric,
$f(\eta,\mathbf y)=F(\eta,r')$ with $r':=|\mathbf y|$, while retaining
smooth time dependence and compact spacetime support. At each fixed
$\eta$, for $\rho$ beyond the outer radius of its spatial support,
the exterior Newton potential gives
\begin{equation}\label{eq:SVT-trace-potential-exterior}
 \Upsilon(\eta,\bx)
 =\frac{3e_{ab}n_an_b}{\rho^3}
       \int_0^\infty\dd r'\,r'^2F(\eta,r').
\end{equation}
This is a nonzero harmonic quadrupole when the radial integral is
nonzero; its boundary values are supplied by the interior. The scalar and tensor fields
in Eqs.~\eqref{eq:SVT-compact-exterior} then need not vanish, although
the exterior metric is zero in its original representative. Their
contributions to the exterior curvature cancel; in particular,
$\partial_i\partial_j\Psi-\tfrac12\ddot\gamma_{ij}=0$.
This is a property of the spatial projection, not an additional
exterior gravitational field.

Equations~\eqref{eq:canonical-SVT-map} give the SVT fields of the
canonical multipole metric continued to the spatial origin as a
distribution. This continuation reproduces the exterior metric but
generally differs from the interior metric of an extended source.
Because the SVT projection contains inverse Laplacians, this interior
difference can change the projected fields even outside the source.
Recovering the whole-space SVT fields of the extended source therefore
requires additional interior information, or equivalent boundary data.

For gravitational measurements confined to the exterior, the physical
information is contained in the exterior metric, up to gauge. Two
solutions with the same exterior geometry give the same such
measurements for corresponding observer worldlines and frames, even
when their individual globally defined SVT fields differ. Within the
solution class considered here, the canonical moments determine this
exterior metric and its cosmic fluid response.

\section{Discussion}
We have constructed the retarded linearized metric and its accompanying
cosmic fluid response for localized sources in an arbitrary smooth,
spatially flat FLRW background. We connected the prescribed source
history to the exterior field through cosmological Green functions
and retarded source moments at every multipole order, retaining both
the light-cone signal and the curvature tail. Compatible past fields
allowed us to retain pre-existing mass and current multipoles. The construction
did not require a slow-motion, sub-Hubble or geometric-optics
approximation. We also obtained analytical representations of both
metric Green functions in $\Lambda$CDM with radiation.
We derived effective multipolar sources by truncating
the source-size expansion of sub-Hubble, slowly varying extended
sources. Our quadrupolar truncation included the cosmic-fluid
response. In de Sitter spacetime, our quadrupolar and octupolar
truncations recovered the earlier
models~\cite{Compere:2023ktn,Compere:2026xoe,Harsh:2026vmh}.

The canonical reduction yielded source expressions for the mass
moments $M_L$, current moments $J_L$ and canonical cosmic fluid moment
$N_L$, from which we reconstructed the exterior metric and matter
perturbations within the prescribed-source class. The independent
fluid response encoded by $N_L$ arises because the gravitational field
disturbs the cosmic fluid supporting the background, including outside
the region occupied by the source. In pure de Sitter spacetime, the
background is supported solely by a cosmological constant and has zero
enthalpy density, so this fluid response vanishes. The
nonzero spacetime curvature nevertheless produces a tail in the
gravitational Green function.

Our comparison with the SVT framework of cosmological perturbation
theory \cite{Bardeen:1980kt,Kodama:1984ziu} showed that exterior
gauge-invariant multipoles and whole-space SVT fields carry different
information. Whole-space SVT
projections involve inverse Laplacians: their values outside an
extended source depend also on the metric inside it, even for
gauge-invariant combinations
\cite{Flanagan:2005yc,Ashtekar:2017wgq}. We exhibited metrics with the
same exterior geometry but different individual SVT fields there,
whose differences cancel in the exterior curvature. By deriving the
canonical moments directly from the metric and fluid equations, we
obtained an exterior characterization without choosing an interior
continuation for a spatial SVT projection. %Such a continuation was needed only when relating this characterization to the whole-space cosmological variables.

The Blanchet--Damour formalism
\cite{Blanchet:1985sp,Blanchet:1989ki}, together with the linear source
analysis of Damour and Iyer \cite{Damour:1990gj}, established how source
multipoles encode the exterior geometry up to coordinate freedom.
Extending this description to FLRW required us to derive the
background-dependent weights that combine different emission times
across the source, together with the metric propagators. We also
imposed a cosmological version of past stationarity, retaining compatible
past fields with their prescribed scale-factor dependence. In the flat
limit, we recovered the canonical metric and past-stationarity
prescription of Blanchet--Damour, together with the exact source-moment
expressions of Damour--Iyer.

We regard the generation of all three families $M_L,J_L,N_L$ by
cosmological processes as a promising area of application. Possible
sources include primordial-black-hole formation from inflationary
curvature perturbations, as in the hybrid-inflation scenario of Clesse
and Garc\'ia-Bellido \cite{ClesseGarciaBellido:2015}, bubble and plasma
dynamics during first-order phase transitions
\cite{Kamionkowski:1994phase}, and the collapse or decay of cosmic
defects reviewed by Vilenkin \cite{Vilenkin:1985defects}. These processes
involve localized stresses and motions together with disturbances of
the cosmic medium. For the subsequent fluid evolution, the acoustic
Green functions of Bashinsky and Bertschinger \cite{Bashinsky:2002vx}
show how an initial perturbation propagates through the coupled
photon--baryon and cold-dark-matter system before recombination. The
canonical moments offer a common description in which these aspects
of a cosmological event can be related. Their physical realization
will depend on the source model, its initial data and the properties
of the cosmic medium; applications involving a strongly gravitating
interior would concern its weak-field exterior and require suitable
matching data.

Lasting imprints of localized events offer another promising connection
with the literature. Cosmological displacement memory
connects localized emission events to lasting changes in the separation
of freely falling observers \cite{Tolish:2016ggo}. Understanding how
this imprint depends on cosmological propagation also involves the
curvature tails, whose contribution to memory has been studied in
constant-equation-of-state backgrounds \cite{Chu:2016ngc}. Our
quadrupolar de Sitter analysis \cite{Compere:2023ktn} gave this
observable effect an asymptotic interpretation, relating it to changes
in the gravitational field at future infinity. In the even-parity
sector, this change is a transition generated by the $\Lambda$-BMS
symmetries, the cosmological extension of BMS symmetry developed in
\cite{Compere:2019bua,Compere:2020lrt}.
The source-based description developed here may be useful in relating
such effects to the geometry and past evolution of cosmological
sources. Their interpretation as memory involves the motion of
specified observers relative to the unperturbed cosmology.

All results derived here concern linearized perturbations. A nonlinear
extension would need to incorporate gravitational self-interaction
and the coupled matter response consistently. The multipolar
post-Minkowskian construction for isolated systems
\cite{Blanchet:2024review} provides an important precedent for this
further development. One possible application is nonlinear
gravitational memory, established in asymptotically flat spacetime
\cite{Blanchet1990,Christodoulou:1991cr,Wiseman:1991ss,Th92,Blanchet:1992br}; Jokela, Kajantie and
Sarkkinen \cite{Jokela:2022rhk} studied its cosmological propagation,
including the curvature tail sourced by gravitational radiation.
Persistent imprints in matter correlations provide another direction.
Masui and Pen
\cite{Masui:2010cz} proposed gravitational-wave fossils as persistent
anisotropies left by primordial tensor perturbations. Schmidt, Pajer
and Zaldarriaga \cite{Schmidt:2013gwa} derived the tidal response of
small-scale matter fluctuations at second order in the combined tensor
and matter amplitudes, including an imprint that survives the decay of
the tensor mode. It would be interesting to investigate
whether gravitational radiation from localized cosmological sources
can leave analogous persistent anisotropies in matter clustering.

\section*{Acknowledgements}
G.C. is Research
Director of the F.R.S.-FNRS. The authors thanks Luc Blanchet for helpful discussions. The work of J.H. is supported in part 
by the Early Career Research Grant (ANRF/ECRG/2025/002233/PMS) from ANRF, Government of India. J.H.  also acknowledges NFSG grant (N8/26/013) from BITS Pilani, Hyderabad campus. \\ {\bf AI disclosure:} During the course of this work, we  interacted with
ChatGPT Astra. Our methodological choices and analytical approaches were challenged and tested by the AI and led us to perform iterative    adjustments. All AI-generated content was entirely reviewed and revised. The authors take
full responsibility of all content.

\appendix

\section{Convergence and numerical evaluation of the retarded Green functions}
\label{app:green-identities}
The characteristic construction of Section~\ref{sec:green-definition}
determines the retarded Green function through $J_V$. We will
establish convergence, uniqueness and truncation bounds for its
Volterra series, including the radial derivative needed to obtain
the tail. We will then describe a characteristic-grid scheme that
computes $J_V$ directly and extracts the tail coefficient $\mathcal V_V$.

\subsection{Convergence and error estimates for the Volterra construction}
\label{app:volterra-estimates}\label{app:green-cone}
We will establish convergence of the iteration of
Eq.~\eqref{eq:volterra} and bound its truncation error.

Throughout this subsection, the functions $J_V$, the iterates $J_n$
and their derivatives use the characteristic arguments $(U,V;\eta')$
of Eq.~\eqref{eq:J-characteristic-arguments}, with $\eta'$ fixed.
We start with $J_0(U,V;\eta'):=1$ and $J_{n+1}$ given by the double integral in
Eq.~\eqref{eq:volterra} applied to $J_n$.
On a compact characteristic rectangle $[0,U]\times[0,V]$ with
$[\eta',\eta'+(U+V)/c]\subset\Eta$, define
$M:=\max_{\zeta\in[\eta',\eta'+(U+V)/c]}|V(\zeta)|$.
This maximum is finite by smoothness of the potential. Iteration gives
\begin{equation}\label{eq:volterra-bound}
 |J_n(U,V;\eta')|\le\frac{M^nU^nV^n}{(n!)^2}.
\end{equation}
The series converges uniformly for every finite bound $M$; no
small-curvature condition $MVU\ll1$ is required. Summing the bounds
on the omitted terms gives the truncation estimate
\begin{equation}\label{eq:volterra-remainder}
 \left|J_V-\sum_{n=0}^N J_n\right|
 \le\sum_{n=N+1}^{\infty}\frac{(MVU)^n}{(n!)^2}.
\end{equation}
To justify the radial derivative in Eq.~\eqref{eq:tail-from-J},
differentiate the upper limits in the recurrence for $J_n$. For
$n\ge1$, this gives
\begin{equation}\label{eq:volterra-derivative-bounds}
 |\partial_VJ_n|\le
 \frac{M^nV^{n-1}U^n}{(n-1)!\,n!},\qquad
 |\partial_UJ_n|\le
 \frac{M^nV^nU^{n-1}}{n!\,(n-1)!}.
\end{equation}
Both derivative series converge uniformly on compact characteristic
rectangles. At fixed $\eta,\eta'$, the coordinate definitions give
$\partial_R=(\partial_V-\partial_U)/2$, so the radial derivative
can be taken term by term. The remainder estimate~\eqref{eq:volterra-remainder}
controls $J_V$; summing the corresponding derivative bounds controls
the differentiated series.

To prove uniqueness, let
$D:=J_V^{(1)}-J_V^{(2)}$ be the difference of two solutions with the
same characteristic data. Their boundary terms cancel, so $D$ obeys
the homogeneous Volterra integral equation. On the compact rectangle,
define $C:=\sup|D|$. Iterating that equation $n$ times gives
\begin{equation}\label{eq:volterra-uniqueness-bound}
 |D(U,V;\eta')|\le C\,\frac{M^nU^nV^n}{(n!)^2}
 \longrightarrow0\qquad(n\to\infty).
\end{equation}
Hence $D=0$ on every such rectangle. Differentiating the integral
equation on compact subsets establishes smoothness.

\subsection{Characteristic numerical evolution}
\label{app:green-numerics}
The Volterra equation also determines the tail by a local numerical
evolution. We begin at fixed source time $\eta'$ and discretize
the characteristic coordinates $(U,V)$ and initialize $J_V=1$ on both
axes. Subtracting the Volterra equation at the four corners of a cell
cancels the integrals over the preceding domain and leaves only the
integral over that cell. Midpoint quadrature, with the midpoint field
estimated from the two known opposite corners, then determines the
fourth corner explicitly. This is the characteristic integration
scheme of \cite[Eq.~(26)]{Gundlach:1993tp}, adapted to our potential
sign and characteristic-coordinate normalization. A sweep through the
grid has a cost proportional to the number of cells and requires no
global linear solve or separate evaluation of the successive terms
$J_n$. The smooth characteristic function $J_V$ is evolved directly; the
light-cone delta distribution remains analytical.

On a finite domain with a smooth solution, the scheme is second-order
accurate once the grid resolves changes in both the background
potential $V(\eta)$ and the propagating response $J_V$.
Comparing solutions at successively finer steps tests convergence and
estimates the discretization error. Additional refinement can be
needed where $V(\eta)$ varies rapidly near the past or future boundary
of the spacetime interval $\Eta$. To recover the tail
coefficient $\mathcal V_V$ from
Eq.~\eqref{eq:tail-from-J}, the evenness of $J_V$ in $R$ gives
$\mathcal V_V(\eta;0;\eta')
=-\left.\partial_R^2J_V(\eta,R;\eta')\right|_{R=0}$;
using this limit avoids division by a small radius.

\section{Fourier dictionary for the retarded fields and source moments}
\label{app:fourier-representations}
This appendix will translate the Green functions, metric reconstruction
and source moments of Sections~\ref{sec:green}--\ref{sec:multipoles}
and~\ref{sec:minkowski} into Fourier representations. Spatial Fourier
modes apply to any smooth expansion history. We will give their
retarded evolution and express the source coefficients through the
Cartesian moments and their STF traces, including the finite
polynomials associated with effective multipolar sources. We will
then identify the light-cone time-average weight with its spherical
Bessel representation and give the corresponding Minkowski metric.
These temporal Fourier formulas describe the light-cone contribution
on a general background and the complete field in the flat limit.

\subsection{Spatial Fourier modes and retarded evolution}
\label{sec:green-fourier}
The Green-function equation~\eqref{eq:green-def} and the retarded
reconstruction~\eqref{eq:retarded-metric-reconstruction} become ordinary
differential equations and source-time integrals at each spatial
wavevector. We start by defining the spatial Fourier transform and
its inverse, with $k:=|\bk|$:
\begin{subequations}\label{eq:spatial-fourier-convention}
\begin{align}
 \widehat f(\eta,\bk)&:=\int\dd^3x\,e^{-i\bk\cdot\bx}f(\eta,\bx),
 \label{eq:spatial-fourier-transform}\\
 f(\eta,\bx)&=\int\frac{\dd^3k}{(2\pi)^3}
                 e^{i\bk\cdot\bx}\widehat f(\eta,\bk).
 \label{eq:spatial-fourier-inverse}
\end{align}
\end{subequations}
Thus $\partial_i$ becomes $ik_i$ and $\Delta$ becomes $-k^2$.
The spatial metric equation~\eqref{eq:spatial} takes the mode form
\begin{subequations}\label{eq:fourier-metric-mode-equations}
\begin{align}
 [\partial_\eta^2+c^2k^2-c^2V^-(\eta)]\widehat{\bar h}_{ij}(\eta,\bk)
 &=\frac{16\pi G}{c^2}a(\eta)\widehat\tau_{ij}(\eta,\bk),
 \label{eq:fourier-rescaled-metric-mode}\\
 [\partial_\eta^2+2\cH(\eta)\partial_\eta+c^2k^2]
 \left[\frac{\widehat{\bar h}_{ij}(\eta,\bk)}{a(\eta)}\right]
 &=\frac{16\pi G}{c^2}\widehat\tau_{ij}(\eta,\bk).
 \label{eq:fourier-damped-metric-mode}
\end{align}
\end{subequations}
The second line follows from $V^-=\ddot a/(c^2a)$ and displays the usual
expansion damping before the field is rescaled by $a$.
These equations act on all six spatial components in generalized
harmonic gauge. The mixed wave equation~\eqref{eq:mixed} has the
form~\eqref{eq:fourier-rescaled-metric-mode} with $ij$ replaced by
$0i$ and $V^-$ by $V^+$.

We denote the retarded Fourier mode of the Green function by
$\widehat G_{V,k}(\eta;\eta')$, keeping the wave number $k$ as a mode
label in the subscript. The semicolon separates observation time from
source time. Taking the spatial Fourier transform of
Eq.~\eqref{eq:green-def}, with the convention~\eqref{eq:spatial-fourier-convention}
and retarded boundary conditions, gives

\begin{subequations}\label{eq:mode}
\begin{align}
 [-c^{-2}\partial_\eta^2-k^2+V(\eta)]\widehat G_{V,k}(\eta;\eta')
 &=-\frac{4\pi}{c}\delta(\eta-\eta'),\label{eq:mode-equation}\\
 \widehat G_{V,k}(\eta'^+;\eta')&=0,\label{eq:mode-initial-value}\\
 \partial_\eta\widehat G_{V,k}(\eta'^+;\eta')&=4\pi c.\label{eq:mode-initial-derivative}
\end{align}
\end{subequations}

Integrating Eq.~\eqref{eq:mode-equation} across $\eta=\eta'$ gives
a derivative jump of $4\pi c$. Together with continuity and retarded
support, this gives the stated initial data. Given two independent homogeneous modes
$y_1(k;\eta),y_2(k;\eta)$, variation of parameters
\cite[Eq.~1.13.10]{DLMF} gives

\begin{equation}\label{eq:fourier-mode-wronskian}
 \widehat G_{V,k}(\eta;\eta')=4\pi  c\,\theta(\eta-\eta')
 \frac{y_1(k;\eta')y_2(k;\eta)-y_2(k;\eta')y_1(k;\eta)}
 {y_1(k;\eta')\dot y_2(k;\eta')-\dot y_1(k;\eta')y_2(k;\eta')}.
\end{equation}

The denominator is the Wronskian. It is constant in $\eta'$ because
the mode equation has no first derivative
\cite[Eq.~1.13.5]{DLMF}. The resulting Green function has the initial
data in Eqs.~\eqref{eq:mode}. Fourier transforming the complete
retarded fields~\eqref{eq:spatial-solution}--\eqref{eq:mixed-solution}
therefore gives
\begin{subequations}\label{eq:fourier-field}
\begin{align}
 \widehat{\bar h}_{ij}(\eta,\bk)
 &=\widehat{\bar h}^{\mathrm{past}}_{ij}(\eta,\bk)
 +\frac{4G}{c^4}\int_{\eta_i}^{\eta}c\,\dd\eta'\,
 \widehat G_{V^-,k}(\eta;\eta')a(\eta')
 \widehat\tau^{\mathrm{dyn}}_{ij}(\eta',\bk),
 \label{eq:fourier-spatial-field}\\
 \widehat{\bar h}_{0i}(\eta,\bk)
 &=\widehat{\bar h}^{\mathrm{past}}_{0i}(\eta,\bk)
 +\frac{4G}{c^4}\int_{\eta_i}^{\eta}c\,\dd\eta'\,
 \widehat G_{V^+,k}(\eta;\eta')a(\eta')
 \widehat\tau^{\mathrm{dyn}}_{0i}(\eta',\bk).
 \label{eq:fourier-mixed-field}
\end{align}
\end{subequations}
The past terms are the spatial transforms of
Eqs.~\eqref{eq:past-reference-metric}; the integrals have zero initial
data and contain only the dynamical sources of
Eq.~\eqref{eq:past-dynamical-source}. The remaining metric components
follow from Eq.~\eqref{eq:scalar-solution}, with
$\partial_i\bar h_{0i}$ replaced by $ik_i\widehat{\bar h}_{0i}$,
and $\widehat{\bar h}_{00}=\widehat{\hat h}-\widehat{\bar h}_{kk}$.

For $R:=|\bx-\bx'|$, angular integration gives
$\int\dd\Omega_k\,e^{i\bk\cdot(\bx-\bx')}=4\pi j_0(kR)$
\cite[Eq.~10.54.2 with $n=0$]{DLMF}. The inverse transform therefore
recovers the retarded Green function in position space:

\begin{equation}\label{eq:fourier-green}
\begin{aligned}
 G_V(\eta;|\bx-\bx'|;\eta')
 &=\int\frac{\dd^3k}{(2\pi)^3}
      e^{i\bk\cdot(\bx-\bx')}\widehat G_{V,k}(\eta;\eta')\\
 &=\frac1{2\pi^2}\int_0^\infty\dd k\,k^2
                  \widehat G_{V,k}(\eta;\eta')j_0(k|\bx-\bx'|).
\end{aligned}
\end{equation}

Here $j_\ell$ denotes the spherical Bessel function of the first
kind. The Fourier inversion is distributional: it reconstructs both
the light-cone singularity and the curvature tail. For smooth,
spatially compact dynamical sources, the field inversion is an
ordinary integral: on compact time intervals, mode evolution
preserves rapid spatial Fourier decay, including after any fixed
number of derivatives \cite[Chap.~2, Sec.~8; Chap.~3, Sec.~3]{Taylor:2011PDE}.
For the {effective stress and momentum sources in
Eqs.~\eqref{eq:quadrupolar-effective-sources}}, the inversion is
instead understood distributionally.

In Minkowski spacetime, $a=1$, $t=\eta$ and $V=0$ give

\begin{equation}\label{eq:flat-fourier-kernel}
 \widehat G_{0,k}(t;t')=4\pi\theta(t-t')\frac{\sin[ck(t-t')]}{k}.
\end{equation}

At $k=0$, the continuous limit is
$4\pi c(t-t')\theta(t-t')$, with the same derivative jump as in
Eq.~\eqref{eq:mode-initial-derivative}. Fourier inversion recovers
the retarded Green function~\eqref{eq:flat-kernel}.

\subsection{Source moments and spatial Fourier coefficients}
\label{sec:bessel-moments}\label{sec:resummed-field}
The source moments~\eqref{eq:source-moment-definitions} and their
STF traces determine the spatial Fourier coefficients of the compact
stress and momentum components. We will express this correspondence
for the wave-equation source $a\tau_{ij}$; the same identities apply
separately to its past and dynamical parts.

We start by expanding the Fourier exponential and using the moment
definition. At each fixed wavevector, spatial compactness gives
\begin{equation}\label{eq:cartesian-fourier}
 a(\eta)\widehat\tau_{ij}(\eta,\bk)
 =\sum_{\ell\ge0}\frac{(-i)^\ell}{\ell!}k_L
       a(\eta)^{-\ell}S_{ij|L}(\eta).
\end{equation}
The factor $a^{-\ell}$ follows from the $a^{\ell+1}$ in the
source-moment definition, leaving the single factor $a$ in the
wave-equation source.

The monomial decomposition~\eqref{eq:STFdecomp} organizes this
series by angular rank. At rank $\ell$, its traces sum to the
spherical Bessel weight \cite[Eq.~10.53.1]{DLMF}
\begin{equation}\label{eq:bessel-weight-series}
 (2\ell+1)!!\frac{j_\ell(z)}{z^\ell}
 =\sum_{p=0}^\infty(-1)^p c_{\ell+1,p}z^{2p},
\end{equation}
where $c_{\ell+1,p}$ is defined in Eq.~\eqref{eq:tracecoefficients}.
Using the radial trace identity~\eqref{eq:spatial-radialmoments-traces}
then gives the equivalent source representations
\begin{subequations}\label{eq:sourcefourier}
\begin{align}
 a(\eta)\widehat\tau_{ij}(\eta,\bk)
 &=a(\eta)\sum_{\ell\ge0}\frac{(-i)^\ell(2\ell+1)!!}{\ell!}k_L
   \int\dd^3x'\,x'_{\langle L\rangle}
   \frac{j_\ell(k\rho')}{(k\rho')^\ell}\tau_{ij}(\eta,\bx'),
 \label{eq:source-fourier-STF-integral}\\
 &=\sum_{\ell\ge0}\frac{(-i)^\ell}{\ell!}k_L
   \sum_{p\ge0}(-1)^p c_{\ell+1,p}k^{2p}a(\eta)^{-\ell-2p}
       \STF_L S_{ij|L j_1j_1\cdots j_pj_p}(\eta).
 \label{eq:source-fourier-STF-traces}
\end{align}
\end{subequations}
The quotient $j_\ell(k\rho')/(k\rho')^\ell$ has limit
$1/(2\ell+1)!!$ at $k\rho'=0$. Thus the leading term at angular
rank $\ell$ is $a^{-\ell}S_{ij|\langle L\rangle}$, while the
higher powers of $k^2$ carry the radial traces defined in
Section~\ref{sec:source-moment-STF}. Replacing
$\tau_{ij},S_{ij|L}$ by $\tau_{0i}/c,P_{i|L}$ gives
$a\widehat\tau_{0i}/c$, by Eq.~\eqref{eq:mixed-radialmoments-traces}.
Thus the mixed field has the prefactor $4G/c^3$ when expressed
through the momentum moments.

{For the stress and momentum components of an effective
multipolar source of order $m$, use the finite delta-function
expansions~\eqref{eq:quadrupolar-effective-sources} with upper limit
$m$. The Fourier transform}
of $\partial'_L\delta^{(3)}(\bx')$ is $i^\ell k_L$.
Equation~\eqref{eq:cartesian-fourier} therefore terminates at
$\ell=m$, and the trace sum~\eqref{eq:source-fourier-STF-traces}
contains precisely the terms $\ell+2p\le m$. This is the Fourier
representation of the effective source, retaining its Cartesian traces.
For a smooth extended source, the infinite series converges at fixed
$\bk$; its individual polynomial terms do not retain the rapid
large-$k$ decay of the full transform. The sum must therefore be
performed before the unbounded wave-number integral unless a separate
justification permits their interchange.

\subsection{Temporal Fourier modes and the light-cone moments}
\label{app:retarded-STF}
The light-cone source moments~\eqref{eq:timemoment} and
fields~\eqref{eq:LCSTF} have a direct frequency-space
representation \cite[Appendix~B]{Blanchet:1989ki},
\cite[Sec.~III]{Damour:1990gj}. We will identify the time-average
weight with a spherical Bessel function and the propagated field
with an outgoing radial wave. This applies to the light-cone part
in any background and to the full dynamical field in Minkowski
spacetime.

To justify temporal Fourier inversion, we multiply the smooth source
product $a\tau^{\mathrm{dyn}}_{\mu\nu}$ by a smooth cutoff with
compact support in $\Eta$, equal to one near all emission times
contributing to a chosen compact exterior observation region.
The formulas below use the resulting product, extended by zero to
$\mathbb R$. Its smooth, compact temporal support ensures rapid
frequency decay, while the light-cone field remains unchanged in
that region.

In this subsection and the next, hats denote temporal Fourier
transforms, with convention
\begin{subequations}\label{eq:temporal-fourier-convention}
\begin{align}
 \widehat f(\omega)&:=\int_{\mathbb R}\dd\eta\,
                       e^{i\omega\eta}f(\eta),
 \label{eq:temporal-fourier-transform}\\
 f(\eta)&=\int_{\mathbb R}\frac{\dd\omega}{2\pi}\,
                       e^{-i\omega\eta}\widehat f(\omega).
 \label{eq:temporal-fourier-inverse}
\end{align}
\end{subequations}
For the weight in Eq.~\eqref{eq:time-weight-definition}, the Poisson
integral \cite[Eq.~10.54.1]{DLMF} gives
\begin{equation}\label{eq:bessel-time-weight}
 \int_{-1}^1\dd z\,\delta_\ell(z)e^{i\omega\rho'z/c}
 =(2\ell+1)!!\frac{j_\ell(\omega\rho'/c)}{(\omega\rho'/c)^\ell}.
\end{equation}
Shifting the time variable in Eq.~\eqref{eq:timemoment} and using
the evenness of $\delta_\ell$ therefore gives
\begin{subequations}\label{eq:temporal-retarded-source-moment}
\begin{align}
 \widehat{\mathcal F}^{\mathrm{LC}}_{ij|L}(\omega;a)
 &=(2\ell+1)!!\int\dd^3x'\,x'_{\langle L\rangle}
   \frac{j_\ell(\omega\rho'/c)}{(\omega\rho'/c)^\ell}
   \widehat{a\tau^{\mathrm{dyn}}}_{ij}(\omega,\bx'),
 \label{eq:temporal-LC-stress-moment}\\
 \widehat{\mathcal F}^{\mathrm{LC}}_{0i|L}(\omega;a)
 &=\frac{(2\ell+1)!!}{c}\int\dd^3x'\,x'_{\langle L\rangle}
   \frac{j_\ell(\omega\rho'/c)}{(\omega\rho'/c)^\ell}
   \widehat{a\tau^{\mathrm{dyn}}}_{0i}(\omega,\bx').
 \label{eq:temporal-LC-momentum-moment}
\end{align}
\end{subequations}
Here $\widehat{a\tau^{\mathrm{dyn}}}_{\mu\nu}$ is the temporal
transform of the cutoff product $a\tau^{\mathrm{dyn}}_{\mu\nu}$;
the scale factor is included because it multiplies the source in
Eq.~\eqref{eq:timemoment}. The quotient $j_\ell(z)/z^\ell$ is even,
so its value is the same as in
Eq.~\eqref{eq:source-fourier-STF-integral} at $k=|\omega|/c$.

The delay $\eta-\rho/c$ in Eqs.~\eqref{eq:LCSTF} supplies the outgoing
phase $e^{i\omega\rho/c}$. Hence
\begin{subequations}\label{eq:frequency-light-cone-field}
\begin{align}
 \widehat{\bar h}^{\mathrm{LC}}_{ij}(\omega,\bx)
 &\overset{\rho>d}{=}\frac{4G}{c^4}\sum_{\ell\ge0}\frac{(-1)^\ell}{\ell!}
   \widehat{\mathcal F}^{\mathrm{LC}}_{ij|L}(\omega;a)
   \partial_L\frac{e^{i\omega\rho/c}}\rho,
 \label{eq:frequency-LC-spatial}\\
 \widehat{\bar h}^{\mathrm{LC}}_{0i}(\omega,\bx)
 &\overset{\rho>d}{=}\frac{4G}{c^3}\sum_{\ell\ge0}\frac{(-1)^\ell}{\ell!}
   \widehat{\mathcal F}^{\mathrm{LC}}_{0i|L}(\omega;a)
   \partial_L\frac{e^{i\omega\rho/c}}\rho.
 \label{eq:frequency-LC-mixed}
\end{align}
\end{subequations}
We now express the STF derivatives in terms of outgoing spherical
waves, making the radial and angular dependence of each multipole
explicit. With $\bn:=\bx/\rho$, $\bn':=\bx'/\rho'$ and
$\mu:=\bn\cdot\bn'$, the outgoing Helmholtz addition theorem
\cite[Eqs.~10.60.1--10.60.2]{DLMF} gives
\begin{equation}\label{eq:helmholtz-addition}
 \frac{e^{i\omega R/c}}R
 \overset{\rho>\rho'}{=}\frac{i\omega}{c}\sum_{\ell\ge0}(2\ell+1)
   j_\ell(\omega\rho'/c)h_\ell^{(1)}(\omega\rho/c)P_\ell(\mu),
 \qquad R:=|\bx-\bx'|.
\end{equation}
Here $h_\ell^{(1)}:=j_\ell+i y_\ell$ is the spherical Hankel
function of the first kind, with $y_\ell$ the spherical Bessel
function of the second kind. The STF derivative identity
\cite[Eq.~(A30)]{Blanchet:1985sp}, \cite[Sec.~10.51]{DLMF} reads
\begin{equation}\label{eq:outgoing-STF-derivative}
 \partial_{\langle L\rangle}\frac{e^{i\omega\rho/c}}\rho
 =i(-1)^\ell(\omega/c)^{\ell+1}
   h_\ell^{(1)}(\omega\rho/c)n_{\langle L\rangle}.
\end{equation}
Since the moments in
Eqs.~\eqref{eq:temporal-retarded-source-moment} are STF in the
source-position indices $L$, only the STF part of the derivatives
contributes to Eqs.~\eqref{eq:frequency-light-cone-field}.
Equation~\eqref{eq:outgoing-STF-derivative} therefore identifies the
outgoing radial function of each multipole. Together with the source
weights in Eqs.~\eqref{eq:temporal-retarded-source-moment} and the STF
representation
$P_\ell(\mu)=(2\ell-1)!!\,n_{\langle L\rangle}
 n'_{\langle L\rangle}/\ell!$
\cite[Eq.~(A25)]{Blanchet:1985sp}, this reproduces the field obtained
by integrating Eq.~\eqref{eq:helmholtz-addition} against the source.
The sign of the frequency $\omega$ is retained in the outgoing phase;
negative frequencies give the complex-conjugate kernel. The products
in Eqs.~\eqref{eq:helmholtz-addition} and~\eqref{eq:outgoing-STF-derivative}
have continuous zero-frequency limits.
Temporal inversion returns the original light-cone fields
\eqref{eq:LCSTF} on the chosen observation region. Angular convergence
is established in Section~\ref{sec:lightconemoments}.
On a curved background, this derivation covers only the light-cone
part, whose Green function is $G^{\mathrm{LC}}=G_0$ by
Eq.~\eqref{eq:light-cone-zero-potential-kernel}.
The curvature tail samples earlier source times and is included
through the full Green functions in the spatial Fourier
reconstruction~\eqref{eq:fourier-field}. Adding the tail and the
prescribed past field to the light-cone field gives the complete
fields in Eqs.~\eqref{eq:full-multipolar-fields}.

\subsection{Cartesian moments at fixed frequency in Minkowski spacetime}
\label{app:Cartesian-convergence}
The Minkowski metric~\eqref{eq:flat-component-metric} can also be
expressed through the temporal transforms of the Cartesian moments
$Q^{(\rho)}_L,Q^{(p)}_L,P_{i|L},S_{ij|L}$. We will give this form
using the convention~\eqref{eq:temporal-fourier-convention}, with
the Cartesian sum performed before the frequency integral.

We start from the retarded field~\eqref{eq:flat-retarded-field},
with $a=1$, $t=\eta$ and $\tau_{\mu\nu}=\delta T_{\mu\nu}$.
The fields and moments in this subsection refer to the full source,
including its prescribed stationary past, as in Section~\ref{sec:minkowski}.
For Fourier inversion on a compact observation region, the same
temporal cutoff can be applied to this full source on all sampled
times. Transforming the retarded kernel gives $e^{i\omega R/c}/R$.
At fixed real frequency, its Cartesian expansion is

\begin{equation}\label{eq:flat-Helmholtz-Cartesian}
 \frac{e^{i\omega|\bx-\bx'|/c}}{|\bx-\bx'|}
 \overset{\rho>\rho'}{=}\sum_{\ell\ge0}\frac{(-1)^\ell}{\ell!}x'_L
           \partial_L\frac{e^{i\omega\rho/c}}\rho.
\end{equation}

At fixed $\omega$, analyticity of the spatial kernel gives uniform
convergence over the compact source $|\bx'|\le d<\rho$. Integrating
against the source and using the Cartesian moment definitions
\eqref{eq:source-moment-definitions}, then undoing trace reversal,
gives the frequency representation of
Eqs.~\eqref{eq:flat-component-metric}:

\begin{subequations}\label{eq:flat-Cartesian-metric}
\begin{align}
 h_{00}(t,\bx)
 &\overset{\rho>d}{=}\frac{2\mathfrak{s}G}{c^2}\int_{\mathbb R}\frac{\dd\omega}{2\pi}e^{-i\omega t}
   \left\{\sum_{\ell\ge0}\frac{(-1)^\ell}{\ell!}
     \left(\widehat Q^{(\rho)}_L+\widehat Q^{(p)}_L\right)(\omega)
     \partial_L\frac{e^{i\omega\rho/c}}\rho\right\},
 \label{eq:flat-Cartesian00}\\
 h_{0i}(t,\bx)
 &\overset{\rho>d}{=}\frac{4\mathfrak{s}G}{c^3}\int_{\mathbb R}\frac{\dd\omega}{2\pi}e^{-i\omega t}
   \left\{\sum_{\ell\ge0}\frac{(-1)^\ell}{\ell!}
     \widehat P_{i|L}(\omega)
     \partial_L\frac{e^{i\omega\rho/c}}\rho\right\},
 \label{eq:flat-Cartesian0i}\\
 h_{ij}(t,\bx)
 &\overset{\rho>d}{=}\frac{4\mathfrak{s}G}{c^4}\int_{\mathbb R}\frac{\dd\omega}{2\pi}e^{-i\omega t}
   \bigg\{\sum_{\ell\ge0}\frac{(-1)^\ell}{\ell!}
     \bigg[\widehat S_{ij|L}(\omega)
       +\frac{c^2\delta_{ij}}2\left(\widehat Q^{(\rho)}_L
                               -\widehat Q^{(p)}_L\right)(\omega)
     \bigg]\partial_L\frac{e^{i\omega\rho/c}}\rho\bigg\}.
 \label{eq:flat-Cartesianij}
\end{align}
\end{subequations}

The braces specify the order of operations. After the Cartesian
sum, the kernel is $e^{i\omega R/c}/R$, bounded by $1/(\rho-d)$ for
real $\omega$. The rapid frequency decay of the smooth compact-time
source then ensures Fourier inversion, including any fixed number
of spacetime derivatives. The cutoff is one on the contributing
source times, so the result equals the full retarded metric on the
chosen observation region. Equivalently, if the stationary past field
is separated before transforming, its transform is
$2\pi\delta(\omega)h^{\mathrm{past}}_{\mu\nu}(\bx)$ and is added
once to the dynamical contribution.

The connection to the STF metric is algebraic at each frequency:
$\Delta(e^{i\omega\rho/c}/\rho)=-(\omega^2/c^2)e^{i\omega\rho/c}/\rho$ for
$\rho>0$, so the Cartesian trace sum gives the Bessel weight
\eqref{eq:bessel-weight-series}. Its time-average form
\eqref{eq:bessel-time-weight} recovers the moments
\eqref{eq:flat-source-STF} and the metric
\eqref{eq:flat-component-metric}.
Inverting each Cartesian term before summation gives spatial
derivatives of moments evaluated at $t-\rho/c$, as in the flat limit
of the finite source-size expansion~\eqref{eq:cartesian-general}.
The resulting infinite time-domain Taylor series requires additional
convergence assumptions.

% Bibliography included for a self-contained pdfLaTeX/Overleaf build.

\end{document}